\documentclass[a4paper,12pt]{amsart}
\usepackage[hmarginratio={1:1},vmarginratio={1:1},heightrounded,textwidth=455pt]{geometry}
\usepackage{amsfonts}
\usepackage{amsmath}
\usepackage[utf8]{inputenc}
\usepackage{amssymb, colonequals}
\usepackage{hyperref}
\usepackage{tikz}
\usetikzlibrary{arrows}
\usetikzlibrary{patterns}
\usetikzlibrary{arrows}
\usetikzlibrary{patterns}
\usetikzlibrary{calc}
\usepackage[colorinlistoftodos,prependcaption,textsize=tiny]{todonotes}
\usepackage{url}
\usepackage{enumitem}
\usepackage{multirow}
\usepackage{chngcntr}
\counterwithin{figure}{section}
\setlist[itemize]{leftmargin=0.3in}
\setlist[enumerate]{leftmargin=0.3in}

\newcommand{\Oh}[1]{\ensuremath{\mathcal{O}(#1)}}

\newcommand{\level}{\mathsf{level}}
\newcommand{\canon}{\mathsf{canon}}
\newcommand{\num}{\mathsf{num}}
\newcommand{\inside}{\mathsf{inside}}
\newcommand{\leftside}{\mathsf{left}}
\newcommand{\rightside}{\mathsf{right}}
\newcommand{\pqsmtree}{\mathbb{T}^*}
\newcommand{\pqstree}{\mathbb{T}}
\newcommand{\pqmtreeinnernodes}{\mathbb{T}}
\newcommand{\dataStructure}{\mathrm{DS}}
\newcommand{\strongModules}{\mathcal{M}}

\newcommand{\MMM}{\mathbb{M}}
\newcommand{\SSS}{\mathbb{S}}
\newcommand{\KKK}{\mathbb{K}}
\newcommand{\LLL}{\mathbb{L}}
\newcommand{\QQQ}{\mathbb{Q}}

\newcommand{\camodules}{\mathcal{S}}
\newcommand{\slots}{\mathcal{S}^*}
\newcommand{\metachords}{\mathcal{MC}}

\let\leq\leqslant
\let\geq\geqslant
\let\setminus\smallsetminus

\let\rho\varrho

\makeatletter
\newcommand\ie{i.e\@ifnextchar.{}{.\@}}
\newcommand\etc{etc\@ifnextchar.{}{.\@}}
\newcommand\etal{et~al\@ifnextchar.{}{.\@}}
\def\namedlabel#1#2{\begingroup
    #2%
    \def\@currentlabel{#2}%
    \phantomsection\label{#1}\endgroup
}
\makeatother

\newcounter{dummy} 
\numberwithin{dummy}{section}

\numberwithin{equation}{dummy}
\newtheorem{theorem}[dummy]{Theorem}
\newtheorem{claim}[dummy]{Claim}
\newtheorem{lemma}[dummy]{Lemma}
\newtheorem{proposition}[dummy]{Proposition}
\newtheorem{observation}[dummy]{Observation}
\newtheorem{definition}[dummy]{Definition}

\newcounter{hackcount}

\title[Normalized models of circular-arc graphs]{On the structure of normalized models of circular-arc graphs -- Hsu's approach revisited}

\thanks{
Tomasz Krawczyk is partially supported by the Polish National Science Center grant UMO-2015/17/B/ST6/01873.
}

\author[T.~Krawczyk]{Tomasz Krawczyk}
\address[T.~Krawczyk]{Faculty of Mathematics and Information Science, Warsaw University of Technology, Warsaw, Poland}
\email{tomasz.krawczyk@pw.edu.pl}
\date{}

\begin{document}

\thispagestyle{empty}

\begin{abstract}
Circular-arc graphs are the intersection graphs of arcs of a circle.
The main result of this work describes the structure of all \emph{normalized intersection models} of circular-arc graphs.
Normalized models of a circular-arc graph reflect the neighborhood relation between its vertices and can be seen as its canonical representations; 
in particular, any intersection model can be made normalized by possibly extending some of its arcs.
We~devise a data-structure, called \emph{PQSM-tree}, that maintains the set of all normalized models of a circular-arc graph.
We show that the PQSM-tree of a circular-arc graph can be computed in linear time.
Finally, basing on PQSM-trees, we provide a linear-time algorithm for the canonization
and the isomorphism problem for circular-arc graphs.

We describe the structure of the normalized models of circular-arc graphs using an approach proposed by 
Hsu~[\emph{SIAM J. Comput. 24(3), 411--439, (1995)}].
In the aforementioned work, Hsu claimed the construction of decomposition trees representing the set of all normalized intersection models of circular-arc graphs and an~$\Oh{nm}$ time isomorphism algorithm for this class of graphs.
However, the counterexample given in~[\emph{Discrete Math. Theor. Comput. Sci., 15(1), 157--182, 2013}] shows that Hsu's isomorphism algorithm is incorrect. 
Also, due to the errors pointed out in~\cite{Kra24appendix}, the decomposition trees proposed by Hsu are not constructed correctly; in particular, we showed that there are circular-arc graphs whose all normalized models do not follow the description given by Hsu.
\end{abstract}

\maketitle

\section{Introduction}
\label{sec:introduction}
\emph{Circular-arc graphs} are the intersection graphs of arcs of a circle. 
Circular-arc graphs generalize interval graphs, 
which are the intersection graphs of intervals on the real line.
Although circular-arc graphs and interval graphs are defined in a quite similar way, they
turn out to have significantly different algorithmic and combinatorial properties. 
A number of problems that are solved (or shown to admit polynomial-time solutions) in the class of interval graphs, in the class of circular-arc graphs are still open (are computationally hard, respectively).
One example is the minimum coloring problem, which admits a simple linear algorithm for
interval graphs, but is NP-complete on circular-arc graphs~\cite{GareyJMP80}.
Another example is concerned with the structure of the intersection models of graphs from these classes.
The structure of all intersection models of interval graphs is well-understood -- all such models are maintained by \emph{PQ-trees}, invented by Booth and Lueker already in 1970's~\cite{BoothLueker76}.
Despite some efforts~\cite{Hsu95}, the corresponding structure for circular-arc graphs has not been devised.
Yet another example is the isomorphism problem.
Already in 1970's Lueker and Booth~\cite{LuekerBooth79} devised a linear-time isomorphism algorithm testing isomorphism of interval graphs.
Their algorithm works on PQ-trees;
to test whether two interval graphs are isomorphic it suffices to 
check the isomorphism between their PQ-trees.
The isomorphism problem for the class of circular-arc graphs has been open so~far\footnote{We posed on arxiv a paper with a polynomial-time isomorphism algorithm for circular-arc graphs already in 2019~\cite{Kra19}, however, we have encountered difficulties in publishing our work (despite that no bugs have been reported by the reviewers). This is the reason why we decided to extend our work with a~detailed comparison of our work and Hsu's work~\cite{Hsu95} and why we decided to write a paper~\cite{Kra24appendix} on errors found in Hsu's work~\cite{Hsu95}.}.

\subsection{Our results}
Our first result is the description of the structure of all \emph{normalized intersection models} of circular-arc graphs.
Normalized models (the formal definition is postponed to Section~\ref{sec:preliminaries}) of a circular-arc graph~$G$ reflect the neighborhood relation between the vertices of $G$ and can be seen as its canonical representations; 
in particular, any intersection model of $G$ can be made normalized by possibly extending some of its arcs.
We introduce a data-structure, called \emph{PQSM-tree}, that represents the set of all normalized models of $G$.
Finally, we show that the PQSM-tree for $G$ can be computed in linear time.

To attain our goal, we follow an approach taken by Hsu in the work~\cite{Hsu95} from~1995.
In~\cite{Hsu95} Hsu claimed a theorem describing the structure of all normalized intersection models of circular-arc graphs and introduced so-called \emph{decomposition trees} supposed to represent all normalized models of circular-arc graphs.
Based on decomposition trees, Hsu claimed an $\Oh{nm}$-time algorithm 
for the isomorphism problem for circular-arc graphs.
However, in 2013 Curtis, Lin, McConnell, Nussbaum, Soulignac, Spinrad, and Szwarcfiter~\cite{counterex13} showed that Hsu's algorithm is not correct.
In~\cite{Kra24appendix} we showed that Hsu's decomposition trees are also constructed incorrectly; 
in particular, we showed that there are circular-arc graphs whose all normalized models do not follow the description given by Hsu.

The \emph{graph isomorphism problem} is the computational problem of determining whether two input graphs are isomorphic.
Clearly, the graph isomorphism problem is in NP, it is unlikely NP-complete~\cite{Babai16}, and it is not known to be in P.
The best currently known algorithm for the isomorphism problem works in super-polynomial time~\cite{Babai16}.
For some restricted classes of graphs, e.g.~for graphs possessing certain geometric representation, the isomorphism problem
can be solved in polynomial or even linear time.
A flurry of research has been devoted to distinguish graph classes admitting such algorithms from those in which the problem remains GI-complete (polynomial time equivalent to the general isomorphism problem).
One of the most known classes, whose status has not been known until now,
is the class of circular-arc graphs.

The isomorphism problem for circular-arc graphs has been open for almost 40 years.
There have been two claimed polynomial time algorithms for 
the isomorphism problem on circular-arc graphs, presented in~\cite{Wu83} and~\cite{Hsu95}, which were shown to be incorrect in~\cite{Eschen97} and~\cite{counterex13}, respectively.
There are known linear time isomorphism algorithms on proper circular-arc graphs~\cite{counterex13, Lin08} and for co-bipartite circular-arc graphs~\cite{Eschen97}.
The isomorphism problem can be solved in linear time~\cite{counterex13} and logarithmic space \cite{Kob16} in the class of Helly circular-arc graphs.
The next main theorem of the paper proves the following:
\begin{theorem}
\label{thm:main_isomorphism_theorem}
The isomorphism problem in the class of circular-arc graphs can be solved in linear time\footnote{We assume the standard word RAM model of computation with words of length $\log{n}$ ($n$~is the size of the vertex set of the input graph), 
in which both arithmetic and bitwise operations can be performed in constant time.}.
\end{theorem}

Instead of explicitly providing a linear-time isomorphism algorithm for circular-arc graphs,
we present a linear-time algorithm that solves the \emph{canonization problem} for this class of graphs.
The \emph{canonization problem} for a graph class~$\mathcal{G}$ consists of computing a \emph{canonical string representation}~$\canon(G)$ of an input graph $G \in \mathcal{G}$, such that a~graph~$H$ 
satisfies $\canon(G) = \canon(H)$ if and only if $H$ is isomorphic to $G$.
Clearly, given a polynomial-time canonization algorithm for $\mathcal{G}$, we can solve the isomorphism problem for~$\mathcal{G}$ in the same time.
In this paper, we present a linear-time canonization algorithm for circular-arc graphs. 
For a given circular-arc graph $G$, the algorithm constructs a tuple $\canon(G)$ containing $\Oh{n}$ entries, where $n$ represents the number of vertices in $G$. 
In summary, we establish the following theorem, which subsequently yields Theorem~\ref{thm:main_isomorphism_theorem} as a byproduct.
\begin{theorem}
\label{thm:main_canonization_theroem}
The canonization problem in the class of circular-arc graphs can be solved in linear time.
\end{theorem}
We mention here that a parameterized logspace algorithm computing a canonical string representation of circular-arc graphs was presented by Chandoo~\cite{Chandoo16}.
The canonization procedure for circular-arc graphs is conducted in a manner similar to that used for interval graphs, permutation graphs (see, for example,~\cite{Yamazaki20}), and circle graphs (see, for example,~\cite{Kalisz19}). 
In all these cases, the method relies on the property that certain trees represent all intersection models of graphs within these classes.

\medskip
Our paper is organized as follows:
\begin{itemize}
 \item In Section~\ref{sec:preliminaries} we introduce notation used throughout the paper.
 \item In Section~\ref{sec:basic_tools} we introduce the basic concepts used to describe the structure of the normalized models of circular-arc graphs.
 \item In Section~\ref{sec:related_work} we bring up the related work which has had an impact on the development of the method used in this paper. 
 In particular, since we follow the approach taken by Hsu~\cite{Hsu95}, we compare and point out the main differences (and their consequences) between our works.
 \item In Section~\ref{sec:data_structure} we describe PQSM-tree, a data structure
 used to represent all normalized intersection models of a circular-arc. 
 This section can be read independently by those interested solely in the structure of the normalized models of circular-arc graphs.
 \item In Section~\ref{sec:proof_sketch}, we provide a concise proof of the correctness of our description. The more detailed proofs of certain statements, which are not essential for understanding the flow of the argument, are deferred to Sections~\ref{sec:modular_decomposition_and_chord_models}, \ref{sec:prime_case_properties}, and \ref{sec:parallel_case_properties}.
 \item In Section~\ref{sec:pqsm_tree_construction} we present a linear-time algorithm that constructs PQSM-trees for circular-arc graphs.
 \item In Section~\ref{sec:canonization} we present a linear-time algorithm for the canonization problem for circular-arc graphs.
\end{itemize}

\section{Preliminaries}
\label{sec:preliminaries}

A \emph{graph} $G$ is a pair~$(V,{\sim})$, where $V$ is a \emph{vertex set} and~${\sim}$ is an \emph{edge set} (irreflexive and symmetric \emph{edge relation} on~$V$). 
The complement of the graph $G=(V,{\sim})$ is the graph $\overline{G}= (V,{\parallel})$ where $x \parallel y \iff x \neq y \text{ and not } x \sim y$. 
A \emph{poset} is a pair $(V,{\prec})$ where ${\prec}$ is a transitive and irreflexive relation on $V$.
A poset $(V,{\prec})$ is a \emph{transitive orientation} of the graph $(V,{\sim})$ if  \mbox{$x \sim y \iff x \prec y \text{ or }
x \succ y$.} 
A graph $(V,{\sim})$ is a \emph{comparability graph} if $(V,{\sim})$ admits a transitive orientation.
A graph $(V,{\sim})$ is a \emph{co-comparability graph} if its complement is a comparability graph.

If ${\star}$ is a binary relation on $V$ and $X, Y \subseteq V$, then $X \star Y$ denotes that
$x \star y$ for all $x \in X$ and $y \in Y$. 
If $(V,{\star})$ is a graph or a poset and $X \subseteq V$, then
$(X, {\star})$ denotes the graph or the poset on $X$ in which $\star$ is restricted to $X \times X$.
The pair $(X,{\star})$ is called the subgraph or the subposet of $(V,{\star})$ \emph{induced} 
by the set $X$.

Let $G=(V,{\sim})$ be a graph.
The \emph{neighborhood} of a vertex $v$, denoted by $N_G(v)$, comprises vertices adjacent
to $v$, i.e., $N_G(v) = \{u \in V : u \sim v\}$, and the \emph{closed neighborhood} of $v$ is $N_G[v] = N_G(v) \cup \{v\}$.
A vertex $u \in V$ is \emph{universal} in $G$ if $N_G[u] =V$.
Vertices $u,v \in V$ are \emph{twins} in $G$ if $N_G[u] = N_G[v]$.

A sequence $\tau$ over an alphabet $\Sigma$ is a \emph{word}. 
A~\emph{circular word} represents the set of words which are cyclical shifts of one another. 
Hence, we represent a circular word by a word from its corresponding set of words.
We use ${\equiv}$ to express equality between two circular words.
We use ${=}$ if the equality holds between simple words.
If we want to emphasize that the (circular) word $\tau$ contains every letter from $\Sigma$ exactly once, 
we say that $\tau$ is a \emph{(circular) order} or a \emph{(circular) permutation} of $\Sigma$.

\begin{figure}[htp!]
\centering
\begin{tikzpicture}[yscale=0.58,xscale=0.58,>=latex,shorten >=-0.4pt,shorten <=-0.4pt]

\coordinate (center) at (0,0) {};
\draw (0,0) circle (2cm);

\coordinate (ls10) at ($(center)+(270:2.45cm)$) {};
\coordinate (ls11) at ($(center)+(90:2.45cm)$) {};

\coordinate (ls20) at ($(center)+(180:2.45cm)$) {};
\coordinate (ls21) at ($(center)+(0:2.45cm)$) {};

\coordinate (ls30) at ($(center)+(35:2.45cm)$) {};
\coordinate (ls31) at ($(center)+(-35:2.45cm)$) {};

\coordinate (lc0) at ($(center)+(225:2.4cm)$) {};
\coordinate (lc1) at ($(center)+(60:2.4cm)$) {};

\coordinate (lp) at ($(center)+(135:2.4cm)$) {};
\coordinate (lq) at ($(center)+(300:2.4cm)$) {};

\coordinate (p) at ($(center)+(135:2cm)$) {};
\coordinate (q) at ($(center)+(300:2cm)$) {};

\tikzstyle{every node}=[inner sep=1pt]
\begin{footnotesize}
\node at (ls30) {$a^0_3$};
\node at (ls31) {$a^1_3$};
\node at (ls10) {$a^0_1$};
\node at (ls11) {$a^1_1$};
\node at (ls20) {$a^0_2$};
\node at (ls21) {$a^1_2$};
\node at (lc0) {$c$};
\node at (lc1) {$c$};
\node at (lp) {$p$};
\node at (lq) {$q$};
\end{footnotesize}

\draw[very thick, red] ([shift=(0:1.92cm)]0,0) arc (0:180:1.92cm);

\draw[very thick, green] ([shift=(90:2.08cm)]0,0) arc (90:270:2.08cm);

\draw[very thick, cyan] ([shift=(-35.5:2.08cm)]0,0) arc (-35:35:2.08cm);

\draw[thick,-] ([shift=(60:2cm)]0,0) -- ([shift=(225:2cm)]0,0);

\draw[ultra thick, black,-] ([shift=(133:2cm)]0,0) arc (133:137:2cm);
\draw[ultra thick, black,-] ([shift=(298:2cm)]0,0) arc (298:302:2cm);

\tikzstyle{every node}=[circle,minimum size=4pt,inner sep=0pt,draw,fill]

\draw[white] (-2.8,-3.2)--(-2.8,-2);
\draw[white] (2.8,2.8)--(2.8,2);
\end{tikzpicture} 
\hspace{0.1cm}
\begin{tikzpicture}[yscale=0.58,xscale=0.58,>=latex,shorten >=-0.4pt,shorten <=-0.4pt]
\tikzstyle{every node}=[inner sep=1pt]
\begin{footnotesize}
\node at (0,-2.6) {$L$};
\end{footnotesize}
\draw[thick, dashed] (0,-2.1)--(0,3);
\end{tikzpicture} 
\hspace{0.1cm}
\begin{tikzpicture}[yscale=0.58,xscale=-0.58,>=latex,shorten >=-0.4pt,shorten <=-0.4pt]

\coordinate (center) at (0,0) {};
\draw (0,0) circle (2cm);

\coordinate (ls10) at ($(center)+(270:2.45cm)$) {};
\coordinate (ls11) at ($(center)+(90:2.45cm)$) {};

\coordinate (ls20) at ($(center)+(180:2.45cm)$) {};
\coordinate (ls21) at ($(center)+(0:2.45cm)$) {};

\coordinate (ls30) at ($(center)+(35:2.45cm)$) {};
\coordinate (ls31) at ($(center)+(-35:2.45cm)$) {};

\coordinate (lc0) at ($(center)+(225:2.4cm)$) {};
\coordinate (lc1) at ($(center)+(60:2.4cm)$) {};

\coordinate (lp) at ($(center)+(135:2.4cm)$) {};
\coordinate (lq) at ($(center)+(300:2.4cm)$) {};

\coordinate (p) at ($(center)+(135:2cm)$) {};
\coordinate (q) at ($(center)+(300:2cm)$) {};

\tikzstyle{every node}=[inner sep=1pt]
\begin{footnotesize}
\node at (ls30) {$a^1_3$};
\node at (ls31) {$a^0_3$};
\node at (ls10) {$a^1_1$};
\node at (ls11) {$a^0_1$};
\node at (ls20) {$a^1_2$};
\node at (ls21) {$a^0_2$};
\node at (lc0) {$c$};
\node at (lc1) {$c$};
\node at (lp) {$p$};
\node at (lq) {$q$};
\end{footnotesize}

\draw[very thick, red] ([shift=(0:1.92cm)]0,0) arc (0:180:1.92cm);

\draw[very thick, green] ([shift=(90:2.08cm)]0,0) arc (90:270:2.08cm);

\draw[very thick, cyan] ([shift=(-35.5:2.08cm)]0,0) arc (-35:35:2.08cm);

\draw[thick,-] ([shift=(60:2cm)]0,0) -- ([shift=(225:2cm)]0,0);

\draw[ultra thick, black,-] ([shift=(133:2cm)]0,0) arc (133:137:2cm);
\draw[ultra thick, black,-] ([shift=(298:2cm)]0,0) arc (298:302:2cm);

\tikzstyle{every node}=[circle,minimum size=4pt,inner sep=0pt,draw,fill]

\draw[white] (-2.8,-3.2)--(-2.8,-2);
\draw[white] (2.8,2.8)--(2.8,2);
\end{tikzpicture} 
\hspace{0.5cm}
\begin{tikzpicture}[yscale=0.58,xscale=0.58,>=latex,shorten >=-0.4pt,shorten <=-0.4pt]

\coordinate (center) at (0,0) {};
\draw (0,0) circle (2cm);

\coordinate (ls10) at ($(center)+(270:2.45cm)$) {};
\coordinate (ls11) at ($(center)+(90:2.45cm)$) {};

\coordinate (ls20) at ($(center)+(180:2.45cm)$) {};
\coordinate (ls21) at ($(center)+(0:2.45cm)$) {};

\coordinate (ls30) at ($(center)+(35:2.45cm)$) {};
\coordinate (ls31) at ($(center)+(-35:2.45cm)$) {};

\coordinate (lc0) at ($(center)+(225:2.4cm)$) {};
\coordinate (lc1) at ($(center)+(60:2.4cm)$) {};

\coordinate (lp) at ($(center)+(135:2.4cm)$) {};
\coordinate (lq) at ($(center)+(300:2.4cm)$) {};

\coordinate (p) at ($(center)+(135:2cm)$) {};
\coordinate (q) at ($(center)+(300:2cm)$) {};

\tikzstyle{every node}=[inner sep=1pt]
\begin{footnotesize}
\node at (ls30) {$a^0_3$};
\node at (ls31) {$a^1_3$};
\node at (ls10) {$a^0_1$};
\node at (ls11) {$a^1_1$};
\node at (ls20) {$a^0_2$};
\node at (ls21) {$a^1_2$};
\node at (lc0) {$c$};
\node at (lc1) {$c$};
\node at (lp) {$p$};
\node at (lq) {$q$};
\end{footnotesize}

\draw[ red] ([shift=(0:1.92cm)]0,0) arc (0:180:1.92cm);

\draw[ green] ([shift=(90:2.08cm)]0,0) arc (90:270:2.08cm);

\draw[very thick, red,<-] ([shift=(0:2cm)]0,0) -- ([shift=(180:2cm)]0,0);

\draw[very thick, green,<-] ([shift=(90:2cm)]0,0) -- ([shift=(270:2cm)]0,0);

\draw[very thick, cyan,<-] ([shift=(-35:2cm)]0,0) -- ([shift=(35:2cm)]0,0);

\draw[thick,-] ([shift=(60:2cm)]0,0) -- ([shift=(225:2cm)]0,0);

\draw[ultra thick, black,-] ([shift=(133:2cm)]0,0) arc (133:137:2cm);
\draw[ultra thick, black,-] ([shift=(298:2cm)]0,0) arc (298:302:2cm);

\tikzstyle{every node}=[circle,minimum size=4pt,inner sep=0pt,draw,fill]

\draw[white] (-2.8,-3.2)--(-2.8,-2);
\draw[white] (2.8,2.8)--(2.8,2);
\end{tikzpicture} 
\hspace{0.1cm}
\begin{tikzpicture}[yscale=0.58,xscale=0.58,>=latex,shorten >=-0.4pt,shorten <=-0.4pt]
\tikzstyle{every node}=[inner sep=1pt]
\begin{footnotesize}
\node at (0,-2.6) {$L$};
\end{footnotesize}
\draw[thick, dashed] (0,-2.1)--(0,3);
\end{tikzpicture} 
\hspace{0.1cm}
\begin{tikzpicture}[yscale=0.58,xscale=-0.58,>=latex,shorten >=-0.4pt,shorten <=-0.4pt]

\coordinate (center) at (0,0) {};
\draw (0,0) circle (2cm);

\coordinate (ls10) at ($(center)+(270:2.45cm)$) {};
\coordinate (ls11) at ($(center)+(90:2.45cm)$) {};

\coordinate (ls20) at ($(center)+(180:2.45cm)$) {};
\coordinate (ls21) at ($(center)+(0:2.45cm)$) {};

\coordinate (ls30) at ($(center)+(35:2.45cm)$) {};
\coordinate (ls31) at ($(center)+(-35:2.45cm)$) {};

\coordinate (lc0) at ($(center)+(225:2.4cm)$) {};
\coordinate (lc1) at ($(center)+(60:2.4cm)$) {};

\coordinate (lp) at ($(center)+(135:2.4cm)$) {};
\coordinate (lq) at ($(center)+(300:2.4cm)$) {};

\coordinate (p) at ($(center)+(135:2cm)$) {};
\coordinate (q) at ($(center)+(300:2cm)$) {};

\tikzstyle{every node}=[inner sep=1pt]
\begin{footnotesize}
\node at (ls30) {$a^1_3$};
\node at (ls31) {$a^0_3$};
\node at (ls10) {$a^1_1$};
\node at (ls11) {$a^0_1$};
\node at (ls20) {$a^1_2$};
\node at (ls21) {$a^0_2$};
\node at (lc0) {$c$};
\node at (lc1) {$c$};
\node at (lp) {$p$};
\node at (lq) {$q$};
\end{footnotesize}

\draw[red] ([shift=(0:1.92cm)]0,0) arc (0:180:1.92cm);

\draw[green] ([shift=(90:2.08cm)]0,0) arc (90:270:2.08cm);

\draw[very thick, red,->] ([shift=(0:2cm)]0,0) -- ([shift=(180:2cm)]0,0);

\draw[very thick, green,->] ([shift=(90:2cm)]0,0) -- ([shift=(270:2cm)]0,0);

\draw[very thick, cyan,->] ([shift=(-35:2cm)]0,0) -- ([shift=(35:2cm)]0,0);

\draw[thick,-] ([shift=(60:2cm)]0,0) -- ([shift=(225:2cm)]0,0);

\draw[ultra thick, black,-] ([shift=(133:2cm)]0,0) arc (133:137:2cm);
\draw[ultra thick, black,-] ([shift=(298:2cm)]0,0) arc (298:302:2cm);

\tikzstyle{every node}=[circle,minimum size=4pt,inner sep=0pt,draw,fill]

\draw[white] (-2.8,-3.2)--(-2.8,-2);
\draw[white] (2.8,2.8)--(2.8,2);
\end{tikzpicture} 

\caption{\label{fig:reflection} To the left: a collection $B = \{a_1,a_2,a_3,c,p,q\}$ consisiting of
three arcs $a_1,a_2,a_3$, a chord $c$, and two points $p, q$ is represented by the circular word $\tau(B) \equiv a_2^0pa_1^1ca^0_3a_2^1a^1_3qa_1^0c$.
The word $\tau(B^R) \equiv ca_1^1qa^0_3a_2^0a^1_3ca_1^0pa_2^1$ is the reflection of
$\tau(B)$.
To the right: the same collection in which the arcs are replaced by the corresponding oriented chords.}
\end{figure}
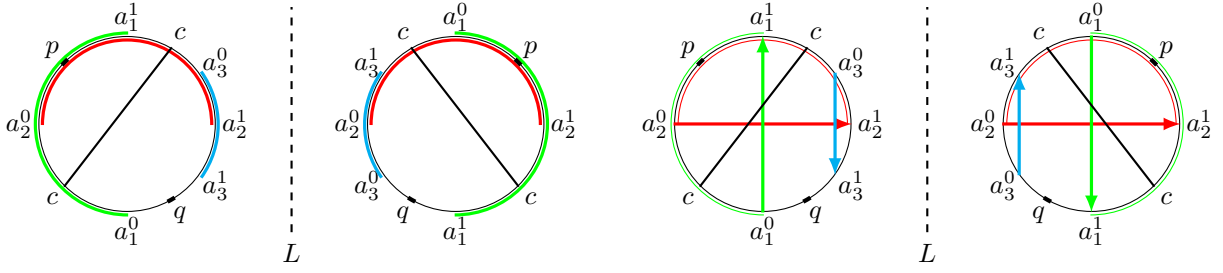

Let $B= A \cup C \cup P$ be a collection of some arcs in the set $A$, some non-oriented chords in the set $C$, 
some points in the set $P$, on some fixed circle such that the endpoints of the objects from $A \cup C$ and the points from $P$ are pairwise different.
We represent the set $B$ by means of a circular word $\tau(B)$ over the set of letters $\Sigma = A^* \cup C \cup P$, where by $A^*$ we denote the set $\{a^0,a^1: a \in A\}$, obtained as follows.
We start with an empty word $\tau(B)$.
We traverse the circle in the clockwise order starting from some arbitrary point and:
\begin{itemize}
\item if we enter/leave the arc $a \in A$, we append the letter $a^0$ ($a^1$, respectively) to $\tau(B)$,
\item if we pass the endpoint of the chord $c \in C$, we append the letter $c$ to $\tau(B)$,
\item if we pass the point $p \in P$, we append the letter $p$ to $\tau(B)$.
\end{itemize}
We make $\tau(B)$ circular when we are back in the starting point.
See Figure~\ref{fig:reflection} to the left.

Let $B=A \cup C \cup P$ be as above and let $\tau(B)$ be a word representation of $B$.
The \emph{reflection $B^R$} of the set $B$ is obtained by mirroring every object from $B$ 
over some fixed line $L$.
Note that the word representation $\tau(B^R)$ of the set $B^R$ is obtained from $\tau(B)$ by reversing the order of the letters in $\tau(B)$ and then by exchanging every superscript $0$~to~$1$ and $1$~to~$0$.
See Figure~\ref{fig:reflection} to the left.

Let $\tau$ be a circular word over the alphabet~$\Sigma$.
The \emph{reflection} $\tau^R$ of $\tau$ is the (circular) word obtained from $\tau$
by reversing the order of the letters in $\tau$ and by exchanging every superscript $0$ to $1$ and
$1$ to $0$.
For $\Sigma' \subseteq \Sigma$, by $\tau \Vert \Sigma'$ we denote a circular word obtained from $\tau$ by restricting to the letters from the set $\Sigma'$.
For example, in Figure~\ref{fig:reflection}, for $\Sigma' = \{a_3^0,a_3^1,p,q,c\}$, 
we have $\tau(B) \Vert  \Sigma' \equiv ca^0_3a^1_3qcp$.

\begin{figure}[htp!]
\centering
\begin{tikzpicture}[xscale=0.7,yscale=0.7,>=latex,shorten >=-0.4pt,shorten <=-0.4pt]
\coordinate (center) at (0,0) {};
\coordinate (Lv1) at ($(center)+(0:2.4cm)$) {};
\coordinate (Lv6) at ($(center)+(60:2.4cm)$) {};
\coordinate (Lv5) at ($(center)+(120:2.4cm)$) {};
\coordinate (Lv4) at ($(center)+(180:2.4cm)$) {};
\coordinate (Lv3) at ($(center)+(240:2.4cm)$) {};
\coordinate (Lv2) at ($(center)+(300:2.5cm)$) {};

\coordinate (Llv1) at ($(center)+(-45:2.5cm)$) {};
\coordinate (Lrv1) at ($(center)+(45:2.5cm)$) {};
\coordinate (Llv6) at ($(center)+(15:2.5cm)$) {};
\coordinate (Lrv6) at ($(center)+(105:2.5cm)$) {};
\coordinate (Llv5) at ($(center)+(75:2.5cm)$) {};
\coordinate (Lrv5) at ($(center)+(165:2.5cm)$) {};
\coordinate (Llv4) at ($(center)+(135:2.5cm)$) {};
\coordinate (Lrv4) at ($(center)+(225:2.5cm)$) {};
\coordinate (Llv3) at ($(center)+(195:2.5cm)$) {};
\coordinate (Lrv3) at ($(center)+(285:2.5cm)$) {};
\coordinate (Llv2) at ($(center)+(255:2.5cm)$) {};
\coordinate (Lrv2) at ($(center)+(345:2.5cm)$) {};

\coordinate (lv1) at ($(center)+(-45:2cm)$) {};
\coordinate (rv1) at ($(center)+(45:2cm)$) {};
\coordinate (lv6) at ($(center)+(15:2cm)$) {};
\coordinate (rv6) at ($(center)+(105:2cm)$) {};
\coordinate (lv5) at ($(center)+(75:2cm)$) {};
\coordinate (rv5) at ($(center)+(165:2cm)$) {};
\coordinate (lv4) at ($(center)+(135:2cm)$) {};
\coordinate (rv4) at ($(center)+(225:2cm)$) {};
\coordinate (lv3) at ($(center)+(195:2cm)$) {};
\coordinate (rv3) at ($(center)+(285:2cm)$) {};
\coordinate (lv2) at ($(center)+(255:2cm)$) {};
\coordinate (rv2) at ($(center)+(345:2cm)$) {};

\tikzstyle{every node}=[inner sep=1pt]
\begin{footnotesize}
\node[red] at (Lrv1) {$v^0_1$};
\node[red] at (Llv1) {$v^1_1$};
\node at (Lrv2) {$v^0_2$};
\node at (Llv2) {$v^1_2$};
\node at (Lrv3) {$v^0_3$};
\node[blue] at (Llv3) {$v^1_3$};
\node[blue] at (Lrv4) {$v^0_4$};
\node at (Llv4) {$v^1_4$};
\node[blue] at (Lrv5) {$v^0_5$};
\node at (Llv5) {$v^1_5$};
\node[red] at (Lrv6) {$v^0_6$};
\node[red] at (Llv6) {$v^1_6$};
\end{footnotesize}

\draw[thick] ([shift=(-45:2cm)]0,0) arc (-45:45:2cm);
\draw[thick] ([shift=(15:2.15cm)]0,0) arc (15:105:2.1cm);
\draw[thick] ([shift=(75:2cm)]0,0) arc (75:165:2cm);
\draw[thick] ([shift=(135:2.15cm)]0,0) arc (135:225:2.1cm);
\draw[thick] ([shift=(195:2cm)]0,0) arc (195:285:2cm);
\draw[thick] ([shift=(255:2.15cm)]0,0) arc (255:345:2.1cm);

\draw[blue,thick,|-|] ([shift=(155:2.8cm)]0,0) arc (155:235:2.8cm);

\draw[white] (-2.7,-3.5)--(-2.5,-3.5);
\draw[white] (2.7,3.5)--(2.5,3.5);

\end{tikzpicture}
\hspace{0.5cm}
\begin{tikzpicture}[xscale=0.7,yscale=0.7,>=latex,shorten >=-0.4pt,shorten <=-0.4pt]
\coordinate (center) at (0,0) {};
\coordinate (Llv1) at ($(center)+(-45:2.4cm)$) {};
\coordinate (Lrv1) at ($(center)+(45:2.4cm)$) {};
\coordinate (Llv6) at ($(center)+(15:2.4cm)$) {};
\coordinate (Lrv6) at ($(center)+(105:2.4cm)$) {};
\coordinate (Llv5) at ($(center)+(75:2.4cm)$) {};
\coordinate (Lrv5) at ($(center)+(165:2.4cm)$) {};
\coordinate (Llv4) at ($(center)+(135:2.4cm)$) {};
\coordinate (Lrv4) at ($(center)+(225:2.4cm)$) {};
\coordinate (Llv3) at ($(center)+(195:2.4cm)$) {};
\coordinate (Lrv3) at ($(center)+(285:2.4cm)$) {};
\coordinate (Llv2) at ($(center)+(255:2.4cm)$) {};
\coordinate (Lrv2) at ($(center)+(345:2.4cm)$) {};

\coordinate (rv1) at ($(center)+(-45:2cm)$) {};
\coordinate (lv1) at ($(center)+(45:2cm)$) {};
\coordinate (rv6) at ($(center)+(15:2cm)$) {};
\coordinate (lv6) at ($(center)+(105:2cm)$) {};
\coordinate (rv5) at ($(center)+(75:2cm)$) {};
\coordinate (lv5) at ($(center)+(165:2cm)$) {};
\coordinate (rv4) at ($(center)+(135:2cm)$) {};
\coordinate (lv4) at ($(center)+(225:2cm)$) {};
\coordinate (rv3) at ($(center)+(195:2cm)$) {};
\coordinate (lv3) at ($(center)+(285:2cm)$) {};
\coordinate (rv2) at ($(center)+(255:2cm)$) {};
\coordinate (lv2) at ($(center)+(345:2cm)$) {};

\tikzstyle{every node}=[inner sep=1pt]
\begin{footnotesize}
\node[red] at (Lrv1) {$v^0_1$};
\node[red] at (Llv1) {$v^1_1$};
\node at (Lrv2) {$v^0_2$};
\node at (Llv2) {$v^1_2$};
\node at (Lrv3) {$v^0_3$};
\node[blue] at (Llv3) {$v^1_3$};
\node[blue] at (Lrv4) {$v^0_4$};
\node at (Llv4) {$v^1_4$};
\node[blue] at (Lrv5) {$v^0_5$};
\node at (Llv5) {$v^1_5$};
\node[red] at (Lrv6) {$v^0_6$};
\node[red] at (Llv6) {$v^1_6$};
\end{footnotesize}

\draw (0,0) circle (2cm);
\draw[->] (lv1)--(rv1);
\draw[->] (lv2)--(rv2);
\draw[->] (lv3)--(rv3);
\draw[->] (lv4)--(rv4);
\draw[->] (lv5)--(rv5);
\draw[->] (lv6)--(rv6);



\draw[blue,thick,|-|] ([shift=(155:2.8cm)]0,0) arc (155:235:2.8cm);

\draw[white] (-2.7,-3.5)--(-2.5,-3.5);
\draw[white] (2.7,3.5)--(2.5,3.5);
\end{tikzpicture}

\caption{\label{fig:examp} To the left: intersection model $\psi$ of a circular-arc graph $G = (V,E)$, where $V = \{v_1,\ldots,v_6\}$ and $E = \{ v_iv_{i+1}: i \in [5]\}\cup \{v_6v_1\}$.
We have $\psi \Vert \{v^0_1,v^1_1, v^0_6,v^1_6\} \equiv v_6^0v_1^0v_6^1v^1_1$ (in red).
The set $\{v^1_3,v^0_4,v^0_5\}$ is contiguous in~$\psi$ (in blue) and we have $\psi|\{v^1_3,v^0_4,v^0_5\} = v^0_4v^1_3v^0_5$.
To the right: the corresponding oriented chord model.}
\end{figure}
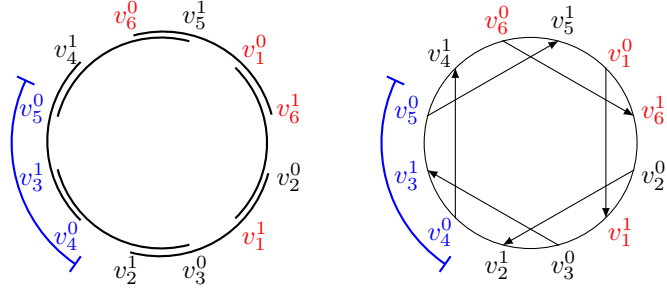

The notation we introduce below can be used to describe relations 
between a set of arcs and points of the circle.
Let $\Sigma$ be a set and let~$\tau$ be a circular permutation of ~$\Sigma$.
A word $\mu'$ is a \emph{contiguous subword} of $\tau$ if
there is $\mu''$ such that $\tau \equiv \mu'\mu''$.
Let $\Sigma' \subseteq \Sigma$.
We say that the letters of $\Sigma'$ are \emph{contiguous} in~$\tau$ if
$\tau \equiv \tau'\tau''$, where 
$\tau'$ is a word in which every letter from $\Sigma'$ occurs exactly one 
and $\tau''$ is a word on $\Sigma \setminus \Sigma'$.
If this is the case, we say that the set $\Sigma'$ forms a \emph{contiguous subword} in~$\tau$, 
and we denote this subword by $\tau|\Sigma'$ (note that $\tau|\Sigma'$ is unique).
For example, in Figure~\ref{fig:examp} the set $\{v^1_3,v^0_4,v^0_5\}$ is contiguous in~$\psi$
and we have $\psi|\{v^1_3,v^0_4,v^0_5\} = v^0_4v^1_3v_5^0$ (depicted in blue in Figure~\ref{fig:examp}).
Let $u'$ and $v'$ be two letters in $\tau$.
A letter $w'$ is \emph{between $u'$ and $v'$ in $\tau$} 
if we pass $w'$ when we traverse $\tau$ in the clockwise order from $u'$ 
to $v'$.
A~letter $w'$ is \emph{strictly between $u'$ and $v'$ in $\tau$} if $w'$ is between $u'$ and $v'$ in $\phi$, $u' \neq w'$, and $u' \neq v'$.
In Figure~\ref{fig:examp} the letters $v_6^0,v_5^1,v_1^0,v_6^1,v_2^0,v_1^1,v_3^0$ are between $v_6^0$ and $v_3^0$ in $\psi$ and 
the letters $v_4^0,v_3^1,v_5^0$ are strictly between $v_2^1$ and $v_4^1$ in $\psi$.

We use analogous notation for simple (non-circular) words.

\subsection{Modular decomposition trees and transitive orientations}
The definitions given below are by Gallai~\cite{Gal67}.

Let $(V,{\sim})$ be a graph and let $(V, \parallel)$ be the complement of~$(V,{\sim})$.
A non-empty set \mbox{$M\subseteq V$} is a \emph{module} in~$(V,{\sim})$ 
if $x\sim M$ or $x \parallel M$ for every $x\in V\smallsetminus M$.
The singleton sets and the whole $V$ are the \emph{trivial} modules of $(V,{\sim})$. 
A module $M$ of $(V,{\sim})$ is \emph{strong} if $M\subseteq N$, $N\subseteq M$, or $M\cap N=\emptyset$ for every other module $N$ in $(V,{\sim})$.
In particular, two strong modules of $(V,{\sim})$ are either nested or disjoint.
The \emph{modular decomposition} of~$(V,{\sim})$, denoted by~$\strongModules(V,{\sim})$, 
consists of all strong modules of~$(V,{\sim})$.
The set $\strongModules(V,{\sim})$, ordered by inclusion, forms a tree in which $V$ is the root, 
the maximal proper subsets from $\strongModules(V,{\sim})$ of $M \in \strongModules(V,{\sim})$ 
are the children of $M$ (the children of $M$ form a partition of $M$), 
and the singleton modules $\{x\}$ for $x\in V$ are the leaves.

A module $M \in \strongModules(V,{\sim})$ is \emph{serial} if $M_1\sim M_2$ for every two children $M_1$ and $M_2$ of~$M$, 
\emph{parallel} if $M_1 \parallel M_2$ for every two children $M_1$ and $M_2$ of~$M$, and \emph{prime} otherwise.
Equivalently, $M \in \strongModules(V,{\sim})$ is serial if $(M,{\parallel})$ is disconnected, 
parallel if $(M,{\sim})$ is disconnected, and prime if both $(M,{\sim})$ and $(M,{\parallel})$ are connected.

Note that the modular decomposition trees~$\strongModules(V,{\sim})$ and~$\strongModules(V,{\parallel})$ are the same;
the only difference is that serial (parallel) module~$M$ in~$\strongModules(V,{\sim})$ is parallel (serial) in~$\strongModules(V,{\parallel})$.

Now assume that $(V,{\sim})$ is a comparability graph.
The relation between the transitive orientations of the graph~$(V,{\sim})$ and the modular decomposition tree of~$(V,{\sim})$ was described by Gallai~\cite{Gal67}.
\begin{theorem}[\cite{Gal67}]
\label{thm:transitive_orientation_of_edges_between_children}
If $M_1,M_2 \in \strongModules(V,{\sim})$ are such that $M_1 \sim M_2$, then every
transitive orientation $(V,{\prec})$ satisfies either $M_1 \prec M_2$ or $M_2 \prec M_1$.
\end{theorem}
For a module $M$ in $\strongModules(V,{\sim})$ let $(M,{\sim_M})$ denote the graph on $M$ whose edge set~${\sim_M}$ contains all the edges from~${\sim}$ that join the vertices from two different children of $M$.
If $x\sim y$ is an edge in $(V,{\sim})$, then $x\sim_M y$ for exactly one strong module $M \in\strongModules(V,{\sim})$.
Hence, the set $\{{\sim_M} : M \in \strongModules(V,{\sim})\}$ forms a partition of the edge set ${\sim}$ of the graph $(V,{\sim})$.
\begin{theorem}[\cite{Gal67}]
\label{thm:transitive_orientations_versus_transitive_orientations_of_strong_modules}
There is one-to-one correspondence between the set of transitive orientations $(V,{\prec})$ of $(V,{\sim})$
and the families $$\{(M,{\prec_M}): M \in \strongModules(V,{\sim}) \text{ and } \prec_M \text{ is a transitive orientation of $(M, \sim_M)$}\}$$
given by $x \prec y \iff x \prec_M y$, where $M$ is the module in $\strongModules(V,{\sim})$ such that $x \sim_M y$.
\end{theorem}
The above theorem asserts that every transitive orientation of $(V,{\sim})$ restricted to the edges of the graph $(M, {\sim_M})$ 
induces a transitive orientation of $(M, {\sim_M})$, for every $M \in \strongModules(V,{\sim})$, and
that every transitive orientation of $(V,{\sim})$ can be obtained 
by independent transitive orientation of the graphs $(M,{\sim_M})$, for $M \in \strongModules(V,{\sim})$.
Gallai~\cite{Gal67} characterized all possible transitive orientations of $(M,{\sim_M})$, where $M$ is a~module of $\strongModules(V,{\sim})$.
\begin{theorem}[\cite{Gal67}]
\label{thm:prime_graph_transitive_orientation}
Let $M$ be a prime module in $\strongModules(V,{\sim})$. 
Then, $(M,{\sim_M})$ has two transitive orientations, one being the reverse of the other.
\end{theorem}
A parallel module $(M,{\sim_M})$ has exactly one (empty) transitive orientation.
The transitive orientations of serial modules $(M,{\sim})$ correspond to the total orderings of its children, that is, every transitive orientation of $(M,{\sim_M})$ is of the form $M_{i_1} \prec \ldots \prec M_{i_k}$, where $i_1 \ldots i_k$ is a permutation of $[k]$ and $M_1, \ldots,M_k$ are the children of $M$ in $\strongModules(V,{\sim})$.

A graph $(V,{\sim})$ is \emph{prime} if every module of $(V,{\sim})$ is trivial.
By Theorem~\ref{thm:prime_graph_transitive_orientation}, every prime comparability graph admits a unique (up to reversal) transitive orientation.

\subsection{Geometric intersection graphs}
For a family $\mathcal{R}$ of geometric objects, the \emph{intersection model} 
(or \emph{$\mathcal{R}$-model}) of a graph $G=(V,E)$ in the family~$\mathcal{R}$ is a
mapping $\phi : V \to \mathcal{R}$ which assigns objects in~$\mathcal{R}$ to the vertices of~$G$ such that $\phi(u) \cap \phi(v) \neq \emptyset \iff uv \in E$ for any two vertices~$u$ and~$v$ of~$G$. 
For this work, we introduce the following classes of intersection graphs:
\begin{itemize}
\item \emph{interval graphs}, which are intersection graphs of intervals on the real line,
\item \emph{circular-arc graphs}, which are intersection graphs of arcs of a circle,
\item \emph{circle graphs}, which are intersection graphs of chords of a circle,
\item \emph{permutation graphs}, which are intersection graphs of chords spanned between two disjoint arcs of a circle.
\end{itemize}
It is known that interval graphs are co-comparability graphs and permutation graphs correspond to the intersection of comparability and co-comparability graphs~\cite{DM41}.
Clearly, permutation graphs form a subclass of circle graphs.

\subsubsection{Permutation graphs}
\begin{figure}[htp!]















\begin{tikzpicture}[xscale=0.85,yscale=0.5,>=latex,shorten >=-0.4pt,shorten <=-0.4pt]
\coordinate (center) at (0,0) {};
\coordinate (lA0) at ($(center)+(120:2.6cm)$) {};
\coordinate (ua) at ($(center)+(105:2cm)$) {};
\coordinate (ub) at ($(center)+(90:2cm)$) {};
\coordinate (uc) at ($(center)+(75:2cm)$) {};
\coordinate (lA1) at ($(center)+(60:2.6cm)$) {};

\coordinate (lua) at ($(center)+(105:2.4cm)$) {};
\coordinate (lub) at ($(center)+(90:2.4cm)$) {};
\coordinate (luc) at ($(center)+(75:2.4cm)$) {};

\coordinate (lB1) at ($(center)+(240:2.6cm)$) {};
\coordinate (bc) at ($(center)+(270:2cm)$) {};
\coordinate (bb) at ($(center)+(255:2cm)$) {};
\coordinate (ba) at ($(center)+(285:2cm)$) {};
\coordinate (lB0) at ($(center)+(300:2.6cm)$) {};

\coordinate (lbc) at ($(center)+(270:2.4cm)$) {};
\coordinate (lbb) at ($(center)+(255:2.4cm)$) {};
\coordinate (lba) at ($(center)+(285:2.4cm)$) {};

\coordinate (tau0) at ($(center)+(55:1.5cm)$) {};
\coordinate (tau1) at ($(center)+(235:1.5cm)$) {};

\tikzstyle{every node}=[inner sep=1pt]
\begin{footnotesize}

\node at (lua) {$a$};
\node at (lba) {$a$};

\node at (lub) {$b$};
\node at (lbb) {$b$};

\node at (luc) {$c$};
\node at (lbc) {$c$};

\node at (tau0) {$\tau^0$};
\node at (tau1) {$\tau^1$};

\node at (lA0) {$A^0$};
\node at (lA1) {$A^1$};
\node at (lB0) {$B^0$};
\node at (lB1) {$B^1$};

\end{footnotesize}

\draw[thick,<-] ([shift=(60:2cm)]0,0) arc (60:120:2cm);
\draw[thick,<-] ([shift=(240:2cm)]0,0) arc (240:300:2cm);

\draw[thick] (ua)--(ba);
\draw[thick] (ub)--(bb);
\draw[thick] (uc)--(bc);

\end{tikzpicture}

\caption{\label{fig:non_oriented_permutation_graph} 
Intersection model $(\tau^0,\tau^1) = (abc,acb)$ of the permutation graph $(\{a,b,c\}, \{a \sim b, a \sim c)\}$ corresponding to the transitive orientations $\{a \prec b, a \prec c\}$ and $\{b < c\}$
of $(V,{\sim})$ and $(V,{\parallel}\}$, respectively.
}
\end{figure}
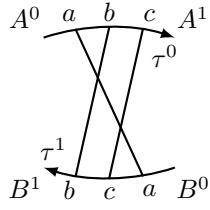

Let $G = (V,{\sim})$ be a permutation graph and let $\psi$ be an intersection model of $(V,{\sim})$ in the set of chords spanned between two disjoint arcs~$A = A^0A^1$ and~$B=B^0B^1$ of the circle.
We represent $\psi$ by means of the pair $(\tau^0, \tau^1)$ of two permutations of $V$, 
where $\tau^0$ ($\tau^1$) is the order of the endpoints of the chords of $\psi$ on the arc $A$ ($B$, respectively) when we traverse it from $A^0$ to $A^1$ (from $B^0$ to $B^1$, respectively).
In particular, for $x,y \in V$ we have
$$
x \sim y \iff x \text{ and } y \text{ occur in the same order in } \tau^0 \text{ and } \tau^1.
$$
See Figure~\ref{fig:non_oriented_permutation_graph}.

Let $(V,{\parallel})$ be the complement of $(V,{\sim})$.
Dushnik and Miller \cite{DM41} showed that both $(V,{\parallel})$ and $(V,{\sim})$ are comparability graphs and that the intersection models of $(V,{\sim})$ can be described by means of transitive orientations of $(V,{\sim})$ and $(V,{\parallel})$, as follows.
Every intersection model $(\tau^0,\tau^1)$ of $(V,{\sim})$ yields transitive orientations ${\prec}$ and ${<}$ of the graphs $(V,{\sim})$ and $(V,{\parallel})$, respectively,
given by:
\begin{equation}
\label{eq:models_of_permutation_graphs_1}
\begin{array}{lll}
x \prec y & \iff & x \text{ occurs before } y \text{ in } \tau^0 \text{ and } x \sim y, \\
x < y & \iff &  x \text{ occurs before } y \text{ in } \tau^0 \text{ and in } x \parallel y. \\
\end{array}
\end{equation}
See Figure~\ref{fig:non_oriented_permutation_graph}.
On the other hand, given transitive orientations ${\prec}$ and ${<}$ of $(V,{\sim})$ and $(V,{\parallel})$, respectively, one can construct a permutation model $(\tau^{0}, \tau^{1})$ of $(V,{\sim})$ such that
\begin{equation}
\label{eq:models_of_permutation_graphs_2}
\begin{array}{lll}
x \text{ occurs before } y \text{ in } \tau^{0} \iff x \prec y \text{ or } x < y,\\
x \text{ occurs before } y \text{ in } \tau^{1} \iff x \prec y \text{ or } y < x.\\
\end{array}
\end{equation}

\begin{theorem}[\cite{DM41}]
\label{thm:permutation_models_transitive_orientations}
Let $(V,{\sim})$ be a permutation graph.
There is one-to-one correspondence between permutation models $(\tau^{0}, \tau^{1})$ of $(V,{\sim})$
and the pairs $({<}, {\prec})$ of transitive orientations of $(V,{\parallel})$ and $(V,{\sim})$, respectively, given by equations \eqref{eq:models_of_permutation_graphs_1} and~\eqref{eq:models_of_permutation_graphs_2}.
\end{theorem}

Theorems~\ref{thm:transitive_orientation_of_edges_between_children} and~\ref{thm:permutation_models_transitive_orientations} yield the following property of strong modules in $(V,{\sim})$ with respect to permutation models of $(V,{\sim})$:
\begin{observation}
\label{obs:strong_modules_are_contiguous_in_permutation_models}
Let $M$ be a module in $\strongModules(V,{\sim})$. 
For every permutation model $(\tau^0,\tau^1)$ of $(V,{\sim})$ the set $M$ is contiguous in both words $\tau^0$ and $\tau^1$.
\end{observation}

\subsubsection{Circle graphs}
Let $G=(V,{\sim})$ be a circle graph and let $\psi$ be a chord model of $G$.
We represent~$\psi$ by means of a circular word on the set~$V$; note that every $v \in V$ appears twice in $\psi$.

The next claim shows that a wide family of subsets $U$ of $V$ induce permutation subgraphs in a circle graph~$(V,{\sim})$.
\begin{claim}
\label{claim:permutation_graphs_in_G_ov}
Let $U \subseteq V$ be such that $x \sim U$ for some $x \in V \setminus U$.
Then, for any chord model $\psi$ of $G$, 
$$\psi \Vert (U \cup \{x\}) \equiv x \tau x \tau',$$
where $(\tau,\tau')$ and $(\tau',\tau)$ are permutation models of $(U,{\sim})$.
In particular, $(U,{\sim})$ is a permutation subgraph of $G$.
\end{claim}
\begin{proof}
Let $\psi$ be a chord model of $G$.
Note that every chord $\psi(u)$ for $u \in U$ has 
its endpoints on different sides of the chord $\psi(x)$.
Thus, $\psi \Vert (U \cup \{x\}) \equiv x \tau x \tau'$, 
where $\tau$ and $\tau'$ are permutations of $U$.
Clearly, since $\psi$ is a chord model of $G$, 
both $(\tau,\tau')$ and $(\tau',\tau)$ are permutation models of $(U,{\sim})$.
\end{proof}

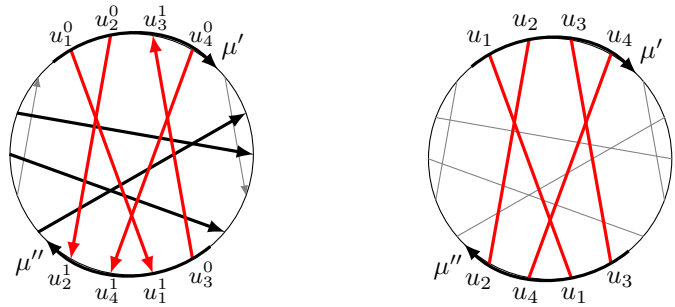
\begin{figure}[htp!]
\centering
\begin{tikzpicture}[scale=0.8,>=latex,shorten >=-0.2pt,shorten <=-0.2pt]
\coordinate (center) at (0,0) {};

\coordinate (lmu') at ($(center)+(47:2.45cm)$) {};
\coordinate (lmu'') at ($(center)+(227:2.45cm)$) {};

\coordinate (u4) at ($(center)+(60:2cm)$) {};
\coordinate (u3) at ($(center)+(80:2cm)$) {};
\coordinate (u2) at ($(center)+(100:2cm)$) {};
\coordinate (u1) at ($(center)+(120:2cm)$) {};

\coordinate (lu4) at ($(center)+(60:2.3cm)$) {};
\coordinate (lu3) at ($(center)+(80:2.3cm)$) {};
\coordinate (lu2) at ($(center)+(100:2.3cm)$) {};
\coordinate (lu1) at ($(center)+(120:2.3cm)$) {};

\coordinate (b4) at ($(center)+(240:2cm)$) {};
\coordinate (b3) at ($(center)+(260:2cm)$) {};
\coordinate (b2) at ($(center)+(280:2cm)$) {};
\coordinate (b1) at ($(center)+(300:2cm)$) {};

\coordinate (lb4) at ($(center)+(240:2.35cm)$) {};
\coordinate (lb3) at ($(center)+(260:2.3cm)$) {};
\coordinate (lb2) at ($(center)+(280:2.3cm)$) {};
\coordinate (lb1) at ($(center)+(300:2.3cm)$) {};

\coordinate (l5) at ($(center)+(140:2cm)$) {};
\coordinate (l4) at ($(center)+(160:2cm)$) {};
\coordinate (l3) at ($(center)+(180:2cm)$) {};
\coordinate (l2) at ($(center)+(200:2cm)$) {};
\coordinate (l1) at ($(center)+(220:2cm)$) {};

\coordinate (rl5) at ($(center)+(140:2cm)$) {};
\coordinate (rl4) at ($(center)+(160:2cm)$) {};
\coordinate (rl3) at ($(center)+(180:2cm)$) {};
\coordinate (rl2) at ($(center)+(200:2cm)$) {};
\coordinate (rl1) at ($(center)+(220:2cm)$) {};

\coordinate (ll4) at ($(center)+(160:2.3cm)$) {};

\coordinate (r5) at ($(center)+(-40:2cm)$) {};
\coordinate (r4) at ($(center)+(-20:2cm)$) {};
\coordinate (r3) at ($(center)+(0:2cm)$) {};
\coordinate (r2) at ($(center)+(20:2cm)$) {};
\coordinate (r1) at ($(center)+(40:2cm)$) {};

\coordinate (lr3) at ($(center)+(0:2.3cm)$) {};

\draw ($(center)$) circle (2cm);

\draw[gray,->] (l2)-- (l5);
\draw[gray,->] (r1)-- (r4);

\draw[black,very thick,->] (l1)-- (r2);
\draw[black,very thick,->] (l3)-- (r5);
\draw[black,very thick,->] (l4)-- (r3);

\draw[very thick,red,->] (u1)-- (b2);
\draw[very thick,red,->] (u2)-- (b4);
\draw[very thick,red,<-] (u3)-- (b1);
\draw[very thick,red,->] (u4)-- (b3);

\draw[very thick, <-] ([shift=(225:2cm)]0,0) arc (225:310:2cm);
\draw[very thick, <-] ([shift=(45:2cm)]0,0) arc (45:130:2cm);

\tikzstyle{every node}=[inner sep=1pt]
\begin{scriptsize}
\node at (lu1) {$u^0_1$};
\node at (lu2) {$u^0_2$};
\node at (lu3) {$u^1_3$};
\node at (lu4) {$u^0_4$};

\node at (lb4) {$u^1_2$};
\node at (lb3) {$u^1_4$};
\node at (lb2) {$u^1_1$};
\node at (lb1) {$u^0_3$};
\end{scriptsize}

\begin{footnotesize}
\node at (lmu') {$\mu'$};
\node at (lmu'') {$\mu''$};
 \end{footnotesize}

\end{tikzpicture} 
\hspace{2cm}
\begin{tikzpicture}[scale=0.8,>=latex,shorten >=-0.2pt,shorten <=-0.2pt]
\coordinate (center) at (0,0) {};

\coordinate (lmu') at ($(center)+(47:2.45cm)$) {};
\coordinate (lmu'') at ($(center)+(227:2.45cm)$) {};

\coordinate (u4) at ($(center)+(60:2cm)$) {};
\coordinate (u3) at ($(center)+(80:2cm)$) {};
\coordinate (u2) at ($(center)+(100:2cm)$) {};
\coordinate (u1) at ($(center)+(120:2cm)$) {};

\coordinate (lu4) at ($(center)+(60:2.3cm)$) {};
\coordinate (lu3) at ($(center)+(80:2.3cm)$) {};
\coordinate (lu2) at ($(center)+(100:2.3cm)$) {};
\coordinate (lu1) at ($(center)+(120:2.3cm)$) {};

\coordinate (b4) at ($(center)+(240:2cm)$) {};
\coordinate (b3) at ($(center)+(260:2cm)$) {};
\coordinate (b2) at ($(center)+(280:2cm)$) {};
\coordinate (b1) at ($(center)+(300:2cm)$) {};

\coordinate (lb4) at ($(center)+(240:2.35cm)$) {};
\coordinate (lb3) at ($(center)+(260:2.3cm)$) {};
\coordinate (lb2) at ($(center)+(280:2.3cm)$) {};
\coordinate (lb1) at ($(center)+(300:2.3cm)$) {};

\coordinate (l5) at ($(center)+(140:2cm)$) {};
\coordinate (l4) at ($(center)+(160:2cm)$) {};
\coordinate (l3) at ($(center)+(180:2cm)$) {};
\coordinate (l2) at ($(center)+(200:2cm)$) {};
\coordinate (l1) at ($(center)+(220:2cm)$) {};

\coordinate (rl5) at ($(center)+(140:2cm)$) {};
\coordinate (rl4) at ($(center)+(160:2cm)$) {};
\coordinate (rl3) at ($(center)+(180:2cm)$) {};
\coordinate (rl2) at ($(center)+(200:2cm)$) {};
\coordinate (rl1) at ($(center)+(220:2cm)$) {};

\coordinate (ll4) at ($(center)+(160:2.3cm)$) {};

\coordinate (r5) at ($(center)+(-40:2cm)$) {};
\coordinate (r4) at ($(center)+(-20:2cm)$) {};
\coordinate (r3) at ($(center)+(0:2cm)$) {};
\coordinate (r2) at ($(center)+(20:2cm)$) {};
\coordinate (r1) at ($(center)+(40:2cm)$) {};

\coordinate (lr3) at ($(center)+(0:2.3cm)$) {};

\draw ($(center)$) circle (2cm);

\draw[gray] (l2)-- (l5);
\draw[gray] (r1)-- (r4);

\draw[gray] (l1)-- (r2);
\draw[gray] (l3)-- (r5);
\draw[gray] (l4)-- (r3);

\draw[very thick,red,-] (u1)-- (b2);
\draw[very thick,red,-] (u2)-- (b4);
\draw[very thick,red,-] (u3)-- (b1);
\draw[very thick,red,-] (u4)-- (b3);

\draw[very thick, <-] ([shift=(225:2cm)]0,0) arc (225:310:2cm);
\draw[very thick, <-] ([shift=(45:2cm)]0,0) arc (45:130:2cm);

\tikzstyle{every node}=[inner sep=1pt]
\begin{footnotesize}
\node at (lu1) {$u_1$};
\node at (lu2) {$u_2$};
\node at (lu3) {$u_3$};
\node at (lu4) {$u_4$};

\node at (lb4) {$u_2$};
\node at (lb3) {$u_4$};
\node at (lb2) {$u_1$};
\node at (lb1) {$u_3$};

\node at (lmu') {$\mu'$};
\node at (lmu'') {$\mu''$};
 \end{footnotesize}

\end{tikzpicture}

\caption{\label{fig:consistent_permutation_model} 
To the left: the set $U=\{u_1,u_2,u_3,u_4\}$ (in red) induces a consistent (oriented) permutation model $(\mu',\mu'')=(u^0_1u^0_2u^1_3u^0_4,u^0_3u^1_1u^1_4u^1_2)$ in some conformal model~$\phi$.
The chord $u_1$ is directed from $\mu'$ to $\mu''$.
The set $B$ (in black) does not induce a consistent permutation model in $\phi$.
To the right: the set $U$ (in red) induces a consistent permutation model $(\mu',\mu'')=(u_1u_2u_3u_4,u_3u_1u_4u_2)$ in some circle model~$\phi$.
}
\end{figure}

Let $U \subseteq V$ and let $\psi$ be a chord model of $G$.
We say the set $U$ induces a \emph{consistent permutation model} in 
$\psi$ if $\psi \equiv \mu'\tau'\mu''\tau''$, where $(\mu',\mu'')$ is a permutation model of $(U,{\sim})$ and $\tau'$ and $\tau''$ are (possibly empty) words on $V$.
See Figure~\ref{fig:consistent_permutation_model} to the right.

\subsubsection{Circular-arc graphs}
\begin{figure}[htp!]
\begin{tikzpicture}[scale=0.5]
\coordinate (center) at (0,0) {};
\coordinate (v) at ($(center)+(90:2cm)$) {};
\coordinate (u) at ($(center)+(270:2cm)$) {};

\coordinate (lv) at ($(center)+(90:2.6cm)$) {};
\coordinate (lu) at ($(center)+(270:2.6cm)$) {};

\tikzstyle{every node}=[inner sep=1pt]
\begin{scriptsize}
\node at (lv) {$\psi(v)$};
\node at (lu) {$\psi(u)$};
\end{scriptsize}

\draw[thick] ([shift=(30:2.1cm)]0,0) arc (30:150:2.1cm);
\draw[thick] ([shift=(210:2.1cm)]0,0) arc (210:330:2.1cm);

\draw[thick, white] (-3.0,-3)--(-3.0,-2.8);
\draw[thick, white] (3.0,3)--(3.0,2.8);

\end{tikzpicture}
\begin{tikzpicture}[scale=0.5]
\coordinate (center) at (0,0) {};
\coordinate (v) at ($(center)+(90:2cm)$) {};
\coordinate (u) at ($(center)+(270:2cm)$) {};

\coordinate (lv) at ($(center)+(90:2.6cm)$) {};
\coordinate (lu) at ($(center)+(90:1.25cm)$) {};

\tikzstyle{every node}=[inner sep=1pt]
\begin{scriptsize}
\node at (lv) {$\psi(v)$};
\node at (lu) {$\psi(u)$};
\end{scriptsize}

\draw[thick] ([shift=(0:2.1cm)]0,0) arc (0:180:2.1cm);
\draw[thick] ([shift=(45:1.8cm)]0,0) arc (45:135:1.8cm);

\draw[thick, white] (-3.0,-3)--(-3.0,-2.8);
\draw[thick, white] (3.0,3)--(3.0,2.8);

\end{tikzpicture}
\begin{tikzpicture}[scale=0.5]
\coordinate (center) at (0,0) {};
\coordinate (v) at ($(center)+(90:2cm)$) {};
\coordinate (u) at ($(center)+(270:2cm)$) {};

\coordinate (lu) at ($(center)+(90:2.6cm)$) {};
\coordinate (lv) at ($(center)+(90:1.25cm)$) {};

\tikzstyle{every node}=[inner sep=1pt]
\begin{scriptsize}
\node at (lv) {$\psi(v)$};
\node at (lu) {$\psi(u)$};
\end{scriptsize}

\draw[thick] ([shift=(0:2.1cm)]0,0) arc (0:180:2.1cm);
\draw[thick] ([shift=(45:1.8cm)]0,0) arc (45:135:1.8cm);

\draw[thick, white] (-3.0,-3)--(-3.0,-2.8);
\draw[thick, white] (3.0,3)--(3.0,2.8);

\end{tikzpicture}
\begin{tikzpicture}[scale=0.5]
\coordinate (center) at (0,0) {};
\coordinate (v) at ($(center)+(90:2cm)$) {};
\coordinate (u) at ($(center)+(270:2cm)$) {};

\coordinate (lv) at ($(center)+(90:2.6cm)$) {};
\coordinate (lu) at ($(center)+(270:2.4cm)$) {};

\tikzstyle{every node}=[inner sep=1pt]
\begin{scriptsize}
\node at (lv) {$\psi(v)$};
\node at (lu) {$\psi(u)$};
\end{scriptsize}

\draw[thick] ([shift=(-20:2.1cm)]0,0) arc (-20:200:2.1cm);
\draw[thick] ([shift=(160:1.9cm)]0,0) arc (160:380:1.9cm);

\draw[thick, white] (-3.0,-3)--(-3.0,-2.8);
\draw[thick, white] (3.0,3)--(3.0,2.8);

\end{tikzpicture}
\begin{tikzpicture}[scale=0.5]
\coordinate (center) at (0,0) {};
\coordinate (v) at ($(center)+(90:2cm)$) {};
\coordinate (u) at ($(center)+(270:2cm)$) {};

\coordinate (lv) at ($(center)+(90:2.6cm)$) {};
\coordinate (lu) at ($(center)+(270:2.4cm)$) {};

\tikzstyle{every node}=[inner sep=1pt]
\begin{scriptsize}
\node at (lv) {$\psi(v)$};
\node at (lu) {$\psi(u)$};
\end{scriptsize}
\draw[thick] ([shift=(70:2.1cm)]0,0) arc (70:200:2.1cm);
\draw[thick] ([shift=(160:1.9cm)]0,0) arc (160:290:1.9cm);

\draw[thick, white] (-3.0,-3)--(-3.0,-2.8);
\draw[thick, white] (3.0,3)--(3.0,2.8);
\end{tikzpicture}

\caption{\label{fig:mutual_arc_position} From left to right:
$\psi(v)$ and $\psi(u)$ are disjoint, $\psi(v)$ contains $\psi(u)$, $\psi(v)$ is contained in $\psi(u)$,
$\psi(v)$ and $\psi(u)$ cover the circle, and $\psi(v)$ and $\psi(u)$ overlap.
}

\end{figure}
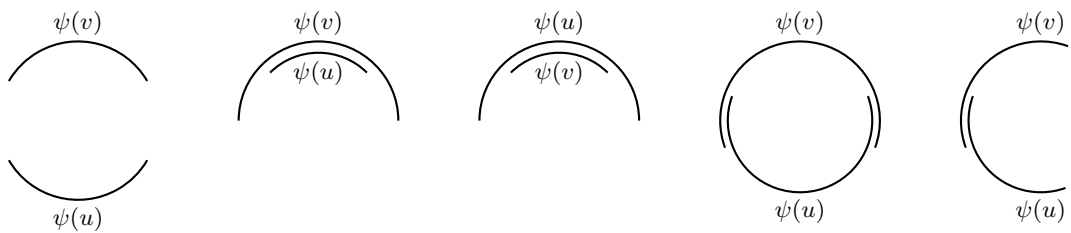

Let $G=(V,E)$ be a circular-arc graph and let $\psi$ be a circular-arc model of $G$ such that all the arcs from $\{\psi(v): v \in V\}$ have distinct endpoints.
We refer to Figure~\ref{fig:mutual_arc_position} which shows possible relations between two arcs from $\{\psi(v): v \in V\}$.

In so-called \emph{normalized models}, defined in \cite{Tucker80, Hsu95}, the relative relation between the arcs reflects the closed neighbourhood relation between the vertices of $G$, as follows.
\begin{definition}
\label{def:normalized_model}
Let $G = (V, E)$ be a circular-arc graph.
A circular-arc model $\psi$ of $G$ is \emph{normalized with respect to $G$} (shortly, \emph{normalized}) if all the arcs from $\{\psi(v): v \in V\}$
have distinct endpoints and for every pair $(v,u)$ of distinct vertices in $G$ the following conditions are satisfied:
\begin{enumerate}
 \item \label{item:v_u_disjoint} if $uv \notin E$, then $\psi(v)$ and $\psi(u)$ are disjoint,
 \item \label{item:v_contains_u} if $N_G[u] \subsetneq N_G[v]$, then $\psi(v)$ contains $\psi(u)$,
 \item \label{item:u_contains_v} if $N_G[v] \subsetneq N_G[u]$, then $\psi(v)$ is contained in $\psi(u)$,
 \item \label{item:v_u_cover_the_circle} if $N_G[v] \cup N_G[u] = V$, $N_G[w] \subsetneq N_G[v]$ for every $w \in N_G[v]\setminus N_G[u]$, and $N_G[w] \subsetneq N_G[u]$ for every $w \in N_G[u]\setminus N_G[v]$, then $\psi(v)$ and $\psi(u)$ cover the circle, 
 \item \label{item:v_u_overlap} If none of the above condition holds, then $\psi(v)$ and $\psi(u)$ overlap.
\end{enumerate}
Furthermore, for a pair $(v,u)$ of distinct vertices from $G$, we say that $v$ and $u$ are \emph{disjoint}, $v$~\emph{contains}~$u$, $v$~\emph{is contained} in~$u$, $v$ and $u$ \emph{cover the circle}, and $v$ and $u$ \emph{overlap} if the pair $(v,u)$ satisfies the assumption of statement
\eqref{item:v_u_disjoint}, \eqref{item:v_contains_u},
\eqref{item:u_contains_v}, \eqref{item:v_u_cover_the_circle}, and \eqref{item:v_u_overlap}, respectively.
\end{definition}
Note that the relation between vertices $u$ and $v$ is symmetric, with one exception:
$u$ is contained in $v$ if and only if $v$ contains $u$.
It is known that, if $G$ has no universal vertices and no twins, every circular-arc
model of $G$ can be turned into a normalized one by possibly extending some arcs of this
model~\cite{Tucker80, Hsu95}.

\section{Normalized models of circular-arc graphs and conformal models of their overlap graphs}
\label{sec:basic_tools}

In this section we introduce basic tools needed to describe 
the set of all normalized intersection models of a circular-arc graph.
As we already said, our approach follows the work of Hsu~\cite{Hsu95}.
Nevertheless, we want to emphasize here that the definitions we introduce below differ from those proposed by Hsu; 
see Section~\ref{sec:related_work} where we discuss how those differences influence our work.

\begin{figure}[htp!]
\begin{tikzpicture}[xscale=0.60,yscale=0.60,>=latex,shorten >=-0.4pt,shorten <=-0.4pt]

\coordinate (lphiv) at (1.5,0) {};
\coordinate (lpsiv) at (-1.5,0) {};

\coordinate (lpsiv) at ($(center)+(160:1.4cm)$) {};
\coordinate (lphiv) at ($(center)+(160:-0.25cm)$) {};

\tikzstyle{every node}=[inner sep=1pt]
\begin{scriptsize}
\node at (lphiv) {$\phi(v)$};
\node at (lpsiv) {$\psi(v)$};
\end{scriptsize}

\draw (0,0) circle (2cm);
\draw[very thick] ([shift=(80:2cm)]0,0) arc (80:240:2cm);
\draw[red,<-,very thick] ([shift=(80:2cm)]0,0)--([shift=(240:2cm)]0,0);
\end{tikzpicture} 
\caption{\label{fig:straightening_bending} The transformation of the arc~$\psi(v)$ into the oriented chord~$\phi(v)$.}
\end{figure}
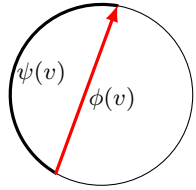

Throughout this section we assume $G=(V,E)$ is a circular-arc graph with no twins and no universal vertices.
\begin{definition}
\label{def:overlap-graph}
The \emph{overlap graph} $G_{ov} = (V,{\sim})$ of $G$ joins with an edge $\sim$ every two 
vertices which overlap in $G$.
\end{definition}
Given a normalized circular-arc model~$\psi$ of~$G$, we transform it into an \emph{oriented chord model} $\phi$ by converting every arc $\psi(v)$ for $v \in V$ into an oriented chord $\phi(v)$ such that the chord $\phi(v)$ has the same endpoints as $\psi(v)$ and $\phi(v)$ is oriented such that it has the arc $\psi(v)$ on its left side -- see Figure~\ref{fig:straightening_bending} for an illustration.
Clearly, for distinct $v,u \in V$, the oriented chords $\phi(v)$ and $\phi(u)$ intersect 
if and only if the arcs $\psi(v)$ and $\psi(u)$ overlap, and 
$\psi(v)$ and $\psi(u)$ overlap if and only if $v \sim u$.
Thus, $\phi$~is an \emph{oriented chord model} of~$G_{ov}$ and hence~$G_{ov}$ is a circle graph.

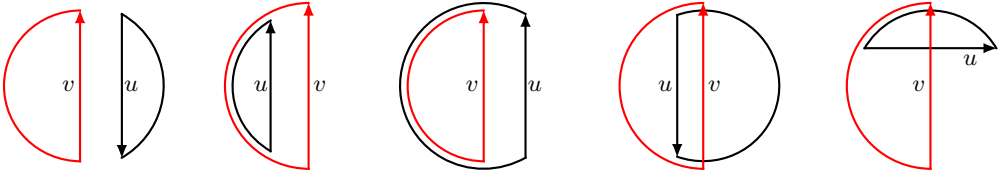
\begin{figure}[htp!]
\begin{tikzpicture}[scale=0.5,>=latex,shorten >=-0.4pt,shorten <=-0.4pt]
\coordinate (label) at (0,-3) {};

\coordinate (lv) at (-0.3,0) {};
\coordinate (lu) at (1.35,0) {};

\tikzstyle{every node}=[inner sep=1pt]
\begin{scriptsize}
\node at (lv) {$v$};
\node at (lu) {$u$};
\end{scriptsize}

\draw[thick] ([shift=(-60:2.2cm)]0,0) arc (-60:60:2.2cm);
\draw[thick,<-] ([shift=(-60:2.2cm)]0,0) -- ([shift=(60:2.2cm)]0,0);

\draw[red, thick] ([shift=(90:2.0cm)]0,0) arc (90:270:2.0cm);
\draw[red, thick,<-] ([shift=(90:2.0cm)]0,0) -- ([shift=(270:2.0cm)]0,0);

\draw[white] (-2.5,-2.5)--(-2.5,-2.3);
\draw[white] (2.5,2.5)--(2.5,2.3);
\end{tikzpicture}
\hspace{0.2cm}
\begin{tikzpicture}[scale=0.5,>=latex,shorten >=-0.4pt,shorten <=-0.4pt]
\coordinate (label) at (0,-3) {};

\coordinate (lv) at (0.3,0) {};
\coordinate (lu) at (-1.25,0) {};

\tikzstyle{every node}=[inner sep=1pt]
\begin{scriptsize}
\node at (lv) {$v$};
\node at (lu) {$u$};
\end{scriptsize}

\draw[thick] ([shift=(120:2.0cm)]0,0) arc (120:240:2.0cm);
\draw[thick,->] ([shift=(240:2.0cm)]0,0) -- ([shift=(120:2.0cm)]0,0);

\draw[red, thick] ([shift=(90:2.2cm)]0,0) arc (90:270:2.2cm);
\draw[red, thick,<-] ([shift=(90:2.2cm)]0,0) -- ([shift=(270:2.2cm)]0,0);

\draw[white] (-2.5,-2.5)--(-2.5,-2.3);
\draw[white] (1.1,2.5)--(1.1,2.3);
\end{tikzpicture}
\hspace{0.2cm}
\begin{tikzpicture}[scale=0.5,>=latex,shorten >=-0.4pt,shorten <=-0.4pt]
\coordinate (label) at (0,-3) {};

\coordinate (lv) at (-0.3,0) {};
\coordinate (lu) at (1.35,0) {};

\tikzstyle{every node}=[inner sep=1pt]
\begin{scriptsize}
\node at (lv) {$v$};
\node at (lu) {$u$};
\end{scriptsize}

\draw[thick] ([shift=(60:2.2cm)]0,0) arc (60:300:2.2cm);
\draw[thick,->] ([shift=(300:2.2cm)]0,0) -- ([shift=(60:2.2cm)]0,0);

\draw[red, thick] ([shift=(90:2.0cm)]0,0) arc (90:270:2.0cm);
\draw[red, thick,<-] ([shift=(90:2.0cm)]0,0) -- ([shift=(270:2.0cm)]0,0);

\draw[white] (-2.5,-2.5)--(-2.5,-2.3);
\draw[white] (2.3,2.5)--(2.3,2.3);
\end{tikzpicture}
\hspace{0.2cm}
\begin{tikzpicture}[scale=0.5,>=latex,shorten >=-0.4pt,shorten <=-0.4pt]
\coordinate (label) at (0,-3) {};

\coordinate (lv) at (0.3,0) {};
\coordinate (lu) at (-1.0,0) {};

\tikzstyle{every node}=[inner sep=1pt]
\begin{scriptsize}
\node at (lv) {$v$};
\node at (lu) {$u$};
\end{scriptsize}

\draw[thick] ([shift=(250:2.0cm)]0,0) arc (250:470:2.0cm);
\draw[thick,<-] ([shift=(250:2.0cm)]0,0) -- ([shift=(470:2.0cm)]0,0);

\draw[red, thick] ([shift=(90:2.2cm)]0,0) arc (90:270:2.2cm);
\draw[red, thick,<-] ([shift=(90:2.2cm)]0,0) -- ([shift=(270:2.2cm)]0,0);

\draw[white] (-2.5,-2.5)--(-2.5,-2.3);
\draw[white] (2.5,2.5)--(2.5,2.3);
\end{tikzpicture}
\hspace{0.2cm}
\begin{tikzpicture}[scale=0.5,>=latex,shorten >=-0.4pt,shorten <=-0.4pt]
\coordinate (label) at (0,-3) {};

\coordinate (lv) at (-0.3,0) {};
\coordinate (lu) at (1.05,0.7) {};

\tikzstyle{every node}=[inner sep=1pt]
\begin{scriptsize}
\node at (lv) {$v$};
\node at (lu) {$u$};
\end{scriptsize}

\draw[thick] ([shift=(30:2.0cm)]0,0) arc (30:150:2.0cm);
\draw[thick,<-] ([shift=(30:2.0cm)]0,0) -- ([shift=(150:2.0cm)]0,0);

\draw[red, thick] ([shift=(90:2.2cm)]0,0) arc (90:270:2.2cm);
\draw[red, thick,<-] ([shift=(90:2.2cm)]0,0) -- ([shift=(270:2.2cm)]0,0);

\draw[white] (-2.5,-2.5)--(-2.5,-2.3);
\draw[white] (2.5,2.5)--(2.5,2.3);
\end{tikzpicture}
\caption{\label{fig:uv_pairs} Relations between the arcs $\psi(v)$ and $\psi(u)$ and the corresponding oriented chords $\phi(v)$ and $\phi(u)$ for the cases: $v$ is disjoint with~$u$, $v$~contains~$u$, $v$~is contained in~$u$, $v$~and $u$~cover the circle, and $v$~and $u$~overlap, respectively.}
\end{figure}

Now, we note some properties of the oriented chord models of~$G_{ov}$ obtained from the normalized models of~$G$.
First, we associate with every vertex $v \in V$ two sets, $\leftside(v)$ and $\rightside(v)$, where:
$$
\begin{array}{rcl}
\leftside(v) &=& \{ u \in V: \quad
\text{$v$ contains $u$} \quad \text{or} \quad \text{$v$ and $u$ cover the circle}\}, 
\\
\rightside(v) &=& \{ u \in V:
\quad \text{$v$ and $u$ are disjoint}  \quad \text{or} \quad \text{$v$ is contained in $u$}\},
\end{array}
$$
and we assume the following definition.
\begin{definition}
\label{def:conformal_models}
An oriented chord model $\phi$ of $G_{ov}$ is \emph{conformal to $G$} (shortly, \emph{conformal}) if for every $v,u \in V$:
\begin{itemize}
\item $u \in \leftside(v)$ if and only if $\phi(u)$ lies on the left side of $\phi(v)$,
\item $u \in \rightside(v)$ if and only if $\phi(u)$ lies on the right side of $\phi(v)$.
\end{itemize}
\end{definition}
Clearly, if $\phi$ is an oriented chord model of $G_{ov}$ obtained from a normalized model $\psi$, 
then $\phi$ is conformal.
See Figure~\ref{fig:uv_pairs} for an illustration.

Now, consider a conformal model $\phi$ of $G_{ov}$ and consider the reverse operation
that transforms every oriented chord $\phi(v)$ for $v \in V$ into the arc $\psi(v)$ with the same endpoints as $\phi(v)$ and placed on the left side of $\phi(v)$.
We leave the reader to check that $\psi$ is a normalized circular-arc model of $G$.
Summing up, we have the following:
\begin{theorem}
\label{thm:normalized_conformal_correspondence}
Let $G$ be a circular-arc graph with no twins and no universal vertices.
There is a one-to-one correspondence between the normalized models of $G$ 
and the conformal models of $G_{ov}$.
\end{theorem}
Our goal is to characterize the oriented conformal models of $G_{ov}$ 
which are conformal.
For this purpose, we exploit the structure of all oriented chord models of $G_{ov}$
and, among them, we identify those that are conformal.
As a byproduct, we also describe the set of operations that allow to 
transform any conformal model into any other conformal model.

Due to the correspondence between the arcs and the oriented chords, we 
use the same notation for the oriented chords as for the arcs.
Note that, to mimic the reflection of the arcs, we need to reverse the orientation of the chords after mirroring them along the line~$L$ -- see Figure~\ref{fig:reflection} to the right.

\subsection{Normalized and conformal models of the induced subgraphs of $G$ and $G_{ov}$}
In the rest of the paper we will require an analogue of Theorem~\ref{thm:normalized_conformal_correspondence} extended on the induced subgraphs of $G=(V,E)$ and $G_{ov}=(V,{\sim})$.
\begin{definition}
Let $U$ be a non-empty subset of $V$.
A circular-arc model $\psi$ of the induced subgraph $(U,E)$ of $G$ is \emph{normalized with respect to $G$} (shortly, \emph{normalized}) if every pair of distinct vertices $(v,u)$ from $U$ satisfies Conditions~\ref{def:normalized_model}.\eqref{item:v_u_disjoint}-\eqref{item:v_u_overlap}.
\end{definition}
Note that the pair $(v,u)$ needs to satisfy conditions \ref{def:normalized_model}.\eqref{item:v_u_disjoint}-\eqref{item:v_u_overlap} with respect to the closed neighbourhoods in $G$ (not with respect to the closed neighbourhoods in $(U,E)$).
In particular, for any $U \subseteq V$, if $\psi$ is a normalized model of~$G$,
then $\psi$ restricted to $U$ is a normalized model of~$(U,E)$ (with respect to~$G$).

\begin{definition}
Let $U$ be a non-empty subset of~$V$.
An oriented chord model $\phi$ of the induced subgraph $(U,{\sim})$ of $G_{ov}$ is \emph{conformal to $G$} (shortly, \emph{conformal}) if for every $v \in U$ the oriented chords $\phi(u)$ for
$u \in \leftside(v) \cap U$ are on the left side of $\phi(v)$ 
and the oriented chords $\phi(u)$ for $u \in \rightside(v) \cap U$ 
are on the right side of~$\phi(v)$.
\end{definition}
Clearly, if $\phi$ is conformal for $(V,{\sim})$, then $\phi$ restricted to $U$ 
is conformal for $(U,{\sim})$.
\begin{theorem}
\label{thm:subgraphs_normalized_conformal_correspondence}
Let $U$ be a non-empty subset of~$V$.
There is a one-to-one correspondence between the normalized models of $(U,E)$ 
and the conformal models of $(U,{\sim})$.
\end{theorem}

\subsection{Permutation subgraphs of $G_{ov}$}
Let $\phi$ be a conformal model of $G_{ov}$ and let $U \subseteq V$.
We say $U$ \emph{induces a consistent (oriented) permutation model $(\mu',\mu'')$} in~$\phi$ if $\phi \equiv \mu' \tau' \mu'' \tau''$, where $(\mu',\mu'')$ is an (oriented) permutation model of $(U,{\sim})$ 
and $\tau'$, $\tau''$ are some (possibly empty) words over~$V^* \setminus U^*$.
In particular, if $U$ induces a consistent permutation model $(\mu',\mu'')$ of $(U,{\sim})$ in~$\phi$, 
$U$ induces also a consistent permutation model $(\mu'',\mu')$ in~$\phi$.

Assume that a set $U \subseteq V$ induces a consistent permutation model $(\mu',\mu'')$ in 
a conformal model $\phi$.
Let $U'$ and $U''$ be the letters occurring in the words $\mu'$ and $\mu''$; 
note that $\{U',U"\}$ is a partition of $U^*$ and both $U'$ and $U''$ are \emph{superscripted copies} of $U$, which means that for every $u \in U$ we have $|\{u^0,u^1\} \cap U'| = |\{u^0,u^1\} \cap U''|=1$.
Let $u \in U$.
We say the vertex~$u$ (the chord~$\phi(u)$) is \emph{oriented} from $\mu'$ to $\mu''$ or from $U'$ to $U''$ (from $\mu''$ to $\mu'$ or from $U''$ to $U'$) if $u^0 \in U'$ and 
$u^1 \in U''$ (if $u^0 \in U''$ and 
$u^1 \in U'$, respectively).

\section{Related work}
\label{sec:related_work}
In this section we describe the work that had an impact 
on the development of the method used to characterize the structure
of the normalized models of circular-arc graphs.

The modular decomposition trees introduced by Gallai~\cite{Gal67} have turned out to be useful to represent the structure of intersection models of certain geometric intersection graphs, especially those related to partial orders.
We have seen that the modular decomposition trees represent the intersection models of permutation graphs.
Similarly, modular decomposition trees can be used to represent intersection models of interval graphs; a~reader familiar with PQ-trees can check that the PQ-tree of an interval graph~$G$ can be obtained from the modular decomposition tree of~$G$.
In particular, Theorems~\ref{thm:prime_graph_transitive_orientation} and \ref{thm:transitive_orientations_versus_transitive_orientations_of_strong_modules} by Gallai have the following impact on the structure of the intersection models of a permutation/interval graph $G=(V,{\sim})$:
\begin{description}
 \item[\namedlabel{prop:M1}{(M1)}] If $(U,{\sim})$ is a prime induced subgraph of $(V,{\sim})$, then $(U,{\sim})$ has a unique intersection model (up to certain normalizations and reflections).
 \item[\namedlabel{prop:M2}{(M2)}] The structure of the intersection models of $(V,{\sim})$ is represented by the modular decomposition tree of $(V,{\sim})$.
\end{description}

As for circular-arc graphs, the work \cite{Spin88} of Spinrad from 1988 is the first that allows to describe, 
in the way given above, the structure of the normalized models of certain graphs in this class, namely those that
are \emph{co-bipartite} (whose vertex set can be partitioned into two cliques).

\begin{figure}[htp!]
\begin{tikzpicture}[scale=0.5,>=latex,shorten >=-0.4pt,shorten <=-0.4pt]

\coordinate (center) at (0,0) {};
\coordinate (A) at ($(center)+(180:2cm)$) {};
\coordinate (B) at ($(center)+(0:2.6cm)$) {};

\coordinate (lA) at ($(center)+(180:2.4cm)$) {};
\coordinate (lB) at ($(center)+(0:2.6cm)$) {};

\draw[very thick] ([shift=(110:1.9cm)]0,0) arc (110:250:1.9cm);
\draw[dashed] ([shift=(110:1.9cm)]0,0) -- ([shift=(250:1.9cm)]0,0);
\draw[red,very thick] ([shift=(70:2.1cm)]0,0) arc (70:290:2.1cm);
\draw[red, dashed] ([shift=(70:2.1cm)]0,0) -- ([shift=(290:2.1cm)]0,0);

\draw ([shift=(180:1.6cm)]0,0) -- ([shift=(180:2.2cm)]0,0);
\draw ([shift=(0:1.6cm)]0,0) -- ([shift=(0:2.2cm)]0,0);

\coordinate (lu) at ($(center)+(125:1.6cm)$) {};
\coordinate (lv) at ($(center)+(80:1.8cm)$) {};

\tikzstyle{every node}=[inner sep=1pt]
\begin{scriptsize}
\node at (lA) {$A$};
\node at (lB) {$B$};
\node at (lu) {$u$};
\node at (lv) {$v$};
\end{scriptsize}

\draw[white] (-3,-2.5)--(-3,2.5);
\draw[white] (3,-2.5)--(3,2.5);
\end{tikzpicture}
\hspace{0.4cm}
\begin{tikzpicture}[scale=0.5,>=latex,shorten >=-0.4pt,shorten <=-0.4pt]

\coordinate (center) at (0,0) {};
\coordinate (A) at ($(center)+(180:2cm)$) {};
\coordinate (B) at ($(center)+(0:2.6cm)$) {};

\coordinate (lA) at ($(center)+(180:2.4cm)$) {};
\coordinate (lB) at ($(center)+(0:2.6cm)$) {};

\draw[very thick] ([shift=(250:2.1cm)]0,0) arc (250:470:2.1cm);
\draw[dashed] ([shift=(250:2.1cm)]0,0) -- ([shift=(470:2.1cm)]0,0);
\draw[red,very thick] ([shift=(290:1.9cm)]0,0) arc (290:430:1.9cm);
\draw[red, dashed] ([shift=(290:1.9cm)]0,0) -- ([shift=(430:1.9cm)]0,0);

\draw ([shift=(180:1.6cm)]0,0) -- ([shift=(180:2.2cm)]0,0);
\draw ([shift=(0:1.6cm)]0,0) -- ([shift=(0:2.2cm)]0,0);

\coordinate (lu) at ($(center)+(105:1.8cm)$) {};
\coordinate (lv) at ($(center)+(55:1.6cm)$) {};

\tikzstyle{every node}=[inner sep=1pt]
\begin{scriptsize}
\node at (lA) {$A$};
\node at (lB) {$B$};
\node at (lu) {$u$};
\node at (lv) {$v$};
\end{scriptsize}

\draw[white] (-3,-2.5)--(-3,2.5);
\draw[white] (3,-2.5)--(3,2.5);
\end{tikzpicture}
\hspace{0.4cm}
\begin{tikzpicture}[scale=0.5,>=latex,shorten >=-0.4pt,shorten <=-0.4pt]

\coordinate (center) at (0,0) {};
\coordinate (A) at ($(center)+(180:2cm)$) {};
\coordinate (B) at ($(center)+(0:2.5cm)$) {};

\coordinate (lA) at ($(center)+(180:2.5cm)$) {};
\coordinate (lB) at ($(center)+(0:2.6cm)$) {};

\draw[very thick] ([shift=(110:2cm)]0,0) arc (110:250:2cm);
\draw[dashed] ([shift=(110:2cm)]0,0) -- ([shift=(250:2cm)]0,0);
\draw[red,very thick] ([shift=(290:2cm)]0,0) arc (290:430:2cm);
\draw[red, dashed] ([shift=(290:2cm)]0,0) -- ([shift=(430:2cm)]0,0);

\draw ([shift=(180:1.8cm)]0,0) -- ([shift=(180:2.2cm)]0,0);
\draw ([shift=(0:1.8cm)]0,0) -- ([shift=(0:2.2cm)]0,0);

\coordinate (lu) at ($(center)+(125:1.7cm)$) {};
\coordinate (lv) at ($(center)+(55:1.7cm)$) {};

\tikzstyle{every node}=[inner sep=1pt]
\begin{scriptsize}
\node at (lA) {$A$};
\node at (lB) {$B$};
\node at (lu) {$u$};
\node at (lv) {$v$};
\end{scriptsize}

\draw[white] (-3,-2.5)--(-3,2.5);
\draw[white] (3,-2.5)--(3,2.5);
\end{tikzpicture}
\hspace{0.4cm}
\begin{tikzpicture}[scale=0.5,>=latex,shorten >=-0.4pt,shorten <=-0.4pt]

\coordinate (center) at (0,0) {};
\coordinate (A) at ($(center)+(180:2cm)$) {};
\coordinate (B) at ($(center)+(0:2.5cm)$) {};

\coordinate (lA) at ($(center)+(180:2.5cm)$) {};
\coordinate (lB) at ($(center)+(0:2.6cm)$) {};

\draw[very thick] ([shift=(250:1.9cm)]0,0) arc (250:470:1.9cm);
\draw[dashed] ([shift=(250:1.9cm)]0,0) -- ([shift=(470:1.9cm)]0,0);

\draw[red,very thick] ([shift=(70:2.1cm)]0,0) arc (70:290:2.1cm);
\draw[red, dashed] ([shift=(70:2.1cm)]0,0) -- ([shift=(290:2.1cm)]0,0);

\draw ([shift=(180:1.8cm)]0,0) -- ([shift=(180:2.2cm)]0,0);
\draw ([shift=(0:1.8cm)]0,0) -- ([shift=(0:2.2cm)]0,0);

\coordinate (lu) at ($(center)+(100:1.6cm)$) {};
\coordinate (lv) at ($(center)+(80:2.4cm)$) {};

\tikzstyle{every node}=[inner sep=1pt]
\begin{scriptsize}
\node at (lA) {$A$};
\node at (lB) {$B$};
\node at (lu) {$u$};
\node at (lv) {$v$};
\end{scriptsize}

\draw[white] (-3,-2.5)--(-3,2.5);
\draw[white] (3,-2.5)--(3,2.5);
\end{tikzpicture}
\caption{\label{fig:Spinrad_relation} Spinrad sets $u < v$ if and only if either $u,v \in C_A$ and $u$ is contained in $v$ (left) or $u,v \in C_B$ and $u$ contains $v$ (middle left) or $u \in C_A$, $v \in C_B$, and $u,v$ are disjoint, or $u \in C_B$, $v \in C_A$, and $u,v$ cover the circle.}
\end{figure}
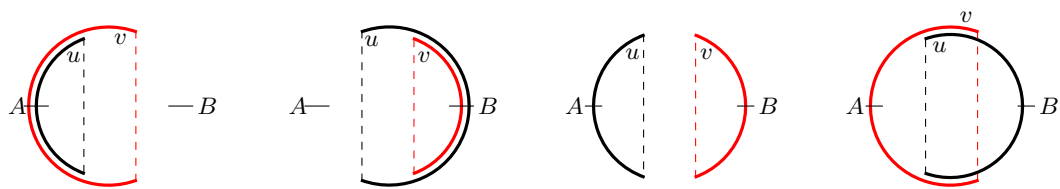

Let us briefly discuss the ideas of Spinrad.
Let $G=(V,E)$ be a \emph{co-bipartite circular-arc} graph whose vertex set can be partitioned into two cliques, say $C_A$ and $C_B$.
We assume~$G$ has no universal vertices and no twins.
Let $G_{ov} = (V,{\sim})$ be the overlap graph of $G$ and let $(V,\parallel)$ 
be the complement of $(V,{\sim})$.
First, Spinrad proved that~$G$ has a circular-arc model $\psi$ such that: 
\begin{equation}
\label{eq:Spinrad_models_A_B}
\begin{array}{c}
\text{all the arcs from $\{\psi(x): x \in C_A\}$ pass through a point $A$ of the circle, and }\\
\text{all the arcs from $\{\psi(x): x \in C_B\}$ pass through a point $B$ of the circle.}
\end{array}
\end{equation}
We assume that $A$ is the leftmost point and $B$ is the rightmost point of the circle --
see Figure~\ref{fig:Spinrad_relation} for an illustration.
Note that no arc from $\{\psi(v): v \in V\}$ contains both the points $A$ and $B$ as $G$ has no universal vertices.
Hence, each arc from $\{\psi(v): v \in V\}$ has one endpoint on the upper and one endpoint on the lower part of the circle.
A~normalized model of $G$ is said to be \emph{strongly normalized} if it additionally satisfies Property~\eqref{eq:Spinrad_models_A_B}.
Given a strongly normalized model~$\psi$ of $G$, let $\tau^0_\psi$ and  $\tau^1_\psi$ be the permutations of $V$ that encode the left-to-right (right-to-left) order of the endpoints of the arcs of $\psi$ on the upper (the lower, respectively) part of the circle.
Observe that $(\tau^0_\psi, \tau^1_\psi)$ is a permutation model of the overlap graph~$(V,{\sim})$ of~$G$.
Let $(V,{<_{\psi}})$ and $(V,{\prec_{\psi}})$ be the transitive orientations of $(V,{\parallel})$ and 
$(V,{\sim})$ that correspond to the permutation model $(\tau^0_\psi, \tau^1_\psi)$ in the way given by Theorem~\ref{thm:permutation_models_transitive_orientations}.
Spinrad observed that the transitive orientation ${<_{\psi}}$ of $(V,\parallel)$ is the same for every strongly normalized model~$\psi$ of~$G$. 
Hence, we can denote it by $(V,{<})$.
In fact, Spinrad showed that $(V,{<})$ can be defined in a purely combinatorial way;
the orientation ${<}$ of $u \parallel v$ depends only on whether $u$ and $v$ belong to $C_A$ or to $C_B$ and on the relation between the closed neighbourhoods of $u$ and $v$ in the graph~$G$ -- see Figure~\ref{fig:Spinrad_relation}.

By the observation of Spinrad, the strongly normalized models of~$G$ are in one-to-one correspondence with the permutation models of~$(V,{\sim})$ in which the transitive orientation of $(V,{\parallel})$ induced by the left-to-right order between the segments of $\phi$ is equal to $(V,{<})$. 
Let us call such permutation models of $(V,{\sim})$ as \emph{conformal} to $(V,{<})$.
Due to Theorem~\ref{thm:permutation_models_transitive_orientations}, permutation models conformal to $(V,{<})$ are in the correspondence with the transitive orientations of $(V,{\sim})$.
Hence, Theorems~\ref{thm:prime_graph_transitive_orientation} and \ref{thm:transitive_orientations_versus_transitive_orientations_of_strong_modules} by Gallai assert that:
\begin{description}
 \item[\namedlabel{prop:S1}{(S1)}] if $(U,{\sim})$ is a prime subgraph of $(V,{\sim})$, 
 then $(U,{\sim})$ admits a unique (up to reflection) permutation model conformal to $(U,{<})$,
 \item[\namedlabel{prop:S2}{(S2)}] the set of all permutation models of $(V,{\sim})$ conformal to $(V,{<})$ is represented by the modular decomposition of $(V,{\sim})$.
\end{description}

In the work~\cite{Hsu95} from 1995 Hsu tries to use Gallai's framework to describe the set of all normalized models of any circular-arc graph.
Given a circular-arc graph~$G=(V,E)$, Hsu introduces the overlap graph $G_{ov}=(V,{\sim})$
and transforms normalized models of $G$ into \textbf{non-oriented} chord models of $G_{ov}$ by 
mapping the arcs into the chords with the same endpoints.
The main difference between our and Hsu's approach lies in the definition of conformal models of~$G_{ov}$.
In fact, Hsu assumes the following definition: a (non-oriented) \emph{chord model $\phi$ of~$G_{ov}$ is conformal 
if for every vertex $v$ of $G$ the chords associated with vertices in $\leftside(v)$ are on one side of $\phi(v)$
and those associated with vertices in $\rightside(v)$ are on the other side of $\phi(v)$} (Section~5.2 in~\cite{Hsu95}).
Then, Hsu tries to describe the structure of all conformal models of $G_{ov}=(V,{\sim})$ by proving it admits the properties analogous to~\ref{prop:S1} and \ref{prop:S2}:
\begin{description}
 \item[\namedlabel{prop:H1}{(H1)}] If $(U,{\sim})$ is a prime induced subgraph of $(V,{\sim})$, then $(U,{\sim})$ has a unique conformal model (up to reflection).
 \item[\namedlabel{prop:H2}{(H2)}] The structure of the conformal models of $(V,{\sim})$ can be described by the modular decomposition tree of $(V,{\sim})$.
\end{description}
Unfortunately, as we have shown in the companion paper~\cite{Kra24appendix}, both of these steps are not accomplished correctly by Hsu~\cite{Hsu95}.

We want to emphesize that Property~\ref{prop:H1} is crucial for the method as
the structure of the conformal models is ``build'' upon the unique conformal models of prime subgraphs of $G_{ov}$
(in fact, its role is similar to the role played by the fact that every prime graph has a unique (up to reversal) transitive orientation in the description of the structure of all transitive orientations of a comparability graph).
Note that $G_{ov}$ is the circle graph, and hence Property~\ref{prop:H1} does not follow by Theorem~\ref{thm:prime_graph_transitive_orientation}.
In~\cite{Kra24appendix} we showed that the ``proof'' of~\ref{prop:H1} proposed by Hsu (Theorem 5.7 in \cite{Hsu95}) is not correct (it is based on two claims, and both of them are not correct).
In this work we show~\ref{prop:H1} using a different approach.

To accomplish~\ref{prop:H2} Hsu divides the description of the structure of the conformal models of $(V,{\sim})$ into three parts, corresponding to the cases when $V$ is serial, prime, and parallel in $\strongModules(V,{\sim})$.
Hsu's work~\cite{Hsu95} correctly deals with the case where $V$ is serial (in this case, any component of $(V,{\parallel})$ induces a co-bipartite circular-arc graph), but it is not correct when $V$ is prime, and 
is incomplete (and also incorrect) when $V$ is parallel.

When $V$ is prime, Hsu first ``proves'' Property~\ref{prop:H1}.
Next, Hsu tries to partition the set $V$ into so-called ``consistent modules''\footnote{In our work we use the name
``CA-modules'' instead of ``consistent modules''.}, which were supposed to satisfy the following properties:
\begin{description}
 \item[\namedlabel{prop:H3}{(H3)}] For every conformal model $\phi$ and every consistent module $M$ the 
set $M$ induces a consistent permutation model $(\mu'_M, \mu''_M)$ of $(M,{\sim})$ in $\phi$;
moreover, there is a unique circular order (up to reflection) in which the words $\mu'_M$ and $\mu''_M$  associated with the consistent modules $M$ may occur in $\phi$. 
\end{description}
In the case when $G$ is co-bipartite, $G_c$ is a permutation graph, and Observation~\ref{obs:strong_modules_are_contiguous_in_permutation_models} and Property~\ref{prop:S1} applied to a set containing a representative vertex from each child $M$ of $V$ 
assert that, at least for this case, we can take the children of $V$ in $\strongModules(V,{\sim})$ 
as the consistent modules.
Hsu rightly noted that this is not the case in all circular-arc graphs.
In fact, Hsu ``proves'' that if a child of $V$ in $\strongModules(V,{\sim})$ does not induce a consistent permutation model in some conformal model $\phi$, then it must be serial (Lemma~6.3 in~\cite{Hsu95}). 
In~\cite{Kra24appendix} we gave a counterexample to this statement, that is, we have shown that parallel children of $V$ might also induce not consistent permutation models in some conformal models of $(V,{\sim})$.
Hsu properly ``refines'' serial children of $V$ into consistent modules and his ideas are used in our paper.
However, due to the error in Lemma~6.3, Hsu's work leaves the cases of prime and parallel children of $V$ unsolved.
Finally, since the consistent modules of $G$ are not properly determined in Hsu's work, 
the theorems from~\cite{Hsu95} ``proving'' Property~\ref{prop:H3} are also not correct -- see~\cite{Kra24appendix} for more details.
Nevertheless, we want to emphasize that the work of Hsu helps to understand the structure of the conformal models of $G_{ov}$ in the case where $V$ is prime; in fact, if we refine prime and parallel children of $V$ accordingly, 
we can prove that the resulted consistent modules satisfy Property~\ref{prop:H3}.

The description of Hsu is incomplete (and also incorrect) in the case when $V$ is parallel.
First of all, the consistent modules of the components of $G_{ov}$ are used to construct the decomposition tree of $G_{ov}$ (which maintains a track of all conformal models of $G_{ov}$). 
However, the consistent modules for the prime components of $G_{ov}$ are not correctly determined (see the errors mentioned above) and the consistent modules for the serial components of $G_{ov}$ are not defined
(we were not able to find their definition in~\cite{Hsu95}).
The main effort of this part of our work is to grasp the role of the ``reflection'' and to understand the differences/similarities between prime and serial children
of $V$ in $\strongModules(V,{\sim})$.
We show that, despite significant differences in the description of
the conformal models of prime and serial children of $V$, 
their role in representing the conformal models of $(V,{\sim})$ is the same.
These important issues, however, are not thoroughly investigated in~\cite{Hsu95}.

Although we report in~\cite{Kra24appendix} a number of errors in Hsu's work, the Hsu's work~\cite{Hsu95} had a great influence on our work, where many of Hsu's ideas, appropriately extended and adopted, have been used.
This includes:
\begin{itemize}
 \item the description of the conformal models in the case when $V$ is serial in $\strongModules(V,{\sim})$,
 \item the refinement of serial children of $V$ in $\strongModules(V,{\sim})$ into the consistent modules in the case when $V$ is prime,
 \item the definition of the \emph{series-parallel} tree $T_{NM}$ in the case when $V$ is parallel,
 \item the description of the role of \emph{maximal NS-collections} (N-nodes) in $T_{NM}$-tree,
 \item the notion of the \emph{conformal models for the c-inseparable components} of $G_{ov}$, which inspired the definition of the extended conformal models for the components of $G_{ov}$.
\end{itemize}

As we said, the main difference between our and Hsu's approach lies in 
the definition of the conformal models.
In Hsu's approach, the chords are non-oriented, which causes 
considerable difficulties in his proofs.

Firstly, Hsu divides all the triples $(u,v,w)$ consisting of pairwise non-adjacent vertices in $(V,{\sim})$ into two categories:
\begin{itemize}
 \item $(u,v,w)$ is \emph{in parallel}, written $u|v|w$, if the vertex $v$ has the vertices $u$ and $w$ on its different sides (that is, either $u \in \leftside(v)$ and $w \in \rightside(v)$ or $w \in \leftside(v)$ and $u \in \rightside(v)$),
 \item $(u,v,w)$ is \emph{in series}, written $u-v-w$, if any vertex from $\{u,v,w\}$ has the remaining two vertices on the same side,
\end{itemize}
(see Section 5.1 of \cite{Hsu95}).
Then, Hsu is searching for chord models $\phi$ of $G_{ov}$ that satisfy 
the conditions:
\begin{itemize}
 \item if $u|v|w$ then the chord $\phi(v)$ has $\phi(u)$ and $\phi(w)$ on its different sides,
 \item if $u-v-w$ then every chord from $\{\phi(u), \phi(v), \phi(w)\}$ has the remaining two chords on the same side.
\end{itemize}
Hsu showed that such chord models correspond to conformal models (see Section 5.2 in \cite{Hsu95}). Consequently, he aims to describe the set of transformations between the circle models of $G_{ov}$ that preserve the relationships among the triples of non-intersecting chords. 
By orienting the chords in the conformal models, we can concentrate solely on pairs of non-intersecting chords. 
Specifically, we seek transformations that maintain the relative relationships between every pair of non-intersecting oriented chords. 
While this difference may seem minor, it significantly simplifies the task of identifying the set of operations that can transform one conformal model into another.

Secondly, if we reflect two intersecting non-oriented chords $a$ and $b$, represented by the circular word $abab$, the result is still two intersecting chords represented by the same word $abab$. 
Therefore, reflection has no effect on two intersecting non-oriented chords; additional chords are needed to observe any change.
In contrast, when reflecting two intersecting oriented chords $a$ and $b$, represented by the word $ a^0b^0a^1b^1$, we obtain intersecting oriented chords represented by the non-equivalent circular word $a^0b^1a^1b^0$. 
This means that reflection alters the orientation of the chords: before reflection, the head of $b$ is to the right of $a$, but after reflection, it shifts to the left side of $a$.
Understanding the role of reflection -- specifically, which components of the conformal models can be reflected independently of others -- is crucial for identifying all conformal models of $G_{ov}$. 
When dealing with oriented chords, the effects of reflection can be easily articulated.



\section{Conformal models of~$G_{ov}$}
\label{sec:data_structure}

Let $G = (V,E)$ be a circular-arc graph with no twins and no universal vertices,
$G_{ov}=(V,{\sim})$ be the overlap graph of~$G$, $\overline{G_{ov}} = (V,{\parallel})$ be the complement of $G_{ov}$, and $\strongModules(G_{ov})$ be the modular decomposition tree of $G_{ov}$.

In this section we describe the structure of all conformal models of the graph $G_{ov}$.
Formally, the set of such models is described by the quadruple $\dataStructure = (\camodules, \slots, \metachords, \Pi)$, 
whose subsequent components denote, respectively, the sets of \emph{CA-modules} $\camodules$, \emph{slots} $\slots$, \emph{metachords} $\metachords$, and \emph{circular orders of the slots} $\Pi$, of the graph~$G$.
The basic properties of these components and the role they play is described in Subsection~\ref{subsec:data_structure_components}. 

As we will see, the components of $\dataStructure$ represent the conformal models of $G_{ov}$ in a~quite simple way. 
However, some components of $\dataStructure$ may have super-linear size, and we need to provide a method to represent them within linear space.
For instance, the set $\Pi$ might have exponentially many members, however, it is represented by a \emph{PQS-tree} $\pqstree$ of $G$, which has a linear size in relation to $G$.
Finally, we show that all the components of $\dataStructure$ can be represented in linear space using a \emph{PQSM-tree} $\pqsmtree$ of $G$, which is a simple extension of the PQS-tree $\pqstree$.

Eventually, in this section we show how to construct all the aforementioned components from a given conformal model of $G_{ov}$. 
In Section~\ref{sec:proof_sketch} we will define these components in a purely combinatorial way and show that they satisfy the properties outlined in this section.

\subsection{The components of $\dataStructure$}
\label{subsec:data_structure_components}

The first component of $\dataStructure$ is the set $\camodules$ of \emph{CA-modules} of~$G$. 
Each CA-module of $G$ is a subset of $M$, where $M$ is a child of $V$ in $\strongModules(G_{ov})$, and all CA-modules of $G$ partition the set $V$. 
The CA-modules satisfy the following property with respect to the conformal models of $G_{ov}$:
\begin{description}
 \item[\namedlabel{prop:CA-modules}{(P1)}] For every CA-module $S$ and every conformal model $\phi$ of $G_{ov}$ the set $S$ induces a consistent permutation model in $\phi$.
 Moreover, in all such models (induced by $S$) the relative orientations of the chords representing the vertices from $S$ is the same.
\end{description}
Assume that $\camodules = \{S_1,\ldots,S_t\}$ is the set of all CA-modules of $G$.
We pick a vertex $s_i$ in every set $S_i$, called the \emph{representant} of $S_i$.

\begin{figure}[h]
\begin{tikzpicture}[scale=0.8,>=latex,shorten >=-0.2pt,shorten <=-0.2pt]
\coordinate (center) at (0.0,0.0) {};

\coordinate (A0) at ($(center)+(240:2cm)$) {};
\coordinate (A1) at ($(center)+(100:2cm)$) {};

\coordinate (B0) at ($(center)+(270:2cm)$) {};
\coordinate (B1) at ($(center)+(90:2cm)$) {};

\coordinate (C0) at ($(center)+(80:2cm)$) {};
\coordinate (C1) at ($(center)+(280:2cm)$) {};

\coordinate (D0) at ($(center)+(250:2cm)$) {};
\coordinate (D1) at ($(center)+(70:2cm)$) {};

\coordinate (E0) at ($(center)+(60:2cm)$) {};
\coordinate (E1) at ($(center)+(260:2cm)$) {};

\coordinate (F0) at ($(center)+(125:2cm)$) {};
\coordinate (F1) at ($(center)+(30:2cm)$) {};

\coordinate (G0) at ($(center)+(215:2cm)$) {};
\coordinate (G1) at ($(center)+(140:2cm)$) {};

\coordinate (H0) at ($(center)+(165:2cm)$) {};
\coordinate (H1) at ($(center)+(350:2cm)$) {};

\coordinate (I0) at ($(center)+(335:2cm)$) {};
\coordinate (I1) at ($(center)+(180:2cm)$) {};

\coordinate (lA0) at ($(center)+(240:2.25cm)$) {};
\coordinate (lA1) at ($(center)+(100:2.25cm)$) {};

\coordinate (lB0) at ($(center)+(270:2.25cm)$) {};
\coordinate (lB1) at ($(center)+(90:2.25cm)$) {};

\coordinate (lC0) at ($(center)+(80:2.25cm)$) {};
\coordinate (lC1) at ($(center)+(280:2.25cm)$) {};

\coordinate (lD0) at ($(center)+(250:2.25cm)$) {};
\coordinate (lD1) at ($(center)+(70:2.25cm)$) {};

\coordinate (lE0) at ($(center)+(60:2.25cm)$) {};
\coordinate (lE1) at ($(center)+(260:2.25cm)$) {};

\coordinate (lF0) at ($(center)+(125:2.25cm)$) {};
\coordinate (lF1) at ($(center)+(30:2.3cm)$) {};

\coordinate (lG0) at ($(center)+(215:2.25cm)$) {};
\coordinate (lG1) at ($(center)+(140:2.25cm)$) {};

\coordinate (lH0) at ($(center)+(165:2.25cm)$) {};
\coordinate (lH1) at ($(center)+(350:2.25cm)$) {};

\coordinate (lI0) at ($(center)+(335:2.25cm)$) {};
\coordinate (lI1) at ($(center)+(180:2.25cm)$) {};

\coordinate (LA0) at ($(center)+(240:3cm)$) {};
\coordinate (LA1) at ($(center)+(100:3cm)$) {};

\coordinate (LB0) at ($(center)+(270:3cm)$) {};
\coordinate (LB1) at ($(center)+(90:3cm)$) {};

\coordinate (LC0) at ($(center)+(80:3cm)$) {};
\coordinate (LC1) at ($(center)+(280:3cm)$) {};

\coordinate (LD0) at ($(center)+(250:3cm)$) {};
\coordinate (LD1) at ($(center)+(70:3cm)$) {};

\coordinate (LE0) at ($(center)+(60:3cm)$) {};
\coordinate (LE1) at ($(center)+(260:3cm)$) {};

\coordinate (LF0) at ($(center)+(125:3cm)$) {};
\coordinate (LF1) at ($(center)+(30:3cm)$) {};

\coordinate (LG0) at ($(center)+(215:3cm)$) {};
\coordinate (LG1) at ($(center)+(140:3cm)$) {};

\coordinate (LH0) at ($(center)+(165:3cm)$) {};
\coordinate (LH1) at ($(center)+(350:3cm)$) {};

\coordinate (LI0) at ($(center)+(335:3cm)$) {};
\coordinate (LI1) at ($(center)+(180:3cm)$) {};

\coordinate (LHI0) at ($(center)+(172.5:3cm)$) {};
\coordinate (LHI1) at ($(center)+(342.5:3cm)$) {};

\coordinate (L) at (3,-3) {};

\draw (center) circle (2cm);
\draw[dashed] (center) circle (2.7cm);

\draw[->] (A0)--(A1);
\draw[->,red, thick] (B0)--(B1);
\draw[->] (C0)--(C1);
\draw[->] (D0)--(D1);
\draw[->] (E0)--(E1);

\draw[->,thick, red] (F0)--(F1);

\draw[->,thick, red] (G0)--(G1);

\draw[->, thick, red] (H0)--(H1);

\draw[->] (I0)--(I1);

\draw[very thick, -] ([shift=(105:2.7cm)]0,0) arc (105:50:2.7cm);
\draw[very thick, -] ([shift=(230:2.7cm)]0,0) arc (230:285:2.7cm);

\draw[very thick, -] ([shift=(120:2.7cm)]0,0) arc (120:130:2.7cm);
\draw[very thick, -] ([shift=(25:2.7cm)]0,0) arc (25:35:2.7cm);

\draw[very thick, -] ([shift=(135:2.7cm)]0,0) arc (135:145:2.7cm);
\draw[very thick, -] ([shift=(210:2.7cm)]0,0) arc (210:220:2.7cm);

\draw[very thick, -] ([shift=(160:2.7cm)]0,0) arc (160:185:2.7cm);
\draw[very thick, -] ([shift=(330:2.7cm)]0,0) arc (330:355:2.7cm);

\draw[very thick, -] ([shift=(105:2.7cm)]0,0) arc (105:55:2.7cm);
\draw[very thick, -] ([shift=(235:2.7cm)]0,0) arc (235:285:2.7cm);

\draw[very thick, -] ([shift=(120:2.7cm)]0,0) arc (120:130:2.7cm);
\draw[very thick, -] ([shift=(25:2.7cm)]0,0) arc (25:35:2cm);

\draw[very thick, -] ([shift=(135:2.7cm)]0,0) arc (135:145:2.7cm);
\draw[very thick, -] ([shift=(210:2.7cm)]0,0) arc (210:220:2.7cm);

\draw[very thick, -] ([shift=(160:2.7cm)]0,0) arc (160:185:2.7cm);
\draw[very thick, -] ([shift=(330:2.7cm)]0,0) arc (330:355:2.7cm);

\tikzstyle{every node}=[inner sep=1pt]
\begin{scriptsize}
\node at (lA0) {$a^0$};
\node at (lA1) {$a^1$};
\node at (lB0) {$b^0$};
\node at (lB1) {$b^1$};
\node at (lC0) {$c^0$};
\node at (lC1) {$c^1$};
\node at (lD0) {$d^0$};
\node at (lD1) {$d^1$};
\node at (lE0) {$e^0$};
\node at (lE1) {$e^1$};
\node at (lF0) {$f^0$};
\node at (lF1) {$f^1$};
\node at (lG0) {$g^0$};
\node at (lG1) {$g^1$};
\node at (lH0) {$h^0$};
\node at (lH1) {$h^1$};
\node at (lI0) {$i^0$};
\node at (lI1) {$i^1$};
\end{scriptsize}

\tikzstyle{every node}=[inner sep=1pt]
\begin{tiny}
\node at (LC0) {$S^1_1$};
\node at (LE1) {$S^0_1$};
\node at (LF0) {$S^0_2$};
\node at (LF1) {$S^1_2$};
\node at (LG0) {$S^0_3$};
\node at (LG1) {$S^1_3$};
\node at (LHI0) {$S^0_4$};
\node at (LHI1) {$S^1_4$};
\end{tiny}
\node at (L) {$\phi$};

\draw[white] (-4,-3)--(-4,-2.5);
\draw[white] (4,3)--(4,2.5);
\end{tikzpicture}
\begin{tikzpicture}[scale=0.8,>=latex,shorten >=-0.2pt,shorten <=-0.2pt]
\coordinate (center) at (0.0,0.0) {};

\coordinate (C0) at ($(center)+(80:2.5cm)$) {};
\coordinate (C1) at ($(center)+(280:2.5cm)$) {};

\coordinate (F0) at ($(center)+(125:2.5cm)$) {};
\coordinate (F1) at ($(center)+(30:2.5cm)$) {};

\coordinate (G0) at ($(center)+(215:2.5cm)$) {};
\coordinate (G1) at ($(center)+(140:2.5cm)$) {};

\coordinate (HI0) at ($(center)+(172.5:2.5cm)$) {};
\coordinate (HI1) at ($(center)+(342.5:2.5cm)$) {};

\coordinate (LC0) at ($(center)+(80:2.8cm)$) {};
\coordinate (LC1) at ($(center)+(280:2.8cm)$) {};

\coordinate (LF0) at ($(center)+(125:2.8cm)$) {};
\coordinate (LF1) at ($(center)+(30:2.8cm)$) {};

\coordinate (LG0) at ($(center)+(215:2.8cm)$) {};
\coordinate (LG1) at ($(center)+(140:2.8cm)$) {};

\coordinate (LHI0) at ($(center)+(172.5:2.8cm)$) {};
\coordinate (LHI1) at ($(center)+(342.5:2.8cm)$) {};

\coordinate (L) at (3,-3) {};

\draw (center) circle (2.5cm);

\draw[->,red,very thick] (C1)--(C0);

\draw[->,red,very thick] (F0)--(F1);

\draw[->,red,very thick] (G0)--(G1);

\draw[->,red,very thick] (HI0)--(HI1);

\tikzstyle{every node}=[inner sep=1pt]
\begin{tiny}
\node at (LC0) {$S^1_1$};
\node at (LC1) {$S^0_1$};
\node at (LF0) {$S^0_2$};
\node at (LF1) {$S^1_2$};
\node at (LG0) {$S^0_3$};
\node at (LG1) {$S^1_3$};
\node at (LHI0) {$S^0_4$};
\node at (LHI1) {$S^1_4$};
\end{tiny}
\node at (L) {$\pi(\phi)$};

\draw[white] (-4,-3)--(-4,-2.5);
\draw[white] (4,3)--(4,2.5);
\end{tikzpicture}

\caption{\label{fig:example_metachords} A conformal model $\phi$ of $G_{ov}$ and the circular order of the slots $\pi(\phi)$ in $\phi$.}
\end{figure}
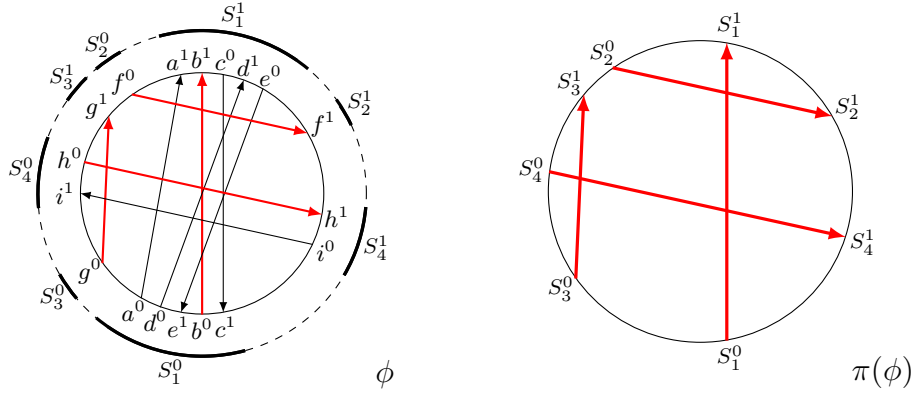

Let $i \in [t]$. 
By \ref{prop:CA-modules}, there is a partition $\{S^0_i,S^1_i\}$ of the set $S^*_i$ such that 
for every conformal model $\phi$ both the sets $S^0_i$ and $S^1_i$ are contiguous in $\phi$ and
$(\phi|S^0_i, \phi|S^1_i)$ is a consistent permutation model induced by $S_i$ in $\phi$.
Note that $S^0_i$ and $S^1_i$ are superscripted copies of $S_i$ and $\{S^0_i, S^1_i\}$ is a partition of $S^*_i$.
We assume the superscripts in $S^0_i,S^1_i$ are chosen such that $s_i$ is oriented from $S^0_i$ to $S^1_i$.
Since for every conformal model $\phi$ the relative orientation of the chords in $(\phi|S^0_i,\phi|S^1_i)$ is the same and since $\phi$ is conformal to $G$ ($\phi$ keeps the left/right relation between every two non-intersecting chords),
the transitive orientations of $(S_i,{\parallel})$ induced by the permutation models $(\phi|S^0_i,\phi|S^1_i)$ are all the same.
We denote this transitive orientation by ${<_{S_i}}$.
The triple $\SSS_i = (S^0_i,S^1_i,{<_{S_i}})$ is called the \emph{metachord} and 
the sets $S^0_i$ and $S^1_i$ are called the \emph{slots} of the CA-module $S_i$.
The set $\slots =\{S^0_1,S^1_1,\ldots,S^0_t,S^1_t\}$ of slots and the set $\metachords = \{ \SSS_1,\ldots,\SSS_t\}$ of metachords are the components of $\dataStructure$.
For the metachord $\SSS_i$ it is convenient to assume the following definition.
\begin{definition}
\label{def:admissible-models-for-metachord}
A permutation model $\tau = (\tau^0,\tau^1)$ of $(S_i,{\sim})$ is \emph{admissible by $\mathbb{S}_i$} if:
\begin{itemize}
 \item $\tau^j$ is a permutation of $S^j_i$ for $j \in \{0,1\}$,
 \item we have ${<_{\tau}} = {<_{S_i}}$ (${\prec_{\tau}}$ is not restricted), where ${<_{\tau}}$ and ${\prec_{\tau}}$ are transitive orientations 
 of $(S_i,{\parallel})$ and $(S_i,{\sim})$, respectively, corresponding to $\tau$.
\end{itemize}
\end{definition}
Then, Property \ref{prop:CA-modules} allows us to treat any conformal model $\phi$ as a collection of 
$t$ permutation models $(\phi|S^0_1, \phi|S^1_1), \ldots, (\phi|S^0_t, \phi|S^1_t)$  admissible by the metachords $\SSS_1,\ldots,\SSS_t$ and spanned between the slots $S^0_1 \leftrightarrow S^1_1, \ldots, S^0_t \leftrightarrow S^1_t$, respectively.
See Figure~\ref{fig:example_metachords} to the left.

Let $\phi$ be a conformal model of~$G$.
Property~\ref{prop:CA-modules} allows us to denote by $\pi(\phi)$ the \emph{circular order of the slots in $\phi$}, that is, the circular order of $\slots$ obtained from $\phi$ by substituting every contiguous subword $\phi|S^j_i$ by the letter $S^j_i$. 
Usually we draw $\pi(\phi)$ in the way shown 
in Figure~\ref{fig:example_metachords}: for every two corresponding slots $S^0_i$ and $S^1_i$ placed on the circle we draw a chord oriented from $S^0_i$ to $S^1_i$.
Since such a chord represents a model admissible for the metachord $\SSS_i$, 
we call it simply the \emph{metachord} of $\SSS_i$.

The last component~$\Pi$ of the data structure~$\dataStructure$ contains the set of \emph{circular orders of the slots} that might appear in the conformal models of~$G_{ov}$.
In particular, $\Pi$ is defined such that:
\begin{description}
 \item [\namedlabel{prop:circular-orders-of-the-slots}{(P2)}] For every conformal model~$\phi$ of~$G_{ov}$ the circular word~$\pi(\phi)$ is a member of~$\Pi$.
\end{description}
Eventually, all the components in $\dataStructure$ are defined such that the following holds:
\begin{description}
\item [\namedlabel{prop:completeness}{(P3)}] We can generate any conformal model of~$G_{ov}$ by:  
\begin{itemize}
 \item picking a circular order of the slots $\pi$ from the set $\Pi$,
 \item replacing the slots $S^0_i$ and $S^1_i$ in $\pi$ by words~$\tau^0_i$ and~$\tau^1_i$, 
 where $(\tau^0_i,\tau^1_i)$ is an oriented permutation model of~$(S_i,{\sim})$ admissible by the metachord $\mathbb{S}_i$.
\end{itemize}
\end{description}

Consider the circular-arc graph $G=(V,E)$ whose conformal model $\phi$ is shown in Figure~\ref{fig:example_metachords}.
We have $V = \{a,b,c,d,e,f,g,h,i\}$, the edges $E$ can be easily read from~$\phi$.
Figure~\ref{fig:example_all_models} shows all non-equivalent conformal models of $G_{ov}$.
The data structure $\dataStructure$ representing the conformal models of $G_{ov}$ consists of:
\begin{itemize}
 \item four CA-modules $S_1 = \{a,b,c,d\}$, $S_2 = \{f\}$, $S_3 = \{g\}$, $S_4 = \{h,i\}$, represented by vertices
 $b$, $f$, $g$, and $h$, respectively,
 \item slots $S^0_1 = \{a^0,b^0,c^1,d^0,e^1\}$, $S^1_1 = \{a^1,b^1,c^0,d^1,e^0\}$,
 $S^0_2 = \{f^0\}$, $S^1_2 = \{f^1\}$, $S^0_3 = \{g^0\}$, $S^1_3 = \{g^1\}$, $S^0_4 = \{h^0,i^1\}$, $S^1_4 = \{h^1,i^0\}$,
 \item metachords $(S^0_1,S^1_1,{<_{S_1}}), \ldots, (S^0_4,S^1_4,{<_{S_4}})$, where ${<_{S_1}}$ consists of the pairs $\{b <_{S_1} a, c <_{S_1} a, d <_{S_1} a, e <_{S_1} a, c <_{S_1} b, e <_{S_1} d\}$,
 ${<_{S_4}}$ consists of the pair $\{i <_{S_4} h\}$, ${<_{S_2}}$ and ${<_{S_3}}$ are empty.
 \item the set $\Pi = \{\pi,\pi^R\}$ of circular order of the slots, where 
 $\pi \equiv S^1_1S^1_2S^1_4S^0_1S^0_3S^0_4S^1_3S^0_2$ and $\pi^R$ is the reflection of $\pi$.
\end{itemize}
We note that:
\begin{itemize}
 \item the metachord $\mathbb{S}_1$ has two admissible models: $\tau = (c^1b^0e^1d^0a^0, a^1b^1c^0d^1e^0)$ and $\mu = (e^1d^0c^1b^0a^0, a^1d^1e^0b^1c^0)$.
 The metachords $\mathbb{S}_2, \mathbb{S}_3, \mathbb{S}_4$ admit one admissible model, respectively, $(f^0,f^1)$, $(g^0,g^1)$, and $(i^1h^0, h^1i^0)$.
\end{itemize}
Note that $\phi_2$ is the reflection of $\phi_1$, $\phi_4$ is the reflection of $\phi_3$, $\phi_3$ is obtained from $\phi_1$ by replacing~$\tau$ by~$\mu$, 
and $\phi_2$ is obtained from $\phi_4$ by replacing $\tau$ by~$\mu$.

\input ./figures/data_structure/example.tex

The rest of this section is organized as follows:
\begin{itemize}
 \item In Subsection~\ref{sub:ca-modules} we show how to read the 
 CA-modules, the slots, and the metachords of~$G$ from a given conformal model of~$G_{ov}$.
 \item In Subsection~\ref{sub:PQ_tree} we describe the PQS-tree of~$G$ and we show 
 how it represents the set~$\Pi$. 
 We also show how to read the PQS-tree from a given conformal model of~$G_{ov}$.
 \item In Subsection~\ref{sub:admissible_models} we describe the structure of all admissible models for 
 a single metachord.
 \item In Subsection~\ref{sub:PQM_tree} we introduce the PQSM-tree $\pqsmtree$ and we show how it represents the conformal models of~$G_{ov}$.
\end{itemize}

\subsection{Determining CA-modules, slots, and metachords of $G$}
\label{sub:ca-modules}
Let $\phi$ be a fixed conformal model of~$G_{ov}$.
The CA-modules of $G$ can be read from $\phi$ using the following rule:
\begin{description}
 \item[\namedlabel{prop:CA-modules-rule}{(R)}] A set $S \subseteq V$ is a CA-module of $G$ if $S$ is an inclusion-wise maximal module (not necessary strong) in $G_{ov}$ such that:
\begin{itemize}
\item $S \subseteq M$ for some child $M$ of $V$ in $\strongModules(G_{ov})$,
\item $S$ induces a consistent permutation model in $\phi$.
\end{itemize}
\end{description}
For a child $M$ of $V$ in $\strongModules(G_{ov})$, let $\camodules(M)$ denote the set of all CA-modules contained in~$M$.
In Section~\ref{sec:proof_sketch} we show that the set $\camodules(M)$
does not depend on~$\phi$ and that $\camodules(M)$ forms a partition of the set $M$.
Given CA-modules of $G$, the slots and the metachords of~$G$ can be easily read from $\phi$.

\subsection{PQS-tree}
\label{sub:PQ_tree}
As we mentioned, the set $\Pi$ may contain exponentially many members, but it has a linear-size representation by means of the PQS-tree~$\pqstree$. 
In this subsection we show how to construct the PQS-tree $\pqstree$ from a given conformal model $\phi$ of $G_{ov}$.
Again, for a combinatorial definition we refer to Section~\ref{sec:proof_sketch}.

\begin{figure}[htp!]
\begin{tikzpicture}[scale=0.8,>=latex,shorten >=-0.4pt,shorten <=-0.4pt]
\coordinate (center) at (0,0) {};
\coordinate (label) at (-2.5,-2.5) {};

\coordinate (s10) at ($(center)+(120:2.0cm)$) {};
\coordinate (s11) at ($(center)+(240:2.0cm)$) {};

\coordinate (s20) at ($(center)+(210:2.0cm)$) {};
\coordinate (s21) at ($(center)+(330:2.0cm)$) {};

\coordinate (m20) at ($(center)+(50:2.0cm)$) {};
\coordinate (m21) at ($(center)+(100:2.0cm)$) {};

\coordinate (m31) at ($(center)+(-10:2.0cm)$) {};
\coordinate (m30) at ($(center)+(40:2.0cm)$) {};

\coordinate (m40) at ($(center)+(260:2.0cm)$) {};
\coordinate (m41) at ($(center)+(310:2.0cm)$) {};

\coordinate (m50) at ($(center)+(140:2.0cm)$) {};
\coordinate (m51) at ($(center)+(190:2.0cm)$) {};

\coordinate (lp1) at ($(center)+(45:0.5cm)$) {};
\coordinate (lp3) at ($(center)+(190:1.5cm)$) {};
\coordinate (lp2) at ($(center)+(260:1.5cm)$) {};

\coordinate (ls10) at ($(center)+(120:2.4cm)$) {};
\coordinate (ls11) at ($(center)+(240:2.4cm)$) {};

\coordinate (ls20) at ($(center)+(210:2.4cm)$) {};
\coordinate (ls21) at ($(center)+(330:2.4cm)$) {};

\coordinate (lm20) at ($(center)+(53:2.4cm)$) {};
\coordinate (lm21) at ($(center)+(100:2.4cm)$) {};

\coordinate (lm31) at ($(center)+(-10:2.4cm)$) {};
\coordinate (lm30) at ($(center)+(37:2.4cm)$) {};

\coordinate (lm40) at ($(center)+(260:2.4cm)$) {};
\coordinate (lm41) at ($(center)+(310:2.4cm)$) {};

\coordinate (lm50) at ($(center)+(140:2.4cm)$) {};
\coordinate (lm51) at ($(center)+(190:2.4cm)$) {};

\tikzstyle{every node}=[inner sep=1pt]
\begin{scriptsize}
\node at (lp1) {$P_1$};
\node at (lp2) {$P_2$};
\node at (lp3) {$P_3$};

\node at (ls10) {$S^0_1$};
\node at (ls11) {$S^1_1$};

\node at (ls20) {$S^0_2$};
\node at (ls21) {$S^1_2$};

\node at (lm20) {$Q^0_2$};
\node at (lm21) {$Q^1_2$};

\node at (lm30) {$Q^0_3$};
\node at (lm31) {$Q^1_3$};

\node at (lm40) {$Q^0_4$};
\node at (lm41) {$Q^1_4$};

\node at (lm50) {$Q^0_5$};
\node at (lm51) {$Q^1_5$};

\end{scriptsize}
\node at (label) {$\pi$};

\draw (0,0) circle (2cm);

\draw[very thick,red,->] (s10) -- (s11);
\draw[very thick,red,->] (s20) -- (s21);
\draw[very thick,->] (m20) -- (m21);
\draw[very thick,->] (m30) -- (m31);
\draw[very thick,->] (m40) -- (m41);
\draw[very thick,->] (m50) -- (m51);

\draw[white] (-2.8,-2.8)--(-2.8,-2);
\draw[white] (2.8,2.8)--(2.8,2);

\end{tikzpicture}
\hspace{1cm}
\begin{tikzpicture}[scale=0.8,>=latex,shorten >=-0.4pt,shorten <=-0.4pt]
\coordinate (label) at (-2.5,-2.5) {};

  \tikzstyle{every node}=[circle,minimum size=10pt,inner sep=0.5,draw];
  \begin{scriptsize}
  \node[red] (m1) at (0,0) {$Q_1$};
  \node (m2) at (2,1.5) {$Q_2$};
  \node (m3) at (2,0.25) {$Q_3$};
  \node (m4) at (2,-1) {$Q_4$};
  \node (m5) at (-2,0.0) {$Q_5$};
  \end{scriptsize}
  \tikzstyle{every node}=[circle,minimum size=10pt,inner sep=0.5,draw];
  \begin{scriptsize}
  \node (p1) at (1,1.0) {$P_1$};
  \node (p2) at (1,-1.0) {$P_2$};
  \node (p3) at (-1.0,0) {$P_3$};
  \end{scriptsize}
\tikzstyle{every node}=[inner sep=1pt]
  \begin{scriptsize}
  \node (m12) at (2,2.5) {$Q^1_2$};
  \node (m02) at (3,1.5) {$Q^0_2$};
  \node (m03) at (3,0.7) {$Q^0_3$};
  \node (m13) at (3,-0.2) {$Q^1_3$};
  \node (m14) at (3,-1) {$Q^1_4$};
  \node (m04) at (2,-2) {$Q^0_4$};

  \node (s01) at (0,1) {$S^0_1$};
  \node (s11) at (0,-1) {$S^1_1$};
  \node (s02) at (-1,-1) {$S^0_2$};
  \node (s12) at (1,0) {$S^1_2$};
  \node (m05) at (-2,1) {$Q^0_5$};
  \node (m15) at (-2,-1) {$Q^1_5$};

  \end{scriptsize}
  \node at (label) {$\pqstree_{\pi}$};

\path (m1) edge (p1); 
\path (m1) edge (p2); 
\path (m1) edge (p3); 
\path (m2) edge (m02); 
\path (m2) edge (m12); 
\path (m3) edge (m03); 
\path (m3) edge (m13); 
\path (m4) edge (m04); 
\path (m4) edge (m14); 
\path (m5) edge (m05); 
\path (m5) edge (m15); 

\path (m1) edge (s01); 
\path (m1) edge (s11); 
\path (m1) edge (s02); 
\path (m1) edge (s12); 

\path (p1) edge (m2); 
\path (p1) edge (m3); 
\path (p2) edge (m4); 
\path (p3) edge (m5); 

\draw[white] (-2.8,-2.8)--(-2.8,-2);
\draw[white] (2.8,2.8)--(2.8,2);
\end{tikzpicture}
\caption{\label{fig:PQS_tree} To the left: the circular order of the slots $\pi$ in some conformal model~$\phi$
for the case when $V$ is parallel in $\strongModules(G_{ov})$.
$Q_1,Q_2,Q_3,Q_4,Q_5$ are the connected components of $G_{ov}$, 
$\camodules(Q_1)=\{S_1,S_2\}$ (metachords of $Q_1$ are drawn in red) 
and $\camodules(Q_i)=\{Q_i\}$ for $i \in [2,5]$.
P-nodes $P_1$, $P_2$, and $P_3$ correspond to 
the regions neighbouring the components $Q_1,Q_2,Q_3$,
$Q_1,Q_4$, and $Q_1,Q_5$, respectively.
We have $\pi_{|Q_1} \equiv S^0_1P_1S^1_2P_2S^1_1S^0_2P_3$ and 
$\pi_{|P_1} \equiv Q_1Q_2Q_3$.
To the right: the ordered PQS-tree $\pqstree_{\pi}$ representing $\pi$.
}
\end{figure}
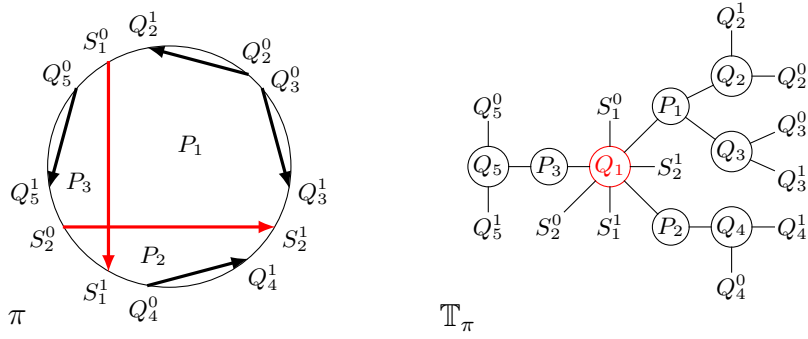

The PQS-tree~$\pqstree$ is an unrooted tree.
The leaf nodes of~$\pqstree$, called also as \emph{S-nodes}, are in the correspondence with the slots of $G$.
The non-leaf nodes of~$\pqstree$ are labelled either by the letter P (\mbox{P-nodes}) 
or the letter Q (\mbox{Q-nodes}).
The \emph{Q-nodes} are in correspondence with the connected components of $G_{ov}$.
We have an edge in $\pqstree$ between the slots $S^0,S^1$ of CA-module~$S$ 
and the Q-node $Q$ if $Q$ is the component of $G_{ov}$ containing~$S$.
Let $\pi \equiv \pi(\phi)$ be the circular order of the slots in $\phi$.
Consider the drawing of $\pi$ -- see Figure~\ref{fig:PQS_tree} to the left.
Clearly, for every component $Q$ the metachords of the component $Q$ (that is, the metachords corresponding to the CA-modules contained in~$Q$) form an arc-wise connected set.
All the metachords of~$\pi$ divide the interior of the circle into arcwise connected regions.
Every region adjacent to metachords from at least two different components
gives rise to a~\emph{P-node}.
In particular, when $V$ is prime or serial in~$\strongModules(G_{ov})$ (the graph $(V,{\sim})$ is then connected), 
$\pqstree$ consists of one inner Q-node~$V$ and the set of slots $\slots$ adjacent to $V$.
When~$V$ is parallel in~$\strongModules(G_{ov})$, we have an edge in $\pqstree$ between a P-node~$P$ 
and a Q-node~$Q$ if the region~$P$ is adjacent to some metachord of the component~$Q$.
See Figure~\ref{fig:PQS_tree} for an illustration.
In Section \ref{sec:proof_sketch} we will define the PQS-tree~$\pqstree$ combinatorially;
in particular, we will show that $\pqstree$ does not depend on the model~$\phi$.

We represent the set $\Pi$ through the sets $\Pi(N)$ of (possible) \emph{orderings} of a node $N$, which are defined for all inner nodes $N$ of $\pqstree$.
For every inner node $N$ of $\pqstree$, the set $\Pi(N)$ contains some circular orders of the set of the neighbours of $N$ in~$\pqstree$ (the precise definition of~$\Pi(N)$ will be given later on).
The PQS-tree~$\pqstree$ is said to be \emph{ordered} if every inner node~$N$ in~$\pqstree$ is assigned 
an ordering from the set~$\Pi(N)$.
Every ordered PQS-tree~$\pqstree$ \emph{represents} a circular order of the set~$\slots$, as follows:
we draw $\pqstree$ in the plane such that the clockwise order of the neighbours of every inner node~$N$ is consistent with the assigned ordering from $\Pi(N)$ and then we list all the slots as they occur when we walk the boundary of~$\pqstree$ in the clockwise order.
The sets $\Pi(\cdot)$ are such that there is the correspondence between the members of~$\Pi$ and the circular orders of $\slots$ represented by the ordered PQS-trees~$\pqstree$.
Hence, for $\pi \in \Pi$ we denote the ordered tree representing $\pi$ by~$\pqstree_{\pi}$ and,
for any inner node~$N$ in $\pqstree$, by $\pi_{|N}$ we denote the ordering (which is in the set $\Pi(N)$) of the node~$N$ in the ordered tree $\pqstree_{\pi}$.
In particular, for every inner node $N$ we have 
$$\Pi(N) = \{\pi_{|N}: \pi \in \Pi\}.$$

Let $\phi$ be a conformal model and let $\pi = \pi(\phi)$ be the circular order of the slots in $\phi$.
We show how to construct the ordered PQS-tree $\pqstree_{\pi}$ corresponding to $\pi$.
To get $\pqstree_{\pi}$, we order every inner node $N$ consistently with~$\pi_{|N}$, where $\pi_{|N}$ can be read from the drawing of $\pi$, as follows:
\begin{itemize}
 \item if $N$ is a P-node, then the word $\pi_{|N}$ contains all Q-nodes adjacent to $N$ in~$\pqstree$
 ordered as they occur when we walk in the clockwise order the boundary of the region $P$,
 \item if $N$ is a Q-node, then the word $\pi_{|N}$ contains all P-nodes and slots adjacent to $N$ in~$\pqstree$ ordered as they occur when we walk in the clockwise order the boundary of the arc-wise connected set consisting of all metachords of the component $N$.
\end{itemize}
See Figure~\ref{fig:PQS_tree} to the right for an illustration.
We leave the reader to verify that the ordered PQS-tree $\pqstree$ in which every inner node $N$ is ordered consistently with $\pi_{|N}$ represents the circular order of the slots $\pi$.

The description of the sets $\Pi(\cdot)$ differs depending on whether~$V$ is serial/prime/parallel in~$\strongModules(G_{ov})$.
Again, in this section we show how to determine the sets $\Pi(\cdot)$ given the conformal model~$\phi$ of~$G_{ov}$.

\subsubsection{$V$ is serial in $\strongModules(G_{ov})$}
In this case we have $\camodules(M)=\{M\}$ for any child $M$ of $V$ in $\strongModules(G_{ov})$ -- see Section~\ref{sec:proof_sketch}.
Hence the set of CA-modules $\camodules$ of $G$ coincides with the set of children of $V$ in $\strongModules(G_{ov})$.
Let $\camodules = \{S_1,\ldots,S_t\}$.
Since $V$ is serial, $(S_i,{\parallel})$ is connected for every $i \in [t]$ and $S_i \sim S_j$ for every two distinct $i,j \in [t]$.
In Section~\ref{sec:proof_sketch} we show that
$$
\Pi(V) =  \left \{ \pi :
\begin{array}{c}
\text{$\pi$ is a circular order of $S^0_1,S^1_1,\ldots,S^0_t,S^1_t$ such that for every } \\
\text{distinct $i,j \in [t]$ the slots associated with $S_i$ and $S_j$ overlap} 
\end{array}
\right \}.
$$
Since $\pqstree$ has only one inner Q-node $V$, we have $\Pi = \Pi(V)$.
See Figure~\ref{fig:Pi-serial} for an illustration.

\begin{center}
\begin{figure}[ht]
\begin{tikzpicture}[xscale=0.6,yscale=0.6,>=latex,shorten >=-0.4pt,shorten <=-0.4pt]
\coordinate (label) at (-2.2,-2.3) {};
\tikzstyle{every node}=[circle,minimum size=15pt,inner sep=0.5,draw];
\begin{scriptsize}
\node (V) at (0.0,0) {$V$};
\end{scriptsize}
\tikzstyle{every node}=[inner sep=1pt]
\begin{scriptsize}
\node (s1) at ($(center)+(0:2.0cm)$) {$S^1_1$};
\node (s2) at ($(center)+(45:2.0cm)$) {$S^0_2$};
\node (s3) at ($(center)+(90:2.0cm)$) {$S^1_3$};
\node (s4) at ($(center)+(135:2.0cm)$) {$S^1_4$};
\node (s5) at ($(center)+(180:2.0cm)$) {$S^0_1$};
\node (s6) at ($(center)+(225:2.0cm)$) {$S^1_2$};
\node (s7) at ($(center)+(270:2.0cm)$) {$S^0_3$};
\node (s8) at ($(center)+(315:2.0cm)$) {$S^1_4$};
\end{scriptsize}
\node at (label) {$\pqstree_{\pi}$};
\path (V) edge (s1); 
\path (V) edge (s2); 
\path (V) edge (s3); 
\path (V) edge (s4); 
\path (V) edge (s5); 
\path (V) edge (s6); 
\path (V) edge (s7); 
\path (V) edge (s8); 
\draw[white] (-2.8,-2.8)--(-2.8,-2);
\draw[white] (2.8,2.8)--(2.8,2);
\end{tikzpicture}
\hspace{0.1cm}
\begin{tikzpicture}[yscale=0.6,xscale=0.6,>=latex,shorten >=-0.4pt,shorten <=-0.4pt]
\coordinate (center) at (0,0) {};
\coordinate (label) at (-2.5,-2.5) {};

\coordinate (s1) at ($(center)+(0:2.0cm)$) {};
\coordinate (s2) at ($(center)+(45:2.0cm)$) {};
\coordinate (s3) at ($(center)+(90:2.0cm)$) {};
\coordinate (s4) at ($(center)+(135:2.0cm)$) {};
\coordinate (s5) at ($(center)+(180:2.0cm)$) {};
\coordinate (s6) at ($(center)+(225:2.0cm)$) {};
\coordinate (s7) at ($(center)+(270:2.0cm)$) {};
\coordinate (s8) at ($(center)+(315:2.0cm)$) {};

\coordinate (ls1) at ($(center)+(0:2.4cm)$) {};
\coordinate (ls2) at ($(center)+(45:2.4cm)$) {};
\coordinate (ls3) at ($(center)+(90:2.4cm)$) {};
\coordinate (ls4) at ($(center)+(135:2.4cm)$) {};
\coordinate (ls5) at ($(center)+(180:2.4cm)$) {};
\coordinate (ls6) at ($(center)+(225:2.4cm)$) {};
\coordinate (ls7) at ($(center)+(270:2.4cm)$) {};
\coordinate (ls8) at ($(center)+(315:2.4cm)$) {};

\tikzstyle{every node}=[inner sep=1pt]
\begin{scriptsize}
\node at (ls1) {$S^1_1$};
\node at (ls2) {$S^0_2$};
\node at (ls3) {$S^1_3$};
\node at (ls4) {$S^1_4$};
\node at (ls5) {$S^0_1$};
\node at (ls6) {$S^1_2$};
\node at (ls7) {$S^0_3$};
\node at (ls8) {$S^0_4$};
\end{scriptsize}
\node at (label) {$\pi$};

\draw (0,0) circle (2cm);

\draw[very thick,red,<-] (s1) -- (s5);
\draw[very thick,blue,->] (s2) -- (s6);
\draw[very thick,green,<-] (s3) -- (s7);
\draw[very thick,<-] (s4) -- (s8);

\draw[white] (-2.8,-2.8)--(-2.8,-2);
\draw[white] (2.8,2.8)--(2.8,2);
\end{tikzpicture}
\hspace{0.3cm}
\begin{tikzpicture}[yscale=0.6,xscale=0.6,>=latex,shorten >=-0.4pt,shorten <=-0.4pt]
\coordinate (center) at (0,0) {};
\coordinate (label) at (-2.5,-2.5) {};

\coordinate (s1) at ($(center)+(0:2.0cm)$) {};
\coordinate (s2) at ($(center)+(45:2.0cm)$) {};
\coordinate (s3) at ($(center)+(90:2.0cm)$) {};
\coordinate (s4) at ($(center)+(135:2.0cm)$) {};
\coordinate (s5) at ($(center)+(180:2.0cm)$) {};
\coordinate (s6) at ($(center)+(225:2.0cm)$) {};
\coordinate (s7) at ($(center)+(270:2.0cm)$) {};
\coordinate (s8) at ($(center)+(315:2.0cm)$) {};

\coordinate (ls1) at ($(center)+(0:2.4cm)$) {};
\coordinate (ls2) at ($(center)+(45:2.4cm)$) {};
\coordinate (ls3) at ($(center)+(90:2.4cm)$) {};
\coordinate (ls4) at ($(center)+(135:2.4cm)$) {};
\coordinate (ls5) at ($(center)+(180:2.4cm)$) {};
\coordinate (ls6) at ($(center)+(225:2.4cm)$) {};
\coordinate (ls7) at ($(center)+(270:2.4cm)$) {};
\coordinate (ls8) at ($(center)+(315:2.4cm)$) {};

\tikzstyle{every node}=[inner sep=1pt]
\begin{scriptsize}
\node at (ls1) {$S^0_2$};
\node at (ls2) {$S^1_1$};
\node at (ls3) {$S^0_3$};
\node at (ls4) {$S^1_4$};
\node at (ls5) {$S^1_2$};
\node at (ls6) {$S^0_1$};
\node at (ls7) {$S^1_3$};
\node at (ls8) {$S^0_4$};
\end{scriptsize}
\node at (label) {$\pi'$};

\draw (0,0) circle (2cm);

\draw[very thick,blue,->] (s1) -- (s5);
\draw[very thick,red,<-] (s2) -- (s6);
\draw[very thick,green,->] (s3) -- (s7);
\draw[very thick,<-] (s4) -- (s8);
\draw[white] (-2.8,-2.8)--(-2.8,-2);
\draw[white] (2.8,2.8)--(2.8,2);
\end{tikzpicture}
\hspace{0.1cm}
\begin{tikzpicture}[xscale=0.6,yscale=0.6,>=latex,shorten >=-0.4pt,shorten <=-0.4pt]
\coordinate (label) at (-2.2,-2.3) {};
\tikzstyle{every node}=[circle,minimum size=15pt,inner sep=0.5,draw];
\begin{scriptsize}
\node (V) at (0.0,0) {$V$};
\end{scriptsize}
\tikzstyle{every node}=[inner sep=1pt]
\begin{scriptsize}
\node (s1) at ($(center)+(0:2.0cm)$) {$S^0_2$};
\node (s2) at ($(center)+(45:2.0cm)$) {$S^1_1$};
\node (s3) at ($(center)+(90:2.0cm)$) {$S^0_3$};
\node (s4) at ($(center)+(135:2.0cm)$) {$S^1_4$};
\node (s5) at ($(center)+(180:2.0cm)$) {$S^1_2$};
\node (s6) at ($(center)+(225:2.0cm)$) {$S^0_1$};
\node (s7) at ($(center)+(270:2.0cm)$) {$S^1_3$};
\node (s8) at ($(center)+(315:2.0cm)$) {$S^1_4$};
\end{scriptsize}
\node at (label) {$\pqstree_{\pi'}$};
\path (V) edge (s1); 
\path (V) edge (s2); 
\path (V) edge (s3); 
\path (V) edge (s4); 
\path (V) edge (s5); 
\path (V) edge (s6); 
\path (V) edge (s7); 
\path (V) edge (s8); 
\draw[white] (-2.8,-2.8)--(-2.8,-2);
\draw[white] (2.8,2.8)--(2.8,2);
\end{tikzpicture}
\caption{\label{fig:Pi-serial} Two members $\pi$ and $\pi'$ of the set $\Pi$ and the ordered PQS-trees $\pqstree_{\pi}$ and $\pqstree_{\pi'}$ representing $\pi$ and $\pi'$ for the case when $V$ is serial in~$\strongModules(G_{ov})$.}
\end{figure}

\end{center}

\subsubsection{$V$ is prime in $\strongModules(G_{ov})$}
In this case, when $M$ is a prime child of $V$ we have $\camodules(M) = \{M\}$, 
and when $M$ is a serial/parallel child of $V$ every CA-module in $\camodules(M)$ is 
the union of some children of $M$ in $\strongModules(G_{ov})$ -- see Section~\ref{sec:proof_sketch} for more details. 
Let $\camodules = \{S_1,\ldots,S_t\}$ and let $\pi = \pi(\phi)$.
In Section~\ref{sec:proof_sketch} we show that
$$\Pi(V) = \{\pi,\pi^R\},$$
where $\pi^R$ is the reflection of $\pi$. 
Since the PQS-tree $\pqstree$ contains only one inner node~$V$,
we have $\Pi = \Pi(V)$.
See Figure~\ref{fig:Pi-prime} for an illustration.
\begin{center}
\begin{figure}[ht]
\begin{tikzpicture}[xscale=0.6,yscale=0.6,>=latex,shorten >=-0.4pt,shorten <=-0.4pt]
\coordinate (label) at (-2.2,-2.3) {};
\tikzstyle{every node}=[circle,minimum size=15pt,inner sep=0.5,draw];
\begin{scriptsize}
\node (V) at (0.0,0) {$V$};
\end{scriptsize}
\tikzstyle{every node}=[inner sep=1pt]
\begin{scriptsize}
\node (s1) at ($(center)+(0:2.0cm)$) {$S^1_1$};
\node (s2) at ($(center)+(45:2.0cm)$) {$S^1_2$};
\node (s3) at ($(center)+(90:2.0cm)$) {$S^1_3$};
\node (s4) at ($(center)+(135:2.0cm)$) {$S^0_2$};
\node (s5) at ($(center)+(180:2.0cm)$) {$S^1_4$};
\node (s6) at ($(center)+(225:2.0cm)$) {$S^0_1$};
\node (s7) at ($(center)+(270:2.0cm)$) {$S^0_4$};
\node (s8) at ($(center)+(315:2.0cm)$) {$S^1_3$};
\end{scriptsize}
\node at (label) {$\pqstree_{\pi}$};
\path (V) edge (s1); 
\path (V) edge (s2); 
\path (V) edge (s3); 
\path (V) edge (s4); 
\path (V) edge (s5); 
\path (V) edge (s6); 
\path (V) edge (s7); 
\path (V) edge (s8); 
\draw[white] (-2.8,-2.8)--(-2.8,-2);
\draw[white] (2.8,2.8)--(2.8,2);
\end{tikzpicture}
\hspace{0.1cm}
\begin{tikzpicture}[yscale=0.6,xscale=0.6,>=latex,shorten >=-0.4pt,shorten <=-0.4pt]
\coordinate (center) at (0,0) {};
\coordinate (label) at (-2.2,-2.3) {};

\coordinate (s1) at ($(center)+(30:2.0cm)$) {};
\coordinate (s2) at ($(center)+(90:2.0cm)$) {};
\coordinate (s3) at ($(center)+(150:2.0cm)$) {};
\coordinate (s4) at ($(center)+(180:2.0cm)$) {};
\coordinate (s5) at ($(center)+(220:2.0cm)$) {};
\coordinate (s6) at ($(center)+(260:2.0cm)$) {};
\coordinate (s7) at ($(center)+(290:2.0cm)$) {};
\coordinate (s8) at ($(center)+(360:2.0cm)$) {};

\coordinate (ls1) at ($(center)+(30:2.4cm)$) {};
\coordinate (ls2) at ($(center)+(90:2.4cm)$) {};
\coordinate (ls3) at ($(center)+(150:2.4cm)$) {};
\coordinate (ls4) at ($(center)+(180:2.4cm)$) {};
\coordinate (ls5) at ($(center)+(220:2.4cm)$) {};
\coordinate (ls6) at ($(center)+(260:2.4cm)$) {};
\coordinate (ls7) at ($(center)+(290:2.4cm)$) {};
\coordinate (ls8) at ($(center)+(360:2.4cm)$) {};

\tikzstyle{every node}=[inner sep=1pt]
\begin{scriptsize}
\node at (ls1) {$S^1_2$};
\node at (ls2) {$S^1_3$};
\node at (ls3) {$S^0_2$};
\node at (ls4) {$S^1_4$};
\node at (ls5) {$S^0_1$};
\node at (ls6) {$S^0_4$};
\node at (ls7) {$S^0_3$};
\node at (ls8) {$S^1_1$};
\end{scriptsize}
\node at (label) {$\pi$};

\draw (0,0) circle (2cm);

\draw[very thick,->] (s3) -- (s1);
\draw[very thick,red,->] (s7) -- (s2);
\draw[very thick,->] (s6) -- (s4);
\draw[very thick,->] (s5) -- (s8);

\draw[thick,red] ([shift=(90:2.0cm)]0,0) arc (90:290:2.0cm);

\draw[white] (-2.8,-2.8)--(-2.8,-2);
\draw[white] (2.8,2.8)--(2.8,2);
\end{tikzpicture}
\begin{tikzpicture}[yscale=0.6,xscale=-0.6,>=latex,shorten >=-0.4pt,shorten <=-0.4pt]
\draw[white] (-0.5,-3)--(-0.5,-3);
\draw[white] (0.5,3)--(0.5,3);
\draw[black, dashed] (0,-2.5)--(0,2.5);
\end{tikzpicture}
\begin{tikzpicture}[yscale=0.6,xscale=-0.6,>=latex,shorten >=-0.4pt,shorten <=-0.4pt]
\coordinate (center) at (0,0) {};
\coordinate (label) at (-2.2,-2.3) {};

\coordinate (s1) at ($(center)+(30:2.0cm)$) {};
\coordinate (s2) at ($(center)+(90:2.0cm)$) {};
\coordinate (s3) at ($(center)+(150:2.0cm)$) {};
\coordinate (s4) at ($(center)+(180:2.0cm)$) {};
\coordinate (s5) at ($(center)+(220:2.0cm)$) {};
\coordinate (s6) at ($(center)+(260:2.0cm)$) {};
\coordinate (s7) at ($(center)+(290:2.0cm)$) {};
\coordinate (s8) at ($(center)+(360:2.0cm)$) {};

\coordinate (ls1) at ($(center)+(30:2.4cm)$) {};
\coordinate (ls2) at ($(center)+(90:2.4cm)$) {};
\coordinate (ls3) at ($(center)+(150:2.4cm)$) {};
\coordinate (ls4) at ($(center)+(180:2.4cm)$) {};
\coordinate (ls5) at ($(center)+(220:2.4cm)$) {};
\coordinate (ls6) at ($(center)+(260:2.4cm)$) {};
\coordinate (ls7) at ($(center)+(290:2.4cm)$) {};
\coordinate (ls8) at ($(center)+(360:2.4cm)$) {};

\tikzstyle{every node}=[inner sep=1pt]
\begin{scriptsize}
\node at (ls1) {$S^0_2$};
\node at (ls2) {$S^0_3$};
\node at (ls3) {$S^1_2$};
\node at (ls4) {$S^0_4$};
\node at (ls5) {$S^1_1$};
\node at (ls6) {$S^1_4$};
\node at (ls7) {$S^1_3$};
\node at (ls8) {$S^0_1$};
\end{scriptsize}
\node at (label) {$\pi^R$};

\draw (0,0) circle (2cm);

\draw[very thick,<-] (s3) -- (s1);
\draw[very thick,red,<-] (s7) -- (s2);
\draw[very thick,<-] (s6) -- (s4);
\draw[very thick,<-] (s5) -- (s8);

\draw[thick,red] ([shift=(90:2.0cm)]0,0) arc (90:290:2.0cm);

\draw[white] (-2.8,-2.8)--(-2.8,-2);
\draw[white] (2.8,2.8)--(2.8,2);
\end{tikzpicture}
\begin{tikzpicture}[xscale=-0.6,yscale=0.6,>=latex,shorten >=-0.4pt,shorten <=-0.4pt]
\coordinate (label) at (-2.2,-2.3) {};
\tikzstyle{every node}=[circle,minimum size=15pt,inner sep=0.5,draw];
\begin{scriptsize}
\node (V) at (0.0,0) {$V$};
\end{scriptsize}
\tikzstyle{every node}=[inner sep=1pt]
\begin{scriptsize}
\node (s1) at ($(center)+(0:2.0cm)$) {$S^0_1$};
\node (s2) at ($(center)+(45:2.0cm)$) {$S^0_2$};
\node (s3) at ($(center)+(90:2.0cm)$) {$S^0_3$};
\node (s4) at ($(center)+(135:2.0cm)$) {$S^1_2$};
\node (s5) at ($(center)+(180:2.0cm)$) {$S^0_4$};
\node (s6) at ($(center)+(225:2.0cm)$) {$S^1_1$};
\node (s7) at ($(center)+(270:2.0cm)$) {$S^1_4$};
\node (s8) at ($(center)+(315:2.0cm)$) {$S^0_3$};
\end{scriptsize}
\node at (label) {$\pqstree_{\pi^R}$};
\path (V) edge (s1); 
\path (V) edge (s2); 
\path (V) edge (s3); 
\path (V) edge (s4); 
\path (V) edge (s5); 
\path (V) edge (s6); 
\path (V) edge (s7); 
\path (V) edge (s8); 
\draw[white] (-2.8,-2.8)--(-2.8,-2);
\draw[white] (2.8,2.8)--(2.8,2);
\end{tikzpicture}
\caption{\label{fig:Pi-prime} Two members, $\pi$ and its reflection~$\pi^R$, of the set $\Pi$ and the ordered PQS-trees $\pqstree_{\pi}$ and $\pqstree_{\pi^R}$ representing $\pi$ and $\pi^R$ in the case when $V$ is prime in $\strongModules(G_{ov})$.}
\end{figure}

\end{center}

\subsubsection{$V$ is parallel in $\strongModules(G_{ov})$}
In this case the children of $V$ in $\strongModules(G_{ov})$ correspond to the 
connected components of $G_{ov}$, which in turn correspond to the Q-nodes in the PQS-tree~$\pqstree$.
We recall that $\camodules(Q)$ forms a partition of $Q$, for every Q-node $Q$ of $\pqstree$.

\begin{figure}[htp!]
\begin{tikzpicture}[xscale=0.8,yscale=0.8,>=latex,shorten >=-0.4pt,shorten <=-0.4pt]
\coordinate (center) at (0,0) {};
\coordinate (label) at (-2.5,-2.5) {};

\coordinate (s10) at ($(center)+(120:2.0cm)$) {};
\coordinate (s11) at ($(center)+(240:2.0cm)$) {};

\coordinate (s20) at ($(center)+(210:2.0cm)$) {};
\coordinate (s21) at ($(center)+(330:2.0cm)$) {};

\coordinate (m20) at ($(center)+(50:2.0cm)$) {};
\coordinate (m21) at ($(center)+(100:2.0cm)$) {};

\coordinate (m30) at ($(center)+(40:2.0cm)$) {};
\coordinate (m31) at ($(center)+(-10:2.0cm)$) {};

\coordinate (m40) at ($(center)+(260:2.0cm)$) {};
\coordinate (m41) at ($(center)+(310:2.0cm)$) {};

\coordinate (m50) at ($(center)+(140:2.0cm)$) {};
\coordinate (m51) at ($(center)+(190:2.0cm)$) {};

\coordinate (lp1) at ($(center)+(45:0.5cm)$) {};
\coordinate (lp3) at ($(center)+(190:1.5cm)$) {};
\coordinate (lp2) at ($(center)+(260:1.5cm)$) {};

\coordinate (ls10) at ($(center)+(120:2.4cm)$) {};
\coordinate (ls11) at ($(center)+(240:2.4cm)$) {};

\coordinate (ls20) at ($(center)+(210:2.4cm)$) {};
\coordinate (ls21) at ($(center)+(330:2.4cm)$) {};

\coordinate (lm20) at ($(center)+(53:2.4cm)$) {};
\coordinate (lm21) at ($(center)+(100:2.4cm)$) {};

\coordinate (lm31) at ($(center)+(-10:2.4cm)$) {};
\coordinate (lm30) at ($(center)+(37:2.4cm)$) {};

\coordinate (lm40) at ($(center)+(260:2.4cm)$) {};
\coordinate (lm41) at ($(center)+(310:2.4cm)$) {};

\coordinate (lm50) at ($(center)+(140:2.4cm)$) {};
\coordinate (lm51) at ($(center)+(190:2.4cm)$) {};

\tikzstyle{every node}=[inner sep=1pt]
\begin{scriptsize}
\node[green] at (lp1) {$P_1$};
\node[blue] at (lp2) {$P_2$};
\node[red] at (lp3) {$P_3$};

\node at (ls10) {$S^0_1$};
\node at (ls11) {$S^1_1$};

\node at (ls20) {$S^0_2$};
\node at (ls21) {$S^1_2$};

\node at (lm20) {$Q^0_2$};
\node at (lm21) {$Q^1_2$};

\node at (lm30) {$Q^0_3$};
\node at (lm31) {$Q^1_3$};

\node at (lm40) {$Q^0_4$};
\node at (lm41) {$Q^1_4$};

\node at (lm50) {$Q^0_5$};
\node at (lm51) {$Q^1_5$};
\end{scriptsize}
\node at (label) {$\pi$};

\draw (0,0) circle (2cm);

\draw[very thick,->] (s10) -- (s11);
\draw[very thick,->] (s20) -- (s21);
\draw[very thick,lightgray,->] (m20) -- (m21);
\draw[very thick,lightgray,->] (m30) -- (m31);
\draw[very thick,lightgray,->] (m40) -- (m41);
\draw[very thick,lightgray,->] (m50) -- (m51);

\draw[very thick,green,|-|] ([shift=(-15:2cm)]0,0) arc (-15:105:2cm);
\draw[very thick,red,|-|] ([shift=(135:2cm)]0,0) arc (135:195:2cm);
\draw[very thick,blue,|-|] ([shift=(255:2cm)]0,0) arc (255:315:2cm);

\draw[white] (-3.3,-2.8)--(-3.3,-2);
\draw[white] (3.3,2.8)--(3.3,2);

\end{tikzpicture}
\begin{tikzpicture}[xscale=0.8,yscale=0.8,>=latex,shorten >=-0.4pt,shorten <=-0.4pt]
\draw[black,dashed] (0,-2.7)--(0,2.7);
\draw[white] (-0.25,-2.7)--(-0.25,-2);
\draw[white] (0.25,2.7)--(0.25,2);
\end{tikzpicture}
\begin{tikzpicture}[xscale=-0.8,yscale=0.8,>=latex,shorten >=-0.4pt,shorten <=-0.4pt]
\coordinate (center) at (0,0) {};
\coordinate (label) at (2.5,-2.5) {};

\coordinate (s10) at ($(center)+(120:2.0cm)$) {};
\coordinate (s11) at ($(center)+(240:2.0cm)$) {};

\coordinate (s20) at ($(center)+(210:2.0cm)$) {};
\coordinate (s21) at ($(center)+(330:2.0cm)$) {};

\coordinate (m20) at ($(center)+(50:2.0cm)$) {};
\coordinate (m21) at ($(center)+(100:2.0cm)$) {};

\coordinate (m30) at ($(center)+(40:2.0cm)$) {};
\coordinate (m31) at ($(center)+(-10:2.0cm)$) {};

\coordinate (m40) at ($(center)+(260:2.0cm)$) {};
\coordinate (m41) at ($(center)+(310:2.0cm)$) {};

\coordinate (m50) at ($(center)+(140:2.0cm)$) {};
\coordinate (m51) at ($(center)+(190:2.0cm)$) {};

\coordinate (lp1) at ($(center)+(45:0.5cm)$) {};
\coordinate (lp3) at ($(center)+(190:1.5cm)$) {};
\coordinate (lp2) at ($(center)+(260:1.5cm)$) {};

\coordinate (ls10) at ($(center)+(120:2.4cm)$) {};
\coordinate (ls11) at ($(center)+(240:2.4cm)$) {};

\coordinate (ls20) at ($(center)+(210:2.4cm)$) {};
\coordinate (ls21) at ($(center)+(330:2.4cm)$) {};

\coordinate (lm20) at ($(center)+(53:2.4cm)$) {};
\coordinate (lm21) at ($(center)+(100:2.4cm)$) {};

\coordinate (lm31) at ($(center)+(-10:2.4cm)$) {};
\coordinate (lm30) at ($(center)+(37:2.4cm)$) {};

\coordinate (lm40) at ($(center)+(260:2.4cm)$) {};
\coordinate (lm41) at ($(center)+(310:2.4cm)$) {};

\coordinate (lm50) at ($(center)+(140:2.4cm)$) {};
\coordinate (lm51) at ($(center)+(190:2.4cm)$) {};

\tikzstyle{every node}=[inner sep=1pt]
\begin{scriptsize}
\node[green] at (lp1) {$P_1$};
\node[blue] at (lp2) {$P_2$};
\node[red] at (lp3) {$P_3$};

\node at (ls10) {$S^1_1$};
\node at (ls11) {$S^0_1$};

\node at (ls20) {$S^1_2$};
\node at (ls21) {$S^0_2$};

\node at (lm20) {$Q^0_3$};
\node at (lm21) {$Q^1_3$};

\node at (lm30) {$Q^0_2$};
\node at (lm31) {$Q^1_2$};

\node at (lm40) {$Q^1_4$};
\node at (lm41) {$Q^0_4$};

\node at (lm50) {$Q^1_5$};
\node at (lm51) {$Q^0_5$};

\end{scriptsize}
\node at (label) {$\pi'$};

\draw (0,0) circle (2cm);

\draw[very thick,<-] (s10) -- (s11);
\draw[very thick,<-] (s20) -- (s21);
\draw[very thick,lightgray,->] (m20) -- (m21);
\draw[very thick,lightgray,->] (m30) -- (m31);
\draw[very thick,lightgray,<-] (m40) -- (m41);
\draw[very thick,lightgray,<-] (m50) -- (m51);

\draw[very thick,green,|-|] ([shift=(-15:2cm)]0,0) arc (-15:105:2cm);
\draw[very thick,red,|-|] ([shift=(135:2cm)]0,0) arc (135:195:2cm);
\draw[very thick,blue,|-|] ([shift=(255:2cm)]0,0) arc (255:315:2cm);

\draw[white] (-3.3,-2.8)--(-3.3,-2);
\draw[white] (3.3,2.8)--(3.3,2);
\end{tikzpicture}

\vspace{0.5cm}
\begin{tikzpicture}[scale=0.8,>=latex,shorten >=-0.4pt,shorten <=-0.4pt]
  \tikzstyle{every node}=[circle,minimum size=10pt,inner sep=0.5,draw];
\coordinate (label) at (-2.5,-2.5) {};
  \begin{scriptsize}
  \node (m1) at (0,0) {$Q_1$};
  \node (m2) at (2,1.5) {$Q_2$};
  \node (m3) at (2,0.25) {$Q_3$};
  \node (m4) at (2,-1) {$Q_4$};
  \node (m5) at (-2,0.0) {$Q_5$};
  \end{scriptsize}
  \tikzstyle{every node}=[circle,minimum size=10pt,inner sep=0.5,draw];
  \begin{scriptsize}
  \node[green] (p1) at (1,1.0) {$P_1$};
  \node[blue] (p2) at (1,-1.0) {$P_2$};
  \node[red] (p3) at (-1.0,0) {$P_3$};
  \end{scriptsize}
\tikzstyle{every node}=[inner sep=1pt]
  \begin{scriptsize}
  \node (m12) at (2,2.5) {$Q^1_2$};
  \node (m02) at (3,1.5) {$Q^0_2$};
  \node (m03) at (3,0.7) {$Q^0_3$};
  \node (m13) at (3,-0.2) {$Q^1_3$};
  \node (m14) at (3,-1) {$Q^1_4$};
  \node (m04) at (2,-2) {$Q^0_4$};

  \node (s01) at (0,1) {$S^0_1$};
  \node (s11) at (0,-1) {$S^1_1$};
  \node (s02) at (-1,-1) {$S^0_2$};
  \node (s12) at (1,0) {$S^1_2$};
  \node (m05) at (-2,1) {$Q^0_5$};
  \node (m15) at (-2,-1) {$Q^1_5$};

  \end{scriptsize}
  \node at (label) {$\pqstree_{\pi}$};

\path (m1) edge (p1); 
\path (m1) edge (p2); 
\path (m1) edge (p3); 
\path (m2) edge (m02); 
\path (m2) edge (m12); 
\path (m3) edge (m03); 
\path (m3) edge (m13); 
\path (m4) edge (m04); 
\path (m4) edge (m14); 
\path (m5) edge (m05); 
\path (m5) edge (m15); 

\path (m1) edge (s01); 
\path (m1) edge (s11); 
\path (m1) edge (s02); 
\path (m1) edge (s12); 

\path (p1) edge (m2); 
\path (p1) edge (m3); 
\path (p2) edge (m4); 
\path (p3) edge (m5); 

\draw[white] (-2.8,-2.8)--(-2.8,-2);
\draw[white] (3.8,2.8)--(3.8,2);
\end{tikzpicture}
\begin{tikzpicture}[xscale=0.8,yscale=0.8,>=latex,shorten >=-0.4pt,shorten <=-0.4pt]
\draw[black,dashed] (0,-2.7)--(0,2.7);
\draw[white] (-0.25,-2.7)--(-0.25,-2);
\draw[white] (0.25,2.7)--(0.25,2);
\end{tikzpicture}
\begin{tikzpicture}[yscale=0.8,xscale=-0.8,>=latex,shorten >=-0.4pt,shorten <=-0.4pt]
\coordinate (label) at (3.5,-2.5) {};
  \tikzstyle{every node}=[circle,minimum size=10pt,inner sep=0.5,draw];
  \begin{scriptsize}
  \node (m1) at (0,0) {$Q_1$};
  \node (m2) at (2,1.5) {$Q_3$};
  \node (m3) at (2,0.25) {$Q_2$};
  \node (m4) at (2,-1) {$Q_4$};
  \node (m5) at (-2,0.0) {$Q_5$};
  \end{scriptsize}
  \tikzstyle{every node}=[circle,minimum size=10pt,inner sep=0.5,draw];
  \begin{scriptsize}
  \node[green] (p1) at (1,1.0) {$P_1$};
  \node[blue] (p2) at (1,-1.0) {$P_2$};
  \node[red] (p3) at (-1.0,0) {$P_3$};
  \end{scriptsize}
\tikzstyle{every node}=[inner sep=1pt]
  \begin{scriptsize}
  \node (m12) at (2,2.5) {$Q^1_3$};
  \node (m02) at (3,1.5) {$Q^0_3$};
  \node (m03) at (3,0.7) {$Q^0_2$};
  \node (m13) at (3,-0.2) {$Q^1_2$};
  \node (m14) at (3,-1) {$Q^0_4$};
  \node (m04) at (2,-2) {$Q^1_4$};

  \node (s01) at (0,1) {$S^1_1$};
  \node (s11) at (0,-1) {$S^0_1$};
  \node (s02) at (-1,-1) {$S^1_2$};
  \node (s12) at (1,0) {$S^0_2$};
  \node (m05) at (-2,1) {$Q^1_5$};
  \node (m15) at (-2,-1) {$Q^0_5$};

  \end{scriptsize}
  \node at (label) {$\pqstree_{\pi'}$};

\path (m1) edge (p1); 
\path (m1) edge (p2); 
\path (m1) edge (p3); 
\path (m2) edge (m02); 
\path (m2) edge (m12); 
\path (m3) edge (m03); 
\path (m3) edge (m13); 
\path (m4) edge (m04); 
\path (m4) edge (m14); 
\path (m5) edge (m05); 
\path (m5) edge (m15); 

\path (m1) edge (s01); 
\path (m1) edge (s11); 
\path (m1) edge (s02); 
\path (m1) edge (s12); 

\path (p1) edge (m2); 
\path (p1) edge (m3); 
\path (p2) edge (m4); 
\path (p3) edge (m5); 

\draw[white] (-2.8,-2.8)--(-2.8,-2);
\draw[white] (3.8,2.8)--(3.8,2);
\end{tikzpicture}

\caption{\label{fig:Q-node-reflection} Reflection of $Q_1$.
Circular orders of the slots $\pi$ and $\pi'$ and the ordered PQS-trees $\pqstree_{\pi}$ and $\pqstree_{\pi'}$ representing $\pi$ and $\pi'$. 
The tree $\pqstree_{\pi'}$ is obtained from $\pqstree_{\pi}$ by reflecting the node $Q_1$:
we have $\pi_{|Q_1} \equiv S^0_1P_1S^1_2P_2S^1_1S^0_2P_3$ and $\pi'_{|Q_1} \equiv (\pi_{|Q_1})^R \equiv P_3S^1_2S^0_1P_2S^0_2P_1S^1_1$.}
\end{figure}
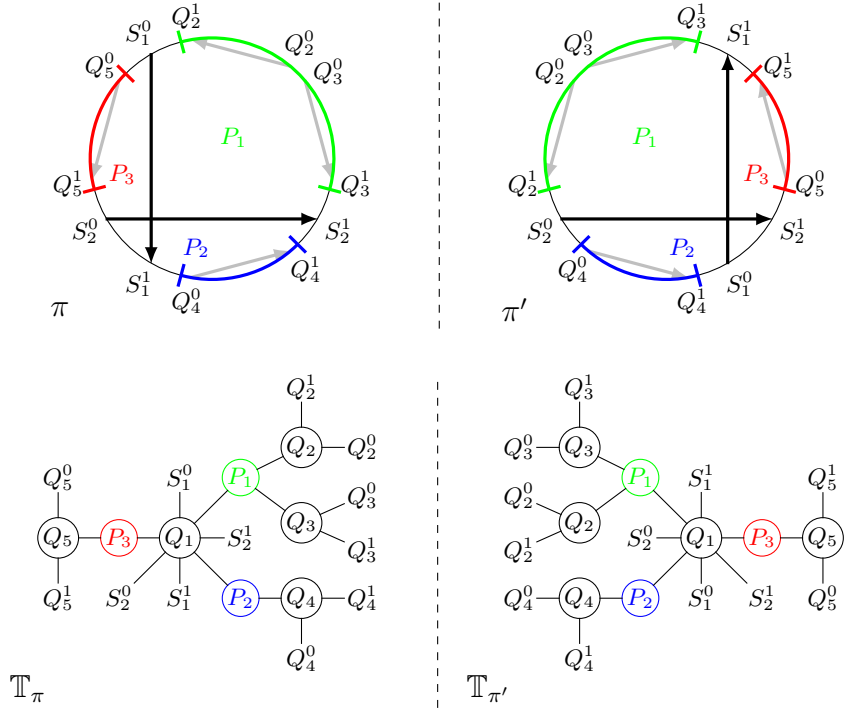

In Section~\ref{sec:proof_sketch} we show that:
\begin{itemize}
 \item for a P-node $P$ the set $\Pi(P)$ contains all circular orders of the set of the neighbours (all are Q-nodes) of~$P$ in~$\pqstree$,
 \item for a Q-node $Q$ the set $\Pi(Q)$ contains two circular permutations of the neighbours of 
 $Q$ in $\pqstree$, one being the reflection of the other.
 In particular, given any $\pi \in \Pi$, the set $\Pi(Q)$ contains $\pi_{|Q}$ and its reflection $(\pi_{|Q})^R$.
\end{itemize}
We note that we might have $\pi_{|Q} \equiv (\pi_{|Q})^R$, which takes place when $\camodules(Q)=\{Q\}$ (such components $Q$ are called \emph{permutation components} -- see the components $Q_2,Q_3,Q_4,Q_5$ in Figure~\ref{fig:PQS_tree}).
Note also that, given $\pi \in \Pi$ and $\pqstree_{\pi}$ representing $\pi$, 
any circular order of the slots in $\Pi$ can be obtained by performing a sequence of operations on the ordered PQS-tree $\pqstree_{\pi}$, where each of them either
\begin{itemize}
 \item reflects a Q-node, or
 \item permutes arbitrarily the neighbours of a P-node.
\end{itemize}
The \emph{reflection of a Q-node} $Q$ transforms $\pqstree_\pi$ into $\pqstree_{\pi'}$, 
where the only difference between $\pqstree_{\pi}$ and $\pqstree_{\pi'}$ is an ordering of the node $Q$:
$\pqstree_{\pi}$ orders the neighbours of $Q$ according to $\pi_{|Q}$ while
$\pqstree_{\pi'}$ orders the neighbours of $Q$ according to $(\pi_{|Q})^R$.
See Figure~\ref{fig:Q-node-reflection} for an illustration.

A \emph{permutation of a P-node} $P$ transforms $\pqstree_\pi$ into $\pqstree_{\pi'}$, 
where the only difference between $\pqstree_{\pi}$ and $\pqstree_{\pi'}$ is an ordering of the node $P$:
$\pqstree_{\pi}$ orders the neighbours of $Q$ consistently with $\pi_{|P}$ and 
$\pqstree_{\pi'}$ orders the neighbours of $Q$ consistently with $\pi'_{|P}$, where $\pi'_{|P}$ is any circular order of the neighbours of $P$.
See Figure~\ref{fig:P-node-permutation} for an illustration.
\begin{figure}[htp!]
\begin{tikzpicture}[xscale=0.8,yscale=0.8,>=latex,shorten >=-0.4pt,shorten <=-0.4pt]
\coordinate (center) at (0,0) {};
\coordinate (label) at (-2.5,-2.5) {};

\coordinate (s10) at ($(center)+(120:2.0cm)$) {};
\coordinate (s11) at ($(center)+(240:2.0cm)$) {};

\coordinate (s20) at ($(center)+(210:2.0cm)$) {};
\coordinate (s21) at ($(center)+(330:2.0cm)$) {};

\coordinate (m20) at ($(center)+(50:2.0cm)$) {};
\coordinate (m21) at ($(center)+(100:2.0cm)$) {};

\coordinate (m30) at ($(center)+(40:2.0cm)$) {};
\coordinate (m31) at ($(center)+(-10:2.0cm)$) {};

\coordinate (m40) at ($(center)+(260:2.0cm)$) {};
\coordinate (m41) at ($(center)+(310:2.0cm)$) {};

\coordinate (m50) at ($(center)+(140:2.0cm)$) {};
\coordinate (m51) at ($(center)+(190:2.0cm)$) {};

\coordinate (lp1) at ($(center)+(45:0.5cm)$) {};
\coordinate (lp3) at ($(center)+(190:1.5cm)$) {};
\coordinate (lp2) at ($(center)+(260:1.5cm)$) {};

\coordinate (ls10) at ($(center)+(120:2.4cm)$) {};
\coordinate (ls11) at ($(center)+(240:2.4cm)$) {};

\coordinate (ls20) at ($(center)+(210:2.4cm)$) {};
\coordinate (ls21) at ($(center)+(330:2.4cm)$) {};

\coordinate (lm20) at ($(center)+(53:2.4cm)$) {};
\coordinate (lm21) at ($(center)+(100:2.4cm)$) {};

\coordinate (lm31) at ($(center)+(-10:2.4cm)$) {};
\coordinate (lm30) at ($(center)+(37:2.4cm)$) {};

\coordinate (lm40) at ($(center)+(260:2.4cm)$) {};
\coordinate (lm41) at ($(center)+(310:2.4cm)$) {};

\coordinate (lm50) at ($(center)+(140:2.4cm)$) {};
\coordinate (lm51) at ($(center)+(190:2.4cm)$) {};

\tikzstyle{every node}=[inner sep=1pt]
\begin{scriptsize}
\node at (lp1) {$P_1$};
\node at (lp2) {$P_2$};
\node at (lp3) {$P_3$};

\node at (ls10) {$S^0_1$};
\node at (ls11) {$S^1_1$};

\node at (ls20) {$S^0_2$};
\node at (ls21) {$S^1_2$};

\node at (lm20) {$Q^0_2$};
\node at (lm21) {$Q^1_2$};

\node at (lm30) {$Q^0_3$};
\node at (lm31) {$Q^1_3$};

\node at (lm40) {$Q^0_4$};
\node at (lm41) {$Q^1_4$};

\node at (lm50) {$Q^0_5$};
\node at (lm51) {$Q^1_5$};
\end{scriptsize}
\node at (label) {$\pi$};

\draw (0,0) circle (2cm);

\draw[very thick,lightgray,->] (s10) -- (s11);
\draw[very thick,lightgray,->] (s20) -- (s21);
\draw[very thick,lightgray,->] (m20) -- (m21);
\draw[very thick,lightgray,->] (m30) -- (m31);
\draw[very thick,lightgray,->] (m40) -- (m41);
\draw[very thick,lightgray,->] (m50) -- (m51);

\draw[very thick,red,|-|] ([shift=(115:2cm)]0,0) arc (115:335:2cm);
\draw[very thick,green,|-|] ([shift=(48:2cm)]0,0) arc (48:105:2cm);
\draw[very thick,blue,|-|] ([shift=(-15:2cm)]0,0) arc (-15:42:2cm);

\draw[white] (-3.3,-2.8)--(-3.3,-2);
\draw[white] (3.3,2.8)--(3.3,2);

\end{tikzpicture}
\begin{tikzpicture}[xscale=0.8,yscale=0.8,>=latex,shorten >=-0.4pt,shorten <=-0.4pt]
\draw[white] (-0.25,-2.7)--(-0.25,-2);
\draw[white] (0.25,2.7)--(0.25,2);
\end{tikzpicture}
\begin{tikzpicture}[xscale=0.8,yscale=0.8,>=latex,shorten >=-0.4pt,shorten <=-0.4pt]
\coordinate (center) at (0,0) {};
\coordinate (label) at (-2.5,-2.5) {};

\coordinate (s10) at ($(center)+(120:2.0cm)$) {};
\coordinate (s11) at ($(center)+(240:2.0cm)$) {};

\coordinate (s20) at ($(center)+(210:2.0cm)$) {};
\coordinate (s21) at ($(center)+(330:2.0cm)$) {};

\coordinate (m20) at ($(center)+(50:2.0cm)$) {};
\coordinate (m21) at ($(center)+(100:2.0cm)$) {};

\coordinate (m30) at ($(center)+(40:2.0cm)$) {};
\coordinate (m31) at ($(center)+(-10:2.0cm)$) {};

\coordinate (m40) at ($(center)+(260:2.0cm)$) {};
\coordinate (m41) at ($(center)+(310:2.0cm)$) {};

\coordinate (m50) at ($(center)+(140:2.0cm)$) {};
\coordinate (m51) at ($(center)+(190:2.0cm)$) {};

\coordinate (lp1) at ($(center)+(45:0.5cm)$) {};
\coordinate (lp3) at ($(center)+(190:1.5cm)$) {};
\coordinate (lp2) at ($(center)+(260:1.5cm)$) {};

\coordinate (ls10) at ($(center)+(120:2.4cm)$) {};
\coordinate (ls11) at ($(center)+(240:2.4cm)$) {};

\coordinate (ls20) at ($(center)+(210:2.4cm)$) {};
\coordinate (ls21) at ($(center)+(330:2.4cm)$) {};

\coordinate (lm20) at ($(center)+(53:2.4cm)$) {};
\coordinate (lm21) at ($(center)+(100:2.4cm)$) {};

\coordinate (lm31) at ($(center)+(-10:2.4cm)$) {};
\coordinate (lm30) at ($(center)+(37:2.4cm)$) {};

\coordinate (lm40) at ($(center)+(260:2.4cm)$) {};
\coordinate (lm41) at ($(center)+(310:2.4cm)$) {};

\coordinate (lm50) at ($(center)+(140:2.4cm)$) {};
\coordinate (lm51) at ($(center)+(190:2.4cm)$) {};

\tikzstyle{every node}=[inner sep=1pt]
\begin{scriptsize}
\node at (lp1) {$P_1$};
\node at (lp2) {$P_2$};
\node at (lp3) {$P_3$};

\node at (ls10) {$S^0_1$};
\node at (ls11) {$S^1_1$};

\node at (ls20) {$S^0_2$};
\node at (ls21) {$S^1_2$};

\node at (lm20) {$Q^1_3$};
\node at (lm21) {$Q^0_3$};

\node at (lm30) {$Q^1_2$};
\node at (lm31) {$Q^0_2$};

\node at (lm40) {$Q^0_4$};
\node at (lm41) {$Q^1_4$};

\node at (lm50) {$Q^0_5$};
\node at (lm51) {$Q^1_5$};
\end{scriptsize}
\node at (label) {$\pi'$};

\draw (0,0) circle (2cm);

\draw[very thick,lightgray,->] (s10) -- (s11);
\draw[very thick,lightgray,->] (s20) -- (s21);
\draw[very thick,lightgray,<-] (m20) -- (m21);
\draw[very thick,lightgray,<-] (m30) -- (m31);
\draw[very thick,lightgray,->] (m40) -- (m41);
\draw[very thick,lightgray,->] (m50) -- (m51);

\draw[very thick,red,|-|] ([shift=(115:2cm)]0,0) arc (115:335:2cm);
\draw[very thick,blue,|-|] ([shift=(48:2cm)]0,0) arc (48:105:2cm);
\draw[very thick,green,|-|] ([shift=(-15:2cm)]0,0) arc (-15:42:2cm);

\draw[white] (-3.3,-2.8)--(-3.3,-2);
\draw[white] (3.3,2.8)--(3.3,2);
\end{tikzpicture}

\vspace{0.5cm}
\begin{tikzpicture}[scale=0.8,>=latex,shorten >=-0.4pt,shorten <=-0.4pt]
  \tikzstyle{every node}=[circle,minimum size=10pt,inner sep=0.5,draw];
\coordinate (label) at (-2.5,-2.5) {};
  \begin{scriptsize}
  \node[red] (m1) at (0,0) {$Q_1$};
  \node[green] (m2) at (2,1.5) {$Q_2$};
  \node[blue] (m3) at (2,0.25) {$Q_3$};
  \node (m4) at (2,-1) {$Q_4$};
  \node (m5) at (-2,0.0) {$Q_5$};
  \end{scriptsize}
  \tikzstyle{every node}=[circle,minimum size=10pt,inner sep=0.5,draw];
  \begin{scriptsize}
  \node (p1) at (1,1.0) {$P_1$};
  \node (p2) at (1,-1.0) {$P_2$};
  \node (p3) at (-1.0,0) {$P_3$};
  \end{scriptsize}
\tikzstyle{every node}=[inner sep=1pt]
  \begin{scriptsize}
  \node (m12) at (2,2.5) {$Q^1_2$};
  \node (m02) at (3,1.5) {$Q^0_2$};
  \node (m03) at (3,0.7) {$Q^0_3$};
  \node (m13) at (3,-0.2) {$Q^1_3$};
  \node (m14) at (3,-1) {$Q^1_4$};
  \node (m04) at (2,-2) {$Q^0_4$};

  \node (s01) at (0,1) {$S^0_1$};
  \node (s11) at (0,-1) {$S^1_1$};
  \node (s02) at (-1,-1) {$S^0_2$};
  \node (s12) at (1,0) {$S^1_2$};
  \node (m05) at (-2,1) {$Q^0_5$};
  \node (m15) at (-2,-1) {$Q^1_5$};

  \end{scriptsize}
  \node at (label) {$\pqstree_{\pi}$};

\path (m1) edge (p1); 
\path (m1) edge (p2); 
\path (m1) edge (p3); 
\path (m2) edge (m02); 
\path (m2) edge (m12); 
\path (m3) edge (m03); 
\path (m3) edge (m13); 
\path (m4) edge (m04); 
\path (m4) edge (m14); 
\path (m5) edge (m05); 
\path (m5) edge (m15); 

\path (m1) edge (s01); 
\path (m1) edge (s11); 
\path (m1) edge (s02); 
\path (m1) edge (s12); 

\path (p1) edge (m2); 
\path (p1) edge (m3); 
\path (p2) edge (m4); 
\path (p3) edge (m5); 

\draw[white] (-2.8,-2.8)--(-2.8,-2);
\draw[white] (3.8,2.8)--(3.8,2);
\end{tikzpicture}
\begin{tikzpicture}[xscale=0.8,yscale=0.8,>=latex,shorten >=-0.4pt,shorten <=-0.4pt]
\draw[white] (-0.25,-2.7)--(-0.25,-2);
\draw[white] (0.25,2.7)--(0.25,2);
\end{tikzpicture}
\begin{tikzpicture}[scale=0.8,>=latex,shorten >=-0.4pt,shorten <=-0.4pt]
  \tikzstyle{every node}=[circle,minimum size=10pt,inner sep=0.5,draw];
\coordinate (label) at (-2.5,-2.5) {};
  \begin{scriptsize}
  \node[red] (m1) at (0,0) {$Q_1$};
  \node[blue] (m2) at (2,1.5) {$Q_3$};
  \node[green] (m3) at (2,0.25) {$Q_2$};
  \node (m4) at (2,-1) {$Q_4$};
  \node (m5) at (-2,0.0) {$Q_5$};
  \end{scriptsize}
  \tikzstyle{every node}=[circle,minimum size=10pt,inner sep=0.5,draw];
  \begin{scriptsize}
  \node (p1) at (1,1.0) {$P_1$};
  \node (p2) at (1,-1.0) {$P_2$};
  \node (p3) at (-1.0,0) {$P_3$};
  \end{scriptsize}
\tikzstyle{every node}=[inner sep=1pt]
  \begin{scriptsize}
  \node (m12) at (2,2.5) {$Q^0_3$};
  \node (m02) at (3,1.5) {$Q^1_3$};
  \node (m03) at (3,0.7) {$Q^1_2$};
  \node (m13) at (3,-0.2) {$Q^0_2$};
  \node (m14) at (3,-1) {$Q^1_4$};
  \node (m04) at (2,-2) {$Q^0_4$};

  \node (s01) at (0,1) {$S^0_1$};
  \node (s11) at (0,-1) {$S^1_1$};
  \node (s02) at (-1,-1) {$S^0_2$};
  \node (s12) at (1,0) {$S^1_2$};
  \node (m05) at (-2,1) {$Q^0_5$};
  \node (m15) at (-2,-1) {$Q^1_5$};

  \end{scriptsize}
  \node at (label) {$\pqstree_{\pi'}$};
\path (m1) edge (p1); 
\path (m1) edge (p2); 
\path (m1) edge (p3); 
\path (m2) edge (m02); 
\path (m2) edge (m12); 
\path (m3) edge (m03); 
\path (m3) edge (m13); 
\path (m4) edge (m04); 
\path (m4) edge (m14); 
\path (m5) edge (m05); 
\path (m5) edge (m15); 

\path (m1) edge (s01); 
\path (m1) edge (s11); 
\path (m1) edge (s02); 
\path (m1) edge (s12); 

\path (p1) edge (m2); 
\path (p1) edge (m3); 
\path (p2) edge (m4); 
\path (p3) edge (m5); 

\draw[white] (-2.8,-2.8)--(-2.8,-2);
\draw[white] (3.8,2.8)--(3.8,2);
\end{tikzpicture}

\caption{\label{fig:P-node-permutation} Permuting the neighbours of $P_1$.
Circular orders of the slots $\pi$ and $\pi'$ and the ordered trees $\pqstree_{\pi}$ and $\pqstree_{\pi'}$ representing $\pi$ and $\pi'$. 
The tree $\pqstree_{\pi'}$ is obtained from $\pqstree_{\pi}$ by permuting the neigbours of $P_1$:
we have $\pi_{|P_1} \equiv Q_1Q_2Q_3$ and $\pi'_{|P_1} \equiv Q_1Q_3Q_2$, where $\camodules(Q_1) = \{S_1,S_2\}$.}
\end{figure}

\subsection{The structure of the admissible models of a metachord}
\label{sub:admissible_models}
Let $S$ be a CA-module of $G$ and let 
$\SSS = (S^0,S^1,{<_{S}})$ be the metachord associated with~$S$.
Clearly, the subgraph of $G$ induced by the set $S$ is co-bipartite. 
Hence, the structure of the admissible models for $S$ can be characterized based on Spinrad's work \cite{Spin88}, as described in Section~\ref{sec:related_work}.
We recall that, due to Theorem~\ref{thm:permutation_models_transitive_orientations}, the models admissible by $\mathbb{S}$ are in the correspondence with the transitive orientations of the permutation graph $(S,{\sim})$.

\begin{figure}[ht]
\begin{tikzpicture}[xscale=0.75,yscale=0.75,>=latex]
\coordinate (a_v1) at (0,3) {};
\coordinate (a_v2) at (1,3) {};
\coordinate (a_v3) at (2,3) {};
\coordinate (a_v4) at (2.5,3) {};
\coordinate (a_v5) at (3.4,3) {};
\coordinate (a_v6) at (3.8,3) {};
\coordinate (a_v7) at (5.2,3) {};
\coordinate (a_v8) at (5.8,3) {};
\coordinate (a_v9) at (7,3) {};

\coordinate (b_v2) at (7,0) {};
\coordinate (b_v4) at (6,0) {};
\coordinate (b_v3) at (5.5,0) {};
\coordinate (b_v6) at (4.5,0) {};
\coordinate (b_v5) at (4.1,0) {};
\coordinate (b_v9) at (2.5,0) {};
\coordinate (b_v1) at (1.5,0) {};
\coordinate (b_v8) at (0.6,0) {};
\coordinate (b_v7) at (0,0) {};

\coordinate (la_v1) at (0,3.3) {};
\coordinate (la_v2) at (1,3.3) {};
\coordinate (la_v3) at (2,3.3) {};
\coordinate (la_v4) at (2.5,3.3) {};
\coordinate (la_v5) at (3.4,3.3) {};
\coordinate (la_v6) at (3.8,3.3) {};
\coordinate (la_v7) at (5.2,3.3) {};
\coordinate (la_v8) at (5.8,3.3) {};
\coordinate (la_v9) at (7,3.3) {};

\coordinate (lb_v2) at (7,-0.3) {};
\coordinate (lb_v4) at (6,-0.3) {};
\coordinate (lb_v3) at (5.5,-0.3) {};
\coordinate (lb_v6) at (4.5,-0.3) {};
\coordinate (lb_v5) at (4.1,-0.3) {};
\coordinate (lb_v9) at (2.5,-0.3) {};
\coordinate (lb_v1) at (1.5,-0.3) {};
\coordinate (lb_v8) at (0.6,-0.3) {};
\coordinate (lb_v7) at (0,-0.3) {};

\coordinate (lS0) at (3.5,5.1) {};

\coordinate (tau0) at (7.9,3);
\coordinate (tau1) at (-0.9,0);

\tikzstyle{every node}=[inner sep=2pt,fill=white]

\coordinate (lS0) at (3.5,5.4) {};
\draw[dotted,|-|] (-0.3,5.1) -- (7.3,5.1);

\coordinate (lS1) at (3.5,-2.4) {};
\draw[dotted,|-|] (-0.3,-2.1) -- (7.3,-2.1);

\draw[fill=gray!30, draw=none] (-0.2,3) -- (0.2,3) -- (1.7,0) -- (1.3,0) -- cycle;
\coordinate (lA01) at (0,4.7) {};
\draw[dotted,|-|] (-0.25,4.4) -- (0.25,4.4);
\coordinate (lA11) at (1.5,-1.75) {};
\draw[dotted,|-|] (1.25,-1.4) -- (1.75,-1.4);

\draw[fill=gray!30, draw=none] (0.7,3) -- (4.1,3) -- (7.3,0) -- (3.8,0) -- cycle;
\coordinate (lA02) at (2.25,4.7) {};
\draw[dotted,thick,|-|] (0.7,4.4) -- (4.1,4.4);
\coordinate (lA12) at (5.55,-1.75) {};
\draw[dotted,thick,|-|] (3.8,-1.4) -- (7.3,-1.4);

\draw[fill=gray!30, draw=none] (4.9,3) -- (6.1,3) -- (0.9,0) -- (-0.3,0) -- cycle;
\coordinate (lA03) at (5.5,4.7) {};
\draw[dotted,|-|] (4.9,4.4) -- (6.1,4.4);
\coordinate (lA13) at (0.3,-1.75) {};
\draw[dotted,|-|] (-0.3,-1.4) -- (0.9,-1.4);

\draw[fill=gray!30, draw=none] (6.7,3) -- (7.3,3) -- (2.8,0) -- (2.2,0) -- cycle;
\coordinate (lA04) at (7,4.7) {};
\draw[dotted,|-|] (6.7,4.4) -- (7.3,4.4);
\coordinate (lA14) at (2.5,-1.75) {};
\draw[dotted,|-|] (2.2,-1.4) -- (2.8,-1.4);

\draw[fill=gray!80, draw=none] (0.7,3) -- (1.3,3) -- (7.3,0) -- (6.7,0) -- cycle;
\coordinate (lB01) at (1,4) {};
\draw[dotted,|-|] (0.7,3.7) -- (1.3,3.7);
\coordinate (lB11) at (7,-1.05) {};
\draw[dotted,|-|] (6.7,-0.7) -- (7.3,-0.7);

\draw[fill=gray!80, draw=none] (1.8,3) -- (2.7,3) -- (6.2,0) -- (5.3,0) -- cycle;
\coordinate (lB02) at (2.25,4) {};
\draw[dotted,thick,|-|] (1.8,3.7) -- (2.7,3.7);
\coordinate (lB12) at (5.75,-1.05) {};
\draw[dotted,thick,|-|] (5.3,-0.7) -- (6.2,-0.7);

\draw[fill=gray!80, draw=none] (3.2,3) -- (4,3) -- (4.7,0) -- (3.9,0) -- cycle;
\coordinate (lB03) at (3.6,4) {};
\draw[dotted,|-|] (3.2,3.7) -- (4,3.7);
\coordinate (lB13) at (4.3,-1.05) {};
\draw[dotted,|-|] (3.9,-0.7) -- (4.7,-0.7);

\draw[<-] (a_v1)--(b_v1);
\draw[->] (a_v2)--(b_v2);
\draw[<-] (a_v3)--(b_v3);
\draw[->] (a_v4)--(b_v4);
\draw[<-] (a_v5)--(b_v5);
\draw[->] (a_v6)--(b_v6);
\draw[<-] (a_v7)--(b_v7);
\draw[->] (a_v8)--(b_v8);
\draw[->] (a_v9)--(b_v9);

\draw[->] (-0.5,3) -- (7.5,3);
\draw[<-] (-0.5,0) -- (7.5,0);

\tikzstyle{every node}=[inner sep=1pt]
\begin{tiny}
\node at (lb_v1) {$v^0_1$};
\node at (lb_v2) {$v^1_2$};
\node at (lb_v3) {$v^0_3$};
\node at (lb_v4) {$v^1_4$};
\node at (lb_v5) {$v^0_5$};
\node at (lb_v6) {$v^1_6$};
\node at (lb_v7) {$v^0_7$};
\node at (lb_v8) {$v^1_8$};
\node at (lb_v9) {$v^1_9$};

\node at (la_v1) {$v^1_1$};
\node at (la_v2) {$v^0_2$};
\node at (la_v3) {$v^1_3$};
\node at (la_v4) {$v^0_4$};
\node at (la_v5) {$v^1_5$};
\node at (la_v6) {$v^0_6$};
\node at (la_v7) {$v^1_7$};
\node at (la_v8) {$v^0_8$};
\node at (la_v9) {$v^0_9$};

\node at (lB01) {$B^0_1$};
\node at (lB02) {$B^0_2$};
\node at (lB03) {$B^0_3$};

\node at (lB11) {$B^1_1$};
\node at (lB12) {$B^1_2$};
\node at (lB13) {$B^1_3$};

\node at (lA01) {$A^0_1$};
\node at (lA02) {$A^0_2$};
\node at (lA03) {$A^0_3$};
\node at (lA04) {$A^0_4$};

\node at (lA11) {$A^1_1$};
\node at (lA12) {$A^1_2$};
\node at (lA13) {$A^1_3$};
\node at (lA14) {$A^1_4$};

\node at (lS0) {$S^0$};
\node at (lS1) {$S^1$};

\node at (tau0) {$\tau^0$};
\node at (tau1) {$\tau^1$};
\end{tiny}
\end{tikzpicture}
\hspace{0.4cm}
\begin{tikzpicture}[xscale=0.75,yscale=0.75,>=latex]
\draw[dashed] (0,-4)--(0,4);
\end{tikzpicture}
\hspace{0.28cm}
\begin{tikzpicture}[xscale=-0.75,yscale=0.75,>=latex]
\coordinate (a_v1) at (0,3) {};
\coordinate (a_v2) at (1,3) {};
\coordinate (a_v3) at (2,3) {};
\coordinate (a_v4) at (2.5,3) {};
\coordinate (a_v5) at (3.4,3) {};
\coordinate (a_v6) at (3.8,3) {};
\coordinate (a_v7) at (5.2,3) {};
\coordinate (a_v8) at (5.8,3) {};
\coordinate (a_v9) at (7,3) {};

\coordinate (b_v2) at (7,0) {};
\coordinate (b_v4) at (6,0) {};
\coordinate (b_v3) at (5.5,0) {};
\coordinate (b_v6) at (4.5,0) {};
\coordinate (b_v5) at (4.1,0) {};
\coordinate (b_v9) at (2.5,0) {};
\coordinate (b_v1) at (1.5,0) {};
\coordinate (b_v8) at (0.6,0) {};
\coordinate (b_v7) at (0,0) {};

\coordinate (la_v1) at (0,3.3) {};
\coordinate (la_v2) at (1,3.3) {};
\coordinate (la_v3) at (2,3.3) {};
\coordinate (la_v4) at (2.5,3.3) {};
\coordinate (la_v5) at (3.4,3.3) {};
\coordinate (la_v6) at (3.8,3.3) {};
\coordinate (la_v7) at (5.2,3.3) {};
\coordinate (la_v8) at (5.8,3.3) {};
\coordinate (la_v9) at (7,3.3) {};

\coordinate (lb_v2) at (7,-0.3) {};
\coordinate (lb_v4) at (6,-0.3) {};
\coordinate (lb_v3) at (5.5,-0.3) {};
\coordinate (lb_v6) at (4.5,-0.3) {};
\coordinate (lb_v5) at (4.1,-0.3) {};
\coordinate (lb_v9) at (2.5,-0.3) {};
\coordinate (lb_v1) at (1.5,-0.3) {};
\coordinate (lb_v8) at (0.6,-0.3) {};
\coordinate (lb_v7) at (0,-0.3) {};

\coordinate (lS0) at (3.5,5.1) {};

\coordinate (tau0) at (7.9,0);
\coordinate (tau1) at (-0.9,3);

\tikzstyle{every node}=[inner sep=2pt,fill=white]

\coordinate (lS0) at (3.5,5.4) {};
\draw[dotted,|-|] (-0.3,5.1) -- (7.3,5.1);

\coordinate (lS1) at (3.5,-2.4) {};
\draw[dotted,|-|] (-0.3,-2.1) -- (7.3,-2.1);

\draw[fill=gray!30, draw=none] (-0.2,3) -- (0.2,3) -- (1.7,0) -- (1.3,0) -- cycle;
\coordinate (lA01) at (0,4.7) {};
\draw[dotted,|-|] (-0.25,4.4) -- (0.25,4.4);
\coordinate (lA11) at (1.5,-1.75) {};
\draw[dotted,|-|] (1.25,-1.4) -- (1.75,-1.4);

\draw[fill=gray!30, draw=none] (0.7,3) -- (4.1,3) -- (7.3,0) -- (3.8,0) -- cycle;
\coordinate (lA02) at (2.25,4.7) {};
\draw[dotted,thick,|-|] (0.7,4.4) -- (4.1,4.4);
\coordinate (lA12) at (5.55,-1.75) {};
\draw[dotted,thick,|-|] (3.8,-1.4) -- (7.3,-1.4);

\draw[fill=gray!30, draw=none] (4.9,3) -- (6.1,3) -- (0.9,0) -- (-0.3,0) -- cycle;
\coordinate (lA03) at (5.5,4.7) {};
\draw[dotted,|-|] (4.9,4.4) -- (6.1,4.4);
\coordinate (lA13) at (0.3,-1.75) {};
\draw[dotted,|-|] (-0.3,-1.4) -- (0.9,-1.4);

\draw[fill=gray!30, draw=none] (6.7,3) -- (7.3,3) -- (2.8,0) -- (2.2,0) -- cycle;
\coordinate (lA04) at (7,4.7) {};
\draw[dotted,|-|] (6.7,4.4) -- (7.3,4.4);
\coordinate (lA14) at (2.5,-1.75) {};
\draw[dotted,|-|] (2.2,-1.4) -- (2.8,-1.4);

\draw[fill=gray!80, draw=none] (0.7,3) -- (1.3,3) -- (7.3,0) -- (6.7,0) -- cycle;
\coordinate (lB01) at (1,4) {};
\draw[dotted,|-|] (0.7,3.7) -- (1.3,3.7);
\coordinate (lB11) at (7,-1.05) {};
\draw[dotted,|-|] (6.7,-0.7) -- (7.3,-0.7);

\draw[fill=gray!80, draw=none] (1.8,3) -- (2.7,3) -- (6.2,0) -- (5.3,0) -- cycle;
\coordinate (lB02) at (2.25,4) {};
\draw[dotted,thick,|-|] (1.8,3.7) -- (2.7,3.7);
\coordinate (lB12) at (5.75,-1.05) {};
\draw[dotted,green,thick,|-|] (5.3,-0.7) -- (6.2,-0.7);

\draw[fill=gray!80, draw=none] (3.2,3) -- (4,3) -- (4.7,0) -- (3.9,0) -- cycle;
\coordinate (lB03) at (3.6,4) {};
\draw[dotted,|-|] (3.2,3.7) -- (4,3.7);
\coordinate (lB13) at (4.3,-1.05) {};
\draw[dotted,|-|] (3.9,-0.7) -- (4.7,-0.7);

\draw[->] (a_v1)--(b_v1);
\draw[<-] (a_v2)--(b_v2);
\draw[->] (a_v3)--(b_v3);
\draw[<-] (a_v4)--(b_v4);
\draw[->] (a_v5)--(b_v5);
\draw[<-] (a_v6)--(b_v6);
\draw[->] (a_v7)--(b_v7);
\draw[<-] (a_v8)--(b_v8);
\draw[<-] (a_v9)--(b_v9);

\draw[<-] (-0.5,3) -- (7.5,3);
\draw[->] (-0.5,0) -- (7.5,0);

\tikzstyle{every node}=[inner sep=1pt]
\begin{tiny}
\node at (lb_v1) {$v^1_1$};
\node at (lb_v2) {$v^0_2$};
\node at (lb_v3) {$v^1_3$};
\node at (lb_v4) {$v^0_4$};
\node at (lb_v5) {$v^1_5$};
\node at (lb_v6) {$v^0_6$};
\node at (lb_v7) {$v^1_7$};
\node at (lb_v8) {$v^0_8$};
\node at (lb_v9) {$v^0_9$};

\node at (la_v1) {$v^0_1$};
\node at (la_v2) {$v^1_2$};
\node at (la_v3) {$v^0_3$};
\node at (la_v4) {$v^1_4$};
\node at (la_v5) {$v^0_5$};
\node at (la_v6) {$v^1_6$};
\node at (la_v7) {$v^0_7$};
\node at (la_v8) {$v^1_8$};
\node at (la_v9) {$v^1_9$};

\node at (lB01) {$B^1_1$};
\node at (lB02) {$B^1_2$};
\node at (lB03) {$B^1_3$};

\node at (lB11) {$B^0_1$};
\node at (lB12) {$B^0_2$};
\node at (lB13) {$B^0_3$};

\node at (lA01) {$A^1_1$};
\node at (lA02) {$A^1_2$};
\node at (lA03) {$A^1_3$};
\node at (lA04) {$A^1_4$};

\node at (lA11) {$A^0_1$};
\node at (lA12) {$A^0_2$};
\node at (lA13) {$A^0_3$};
\node at (lA14) {$A^0_4$};

\node at (lS0) {$S^1$};
\node at (lS1) {$S^0$};

\node at (tau0) {$\mu^0$};
\node at (tau1) {$\mu^1$};
\end{tiny}
\end{tikzpicture}
\caption{\label{fig:admissible-model-reflection} 
Admissible model $\tau = (\tau^0, \tau^1)$ for $\SSS$ (to the left) and its reflection $\mu = (\mu^0, \mu^1)$ (to the right).
We have $\tau_{|S} = (A^0_1A^0_2A^0_3A^0_4, A^1_2A^1_4A^1_1A^1_3)$ and $\mu_{|S}= (A^0_3A^0_1A^0_4A^0_2,A^1_4A^1_3A^1_2A^1_1)$, $\tau_{|A_2} = (B^0_1B^0_2B^0_3, B^1_1B^1_2B^1_3)$ and $\mu_{|A_2}= (B^0_3B^0_2B^0_1,B^1_3B^1_2B^1_1)$
}
\end{figure}
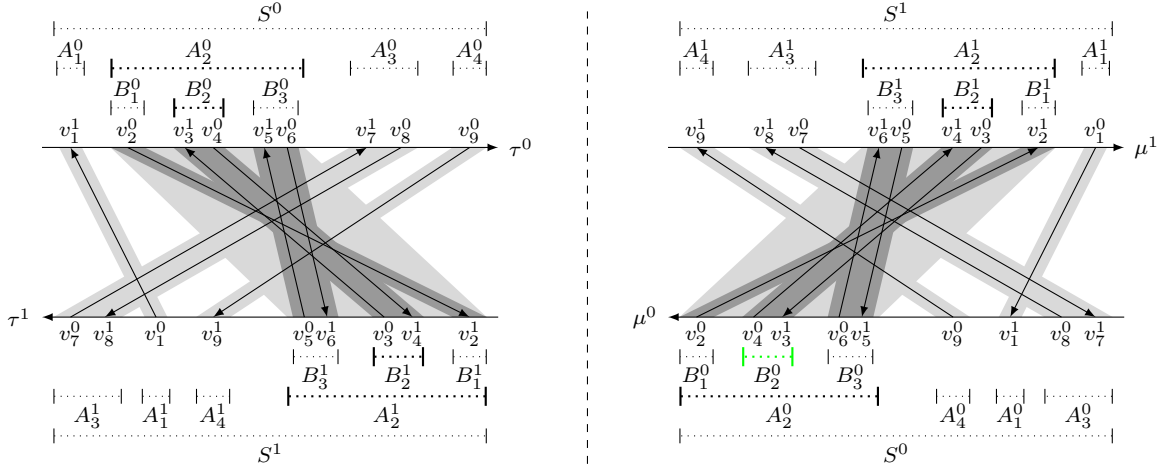

First, note that for any admissible model $\tau = (\tau^0,\tau^1)$ for $\SSS$
its reflection $\mu = (\mu^0,\mu^1)$ is also admissible by $\SSS$.
Note that: 
\begin{itemize}
 \item if $\tau$ and $\mu$ correspond to the transitive orientations 
${\prec_{\tau}}$ and ${\prec_{\mu}}$ of $(S,{\sim})$, then $\prec_{\mu}$ is the reverse of $\prec_{\tau}$,
 \item $\mu^0$ is the reflection of~$\tau^1$ and $\mu^1$ is the reflection of~$\tau^0$.
\end{itemize}
See Figure~\ref{fig:admissible-model-reflection} for an illustration.

From now, we call the modules in $\strongModules(S,{\sim})$ as \emph{nodes}   
and the inner nodes in $\strongModules(S,{\sim})$ as \emph{M-nodes} of the tree in $\strongModules(S,{\sim})$.
Since $(S,{<_S})$ might have a quadratic size, we need to find a way to represent the admissible way of $\SSS$ in linear space.
First, for a node $M \in \strongModules(S,{\sim})$ we define the \emph{metachord $\MMM$} of $M$ as the triple $(M^0,M^1,{<_M})$, where $M^0 = M^* \cap S^0$, $M^1 = M^* \cap S^1$, and ${<_M}$ equals to ${<_{S}}$ restricted to $M$.
Recall that for every admissible model $(\tau^0,\tau^1)$ for $\SSS$ and for every $j \in \{0,1\}$ 
the set $M^j$ is contiguous in~$\tau^j$, and for any child $K$ of $M$ in~$\strongModules(S,{\sim})$ the set $K^j$ is contiguous in the word~$\tau^j|M^j$.
Then, for every M-node~$M$ we define the set $\Pi(M)$ of possible \emph{orderings} of the children of 
$M$ in the words $\tau^0|M^0$ and $\tau^1|M^1$, that is, each member~$(\pi^0,\pi^1)$ of~$\Pi(M)$ 
determines the order in which the contiguous sets $K^0$ and $K^1$ associated with the children $K$ of $M$ 
may occur in the words $\tau^0|M^0$ and $\tau^1|M^1$, respectively.
Formally, for every transitive orientation ${\prec_M}$ of $(M,{\sim}_M)$ we have an ordering $\pi_M=(\pi^0, \pi^1)$ in the set $\Pi(M)$, where the word $\pi^j$ for $j \in \{0,1\}$ is a permutation of the set $\{K^j: K \text{ is a child of $M$ in } \strongModules(S,{\sim})\}$ such that for every two distinct children $K,L$ of $M$:
\begin{equation}
\begin{array}{lll}
K^0 \text{ occurs before } L^0 \text{ in } \pi^{0} \iff K \prec_M L \text{ or } K <_{M} L,\\
K^1 \text{ occurs before } L^1 \text{ in } \pi^{1} \iff K \prec_M L \text{ or } L <_{M} K.\\
\end{array}
\end{equation}
Figure~\ref{fig:admissible-model-structure} shows a modular decomposition tree $\strongModules(S,{\sim})$ for some CA-module $S$ and its admissible model $\tau = (\tau^0,\tau^1)$.
The M-node $S$ is prime, $(S,{\sim_S})$ has two transitive orientations, one being the reverse of the other,
and the set $\Pi(S)$ has two orderings: $(A^0_1A^0_2A^0_3A^0_4, A^1_2A^1_4A^1_1A^1_3)$ and its reflection $(A^0_3A^0_1A^0_4A^0_2, A^1_4A^1_3A^1_2A^1_1)$.
The M-node $A_2$ is serial with three children, $(A_2,{\sim_{A_2}})$ 
has $3!$ transitive orientations corresponding to the linear orders of its children $B_1,B_2,B_3$, 
and the set $\Pi(A_2)$ has $3!$ orderings $(B^0_iB^0_jB^0_k, B^1_iB^1_jB^1_k)$ corresponding to all permutations $i,j,k$ of the set $[3]$.
The remaining M-nodes $A_3, B_2, B_3$ are parallel, each graph $(A_3,{\sim_{A_3}})$, $(B_2,{\sim_{B_2}})$, $(B_3,{\sim_{B_3}})$ has one (empty) transitive orientation, and each set $\Pi(A_3)$, $\Pi(B_2)$, and $\Pi(B_3)$
has one possible ordering.
For example, we have $\Pi(A_3) = \big{\{} (B^0_4 B^0_5, B^1_5 B^1_4) \big{\}}$.

For any $M$-node $M$ we can represent the set $\Pi(M)$ in space linear in the number of children of $M$ in $\strongModules(S,{\sim})$
(for a serial M-node $M$ it suffices to represent the type of $M$ and one member of $\Pi(M)$).
Hence, we can represents all the sets $\Pi(M)$ for all $M$-nodes in space linear in the size of $\strongModules(S,{\sim})$, and hence linear in the size of $(S,{\sim})$.

Following our convention, the tree $\strongModules(S,{\sim})$ is said to be \emph{ordered} if every M-node in $\strongModules(S,{\sim})$ is assigned an ordering from the set~$\Pi(M)$.
Theorems~\ref{thm:transitive_orientations_versus_transitive_orientations_of_strong_modules} and~\ref{thm:permutation_models_transitive_orientations} assert one-to-one correspondence between admissible models~$\tau$ for~$\SSS$ and the ordered trees~$\strongModules(S,{\sim})$.
For every admissible model $\tau$ for $\SSS$, by $\tau_{|M}$ we denote the ordering of the node $M$ in the ordered tree $\strongModules(S,{\sim})_{\tau}$ corresponding to $\tau$ -- see Figure~\ref{fig:admissible-model-structure} for an illustration.

\begin{figure}[ht]
\begin{tikzpicture}[scale=0.7,>=latex]
\tikzstyle{every node}=[rectangle,minimum size=15pt,inner sep=0.5,draw];
  \begin{scriptsize}
  \node (M) at (4.0,7.5) {$S$};
  \node (v1) at (0.5,6) {$A_1 = \{v_1\}$};
  \node (A) at (3,6) {$A_2$};
  \node (B) at (5.5,6) {$A_3$};
  \node (v9) at (8,6) {$A_4 = \{v_9\}$};
  \node (v2) at (0.5,4.5) {$B_1 = \{v_2\}$};
  \node (C) at (2.5,4.5) {$B_2$};
  \node (D) at (3.5,4.5) {$B_3$};
  \node (v7) at (5.75,4.5) {$B_4=\{v_7\}$};
  \node (v8) at (8,4.5) {$B_5 = \{v_8\}$};
  \node (v3) at (0.5,3) {$C_1 = \{v_3\}$};
  \node (v4) at (2.75,3) {$C_2 = \{v_4\}$};
  \node (v5) at (5,3) {$C_3 = \{v_5\}$};
  \node (v6) at (7.25,3) {$C_4 = \{v_6\}$};
 
\end{scriptsize}
\tikzstyle{every node}=[square,minimum size=15pt,inner sep=0.5,draw];
\path (M) edge (v1); 
\path (M) edge (A); 
\path (M) edge (B); 
\path (M) edge (v9); 
\path (A) edge (v2); 
\path (A) edge (C); 
\path (A) edge (D); 
\path (B) edge (v7); 
\path (B) edge (v8); 
\path (C) edge (v3); 
\path (C) edge (v4); 
\path (D) edge (v5); 
\path (D) edge (v6); 

\draw[-,white] (-0.5,1) -- (0,1);
\end{tikzpicture}
\hspace{1cm}
\begin{tikzpicture}[xscale=0.75,yscale=0.75,>=latex]
\coordinate (a_v1) at (0,3) {};
\coordinate (a_v2) at (1,3) {};
\coordinate (a_v3) at (2,3) {};
\coordinate (a_v4) at (2.5,3) {};
\coordinate (a_v5) at (3.4,3) {};
\coordinate (a_v6) at (3.8,3) {};
\coordinate (a_v7) at (5.2,3) {};
\coordinate (a_v8) at (5.8,3) {};
\coordinate (a_v9) at (7,3) {};

\coordinate (b_v2) at (7,0) {};
\coordinate (b_v4) at (6,0) {};
\coordinate (b_v3) at (5.5,0) {};
\coordinate (b_v6) at (4.5,0) {};
\coordinate (b_v5) at (4.1,0) {};
\coordinate (b_v9) at (2.5,0) {};
\coordinate (b_v1) at (1.5,0) {};
\coordinate (b_v8) at (0.6,0) {};
\coordinate (b_v7) at (0,0) {};

\coordinate (la_v1) at (0,3.3) {};
\coordinate (la_v2) at (1,3.3) {};
\coordinate (la_v3) at (2,3.3) {};
\coordinate (la_v4) at (2.5,3.3) {};
\coordinate (la_v5) at (3.4,3.3) {};
\coordinate (la_v6) at (3.8,3.3) {};
\coordinate (la_v7) at (5.2,3.3) {};
\coordinate (la_v8) at (5.8,3.3) {};
\coordinate (la_v9) at (7,3.3) {};

\coordinate (lb_v2) at (7,-0.3) {};
\coordinate (lb_v4) at (6,-0.3) {};
\coordinate (lb_v3) at (5.5,-0.3) {};
\coordinate (lb_v6) at (4.5,-0.3) {};
\coordinate (lb_v5) at (4.1,-0.3) {};
\coordinate (lb_v9) at (2.5,-0.3) {};
\coordinate (lb_v1) at (1.5,-0.3) {};
\coordinate (lb_v8) at (0.6,-0.3) {};
\coordinate (lb_v7) at (0,-0.3) {};

\coordinate (lS0) at (3.5,5.1) {};

\coordinate (tau0) at (7.9,3);
\coordinate (tau1) at (-0.9,0);

\tikzstyle{every node}=[inner sep=2pt,fill=white]

\coordinate (lS0) at (3.5,5.4) {};
\draw[dotted,|-|] (-0.3,5.1) -- (7.3,5.1);

\coordinate (lS1) at (3.5,-2.4) {};
\draw[dotted,|-|] (-0.3,-2.1) -- (7.3,-2.1);

\draw[fill=gray!30, draw=none] (-0.2,3) -- (0.2,3) -- (1.7,0) -- (1.3,0) -- cycle;
\coordinate (lA01) at (0,4.7) {};
\draw[dotted,|-|] (-0.25,4.4) -- (0.25,4.4);
\coordinate (lA11) at (1.5,-1.75) {};
\draw[dotted,|-|] (1.25,-1.4) -- (1.75,-1.4);

\draw[fill=gray!30, draw=none] (0.7,3) -- (4.1,3) -- (7.3,0) -- (3.8,0) -- cycle;
\coordinate (lA02) at (2.25,4.7) {};
\draw[dotted,thick,|-|] (0.7,4.4) -- (4.1,4.4);
\coordinate (lA12) at (5.55,-1.75) {};
\draw[dotted,thick,|-|] (3.8,-1.4) -- (7.3,-1.4);

\draw[fill=gray!30, draw=none] (4.9,3) -- (6.1,3) -- (0.9,0) -- (-0.3,0) -- cycle;
\coordinate (lA03) at (5.5,4.7) {};
\draw[dotted,|-|] (4.9,4.4) -- (6.1,4.4);

\coordinate (lB04) at (5.15,4) {};
\draw[dotted,|-|] (4.9,3.7) -- (5.4,3.7);
\coordinate (lB05) at (5.85,4) {};
\draw[dotted,|-|] (5.6,3.7) -- (6.1,3.7);

\coordinate (lA13) at (0.3,-1.75) {};
\draw[dotted,|-|] (-0.3,-1.4) -- (0.9,-1.4);

\coordinate (lB14) at (-0.05,-1) {};
\draw[dotted,|-|] (-0.3,-0.7) -- (0.2,-0.7);
\coordinate (lB15) at (0.65,-1) {};
\draw[dotted,|-|] (0.4,-0.7) -- (0.9,-0.7);

\draw[fill=gray!30, draw=none] (6.7,3) -- (7.3,3) -- (2.8,0) -- (2.2,0) -- cycle;
\coordinate (lA04) at (7,4.7) {};
\draw[dotted,|-|] (6.7,4.4) -- (7.3,4.4);
\coordinate (lA14) at (2.5,-1.75) {};
\draw[dotted,|-|] (2.2,-1.4) -- (2.8,-1.4);

\draw[fill=gray!80, draw=none] (0.7,3) -- (1.3,3) -- (7.3,0) -- (6.7,0) -- cycle;
\coordinate (lB01) at (1,4) {};
\draw[dotted,|-|] (0.7,3.7) -- (1.3,3.7);
\coordinate (lB11) at (7,-1.05) {};
\draw[dotted,|-|] (6.7,-0.7) -- (7.3,-0.7);

\draw[fill=gray!80, draw=none] (1.8,3) -- (2.7,3) -- (6.2,0) -- (5.3,0) -- cycle;
\coordinate (lB02) at (2.25,4) {};
\draw[dotted,thick,|-|] (1.8,3.7) -- (2.7,3.7);
\coordinate (lB12) at (5.75,-1.05) {};
\draw[dotted,thick,|-|] (5.3,-0.7) -- (6.2,-0.7);

\draw[fill=gray!80, draw=none] (3.2,3) -- (4,3) -- (4.7,0) -- (3.9,0) -- cycle;
\coordinate (lB03) at (3.6,4) {};
\draw[dotted,|-|] (3.2,3.7) -- (4,3.7);
\coordinate (lB13) at (4.3,-1.05) {};
\draw[dotted,|-|] (3.9,-0.7) -- (4.7,-0.7);

\draw[<-] (a_v1)--(b_v1);
\draw[->] (a_v2)--(b_v2);
\draw[<-] (a_v3)--(b_v3);
\draw[->] (a_v4)--(b_v4);
\draw[<-] (a_v5)--(b_v5);
\draw[->] (a_v6)--(b_v6);
\draw[<-] (a_v7)--(b_v7);
\draw[->] (a_v8)--(b_v8);
\draw[->] (a_v9)--(b_v9);

\draw[->] (-0.5,3) -- (7.5,3);
\draw[<-] (-0.5,0) -- (7.5,0);

\tikzstyle{every node}=[inner sep=1pt]
\begin{tiny}
\node at (lb_v1) {$v^0_1$};
\node at (lb_v2) {$v^1_2$};
\node at (lb_v3) {$v^0_3$};
\node at (lb_v4) {$v^1_4$};
\node at (lb_v5) {$v^0_5$};
\node at (lb_v6) {$v^1_6$};
\node at (lb_v7) {$v^0_7$};
\node at (lb_v8) {$v^1_8$};
\node at (lb_v9) {$v^1_9$};

\node at (la_v1) {$v^1_1$};
\node at (la_v2) {$v^0_2$};
\node at (la_v3) {$v^1_3$};
\node at (la_v4) {$v^0_4$};
\node at (la_v5) {$v^1_5$};
\node at (la_v6) {$v^0_6$};
\node at (la_v7) {$v^1_7$};
\node at (la_v8) {$v^0_8$};
\node at (la_v9) {$v^0_9$};

\node at (lB01) {$B^0_1$};
\node at (lB02) {$B^0_2$};
\node at (lB03) {$B^0_3$};
\node at (lB04) {$B^0_4$};
\node at (lB05) {$B^0_5$};

\node at (lB11) {$B^1_1$};
\node at (lB12) {$B^1_2$};
\node at (lB13) {$B^1_3$};
\node at (lB14) {$B^1_4$};
\node at (lB15) {$B^1_5$};

\node at (lA01) {$A^0_1$};
\node at (lA02) {$A^0_2$};
\node at (lA03) {$A^0_3$};
\node at (lA04) {$A^0_4$};

\node at (lA11) {$A^1_1$};
\node at (lA12) {$A^1_2$};
\node at (lA13) {$A^1_3$};
\node at (lA14) {$A^1_4$};

\node at (lS0) {$S^0$};
\node at (lS1) {$S^1$};

\node at (tau0) {$\tau^0$};
\node at (tau1) {$\tau^1$};
\end{tiny}
\end{tikzpicture}
\caption{\label{fig:admissible-model-structure} Modular decomposition tree
$\strongModules(S,{\sim})$ of CA-module $S$ (to the left) and an admissible model $\tau = (\tau^0,\tau^1)$ for $\SSS$.
We have $\tau_{|S} = (A^0_1A^0_2A^0_3A^0_4, A^1_2A^1_4A^1_1A^1_3)$ and $\tau_{|A_2} = (B^0_1B^0_2B^0_3,B^1_1B^1_2B^1_3)$.}
\end{figure}

\subsection{PQSM-tree of $G$}
\label{sub:PQM_tree}
We obtain the \emph{PQSM-tree} $\pqsmtree$ of $G$ by attaching the root $S$ of the modular decomposition tree $\strongModules(S,{\sim})$ to the leaf slot $S^0$ of the PQS-tree $\pqstree$ of $G$, for every $S \in \mathcal{S}$.
A node $N$ of $\pqsmtree$ is a \emph{PQM-node} in $\pqsmtree$ if $N$ is either a P-node, or a Q-node, or an M-node.
Note that the sets~$\Pi(\cdot)$ are defined for all PQM-nodes in~$\pqsmtree$.
Following our convention, we say PQSM-tree~$\pqsmtree$ is \emph{ordered} if every PQM-node~$N$ in~$\pqsmtree$ is assigned an ordering in the set~$\Pi(N)$.
Clearly, there is one-to-one correspondence between the ordered PQSM-trees
and the conformal models of~$G_{ov}$.
For a conformal model~$\phi$ of~$G_{ov}$ and a PQM-node~$N$ in~$\pqsmtree$ 
by~$\phi_{|N}$ we denote the ordering of the node~$N$ in the ordered PQSM-tree~$\pqsmtree_{\phi}$ corresponding to $\phi$.

\section{Sketch of the proof}
\label{sec:proof_sketch}
Let $G=(V,E)$ be a circular-arc graph with no twins and no universal vertices and
let $G_{ov}=(V,{\sim})$ be the overlap graph of~$G$.

In order to prove that the data structure $\dataStructure$ introduced in the previous section represents all conformal models of $G_{ov}$, it is convenient to assume the following definition.
\begin{definition}
A circular word $\phi$ on $V^*$ is \emph{admissible by $\dataStructure = (\camodules, \slots, \metachords, \Pi)$} if
$\phi$ is obtained from some $\pi \in \Pi$ by substituting the slots 
$S^0_i$ and $S^1_i$ by words $\tau^0_i$ and $\tau^1_i$ for $i \in [t]$, 
where  $(\tau^0_i,\tau^1_i)$ is a permutation model of $(S_i,{\sim})$ admissible by $\SSS_i$. 
\end{definition}
Then, it suffices to show:
\begin{theorem}
\label{thm:main_theorem}
Let $\phi$ be a circular word over $V^*$.
The word $\phi$ is a conformal model of $G$ if and only if 
the word $\phi$ is admissible by $\dataStructure$.
\end{theorem}
We prove Theorem~\ref{thm:main_theorem} as follows.
The more difficult part is to prove the necessity: we need to show that any conformal model $\phi$ of $G_{ov}$ is admissible by $\dataStructure$.
Given the necessity, we can easily prove the sufficiency.
Since $G$ is a circular-arc graph, $G_{ov}$ admits a conformal model $\phi$, 
which is admissible by $\dataStructure$ (here we use the necessity).
Then, we observe that we can transform $\phi$ into any other admissible 
model for $\dataStructure$ by:
\begin{itemize}
 \item replacing admissible model $(\phi|S^0_i,\phi|S^1_i)$ with some other admissible model for $\SSS_i$, for any $i \in [t]$,
 \item replacing circular ordering of the slots $\pi(\phi)$ by any other circular order of the slots from~$\Pi$. 
 This operation might be performed in a few steps, by reflecting some Q-nodes (prime and parallel case), permuting some P-nodes (parallel case), or by reordering the slots arbitrarily as long as they overlap (serial case). 
\end{itemize}
We leave the reader to check that all those transformations keep the left/right relation between 
non-intersecting chords (operations of the first type) and non-intersecting metachords (operations of the second type), and hence transform one conformal model into another.
This completes the proof of the sufficiency in Theorem~\ref{thm:main_theorem}.

We are left to prove that every conformal model of $G_{ov}$ is admissible for $\dataStructure$.
We split into cases depending on the type of $V$ in $\strongModules(G_{ov})$.

\subsection{Serial case}
\label{subsec:conformal_models_serial}

The results of this section are inspired by the work of Hsu~\cite{Hsu95}.

In this subsection we describe the conformal models of $G_{ov}=(V,{\sim})$ for the case when
$V$ is serial in $\strongModules(G_{ov})$.
In passing, we also describe the structure of the conformal models of $(Q,{\sim})$, where
$Q$ is a serial child of parallel $V$.

Suppose $Q$ is a serial module in $\strongModules(G_{ov})$ such that either $Q=V$ or $Q$ is a child of parallel~$V$.
Suppose $M_1,\ldots,M_t$ are the children of $Q$ in $\strongModules(G_{ov})$.
Since $Q$ is serial, we have $M_i \sim M_j$ for every two distinct $i,j \in [t]$, 
$(M_i, \sim)$ is a permutation graph by Claim~\ref{claim:permutation_graphs_in_G_ov},
and $(M_i,{\parallel})$ is connected for every $i \in [t]$.
We pick a representant $r_i$ in every set $M_i$.

\begin{figure}[htp!]
\begin{tikzpicture}[scale=0.7]
    \coordinate (M1_center) at (-1.5,1.5) {};
    \coordinate (lM1) at ($(M1_center)+(180:1.5)$) {};
    
    \coordinate (M1r1) at ($(M1_center)+(15:1)$) {};
    \coordinate (M1r2) at ($(M1_center)+(0:1)$) {};
    \coordinate (M1r3) at ($(M1_center)+(-15:1)$) {};

    \coordinate (M1b1) at ($(M1_center)+(285:1)$) {};
    \coordinate (M1b2) at ($(M1_center)+(270:1)$) {};
    \coordinate (M1b3) at ($(M1_center)+(255:1)$) {};
    
    \coordinate (M1ul1) at ($(M1_center)+(150:1)$) {};
    \coordinate (M1ul2) at ($(M1_center)+(135:1)$) {};
    \coordinate (M1ul3) at ($(M1_center)+(120:1)$) {};

    \coordinate (M1br1) at ($(M1_center)+(330:1)$) {};
    \coordinate (M1br2) at ($(M1_center)+(315:1)$) {};
    \coordinate (M1br3) at ($(M1_center)+(300:1)$) {};

    \coordinate (M2_center) at (1.5,1.5) {};
    \coordinate (lM2) at ($(M2_center)+(0:1.5)$) {};
    
    \coordinate (M2l1) at ($(M2_center)+(195:1)$) {};
    \coordinate (M2l2) at ($(M2_center)+(180:1)$) {};
    \coordinate (M2l3) at ($(M2_center)+(165:1)$) {};

    \coordinate (M2b1) at ($(M2_center)+(285:1)$) {};
    \coordinate (M2b2) at ($(M2_center)+(270:1)$) {};
    \coordinate (M2b3) at ($(M2_center)+(255:1)$) {};
    
    \coordinate (M2ur1) at ($(M2_center)+(60:1)$) {};
    \coordinate (M2ur2) at ($(M2_center)+(45:1)$) {};
    \coordinate (M2ur3) at ($(M2_center)+(30:1)$) {};

    \coordinate (M2bl1) at ($(M2_center)+(240:1)$) {};
    \coordinate (M2bl2) at ($(M2_center)+(225:1)$) {};
    \coordinate (M2bl3) at ($(M2_center)+(210:1)$) {};

    \coordinate (M3_center) at (1.5,-1.5) {};
    \coordinate (lM3) at ($(M3_center)+(0:1.5)$) {};
    
    \coordinate (M3l1) at ($(M3_center)+(195:1)$) {};
    \coordinate (M3l2) at ($(M3_center)+(180:1)$) {};
    \coordinate (M3l3) at ($(M3_center)+(165:1)$) {};

    \coordinate (M3u1) at ($(M3_center)+(105:1)$) {};
    \coordinate (M3u2) at ($(M3_center)+(90:1)$) {};
    \coordinate (M3u3) at ($(M3_center)+(75:1)$) {};
    
    \coordinate (M3ul1) at ($(M3_center)+(150:1)$) {};
    \coordinate (M3ul2) at ($(M3_center)+(135:1)$) {};
    \coordinate (M3ul3) at ($(M3_center)+(120:1)$) {};
    
    \coordinate (M3br1) at ($(M3_center)+(330:1)$) {};
    \coordinate (M3br2) at ($(M3_center)+(315:1)$) {};
    \coordinate (M3br3) at ($(M3_center)+(300:1)$) {};
    
    \coordinate (M4_center) at (-1.5,-1.5) {};
    \coordinate (lM4) at ($(M4_center)+(180:1.5)$) {};
    
    \coordinate (M4r1) at ($(M4_center)+(15:1)$) {};
    \coordinate (M4r2) at ($(M4_center)+(0:1)$) {};
    \coordinate (M4r3) at ($(M4_center)+(-15:1)$) {};

    \coordinate (M4u1) at ($(M4_center)+(105:1)$) {};
    \coordinate (M4u2) at ($(M4_center)+(90:1)$) {};
    \coordinate (M4u3) at ($(M4_center)+(75:1)$) {};
    
    \coordinate (M4ur1) at ($(M4_center)+(60:1)$) {};
    \coordinate (M4ur2) at ($(M4_center)+(45:1)$) {};
    \coordinate (M4ur3) at ($(M4_center)+(30:1)$) {};
    
    \coordinate (M4bl1) at ($(M4_center)+(240:1)$) {};
    \coordinate (M4bl2) at ($(M4_center)+(225:1)$) {};
    \coordinate (M4bl3) at ($(M4_center)+(210:1)$) {};
    
    \draw[fill=gray!60] (M1_center) ellipse (1 and 1);
    \draw[fill=gray!60] (M2_center) ellipse (1 and 1);
    \draw[fill=gray!60] (M3_center) ellipse (1 and 1);
    \draw[fill=gray!60] (M4_center) ellipse (1 and 1);

    \draw[-] (M1r1)--(M2l1);
    \draw[-] (M1r1)--(M2l2);
    \draw[-] (M1r1)--(M2l3);
    \draw[-] (M1r2)--(M2l1);
    \draw[-] (M1r2)--(M2l2);
    \draw[-] (M1r2)--(M2l3);
    \draw[-] (M1r3)--(M2l1);
    \draw[-] (M1r3)--(M2l2);
    \draw[-] (M1r3)--(M2l3);

    \draw[-] (M4r1)--(M3l1);
    \draw[-] (M4r1)--(M3l2);
    \draw[-] (M4r1)--(M3l3);
    \draw[-] (M4r2)--(M3l1);
    \draw[-] (M4r2)--(M3l2);
    \draw[-] (M4r2)--(M3l3);
    \draw[-] (M4r3)--(M3l1);
    \draw[-] (M4r3)--(M3l2);
    \draw[-] (M4r3)--(M3l3);
    
    \draw[-] (M1b1)--(M4u1);
    \draw[-] (M1b1)--(M4u2);
    \draw[-] (M1b1)--(M4u3);
    \draw[-] (M1b2)--(M4u1);
    \draw[-] (M1b2)--(M4u2);
    \draw[-] (M1b2)--(M4u3);
    \draw[-] (M1b3)--(M4u1);
    \draw[-] (M1b3)--(M4u2);
    \draw[-] (M1b3)--(M4u3);

    \draw[-] (M2b1)--(M3u1);
    \draw[-] (M2b1)--(M3u2);
    \draw[-] (M2b1)--(M3u3);
    \draw[-] (M2b2)--(M3u1);
    \draw[-] (M2b2)--(M3u2);
    \draw[-] (M2b2)--(M3u3);
    \draw[-] (M2b3)--(M3u1);
    \draw[-] (M2b3)--(M3u2);
    \draw[-] (M2b3)--(M3u3);
    
    \draw[-] (M1br1)--(M3ul1);
    \draw[-] (M1br1)--(M3ul2);
    \draw[-] (M1br1)--(M3ul3);
    \draw[-] (M1br2)--(M3ul1);
    \draw[-] (M1br2)--(M3ul2);
    \draw[-] (M1br2)--(M3ul3);
    \draw[-] (M1br3)--(M3ul1);
    \draw[-] (M1br3)--(M3ul2);
    \draw[-] (M1br3)--(M3ul3);
    
    \draw[-] (M2bl1)--(M4ur1);
    \draw[-] (M2bl1)--(M4ur2);
    \draw[-] (M2bl1)--(M4ur3);
    \draw[-] (M2bl2)--(M4ur1);
    \draw[-] (M2bl2)--(M4ur2);
    \draw[-] (M2bl2)--(M4ur3);
    \draw[-] (M2bl3)--(M4ur1);
    \draw[-] (M2bl3)--(M4ur2);
    \draw[-] (M2bl3)--(M4ur3);

    \tikzstyle{every node}=[inner sep=1pt]
    \node at (M1_center) {$M_1$};
    \node at (M2_center) {$M_2$};
    \node at (M3_center) {$M_3$};
    \node at (M4_center) {$M_4$};

\draw[white,-] (-3,-4)--(-3,-3.8);
\draw[white,-] (3,3)--(3,2.8);
\end{tikzpicture}
\hspace{0.2cm}
\begin{tikzpicture}[scale=1,>=latex]
\coordinate (center) at (0.0,0.0) {};
\coordinate (tau111) at ($(center)+(102:2cm)$) {};
\coordinate (tau112) at ($(center)+(96:2cm)$) {};
\coordinate (tau113) at ($(center)+(90:2cm)$) {};
\coordinate (tau114) at ($(center)+(84:2cm)$) {};
\coordinate (tau115) at ($(center)+(78:2cm)$) {};
\coordinate (tau11) at ($(center)+(90:2.35cm)$) {};

\coordinate (tau011) at ($(center)+(282:2cm)$) {};
\coordinate (tau012) at ($(center)+(276:2cm)$) {};
\coordinate (tau013) at ($(center)+(270:2cm)$) {};
\coordinate (tau014) at ($(center)+(264:2cm)$) {};
\coordinate (tau015) at ($(center)+(258:2cm)$) {};
\coordinate (tau01) at ($(center)+(270:2.35cm)$) {};

\coordinate (tau124) at ($(center)+(32:2cm)$) {};
\coordinate (tau123) at ($(center)+(38:2cm)$) {};
\coordinate (tau122) at ($(center)+(44:2cm)$) {};
\coordinate (tau121) at ($(center)+(50:2cm)$) {};
\coordinate (tau12) at ($(center)+(41:2.35cm)$) {};

\coordinate (tau024) at ($(center)+(212:2cm)$) {};
\coordinate (tau023) at ($(center)+(218:2cm)$) {};
\coordinate (tau022) at ($(center)+(224:2cm)$) {};
\coordinate (tau021) at ($(center)+(230:2cm)$) {};
\coordinate (tau02) at ($(center)+(221:2.35cm)$) {};

\coordinate (tau131) at ($(center)+(5:2cm)$) {};
\coordinate (tau132) at ($(center)+(355:2cm)$) {};
\coordinate (tau031) at ($(center)+(185:2cm)$) {};
\coordinate (tau032) at ($(center)+(175:2cm)$) {};
\coordinate (tau13) at ($(center)+(0:2.35cm)$) {};
\coordinate (tau03) at ($(center)+(180:2.35cm)$) {};

\coordinate (tau141) at ($(center)+(144:2cm)$) {};
\coordinate (tau142) at ($(center)+(134:2cm)$) {};
\coordinate (tau041) at ($(center)+(324:2cm)$) {};
\coordinate (tau042) at ($(center)+(314:2cm)$) {};
\coordinate (tau14) at ($(center)+(139:2.35cm)$) {};
\coordinate (tau04) at ($(center)+(319:2.35cm)$) {};

\draw (center) circle (2cm);

\draw[<-] (tau111)--(tau015);
\draw[<-,red, thick] (tau112)--(tau012);
\draw[->] (tau113)--(tau011);
\draw[<-] (tau114)--(tau014);
\draw[->] (tau115)--(tau013);

\draw[very thick, ->] ([shift=(112:2cm)]0,0) arc (112:68:2cm);
\draw[very thick, ->] ([shift=(292:2cm)]0,0) arc (292:248:2cm);

\draw[<-] (tau121)--(tau023);
\draw[->] (tau122)--(tau021);
\draw[<-] (tau123)--(tau024);
\draw[->,red, thick] (tau124)--(tau022);

\draw[very thick, ->] ([shift=(60:2cm)]0,0) arc (60:22:2cm);
\draw[very thick, ->] ([shift=(240:2cm)]0,0) arc (240:202:2cm);

\draw[<-,red, thick] (tau031)--(tau132);
\draw[->] (tau032)--(tau131);

\draw[very thick, ->] ([shift=(15:2cm)]0,0) arc (15:-15:2cm);
\draw[very thick, ->] ([shift=(195:2cm)]0,0) arc (195:165:2cm);

\draw[<-,red, thick] (tau041)--(tau142);
\draw[->] (tau042)--(tau141);

\draw[very thick, ->] ([shift=(154:2cm)]0,0) arc (154:124:2cm);
\draw[very thick, ->] ([shift=(334:2cm)]0,0) arc (334:304:2cm);

\tikzstyle{every node}=[inner sep=1pt]
\begin{footnotesize}
\node at (tau11) {$\tau^1_{1,\phi}$};
\node at (tau01) {$\tau^0_{1,\phi}$};
\node at (tau12) {$\tau^0_{2,\phi}$};
\node at (tau02) {$\tau^1_{2,\phi}$};
\node at (tau13) {$\tau^0_{3,\phi}$};
\node at (tau03) {$\tau^1_{3,\phi}$};
\node at (tau14) {$\tau^0_{4,\phi}$};
\node at (tau04) {$\tau^1_{4,\phi}$};
\end{footnotesize}
\draw[white] (-2.7,-2.7)--(-2.7,-2.3);
\draw[white] (2.7,2.7)--(2.7,2.3);
\end{tikzpicture}

\caption{\label{fig:serial_case_conformal_models} 
To the left: serial $Q$ has four children $M_1, M_2, M_3, M_4$.
To the right: an exemplary conformal model $\phi$ of $(Q,{\sim})$.
}
\end{figure}
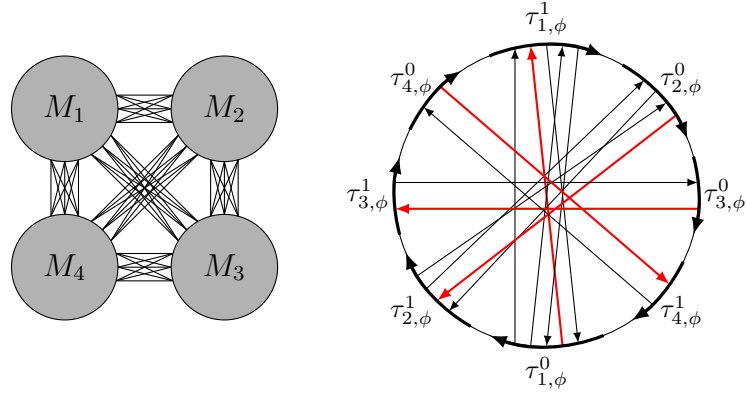

Suppose $\phi$ is a conformal model of $G_{ov}$.
The following properties of~$\phi$, proved by Lemma~\ref{lemma:circle_models_of_proper_modules_N_M} in Section~\ref{sec:modular_decomposition_and_chord_models},
follow by the facts that $Q$ is serial in $\strongModules(G_{ov})$ and $\phi$ is an oriented chord model of $G_{ov}$:
\begin{description}
\item[\namedlabel{prop_serial:contiguous_subwords}{(M1)}]
For every $i \in [t]$ the set $M_i$ induces a consistent permutation model $(\tau^0_{i,\phi}, \tau^1_{i,\phi})$ in~$\phi$, where the superscript in $\tau^0_{i,\phi}$ and $\tau^1_{i,\phi}$ are chosen such that $r_i^0 \in \tau^0_{i,\phi}$.
\item[\namedlabel{prop_serial:contiguous_subwords_overlap}{(M2)}]
For every distinct $i,j \in [t]$ the words $\tau^0_{i,\phi}, \tau^1_{i,\phi}$ and the words $\tau^0_{j,\phi}, \tau^1_{j,\phi}$ overlap in $\phi$.
\end{description}
See Figure \ref{fig:serial_case_conformal_models} for an illustration.

Suppose that $M^{0}_{i,\phi}$ and $M^{1}_{i,\phi}$ are the sets containing all the letters from the words 
$\tau^{0}_{i,\phi}$ and $\tau^{1}_{i,\phi}$, respectively.
Note that $M^{0}_{i,\phi}$ and $M^{1}_{i,\phi}$ are superscripted copies of $M_i$ and 
$\{M_{i,\phi}^{0},M_{i,\phi}^{1}\}$ forms a partition of $M^{*}_i$.
Assume the oriented permutation model $(\tau^0_{i,\phi}, \tau^1_{i,\phi})$ corresponds 
to the pair of transitive orientations $({<_{i,\phi}},{\prec_{i,\phi}})$ 
of $(M_i,{\parallel})$ and $(M_i,{\sim})$, respectively.
It turns out that the transitive orientation ${<_{i,\phi}}$ and the sets $M^{0}_{i,\phi}, M^{1}_{i,\phi}$ 
are independent on the choice of a conformal model $\phi$ of $(Q,{\sim})$.
\begin{claim}
\label{claim:metachords_for_children_of_serial}
For every two conformal models $\phi$ and $\phi'$ of $(Q,{\sim})$ we have:
$$(M^{0}_{i,\phi},M^{1}_{i,\phi},{<_{i,\phi}}) =  (M^{0}_{i,\phi'},M^{1}_{i,\phi'},{<_{i,\phi'}})\quad \text{for every } i \in [t].$$
\end{claim}
\begin{proof}
Suppose $u \in M_i$ and suppose $\phi(u)$ is oriented from $M_{i,\phi}^{0}$ to $M^{1}_{i,\phi}$ 
in every conformal model $\phi$.
Let $v \in M_i$ be such that $v \parallel u$.
Note that for every conformal model $\phi$:
\begin{itemize}
 \item if ($v \in \leftside(u)$ and $u \in \rightside(v)$) or ($v \in \rightside(u)$ and $u \in \leftside(v)$), 
 then $\phi(v)$ and $\phi(u)$ have the same orientations in the consistent permutation model induced by $M_i$ in $\phi$,
   \item if ($v \in \leftside(u)$ and $u \in \leftside(v)$) and ($v \in \rightside(u)$ and $u \in \rightside(v)$), 
   then $\phi(v)$ and $\phi(u)$ have different orientations in the consistent permutation model induced by $M_i$ in $\phi$.
\end{itemize}
See Figure~\ref{fig:spinrad_algorithm}.
Hence, either $u <_{i,\phi} v$ for all conformal models~$\phi$ or 
$v <_{i,\phi} u$ for all conformal models~$\phi$.

\begin{figure}[htp!]
\begin{tikzpicture}[scale=0.65,>=latex]

\coordinate (lu) at (0.6,0) {};
\coordinate (lv) at (-1.25,0) {};
\coordinate (ltau1) at (0,2.4) {};
\coordinate (ltau0) at (0,-2.4) {};

\tikzstyle{every node}=[inner sep=1pt]
\begin{scriptsize}
\node at (lu) {$\phi(u)$};
\node at (lv) {$\phi(v)$};
\node at (ltau0) {$M^0_{i,\phi}$};
\node at (ltau1) {$M^1_{i,\phi}$};
\end{scriptsize}
\draw (0,0) circle (2cm);
\draw[red,<-,thick] ([shift=(90:2cm)]0,0)--([shift=(270:2cm)]0,0);
\draw[black,->,thick] ([shift=(110:2cm)]0,0)--([shift=(250:2cm)]0,0);

\draw[very thick,->] ([shift=(135:2cm)]0,0) arc (135:45:2cm);
\draw[very thick,<-] ([shift=(235:2cm)]0,0) arc (235:315:2cm);

\end{tikzpicture}
\hspace{0.2cm}
\begin{tikzpicture}[scale=0.65,>=latex]

\coordinate (lu) at (0.6,0) {};
\coordinate (lv) at (-1.25,0) {};
\coordinate (ltau1) at (0,2.4) {};
\coordinate (ltau0) at (0,-2.4) {};

\tikzstyle{every node}=[inner sep=1pt]
\begin{scriptsize}
\node at (lu) {$\phi(u)$};
\node at (lv) {$\phi(v)$};
\node at (ltau0) {$M^0_{i,\phi}$};
\node at (ltau1) {$M^1_{i,\phi}$};
\end{scriptsize}
\draw (0,0) circle (2cm);
\draw[red,<-,thick] ([shift=(90:2cm)]0,0)--([shift=(270:2cm)]0,0);
\draw[black,<-,thick] ([shift=(110:2cm)]0,0)--([shift=(250:2cm)]0,0);

\draw[very thick,->] ([shift=(135:2cm)]0,0) arc (135:45:2cm);
\draw[very thick,<-] ([shift=(235:2cm)]0,0) arc (235:315:2cm);
\end{tikzpicture}
\hspace{0.2cm}
\begin{tikzpicture}[scale=0.65,>=latex]

\coordinate (lu) at (-0.6,0) {};
\coordinate (lv) at (1.25,0) {};
\coordinate (ltau1) at (0,2.4) {};
\coordinate (ltau0) at (0,-2.4) {};

\tikzstyle{every node}=[inner sep=1pt]
\begin{scriptsize}
\node at (lu) {$\phi(u)$};
\node at (lv) {$\phi(v)$};
\node at (ltau0) {$M^0_{i,\phi}$};
\node at (ltau1) {$M^1_{i,\phi}$};
\end{scriptsize}
\draw (0,0) circle (2cm);
\draw[red,<-,thick] ([shift=(90:2cm)]0,0)--([shift=(270:2cm)]0,0);
\draw[black,->,thick] ([shift=(70:2cm)]0,0)--([shift=(290:2cm)]0,0);

\draw[very thick,->] ([shift=(135:2cm)]0,0) arc (135:45:2cm);
\draw[very thick,<-] ([shift=(235:2cm)]0,0) arc (235:315:2cm);
\end{tikzpicture}
\hspace{0.2cm}
\begin{tikzpicture}[scale=0.65,>=latex]

\coordinate (lu) at (-0.6,0) {};
\coordinate (lv) at (1.25,0) {};
\coordinate (ltau1) at (0,2.4) {};
\coordinate (ltau0) at (0,-2.4) {};

\tikzstyle{every node}=[inner sep=1pt]
\begin{scriptsize}
\node at (lu) {$\phi(u)$};
\node at (lv) {$\phi(v)$};
\node at (ltau0) {$M^0_{i,\phi}$};
\node at (ltau1) {$M^1_{i,\phi}$};
\end{scriptsize}
\draw (0,0) circle (2cm);
\draw[red,<-,thick] ([shift=(90:2cm)]0,0)--([shift=(270:2cm)]0,0);
\draw[black,<-,thick] ([shift=(70:2cm)]0,0)--([shift=(290:2cm)]0,0);

\draw[very thick,->] ([shift=(135:2cm)]0,0) arc (135:45:2cm);
\draw[very thick,<-] ([shift=(235:2cm)]0,0) arc (235:315:2cm);
\end{tikzpicture}

\caption{\label{fig:spinrad_algorithm}}
\end{figure}

Note that $\phi(r_i)$ is oriented from $M^0_{i,\phi}$ to $M^1_{i,\phi}$ for every conformal model~$\phi$.
Now, the claim follows from the fact that $(M_i, \parallel)$ is connected.
\end{proof}
Claim~\ref{claim:metachords_for_children_of_serial} allows us to define the metachord $\MMM_i$ of $M_i$:
$$
\MMM_i = (M^0_i,M^1_i,{<_{M_i}}) = (M^0_{i,\phi},M^1_{i,\phi},{<_{M_{i,\phi}}}),\quad \text{where $\phi$ is any conformal model of $(M,{\sim})$}.
$$
Eventually, we set 
$$
\Pi(Q) = \left \{ \pi :
\begin{array}{c}
\text{$\pi$ is a circular order of $M^0_1,M^1_1,\ldots,M^0_t,M^1_t$ such that for every two} \\
\text{distinct $i,j \in [t]$ the slots $M^0_i, M^1_i$ overlap with the slots $M^0_j,M^1_j$} 
\end{array}
\right \}
$$
We say a circular word $\phi$ on $Q^*$ is \emph{admissible for $\Pi(Q)$} if $\phi$ arises from some member of $\Pi(Q)$ by replacing $M^0_i$ and $M^1_i$ by $\tau^0_i$ and $\tau^1_i$, respectively, where $(\tau^0_i,\tau^1_i)$ is a model admissible for $\MMM_i$ for $i \in [t]$.
\begin{theorem}
\label{thm:conformal_models_serial_component}
Let $G$ be a circular-arc graph with no twins and no universal vertices,
let $G_{ov}$ be the overlap graph of $G$, and let $Q$ be a serial module of $\strongModules(G_{ov})$
such that $Q=V$ or $Q$ is a child of parallel $V$.

A circular word $\phi$ on $Q^*$ is a conformal model for $(Q,{\sim})$ if and only if $\phi$ is admissible for $\Pi(Q)$.
\end{theorem}
\begin{proof}
Necessity. Properties~\ref{prop_serial:contiguous_subwords}--\ref{prop_serial:contiguous_subwords_overlap} and Claim~\ref{claim:metachords_for_children_of_serial} prove that
any conformal model of $(Q,{\sim})$ is admissible for $\Pi(Q)$.

Sufficiency. We prove the sufficiency the same way as in Theorem~\ref{thm:main_theorem}.
\end{proof}

Note that Theorem~\ref{thm:conformal_models_serial_component} proves Theorem~\ref{thm:main_theorem} for the case when $V$ is serial in $\strongModules(G_{ov})$.

\subsection{Prime case}
\label{subsec:conformal_models_prime}
In this subsection we describe the conformal models of $G_{ov}=(V,{\sim})$ for the case when
$V$ is prime in $\strongModules(G_{ov})$.
In passing, we also describe the structure of the conformal models of~$(Q,{\sim})$ for the case when~$Q$ is a prime child of parallel~$V$.

Let $Q$ be a prime module in $\strongModules(G_{ov})$ such that either~$Q=V$ or~$Q$ is a child of parallel~$V$.
Suppose $M_1,\ldots,M_k$ are the children of~$Q$ in~$\strongModules(G_{ov})$.
Since $Q$ is prime in~$\strongModules(G_{ov})$, Claim~\ref{claim:permutation_graphs_in_G_ov} shows that~$(M_i,{\sim})$ is a permutation graph for every~$i \in [k]$.

\begin{figure}[htp!]
\begin{tikzpicture}[scale=0.5]
    \coordinate (M1_center) at (0,0) {};
    \coordinate (lM1) at ($(M1_center)+(180:2.6)$) {};
    
    \coordinate (M1r1) at ($(M1_center)+(20:2)$) {};
    \coordinate (M1r2) at ($(M1_center)+(0:2)$) {};
    \coordinate (M1r3) at ($(M1_center)+(-20:2)$) {};

    \coordinate (M11_center) at ($(M1_center)+(30:1)$) {};
    \coordinate (M12_center) at ($(M1_center)+(150:1)$) {};
    \coordinate (M13_center) at ($(M1_center)+(270:1)$) {};

    \coordinate (M2_center) at (4,0) {};
    
    \coordinate (M2l1) at ($(M2_center)+(200:1)$) {};
    \coordinate (M2l2) at ($(M2_center)+(180:1)$) {};
    \coordinate (M2l3) at ($(M2_center)+(160:1)$) {};

    \coordinate (M2r1) at ($(M2_center)+(20:1)$) {};
    \coordinate (M2r2) at ($(M2_center)+(0:1)$) {};
    \coordinate (M2r3) at ($(M2_center)+(-20:1)$) {};
    
    \coordinate (M2u1) at ($(M2_center)+(90:1)$) {};
    \coordinate (M2u2) at ($(M2_center)+(75:1)$) {};
    \coordinate (M2u3) at ($(M2_center)+(60:1)$) {};
       
    \coordinate (M3_center) at (7,0) {};

    \coordinate (M3l1) at ($(M3_center)+(200:1)$) {};
    \coordinate (M3l2) at ($(M3_center)+(180:1)$) {};
    \coordinate (M3l3) at ($(M3_center)+(160:1)$) {};

    \coordinate (M3r1) at ($(M3_center)+(20:1)$) {};
    \coordinate (M3r2) at ($(M3_center)+(0:1)$) {};
    \coordinate (M3r3) at ($(M3_center)+(-20:1)$) {};

    \coordinate (M3u1) at ($(M3_center)+(90:1)$) {};
    \coordinate (M3u2) at ($(M3_center)+(105:1)$) {};
    \coordinate (M3u3) at ($(M3_center)+(120:1)$) {};
        
    \coordinate (M4_center) at (10,0) {};

    \coordinate (M4l1) at ($(M4_center)+(200:1)$) {};
    \coordinate (M4l2) at ($(M4_center)+(180:1)$) {};
    \coordinate (M4l3) at ($(M4_center)+(160:1)$) {};
    
    \coordinate (M5_center) at (5.5,4) {};
    \coordinate (lM5) at ($(M5_center)+(0:2.6)$) {};
    \coordinate (M5l1) at ($(M5_center)+(230:2)$) {};
    \coordinate (M5l2) at ($(M5_center)+(250:2)$) {};
    \coordinate (M5l3) at ($(M5_center)+(270:2)$) {};
    \coordinate (M5r1) at ($(M5_center)+(270:2)$) {};
    \coordinate (M5r2) at ($(M5_center)+(290:2)$) {};
    \coordinate (M5r3) at ($(M5_center)+(310:2)$) {};

    \coordinate (M51_center) at ($(M5_center)+(45:1.2)$) {};
    \coordinate (M51l1) at ($(M51_center)+(200:0.6)$) {};
    \coordinate (M51l2) at ($(M51_center)+(180:0.6)$) {};
    \coordinate (M51l3) at ($(M51_center)+(160:0.6)$) {};

    \coordinate (M51b1) at ($(M51_center)+(290:0.6)$) {};
    \coordinate (M51b2) at ($(M51_center)+(270:0.6)$) {};
    \coordinate (M51b3) at ($(M51_center)+(250:0.6)$) {};
    
    \coordinate (M51bl1) at ($(M51_center)+(245:0.6)$) {};
    \coordinate (M51bl2) at ($(M51_center)+(225:0.6)$) {};
    \coordinate (M51bl3) at ($(M51_center)+(205:0.6)$) {};
    
    \coordinate (M52_center) at ($(M5_center)+(135:1.2)$) {};
    \coordinate (M52r1) at ($(M52_center)+(20:0.6)$) {};
    \coordinate (M52r2) at ($(M52_center)+(0:0.6)$) {};
    \coordinate (M52r3) at ($(M52_center)+(-20:0.6)$) {};

    \coordinate (M52b1) at ($(M52_center)+(290:0.6)$) {};
    \coordinate (M52b2) at ($(M52_center)+(270:0.6)$) {};
    \coordinate (M52b3) at ($(M52_center)+(250:0.6)$) {};
    
    \coordinate (M52br1) at ($(M52_center)+(335:0.6)$) {};
    \coordinate (M52br2) at ($(M52_center)+(315:0.6)$) {};
    \coordinate (M52br3) at ($(M52_center)+(305:0.6)$) {};

    \coordinate (M53_center) at ($(M5_center)+(225:1.2)$) {};

    \coordinate (M53r1) at ($(M53_center)+(20:0.6)$) {};
    \coordinate (M53r2) at ($(M53_center)+(0:0.6)$) {};
    \coordinate (M53r3) at ($(M53_center)+(-20:0.6)$) {};

    \coordinate (M53u1) at ($(M53_center)+(70:0.6)$) {};
    \coordinate (M53u2) at ($(M53_center)+(90:0.6)$) {};
    \coordinate (M53u3) at ($(M53_center)+(110:0.6)$) {};
    
    \coordinate (M53ur1) at ($(M53_center)+(25:0.6)$) {};
    \coordinate (M53ur2) at ($(M53_center)+(45:0.6)$) {};
    \coordinate (M53ur3) at ($(M53_center)+(65:0.6)$) {};
    
    \coordinate (M54_center) at ($(M5_center)+(315:1.2)$) {};
    
    \coordinate (M54l1) at ($(M54_center)+(200:0.6)$) {};
    \coordinate (M54l2) at ($(M54_center)+(180:0.6)$) {};
    \coordinate (M54l3) at ($(M54_center)+(160:0.6)$) {};

    \coordinate (M54u1) at ($(M54_center)+(70:0.6)$) {};
    \coordinate (M54u2) at ($(M54_center)+(90:0.6)$) {};
    \coordinate (M54u3) at ($(M54_center)+(110:0.6)$) {};

    \coordinate (M54ul1) at ($(M54_center)+(115:0.6)$) {};
    \coordinate (M54ul2) at ($(M54_center)+(135:0.6)$) {};
    \coordinate (M54ul3) at ($(M54_center)+(155:0.6)$) {};
    
    \draw[fill=gray!30] (M1_center) ellipse (2 and 2);
    \draw[fill=gray!60] (M11_center) ellipse (0.75 and 0.75);
    \draw[fill=gray!60] (M12_center) ellipse (0.75 and 0.75);
    \draw[fill=gray!60] (M13_center) ellipse (0.75 and 0.75);

    \draw[fill=gray!30] (M2_center) ellipse (1 and 1);

    \draw[fill=gray!30] (M3_center) ellipse (1 and 1);

    \draw[fill=gray!30] (M4_center) ellipse (1 and 1);

    \draw[fill=gray!30] (M5_center) ellipse (2 and 2);
    \draw[fill=gray!60] (M51_center) ellipse (0.6 and 0.6);
    \draw[fill=gray!60] (M52_center) ellipse (0.6 and 0.6);
    \draw[fill=gray!60] (M53_center) ellipse (0.6 and 0.6);
    \draw[fill=gray!60] (M54_center) ellipse (0.6 and 0.6);

    \draw[-] (M1r1)--(M2l1);
    \draw[-] (M1r1)--(M2l2);
    \draw[-] (M1r1)--(M2l3);
    \draw[-] (M1r2)--(M2l1);
    \draw[-] (M1r2)--(M2l2);
    \draw[-] (M1r2)--(M2l3);
    \draw[-] (M1r3)--(M2l1);
    \draw[-] (M1r3)--(M2l2);
    \draw[-] (M1r3)--(M2l3);

    \draw[-] (M2r1)--(M3l1);
    \draw[-] (M2r1)--(M3l2);
    \draw[-] (M2r1)--(M3l3);
    \draw[-] (M2r2)--(M3l1);
    \draw[-] (M2r2)--(M3l2);
    \draw[-] (M2r2)--(M3l3);
    \draw[-] (M2r3)--(M3l1);
    \draw[-] (M2r3)--(M3l2);
    \draw[-] (M2r3)--(M3l3);
    
    \draw[-] (M3r1)--(M4l1);
    \draw[-] (M3r1)--(M4l2);
    \draw[-] (M3r1)--(M4l3);
    \draw[-] (M3r2)--(M4l1);
    \draw[-] (M3r2)--(M4l2);
    \draw[-] (M3r2)--(M4l3);
    \draw[-] (M3r3)--(M4l1);
    \draw[-] (M3r3)--(M4l2);
    \draw[-] (M3r3)--(M4l3);
    
    \draw[-] (M2u1)--(M5l1);
    \draw[-] (M2u1)--(M5l2);
    \draw[-] (M2u1)--(M5l3);
    \draw[-] (M2u2)--(M5l1);
    \draw[-] (M2u2)--(M5l2);
    \draw[-] (M2u2)--(M5l3);
    \draw[-] (M2u3)--(M5l1);
    \draw[-] (M2u3)--(M5l2);
    \draw[-] (M2u3)--(M5l3);
    
    \draw[-] (M3u1)--(M5r1);
    \draw[-] (M3u1)--(M5r2);
    \draw[-] (M3u1)--(M5r3);
    \draw[-] (M3u2)--(M5r1);
    \draw[-] (M3u2)--(M5r2);
    \draw[-] (M3u2)--(M5r3);
    \draw[-] (M3u3)--(M5r1);
    \draw[-] (M3u3)--(M5r2);
    \draw[-] (M3u3)--(M5r3);
    
    \draw[-] (M51l1)--(M52r1);
    \draw[-] (M51l1)--(M52r2);
    \draw[-] (M51l1)--(M52r3);
    \draw[-] (M51l2)--(M52r1);
    \draw[-] (M51l2)--(M52r2);
    \draw[-] (M51l2)--(M52r3);
    \draw[-] (M51l3)--(M52r1);
    \draw[-] (M51l3)--(M52r2);
    \draw[-] (M51l3)--(M52r3);

    \draw[-] (M53r1)--(M54l1);
    \draw[-] (M53r1)--(M54l2);
    \draw[-] (M53r1)--(M54l3);
    \draw[-] (M53r2)--(M54l1);
    \draw[-] (M53r2)--(M54l2);
    \draw[-] (M53r2)--(M54l3);
    \draw[-] (M53r3)--(M54l1);
    \draw[-] (M53r3)--(M54l2);
    \draw[-] (M53r3)--(M54l3);

    \draw[-] (M51b1)--(M54u1);
    \draw[-] (M51b1)--(M54u2);
    \draw[-] (M51b1)--(M54u3);
    \draw[-] (M51b2)--(M54u1);
    \draw[-] (M51b2)--(M54u2);
    \draw[-] (M51b2)--(M54u3);
    \draw[-] (M51b3)--(M54u1);
    \draw[-] (M51b3)--(M54u2);
    \draw[-] (M51b3)--(M54u3);
    
    \draw[-] (M52b1)--(M53u1);
    \draw[-] (M52b1)--(M53u2);
    \draw[-] (M52b1)--(M53u3);
    \draw[-] (M52b2)--(M53u1);
    \draw[-] (M52b2)--(M53u2);
    \draw[-] (M52b2)--(M53u3);
    \draw[-] (M52b3)--(M53u1);
    \draw[-] (M52b3)--(M53u2);
    \draw[-] (M52b3)--(M53u3);
    
    \draw[-] (M51bl1)--(M53ur1);
    \draw[-] (M51bl1)--(M53ur2);
    \draw[-] (M51bl1)--(M53ur3);
    \draw[-] (M51bl2)--(M53ur1);
    \draw[-] (M51bl2)--(M53ur2);
    \draw[-] (M51bl2)--(M53ur3);
    \draw[-] (M51bl3)--(M53ur1);
    \draw[-] (M51bl3)--(M53ur2);
    \draw[-] (M51bl3)--(M53ur3);

    \draw[-] (M52br1)--(M54ul1);
    \draw[-] (M52br1)--(M54ul2);
    \draw[-] (M52br1)--(M54ul3);
    \draw[-] (M52br2)--(M54ul1);
    \draw[-] (M52br2)--(M54ul2);
    \draw[-] (M52br2)--(M54ul3);
    \draw[-] (M52br3)--(M54ul1);
    \draw[-] (M52br3)--(M54ul2);
    \draw[-] (M52br3)--(M54ul3);

    \tikzstyle{every node}=[inner sep=1pt]
    \begin{footnotesize}
    \node at (lM1) {$M_1$};
    \node at (M11_center) {$L_1^1$};
    \node at (M12_center) {$L_1^2$};
    \node at (M13_center) {$L_1^3$};
    \node at (M2_center) {$M_2$};
    \node at (M3_center) {$M_3$};
    \node at (M4_center) {$M_4$};
    \node at (lM5) {$M_5$};
    \end{footnotesize}
    \begin{tiny}
    \node at (M51_center) {$L_5^1$};
    \node at (M52_center) {$L_5^2$};
    \node at (M53_center) {$L_5^3$};
    \node at (M54_center) {$L_5^4$};
     
    \end{tiny}
\draw[white] (-3.4,-4.2)--(-3.4,-3.0);
\draw[white] (11,9)--(11,7.0);
\end{tikzpicture}
\hspace{1cm}
\begin{tikzpicture}[scale=1.1,rotate=0,>=latex]
\coordinate (center) at (0.0,0.0) {};
\draw (center) circle (2cm);

\coordinate (m110) at ($(center)+(160:2cm)$) {};
\coordinate (m111) at ($(center)+(200:2cm)$) {};
\coordinate (lm110) at ($(center)+(160:2.2cm)$) {};
\coordinate (lm111) at ($(center)+(200:2.2cm)$) {};

\coordinate (m120) at ($(center)+(145:2cm)$) {};
\coordinate (m121) at ($(center)+(215:2cm)$) {};
\coordinate (lm120) at ($(center)+(145:2.2cm)$) {};
\coordinate (lm121) at ($(center)+(215:2.2cm)$) {};

\draw[very thick,-] ([shift=(140:2.5cm)]0,0) arc (140:165:2.5cm);
\coordinate (lk11) at ($(center)+(152.5:2.75cm)$) {};
\draw[very thick,-] ([shift=(195:2.5cm)]0,0) arc (195:220:2.5cm);
\coordinate (lk12) at ($(center)+(207.5:2.75cm)$) {};


\coordinate (m130) at ($(center)+(30:2cm)$) {};
\coordinate (m131) at ($(center)+(330:2cm)$) {};
\coordinate (lm130) at ($(center)+(30:2.25cm)$) {};
\coordinate (lm131) at ($(center)+(330:2.25cm)$) {};
\coordinate (lk2) at ($(center)+(330:2.2cm)$) {};

\draw[very thick,-] ([shift=(25:2.5cm)]0,0) arc (25:35:2.5cm);
\coordinate (lk21) at ($(center)+(30:2.75cm)$) {};
\draw[very thick,-] ([shift=(325:2.5cm)]0,0) arc (325:335:2.5cm);
\coordinate (lk22) at ($(center)+(330:2.75cm)$) {};


\coordinate (m510) at ($(center)+(45:2cm)$) {};
\coordinate (lm510) at ($(center)+(45:2.2cm)$) {};
\coordinate (m511) at ($(center)+(235:2cm)$) {};
\coordinate (lm511) at ($(center)+(235:2.2cm)$) {};
\coordinate (lk6) at ($(center)+(235:2.2cm)$) {};

\draw[very thick,-] ([shift=(40:2.5cm)]0,0) arc (40:50:2.5cm);
\coordinate (lk61) at ($(center)+(45:2.75cm)$) {};
\draw[very thick,-] ([shift=(230:2.5cm)]0,0) arc (230:240:2.5cm);
\coordinate (lk62) at ($(center)+(235:2.75cm)$) {};

\coordinate (m520) at ($(center)+(110:2cm)$) {};
\coordinate (lm520) at ($(center)+(110:2.2cm)$) {};
\coordinate (m521) at ($(center)+(290:2cm)$) {};
\coordinate (lm521) at ($(center)+(290:2.2cm)$) {};

\coordinate (m530) at ($(center)+(120:2cm)$) {};
\coordinate (lm530) at ($(center)+(120:2.2cm)$) {};
\coordinate (m531) at ($(center)+(300:2cm)$) {};
\coordinate (lm531) at ($(center)+(300:2.2cm)$) {};

\coordinate (m540) at ($(center)+(130:2cm)$) {};
\coordinate (lm540) at ($(center)+(130:2.2cm)$) {};
\coordinate (m541) at ($(center)+(310:2cm)$) {};
\coordinate (lm541) at ($(center)+(310:2.2cm)$) {};
\coordinate (lk7) at ($(center)+(300:2.2cm)$) {};

\draw[very thick,-] ([shift=(105:2.5cm)]0,0) arc (105:135:2.5cm);
\coordinate (lk71) at ($(center)+(120:2.75cm)$) {};
\draw[very thick,-] ([shift=(285:2.5cm)]0,0) arc (285:315:2.5cm);
\coordinate (lk72) at ($(center)+(300:2.75cm)$) {};


\coordinate (m20) at ($(center)+(0:2cm)$) {};
\coordinate (lm20) at ($(center)+(0:2.2cm)$) {};
\coordinate (m21) at ($(center)+(180:2cm)$) {};
\coordinate (lm21) at ($(center)+(180:2.2cm)$) {};
\coordinate (lk3) at ($(center)+(0:2.2cm)$) {};

\draw[very thick,-] ([shift=(-10:2.5cm)]0,0) arc (-10:10:2.5cm);
\coordinate (lk31) at ($(center)+(0:2.75cm)$) {};
\draw[very thick,-] ([shift=(170:2.5cm)]0,0) arc (170:190:2.5cm);
\coordinate (lk32) at ($(center)+(180:2.75cm)$) {};

\coordinate (m30) at ($(center)+(77:2cm)$) {};
\coordinate (lm30) at ($(center)+(77:2.2cm)$) {};
\coordinate (m31) at ($(center)+(265:2cm)$) {};
\coordinate (lm31) at ($(center)+(265:2.2cm)$) {};
\coordinate (lk4) at ($(center)+(265:2.2cm)$) {};

\draw[very thick,-] ([shift=(72:2.5cm)]0,0) arc (72:82:2.5cm);
\coordinate (lk41) at ($(center)+(77:2.75cm)$) {};
\draw[very thick,-] ([shift=(260:2.5cm)]0,0) arc (260:270:2.5cm);
\coordinate (lk42) at ($(center)+(265:2.75cm)$) {};

\coordinate (m40) at ($(center)+(95:2cm)$) {};
\coordinate (lm40) at ($(center)+(95:2.2cm)$) {};
\coordinate (m41) at ($(center)+(60:2cm)$) {};
\coordinate (lm41) at ($(center)+(60:2.2cm)$) {};
\coordinate (lk5) at ($(center)+(60:2.2cm)$) {};

\draw[very thick,-] ([shift=(90:2.5cm)]0,0) arc (90:100:2.5cm);
\coordinate (lk51) at ($(center)+(95:2.75cm)$) {};
\draw[very thick,-] ([shift=(55:2.5cm)]0,0) arc (55:65:2.5cm);
\coordinate (lk52) at ($(center)+(60:2.75cm)$) {};

\tikzstyle{every node}=[inner sep=1pt]
\begin{tiny}
\node at (lk11) {$K_1$};
\node at (lk12) {$K_1$};
\node at (lk21) {$K_2$};
\node at (lk22) {$K_2$};
\node at (lk31) {$K_3$};
\node at (lk32) {$K_3$};
\node at (lk41) {$K_4$};
\node at (lk42) {$K_4$};
\node at (lk51) {$K_5$};
\node at (lk52) {$K_5$};
\node at (lk61) {$K_6$};
\node at (lk62) {$K_6$};
\node at (lk71) {$K_7$};
\node at (lk72) {$K_7$};
\node at (lm110) {$L_{1}^{1}$};
\node at (lm111) {$L_{1}^{1}$};
\node at (lm120) {$L_{1}^{3}$};
\node at (lm121) {$L_{1}^{3}$};
\node at (lm130) {$L_{1}^{2}$};
\node at (lm131) {$L_{1}^{2}$};
\node at (lm20) {$M_{2}$};
\node at (lm21) {$M_{2}$};
\node at (lm30) {$M_{3}$};
\node at (lm31) {$M_{3}$};
\node at (lm40) {$M_{4}$};
\node at (lm41) {$M_{4}$};
\node at (lm510) {$L_{5}^{1}$};
\node at (lm511) {$L_{5}^{1}$};
\node at (lm520) {$L_{5}^{2}$};
\node at (lm521) {$L_{5}^{2}$};
\node at (lm530) {$L_{5}^{3}$};
\node at (lm531) {$L_{5}^{3}$};
\node at (lm540) {$L_{5}^{4}$};
\node at (lm541) {$L_{5}^{4}$};
\end{tiny}
\begin{footnotesize}
 
\end{footnotesize}

\draw[line width=0.4mm,-] (m20)--(m21);

\draw[line width=0.4mm,-] (m30)--(m31);

\draw[line width=0.4mm,-] (m40)--(m41);

\draw[red,line width=0.4mm,-] (m110)--(m111);
\draw[red,line width=0.4mm,-] (m120)--(m121);
\draw[red,line width=0.4mm,-] (m130)--(m131);

\draw[blue,line width=0.4mm,-] (m510)--(m511);
\draw[blue,line width=0.4mm,-] (m520)--(m521);
\draw[blue,line width=0.4mm,-] (m530)--(m531);
\draw[blue,line width=0.4mm,-] (m540)--(m541);

\draw[white] (-2.7,-2.7)--(-2.7,-2.3);
\draw[white] (2.7,2.7)--(2.7,2.3);
\end{tikzpicture}
\caption{\label{fig:prime_case_consistent_decomposition} 
To the left: prime $Q$ has five children $M_1, \ldots, M_5$;
$M_1$ is parallel and has three children $L^1_1,L^2_1,L^3_1$,
$M_5$ is serial and has four children $L^1_5,L^2_5,L^3_5,L^4_5$,
and $M_2,M_3,M_4$ are prime.
To the right: schematic view of a conformal model $\phi$ of $(M,{\sim})$.
Oriented permutation models of $(M_2,{\sim})$, $(M_3,{\sim})$, $(M_4,{\sim})$ and of $(L^i_j,{\sim})$ occuring in $\phi$ are represented by bold chords.
}
\end{figure}
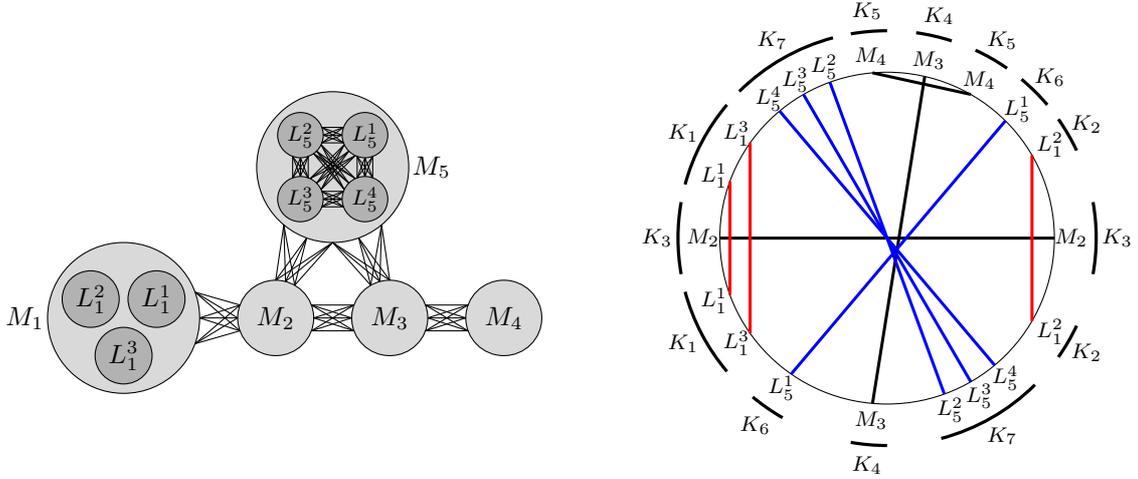

Let $\phi$ be any conformal model of $(Q,{\sim})$.
The properties of $\phi$ listed below follow from the facts that $\phi$ is a chord model of $(Q,{\sim})$ 
and $Q$ is prime in $\strongModules(G_{ov})$. 
Lemma~\ref{lemma:circle_models_of_a_proper_prime_module} in Section~\ref{sec:modular_decomposition_and_chord_models} proves that (see Figure~\ref{fig:prime_case_consistent_decomposition} for an illustration):
\begin{description}
 \item[\namedlabel{prop_prime:contiguous_subwords_prime}{(M3)}] If $M_i$ is prime child of $Q$, then $M_i$ induces a consistent permutation model $(\mu_i,\mu'_i)$ in~$\phi$.
\end{description}
The above property does not necessarily hold when $M_i$ is a parallel or a serial child of~$Q$.
However, Lemmas \ref{lemma:circle_models_of_a_parallel_module} and \ref{lemma:circle_models_of_a_serial_module} 
show that (see Figure~\ref{fig:prime_case_consistent_decomposition} for an illustration):
\begin{description}
 \item[\namedlabel{prop_prime:contiguous_subwords_serial}{(M4)}] If $M_i$ is a serial child of $Q$ and $L$ is a child of $M_i$, then the set
 $L$ induces a consistent permutation model $(\lambda,\lambda')$ in $\phi$.
 Moreover, if $L_1,L_2$ are two children of $M_i$ inducing permutation models $(\lambda_1,\lambda'_1)$ and $(\lambda_2,\lambda'_2)$ then the words $\lambda_1, \lambda'_1$ overlap with the words $\lambda_2,\lambda'_2$ in $\phi$.
 \item[\namedlabel{prop_prime:contiguous_subwords_parallel}{(M5)}] If $M_i$ is a parallel child of $Q$ and $L$ is a child of $M_i$, then the set
 $L$ induces a consistent permutation model $(\lambda,\lambda')$ in $\phi$. 
 Moreover, if $L_1,L_2$ are two children of $M_i$ inducing permutation models $(\lambda_1,\lambda'_1)$ and $(\lambda_2,\lambda'_2)$ then the words $\lambda_1, \lambda'_1$ do not overlap with the words $\lambda_2,\lambda'_2$ in $\phi$.
\end{description}

Next we introduce in every module $M_i$ an equivalence relation $K$,
used to define the CA-modules of $Q$ when $Q =V$.
\begin{definition}
\label{def:prime-K-relation}
Let $M_i$ be a child of $Q$ in $\strongModules(G_{ov})$. 
The $K$-relation in the set $M_i$ is defined as follows:
\begin{itemize}
\item If $M_i$ is prime, then $v K v'$ for every $v,v' \in M_i$.
\item If $M_i$ is parallel, then for every $v,v' \in M_i$:
$$
v K v'  \quad \text{if} \quad
\begin{array}{c}
\vspace{1pt}
\text{either $\{v,v'\} \subseteq \leftside(u)$ or $\{v,v'\} \subseteq \rightside(u)$,} \\
\text{for every $u \in Q \setminus M_i$ such that $u \parallel M_i$.}
\end{array}
$$
\item If $M_i$ is serial, then for every $v,v' \in M_i$:
$$
v K v'  \quad \text{if} \quad 
\begin{array}{l}
\vspace{1pt}
\{ \leftside(v) \cap (Q \setminus M_i), \rightside(v) \cap (Q \setminus M_i) \} = \\
\{ \leftside(v') \cap (Q \setminus M_i), \rightside(v') \cap (Q \setminus M_i)\}.
\end{array}
$$
\end{itemize}
We denote by $K(M_i)$ the equivalence classes of $K$-relation in $M_i$ and 
by $K(Q)$ the set $\bigcup_{i=1}^n K(M_i)$.
\end{definition}
Note that Properties~\ref{prop_prime:contiguous_subwords_serial} and \ref{prop_prime:contiguous_subwords_parallel} 
of the conformal models of $(Q,{\sim})$ assert that every set in $K(M_i)$ is the union of some children of $M_i$ when $M_i$ is serial or parallel.

We refer to Figure~\ref{fig:prime_case_consistent_decomposition} for an example.
Clearly, $K(M_2) = \{M_2\}$, $K(M_3) = \{M_3\}$, $K(M_4) = \{M_4\}$ as $M_2,M_3,M_4$ are prime.
The module $M_1$ is parallel.
Note that the vertices from $L_1^1 \cup L_1^3$ are on the same side ($\leftside$ or $\rightside$) of every vertex from $M_3 \cup M_4 \cup M_5$, 
similarly the vertices from $L_1^2$ are on the same side of every vertex from $M_3 \cup M_4 \cup M_5$, 
but every vertex from $M_3$ has the vertices from $L_1^3 \cup L^1_1$ and the vertices from $L^2_1$ on different sides.
This means that $K(M_1) = \{L_1^1 \cup L_1^3, L^2_1\}$.
The module $M_5$ is serial.
Every vertex from $L^2_5 \cup L^3_5 \cup L^4_5$ has the set $M_4 \cup L^2_1$ on one side 
and the set $L^1_1 \cup L^3_1$ on the other side.
Every vertex from $L^1_5$ has the set $M_4 \cup L^1_1 \cup L^3_1$ on one side 
and the set $L^2_1$ on the second side.
In particular, it means that $K(M_5) = \{L^1_5,L^2_5 \cup L^3_5 \cup L^4_5\}$.
So, $K_1,K_2,K_3,K_4,K_5,K_6,K_7$, equal to $L_1^1 \cup L_1^3, L^2_1, M_2,M_3,M_4,L^1_5,L^2_5 \cup L^3_5 \cup L^4_5$, respectively,
are the members of~$K(Q)$.

Suppose $K_1,\ldots,K_t$ are the members of $K(M)$.
For every $i \in [t]$ we fix a representant~$r_i$ in the set $K_i$ and we let $R = \{r_1,\ldots,r_t\}$.
Lemma~\ref{lemma:prime_consistent_modules}.\eqref{item:prime_consistent_modules_skeleton} of Section~\ref{sec:prime_case_properties} proves the following property of $(R,{\sim})$:
\begin{description}
 \item[\namedlabel{prop:prime_skeleton}{(P1)}] 
 The graph $(R,{\sim})$ has exactly two conformal model, $\phi^0_R$ and $\phi^1_R$, one being the reflection of the other.
\end{description}
The proof of Property~\ref{prop:prime_skeleton} is done in two steps.
First we show that every prime subgraph of~$G_{ov}$
has exactly two conformal models, one being the reflection of the other (Lemma~\ref{lemma:two_models_of_a_prime_graph}).
This shows that every set containing one vertex from every $M_i$
has two conformal models, one being the reflection of the other.
Afterwards we extend this property on every set containing one vertex from every set in $K(Q)$
(Lemma~\ref{lemma:prime_consistent_modules} in Section~\ref{sec:prime_case_properties}).

Lemma~\ref{lemma:prime_consistent_modules}.\eqref{item:prime_consistent_modules_contiguous_subwords} of Section~\ref{sec:prime_case_properties} 
shows the following property of a conformal model~$\phi$ of $(Q,{\sim})$ with respect to the sets in $K(Q)$:
\begin{description} 
 \item[\namedlabel{prop:prime_contiguous_subwords}{(P2)}] For every $i \in [t]$ the set $K_i$ induces a consistent permutation model $(\tau^0_{i,\phi}, \tau^1_{i,\phi})$ in~$\phi$, where $\tau^0_{i,\phi}, \tau^1_{i,\phi}$ are enumerated such that $r^0_i \in K^0_i$ and $r^1_i \in K^1_i$.
\end{description}
We also note that $\phi$ satisfies either $\phi \Vert R^* = \phi^0_R$ or $\phi \Vert R^* = \phi^1_R$, as $\phi \Vert R^*$ is a conformal model of $(R,{\sim})$ 
and $(R,{\sim})$ has two conformal models $\phi^0_R$ and $\phi^1_R$.
Having Properties~\ref{prop:prime_skeleton} and~\ref{prop:prime_contiguous_subwords} in mind, we proceed in the similar way as in the previous case.
Let $K^{0}_{i,\phi}$ and $K^{1}_{i,\phi}$ be the sets containing all the letters occurring in the words 
$\tau^{0}_{i,\phi}$ and $\tau^{1}_{i,\phi}$, respectively, 
and let the oriented permutation model $(\tau^0_{i,\phi}, \tau^1_{i,\phi})$ of $(K_i,{\sim})$ correspond 
to the pair of transitive orientations $({<_{i,\phi}},{\prec_{i,\phi}})$ 
of $(K_i,{\parallel})$ and $(K_i,{\sim})$, respectively.
It turns out that the transitive orientation ${<_{i,\phi}}$ and the sets $K^{0}_{i,\phi}, K^{1}_{i,\phi}$ 
are independent on the choice of a conformal model $\phi$ of $(Q,{\sim})$.
\begin{claim}
\label{claim:metachords_for_children_of_prime}
For every two conformal models $\phi$ and $\phi'$ of $(Q,{\sim})$ and every $i \in [t]$ we have:
$$(K^{0}_{i,\phi},K^{1}_{i,\phi},{<_{i,\phi}}) =  (K^{0}_{i,\phi'},K^{1}_{i,\phi'},{<_{i,\phi'}}).$$
\end{claim}
\begin{proof}
The claim is proved similarly as Claim~\ref{claim:metachords_for_children_of_serial}.
Suppose $s$ is a vertex in $Q$ such that $s \parallel K_i$ --
such a vertex exists as $Q$ is prime.
Since~$K$ induces a consistent permutation model in $\phi$, 
for every $u \in K_i$ and every conformal model~$\phi$ of $(Q,{\sim})$:
\begin{itemize}
 \item $u^0 \in K^0_{i,\phi}$ if the vertices $r_i$ and $u$ have the vertex $s$ on the same side,
 \item $u^0 \in K^1_{i,\phi}$, otherwise.
\end{itemize}
So, the orientation of $\phi(u)$ is the same in every conformal model $\phi$ of $(Q,{\sim})$.
Hence, for every $u,v \in K_i$ such that $u \parallel v$ we have either  $u <_{i, \phi} v$ for all conformal models $\phi$
or $v <_{i, \phi} u$ for all conformal models $\phi$.
This completes the proof.
\end{proof}
\noindent Claim~\ref{claim:metachords_for_children_of_prime} allows to define the metachord $\mathbb{K}_i$ for every $i \in [t]$:
$$
(K^0_i,K^1_i,{<_{K_i}}) = (K^0_{i,\phi},K^1_{i,\phi},{<_{K_{i,\phi}}}),\quad \text{where $\phi$ is any conformal model of $(Q,{\sim})$.}
$$
Now, let $\pi^0(Q)$ be the circular order of $\{K^0_1,K^1_1,\ldots,K^0_t,K^1_t\}$ 
obtained from $\phi^0_R$ by replacing every letter $r^j_i$ by $K^j_i$
and let $\pi^1(Q)$ be the circular order of $\{K^0_1,K^1_1,\ldots,K^0_t,K^1_t\}$ 
obtained from $\phi^1_R$ by replacing every letter $r^j_i$ by $K^j_i$.
In particular, $\pi^1(Q)$ is the reflection of $\pi^0(Q)$ as $\phi^1_R$ is the reflection of $\phi^0_R$.
We set 
$$\Pi(Q) = \{\pi^0(Q), \pi^1(Q)\}.$$
We say a circular word $\phi$ on $Q^*$ is \emph{admissible for $\Pi(Q)$} if 
$\phi$ arises from some member of $\Pi(Q)$ by replacing 
$K^0_i$ and $K^1_i$ by $\tau^0_i$ and $\tau^1_i$, respectively, where $(\tau^0_i,\tau^1_i)$ is a model admissible for~$\KKK_i$ for $i \in [t]$.
Now we can state the main theorem of this subsection.
\begin{theorem}
\label{thm:conformal_models_prime_component}
Let $G$ be a circular-arc graph with no twins and no universal vertices,
let $G_{ov}$ be the overlap graph of $G$, and let $Q$ be a prime module of $\strongModules(G_{ov})$ such that $Q=V$ or $Q$ is a child of parallel $V$.

A circular word $\phi$ on $Q^*$ is a conformal model for $(Q,{\sim})$ if and only if $\phi$ is admissible for $\Pi(Q)$.
\end{theorem}
\begin{proof}
Necessity. Suppose $\phi$ is a conformal model of $(Q,{\sim})$.
Properties \ref{prop:prime_skeleton} and \ref{prop:prime_contiguous_subwords} of~$\phi$
and Claim~\ref{claim:metachords_for_children_of_prime} assert that:
\begin{itemize}
 \item $\phi$ is admissible for $\pi^0(Q)$ if $\phi \Vert R^* = \phi^0_R$,
 \item $\phi$ is admissible for $\pi^1(Q)$ if $\phi \Vert R^* = \phi^1_R$.
\end{itemize}
In particular, $\phi$ is admissible for $\Pi(Q)$.

Sufficiency. We prove the sufficiency the same way as in Theorem~\ref{thm:main_theorem}.
\end{proof}
Note that Theorem~\ref{thm:conformal_models_prime_component} proves Theorem~\ref{thm:main_theorem} for the case when $V$ is prime.
In this case the CA-modules of $V$ are the members $K_1,\ldots,K_t$ of $K(V)$ and
$\Pi = \Pi(V)$ as $V$ is the only connected component of $G_{ov}$.

\subsection{Parallel case}
\label{subsec:conformal_models_parallel}
Suppose $V$ is a parallel module in $\strongModules(G_{ov})$. 
Let $\mathcal{Q}$ be the set of the children of $V$ in $\strongModules(G_{ov})$.
Clearly, $|\mathcal{Q}| \geq 2$ and the members of $\mathcal{Q}$ are the connected components of $G_{ov}$.

First, we define combinatorially the PQS-tree $\pqstree$ of $G$;
We will accomplish our task gradually; we first construct $\pqstree$ restricted to P-nodes and Q-nodes (for this part we follow the work of Hsu~\cite{Hsu95}), and then
we extend this tree by the slots and the sets $\Pi(\cdot)$.

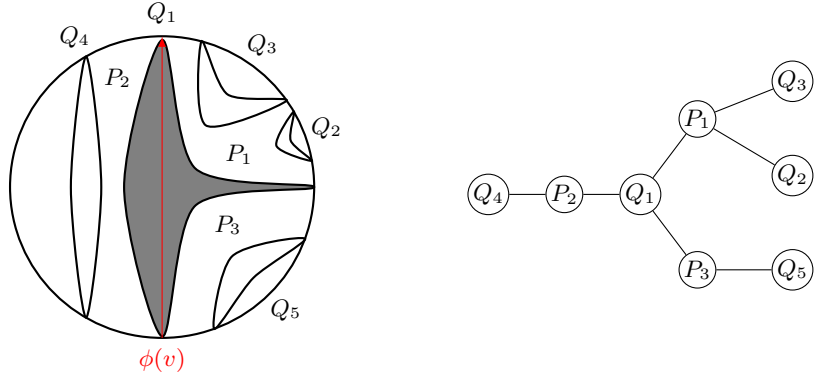
\begin{figure}[htp!]
\begin{tikzpicture}[scale=1,>=latex]
\coordinate (center) at (0,0) {};
\coordinate (label) at (0,-3) {};

\coordinate (m11) at ($(center)+(90:1.95cm)$) {};
\coordinate (m112) at ($(center)+(180:0.5cm)$) {};
\coordinate (m12) at ($(center)+(270:1.98cm)$) {};
\coordinate (m13) at ($(center)+(330:0.5cm)$) {};
\coordinate (m14) at ($(center)+(0:2cm)$) {};
\coordinate (m15) at ($(center)+(30:0.5cm)$) {};
\coordinate (lm1) at ($(center)+(90:2.3cm)$) {};

\coordinate (m21) at ($(center)+(10:2cm)$) {};
\coordinate (m22) at ($(center)+(20:1.8cm)$) {};
\coordinate (m23) at ($(center)+(30:2cm)$) {};
\coordinate (m24) at ($(center)+(20:1.6cm)$) {};
\coordinate (lm2) at ($(center)+(20:2.3cm)$) {};

\coordinate (m31) at ($(center)+(35:2cm)$) {};
\coordinate (m32) at ($(center)+(55:1.5cm)$) {};
\coordinate (m33) at ($(center)+(75:2cm)$) {};
\coordinate (m34) at ($(center)+(55:1cm)$) {};
\coordinate (lm3) at ($(center)+(55:2.3cm)$) {};

\coordinate (m41) at ($(center)+(120:2cm)$) {};
\coordinate (m42) at ($(center)+(0:-1.2cm)$) {};
\coordinate (m43) at ($(center)+(240:2cm)$) {};
\coordinate (m44) at ($(center)+(0:-0.8cm)$) {};
\coordinate (lm4) at ($(center)+(120:2.3cm)$) {};

\coordinate (m51) at ($(center)+(290:2cm)$) {};
\coordinate (m52) at ($(center)+(315:1.7cm)$) {};
\coordinate (m53) at ($(center)+(340:2cm)$) {};
\coordinate (m54) at ($(center)+(315:1.3cm)$) {};
\coordinate (lm5) at ($(center)+(315:2.3cm)$) {};

\coordinate (ln1) at ($(center)+(23:1.1cm)$) {};
\coordinate (ln2) at ($(center)+(112:1.56cm)$) {};
\coordinate (ln3) at ($(center)+(330:1.0cm)$) {};

\coordinate (v) at ($(center)+(270:2.3cm)$) {};

\begin{scriptsize}
\tikzstyle{every node}=[inner sep=1pt]
\node at (lm1) {$Q_1$};
\node at (lm2) {$Q_2$};
\node at (lm3) {$Q_3$};
\node at (lm4) {$Q_4$};
\node at (lm5) {$Q_5$};

\node at (ln1) {$P_1$};
\node at (ln2) {$P_2$};
\node at (ln3) {$P_3$};

\node[red] at (v) {$\phi(v)$};
\end{scriptsize}

\draw[thick, fill = gray] plot [smooth cycle] coordinates {(m11) (m112) (m12) (m13) (m14) (m15)};

\draw[thick] plot [smooth cycle] coordinates {(m21) (m22) (m23) (m24)};

\draw[thick] plot [smooth cycle] coordinates {(m31) (m32) (m33) (m34)};

\draw[thick] plot [smooth cycle] coordinates {(m41) (m42) (m43) (m44)};

\draw[thick] plot [smooth cycle] coordinates {(m51) (m52) (m53) (m54)};

\draw[thick] (0,0) circle (2cm);

\draw[<-, red] ($(center)+(90:2cm)$)--($(center)+(270:2cm)$);

\draw[white] (-2.5,-2.5)--(-2.5,-2.3);
\draw[white] (2.5,2.5)--(2.5,2.3);

\end{tikzpicture} 
\hspace{1cm}
\begin{tikzpicture}[scale=1,>=latex]
  \tikzstyle{every node}=[circle,minimum size=10pt,inner sep=0.5,draw];
  \begin{scriptsize}
  \node (m1) at (0.0,0) {$Q_1$};
  \node (m4) at (-2,0.0) {$Q_4$};
  \node (m3) at (2,1.5) {$Q_3$};
  \node (m2) at (2,0.25) {$Q_2$};
  \node (m5) at (2,-1) {$Q_5$};
  \end{scriptsize}
  \tikzstyle{every node}=[circle,minimum size=10pt,inner sep=0.5,draw];
  \begin{scriptsize}
  \node (n1) at (0.75,1.0) {$P_1$};
  \node (n3) at (0.75,-1.0) {$P_3$};
  \node (n2) at (-1.0,0) {$P_2$};
  \end{scriptsize}
\path (m1) edge (n1); 
\path (m1) edge (n2); 
\path (m1) edge (n3); 

\path (n1) edge (m3); 
\path (n1) edge (m2); 
\path (n3) edge (m5); 
\path (m4) edge (n2); 

\draw[white] (-2.5,-2.4)--(-2.5,-1.8);
\draw[white] (2.5,2.6)--(2.5,2.3);
\end{tikzpicture}
\caption{\label{fig:PQ_tree_inner_nodes} To the left: a schematic view of 
a conformal model $\phi$ of~$G$ in which every component $Q$ is represented by a closed curve 
encompassing all the chords from $\phi(Q)$. 
To the right: the tree $\pqmtreeinnernodes$ with the P-nodes
$P_1,P_2,P_3$ and Q-nodes $Q_1,Q_2,Q_3,Q_4,Q_5$.
Vertex $v$ separates $Q_4$ from $Q_2,Q_3,Q_5$.
The maximal subsets consisting of pairwise neighbouring components are $\{Q_1,Q_2,Q_3\}$, $\{Q_1,Q_4\}$, and $\{Q_1,Q_5\}$,
which correspond to P-nodes $P_1$, $P_2$, and $P_3$, respectively.
}
\end{figure}

Since $G$ is a circular-arc graph, for every component $Q \in \mathcal{Q}$ and every vertex $v \in V \setminus Q$ we have either $Q \subseteq \leftside(v)$ (then we say that $Q$ is on the left side of $v$) 
or $Q \subseteq \rightside(v)$ (then we say $Q$ is on the right side of $v$).
Let $Q,Q' \in \mathcal{Q}$.
We say $v \in V \setminus (Q \cup Q')$ \emph{separates} $Q$ and $Q'$
if $v$ has $Q$ and $Q'$ on different sides.
Finally, $Q$ and $Q'$ are \emph{separated} if there is $v \in V\setminus (Q \cup Q')$ that separates $Q$ and $Q'$;
otherwise $Q$ and $Q'$ are \emph{neighbouring} -- see Figure~\ref{fig:PQ_tree_inner_nodes} for an illustration.

A \emph{P-node} is an inclusion-wise maximal subset of $\mathcal{Q}$ consisting of pairwise neighbouring components.
We denote the set of all P-nodes by $\mathcal{P}$.
Let $\pqstree$ denote a bipartite graph with the bipartition classes $\mathcal{P}$ and $\mathcal{Q}$:
we join a Q-node $Q$ and a P-node $P$ with an edge in $\pqstree$ if $Q \in P$.
Claim~\ref{claim:PQ_tree} of Section~\ref{sec:parallel_case_properties} proves that:
\begin{description}
\item[\namedlabel{prop:PQ_tree}{(T1)}] $\pqstree$ is a tree.
\end{description}
See Figure~\ref{fig:PQ_tree_inner_nodes} for an illustration.

Let $N_{\pqstree}(N)$ denote the neighbours of a node $N$ in $\pqstree$.
Let $Q$ be a Q-node and $P$ be a P-node adjacent to $Q$ in $\pqstree$. 
Let $\pqstree - Q$ be the forest obtained from $\pqstree$ by deleting the node~$Q$ 
and let $V_{\pqstree - Q}(P)$ be the set of vertices contained in the components 
from the Q-nodes of $\pqstree - Q$ 
containing the node~$P$.
Similarly, let $\pqstree - P$ be the forest obtained from $\pqstree$ by deleting the node $P$ and let
$V_{\pqstree - P}(Q)$ be the set of vertices contained in the P-nodes from the Q-nodes of $\pqstree - P$ containing the component $Q$.
In the example shown in Figure~\ref{fig:PQ_tree_inner_nodes}, we have
$V_{\pqstree - Q_1}(P_1) = Q_2 \cup Q_3$, $V_{\pqstree - Q_1}(P_2) = Q_4$,
$V_{\pqstree - P_3}(Q_1) = Q_1 \cup Q_2 \cup Q_3 \cup Q_4$.

Suppose $Q \in \mathcal{Q}$ and $P \in \mathcal{P}$ are adjacent in $\pqstree$.
Claim~\ref{claim:P_nodes_in_a_component} of Section~\ref{sec:parallel_case_properties} proves that:
\begin{description}
\item[\namedlabel{prop:pqtree_P_left_right}{(T2)}] For every $v \in Q$ either $V_{\pqstree-Q}(P) \subseteq \leftside(v)$ or $V_{\pqstree-Q}(P) \subseteq \rightside(v)$.
\end{description}
Hence, we write $P \in \leftside(v)$ if $V_{\pqstree-Q}(P) \subseteq \leftside(v)$ and $P \in \rightside(v)$ if $V_{\pqstree-Q}(P) \subseteq \rightside(v)$.

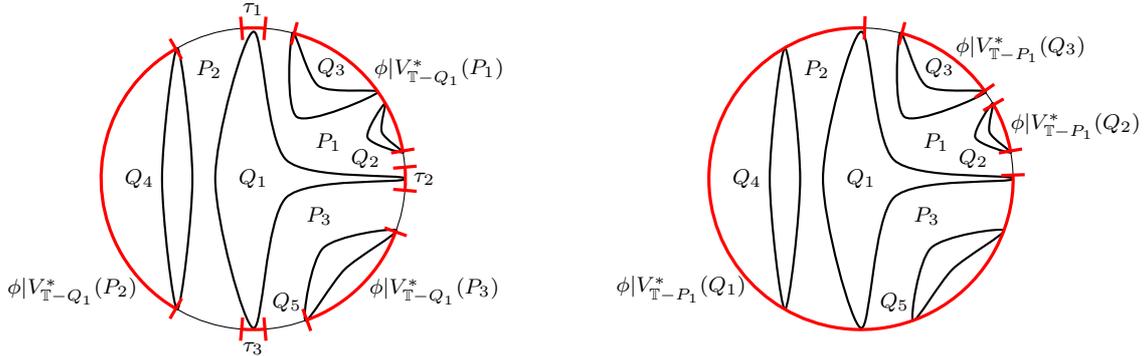
\begin{figure}[htp!]
\begin{tikzpicture}[scale=1,>=latex]
\coordinate (label) at (0,-3) {};

\coordinate (m11) at ($(center)+(90:1.95cm)$) {};
\coordinate (m112) at ($(center)+(180:0.5cm)$) {};
\coordinate (m12) at ($(center)+(270:1.98cm)$) {};
\coordinate (m13) at ($(center)+(330:0.5cm)$) {};
\coordinate (m14) at ($(center)+(0:2cm)$) {};
\coordinate (m15) at ($(center)+(30:0.5cm)$) {};
\coordinate (lm1) at ($(center)+(90:0.0cm)$) {};
\coordinate (phi_m1) at ($(center)+(225:2.07cm)$) {};

\coordinate (m21) at ($(center)+(10:2cm)$) {};
\coordinate (m22) at ($(center)+(20:1.8cm)$) {};
\coordinate (m23) at ($(center)+(30:2cm)$) {};
\coordinate (m24) at ($(center)+(20:1.6cm)$) {};
\coordinate (lm2) at ($(center)+(10:1.5cm)$) {};
\coordinate (phi_m2) at ($(center)+(20:2.05cm)$) {};

\coordinate (m31) at ($(center)+(35:2cm)$) {};
\coordinate (m32) at ($(center)+(55:1.5cm)$) {};
\coordinate (m33) at ($(center)+(75:2cm)$) {};
\coordinate (m34) at ($(center)+(55:1cm)$) {};
\coordinate (lm3) at ($(center)+(55:1.8cm)$) {};

\coordinate (m41) at ($(center)+(120:2cm)$) {};
\coordinate (m42) at ($(center)+(0:-1.2cm)$) {};
\coordinate (m43) at ($(center)+(240:2cm)$) {};
\coordinate (m44) at ($(center)+(0:-0.8cm)$) {};
\coordinate (lm4) at ($(center)+(180:1.5cm)$) {};

\coordinate (m51) at ($(center)+(290:2cm)$) {};
\coordinate (m52) at ($(center)+(315:1.7cm)$) {};
\coordinate (m53) at ($(center)+(340:2cm)$) {};
\coordinate (m54) at ($(center)+(315:1.3cm)$) {};
\coordinate (lm5) at ($(center)+(285:1.7cm)$) {};

\coordinate (ln1) at ($(center)+(26:1.1cm)$) {};
\coordinate (ln2) at ($(center)+(112:1.56cm)$) {};
\coordinate (ln3) at ($(center)+(330:1.0cm)$) {};

\coordinate (phi_n1) at ($(center)+(42.5:2.1cm)$) {};

\coordinate (phi_tau_1) at ($(center)+(90:2.25cm)$) {};

\coordinate (phi_n2) at ($(center)+(225:2.1cm)$) {};

\coordinate (phi_tau_2) at ($(center)+(270:2.25cm)$) {};

\coordinate (phi_n3) at ($(center)+(315:2.1cm)$) {};

\coordinate (phi_tau_3) at ($(center)+(0:2.25cm)$) {};

\begin{tiny}
\tikzstyle{every node}=[inner sep=1pt]
\node[anchor=west] at (phi_n1) {$\phi|V^*_{\pqmtreeinnernodes - Q_1}(P_1)$};
\node[anchor=east] at (phi_n2) {$\phi|V^*_{\pqmtreeinnernodes - Q_1}(P_2)$};
\node[anchor=west] at (phi_n3) {$\phi|V^*_{\pqmtreeinnernodes - Q_1}(P_3)$};
\node at (phi_tau_1) {$\tau_1$};
\node at (phi_tau_2) {$\tau_3$};
\node at (phi_tau_3) {$\tau_2$};

\node at (lm1) {$Q_1$};
\node at (lm2) {$Q_2$};

\node at (lm3) {$Q_3$};

\node at (lm4) {$Q_4$};
\node at (lm5) {$Q_5$};

\node at (ln1) {$P_1$};
\node at (ln2) {$P_2$};
\node at (ln3) {$P_3$};
\end{tiny}

\draw[thick] plot [smooth cycle] coordinates {(m11) (m112) (m12) (m13) (m14) (m15)};

\draw[thick] plot [smooth cycle] coordinates {(m21) (m22) (m23) (m24)};

\draw[thick] plot [smooth cycle] coordinates {(m31) (m32) (m33) (m34)};

\draw[thick] plot [smooth cycle] coordinates {(m41) (m42) (m43) (m44)};

\draw[thick] plot [smooth cycle] coordinates {(m51) (m52) (m53) (m54)};

\draw (0,0) circle (2cm);

\draw[very thick,red,|-|] ([shift=(10:2cm)]0,0) arc (10:75:2cm);

\draw[very thick,red,|-|] ([shift=(85:2cm)]0,0) arc (85:95:2cm);

\draw[very thick,red,|-|] ([shift=(120:2cm)]0,0) arc (120:240:2cm);

\draw[very thick,red,|-|] ([shift=(265:2cm)]0,0) arc (265:275:2cm);

\draw[very thick,red,|-|] ([shift=(290:2cm)]0,0) arc (290:340:2cm);

\draw[very thick,red,|-|] ([shift=(-5:2cm)]0,0) arc (-5:5:2cm);

\draw[white] (-3.4,-2.5)--(-3.4,-2.3);
\draw[white] (3.8,2.5)--(3.8,2.3);

\end{tikzpicture} 
\hspace{0.5cm}
\begin{tikzpicture}[scale=1,>=latex]
\coordinate (label) at (0,-3) {};

\coordinate (m11) at ($(center)+(90:1.95cm)$) {};
\coordinate (m112) at ($(center)+(180:0.5cm)$) {};
\coordinate (m12) at ($(center)+(270:1.98cm)$) {};
\coordinate (m13) at ($(center)+(330:0.5cm)$) {};
\coordinate (m14) at ($(center)+(0:2cm)$) {};
\coordinate (m15) at ($(center)+(30:0.5cm)$) {};
\coordinate (lm1) at ($(center)+(90:0.0cm)$) {};
\coordinate (phi_m1) at ($(center)+(225:2.07cm)$) {};

\coordinate (m21) at ($(center)+(10:2cm)$) {};
\coordinate (m22) at ($(center)+(20:1.8cm)$) {};
\coordinate (m23) at ($(center)+(30:2cm)$) {};
\coordinate (m24) at ($(center)+(20:1.6cm)$) {};
\coordinate (lm2) at ($(center)+(10:1.5cm)$) {};
\coordinate (phi_m2) at ($(center)+(20:2.05cm)$) {};

\coordinate (m31) at ($(center)+(35:2cm)$) {};
\coordinate (m32) at ($(center)+(55:1.5cm)$) {};
\coordinate (m33) at ($(center)+(75:2cm)$) {};
\coordinate (m34) at ($(center)+(55:1cm)$) {};
\coordinate (lm3) at ($(center)+(55:1.8cm)$) {};
\coordinate (phi_m3) at ($(center)+(55:2.1cm)$) {};

\coordinate (m41) at ($(center)+(120:2cm)$) {};
\coordinate (m42) at ($(center)+(0:-1.2cm)$) {};
\coordinate (m43) at ($(center)+(240:2cm)$) {};
\coordinate (m44) at ($(center)+(0:-0.8cm)$) {};
\coordinate (lm4) at ($(center)+(180:1.5cm)$) {};

\coordinate (m51) at ($(center)+(290:2cm)$) {};
\coordinate (m52) at ($(center)+(315:1.7cm)$) {};
\coordinate (m53) at ($(center)+(340:2cm)$) {};
\coordinate (m54) at ($(center)+(315:1.3cm)$) {};
\coordinate (lm5) at ($(center)+(285:1.7cm)$) {};

\coordinate (ln1) at ($(center)+(26:1.1cm)$) {};
\coordinate (ln2) at ($(center)+(112:1.56cm)$) {};
\coordinate (ln3) at ($(center)+(330:1.0cm)$) {};

\begin{tiny}
\tikzstyle{every node}=[inner sep=1pt]
\node at (lm1) {$Q_1$};
\node[anchor=east] at (phi_m1) {$\phi|V^*_{\pqmtreeinnernodes - P_1}(Q_1)$};
\node at (lm2) {$Q_2$};
\node[anchor=west] at (phi_m2) {$\phi|V^*_{\pqmtreeinnernodes - P_1}(Q_2)$};

\node at (lm3) {$Q_3$};
\node[anchor=west] at (phi_m3) {$\phi|V^*_{\pqmtreeinnernodes - P_1}(Q_3)$};

\node at (lm4) {$Q_4$};
\node at (lm5) {$Q_5$};

\node at (ln1) {$P_1$};
\node at (ln2) {$P_2$};
\node at (ln3) {$P_3$};
\end{tiny}

\draw[thick] plot [smooth cycle] coordinates {(m11) (m112) (m12) (m13) (m14) (m15)};

\draw[thick] plot [smooth cycle] coordinates {(m21) (m22) (m23) (m24)};

\draw[thick] plot [smooth cycle] coordinates {(m31) (m32) (m33) (m34)};

\draw[thick] plot [smooth cycle] coordinates {(m41) (m42) (m43) (m44)};

\draw[thick] plot [smooth cycle] coordinates {(m51) (m52) (m53) (m54)};

\draw (0,0) circle (2cm);

\draw[very thick,red,|-|] ([shift=(10:2cm)]0,0) arc (10:30:2cm);
\draw[very thick,red,|-|] ([shift=(35:2cm)]0,0) arc (35:75:2cm);
\draw[very thick,red,|-|] ([shift=(88:2cm)]0,0) arc (88:362:2cm);

\draw[white] (-3.4,-2.5)--(-3.4,-2.3);
\draw[white] (3.8,2.5)--(3.8,2.3);

\end{tikzpicture} 
\caption{\label{fig:pqtree_subwords} 
A conformal model $\phi$ of $G_{ov}$ and its contiguous subwords: $\phi|V^*_{\pqmtreeinnernodes - Q_1}(P_1)$, 
$\phi|V^*_{\pqmtreeinnernodes - Q_1}(P_2)$, and $\phi|V^*_{\pqmtreeinnernodes - Q_1}(P_3)$ (to the left), 
and $\phi|V^*_{\pqmtreeinnernodes - P_1}(Q_1)$, $\phi|V^*_{\pqmtreeinnernodes - P_1}(Q_2)$, and 
$\phi|V^*_{\pqmtreeinnernodes - P_1}(Q_3)$ (to the right). 
We have $\phi[Q_1] \equiv \tau_1P_1\tau_2P_3\tau_3P_2$ and $\phi[P_1] \equiv Q_1Q_3Q_2$.}
\end{figure}

Claim \ref{claim:pq_tree_conformal_models_properties_nodes} of Section~\ref{sec:parallel_case_properties} proves the following properties of a conformal model $\phi$ of $G_{ov}$ with respect to the nodes of $\pqstree$:
\begin{description}
\item[\namedlabel{prop:pqtree_Q_property}{(T3)}] For every $Q \in \mathcal{Q}$ and every $P \in N_{\pqstree}(Q)$ the set $V^*_{\pqstree - Q}(P)$
is contiguous in $\phi$.
Moreover, for every two different nodes $P,P' \in N_{\pqstree}(Q)$ there is $v \in Q$ 
such that $\phi(v)$ separates the words $\phi|V^*_{\pqstree - Q}(P)$
and $\phi|V^*_{\pqstree - Q}(P')$.
\item[\namedlabel{prop:pqtree_P_property}{(T4)}] For every $P \in \mathcal{P}$ and every $Q \in N_{\pqstree}(P)$ the set $V^*_{\pqstree - P}(Q)$ is 
contiguous in $\phi$.
\end{description}
See Figure~\ref{fig:pqtree_subwords} for an illustration.

Let $\phi$ be a conformal model of $G_{ov}$.
For a Q-node $Q$ in $\pqstree$ by $\phi[Q]$ we denote a circular word obtained from $\phi$ by replacing the contiguous subword $\phi|V^*_{\pqstree - Q}(P)$ in $\phi$ by the letter~$P$, for every $P \in N_\pqstree(Q)$.
Note that $\phi[Q]$ is a circular word on $Q^* \cup N_\pqstree(Q)$ and that $\phi[Q]$ extends the conformal model $\phi \Vert Q^*$ of $(Q,{\sim})$ by the letters from $N_{\pqstree}(Q)$.
Hence, $\phi[Q]$ is called \emph{the extended conformal model of $(Q,{\sim})$ induced by $\phi$}.
Observe that, by property~\ref{prop:pqtree_Q_property}, no two letters from $N_\pqstree(Q)$ appear next to each other in $\phi[Q]$.
Similarly, for a P-node $P$ in $\pqstree$ by $\phi[P]$ we denote a circular word obtained from $\phi$ by replacing the contiguous subword $\phi|V^*_{\pqstree - P}(Q)$ by the letter~$Q$, for every $Q \in N_\pqstree(P)$.
Note that $\phi[P]$ is a circular order of the Q-nodes from the set $N_\pqstree(P)$.
See Figure~\ref{fig:pqtree_subwords} for an illustration.

Now, our goal is to characterize the extended conformal models $\phi[Q]$
induced by the conformal models $\phi$ of $G_{ov}$.
We recall that $\phi[Q]$ is an extension of the conformal model $\phi \Vert Q^*$ of $(Q,{\sim})$ by the letters from the set $N_{\pqstree}(Q)$ 
such that: 
\begin{itemize}
\item For every $P \in N_{\pqstree}(Q)$ and every $v \in Q$:
\begin{itemize}
\item $P$ is to the left of the chord of $v$ if $P \in \leftside(v)$,
\item $P$ is to the right of the chord of $v$ if $P \in \rightside(v)$.
\end{itemize}
\end{itemize}
A conformal model $\psi^Q$ of $(Q,{\sim})$ is said to be \emph{extendable by the set $N_{\pqstree}(Q)$} if we can insert the letters from $N_{\pqstree}(Q)$ into $\psi^Q$ 
so as the above condition holds.
Observe that such extension, if exists, is unique, and then we denote it by $\psi_Q$.
Eventually, note that the conformal models of $(Q,{\sim})$ extendable by $N_{\pqstree}(Q)$ are in the correspondence 
with the extended conformal models $\phi[Q]$ induced by the conformal models $\phi$ of $G_{ov}$.
Indeed, given a conformal model $\psi^Q$ of $(Q,{\sim})$ and its extension $\psi_Q$, we obtain a conformal model $\phi$ of $G_{ov}$ satisfying $\phi[Q] \equiv \psi_Q$ as follows:
we replace every $P \in N_{\pqstree}(Q)$ in $\psi_Q$ by the word $\phi'|V^*_{\pqstree -Q}(P)$, 
where $\phi'$ is any conformal model of $G_{ov}$.

Clearly, not every conformal model of $(Q,{\sim})$ can be extended by $N_{\pqstree}(Q)$.
Figure~\ref{fig:inside_notation} shows an extended conformal model $\phi_Q$ of $(Q,\sim)$
for some prime node $Q$.
Let $\phi^Q \equiv \phi_Q \Vert Q^*$.
We may swap the chords $\phi^Q(e), \phi^Q(f)$ with the chords $\phi^Q(g), \phi^Q(h)$ in $\phi^Q$ to get a conformal model of $(Q,{\sim})$, which is still extendable by $N_{\pqstree}(Q)$.
We can also swap the chords $\phi^Q(a), \phi^Q(b)$ with the chords $\phi^Q(c),\phi^Q(d)$ to get another conformal model of $(Q,{\sim})$, however, this model is not extendable by $N_{\pqstree}(Q)$ (as there is no valid place for~$P_1$).  

Now, our goal is to characterize the conformal models of $(Q,{\sim})$ extendable by $N_{\pqstree}(Q)$, for a fixed Q-node $Q$.
We first consider the case when the set $Q$ induces a consistent permutation model in some conformal model of $G_{ov}$; such a component $Q$ is called a~\emph{permutation component}.
Then we examine the case when $Q$ is a non-permutation component;  
we consider two cases depending on whether $Q$ is serial or prime in $\strongModules(G_{ov})$.
For the rest of this section, given an extended conformal model $\phi_Q$ of $(Q,{\sim})$, by $\phi^Q$ we denote the conformal model $\phi_Q \Vert Q^*$ of $(Q,{\sim})$. 

\subsubsection{Permutation components}
Suppose $Q \in \mathcal{Q}$ is a permutation component, that is, suppose $Q$ induces a consistent permutation model $(\tau^0_{\phi},\tau^1_{\phi})$ in some conformal model~$\phi$ of $G_{ov}$.
We assume $\tau^0_{\phi},\tau^1_{\phi}$ are enumerated such that $r^0 \in \tau^0_{\phi}$, where 
$r$ is some fixed representant of the set $Q$.
Suppose $Q^0_{\phi}$ and $Q^1_{\phi}$ are the letters in the words~$\tau^0_\phi$ and~$\tau^1_\phi$, respectively, and suppose $({<_{\phi}},{\prec_{\phi}})$ are the transitive orientations of $(Q,{\parallel})$ and $(Q,{\sim})$ corresponding to the permutation model $(\tau^0_{\phi}, \tau^1_{\phi})$ of $(Q,{\sim})$.
Clearly, since $(\tau^0_\phi,\tau^1_\phi)$ is a consistent permutation model of $(Q,{\sim})$, 
we have $\phi \equiv \tau_{\phi,L} \cdot \tau^{0}_{\phi} \cdot \tau_{\phi,R} \cdot \tau^{1}_{\phi}$ for some two words $\tau_{\phi,L}$ and $\tau_{\phi,R}$.
Since $V$ is parallel, at least one among the words $\tau_{\phi,L}, \tau_{\phi,R}$ is non-empty. 
If $\tau_{\phi,L}$ is non-empty, $\tau_{\phi,L}$ is a permutation 
of the set $V^{*}_{\pqstree - Q}(P_L)$ for some P-node $P_L$ adjacent to $Q$ in $\pqstree$.
Similarly, if $\tau_{\phi,R}$ is non-empty, $\tau_{\phi,R}$ is a permutation 
of the set $V^{*}_{\pqstree - Q}(P_R)$ for some P-node $P_R$ adjacent to $Q$ in $\pqstree$.
Suppose for a while that both the words $\tau_{\phi,L}$ and $\tau_{\phi,R}$ are non-empty (the other cases are handled similarly).
Then, $\phi[Q]$ has the form 
$P_L \tau^0_{\phi} P_R \tau^1_{\phi}$.
Since for every conformal model $\phi'$ of $G_{ov}$ and every $P \in N_{\pqstree}(Q)$ 
the set $V^{*}_{\pqstree - Q}(P)$ is contiguous in $\phi'$,
arguing similarly as in the proof of Claim~\ref{claim:metachords_for_children_of_prime}, 
we deduce that $\phi'[Q]$ has the form $P_L \tau^0_{\phi'} P_R \tau^1_{\phi'}$,
where $\tau^j_{\phi'}$ is a permutation of $\tau^j_{\phi}$ for $j \in \{0,1\}$ and $(\tau^0_{\phi'},\tau^1_{\phi'})$ is a consistent permutation model of $(Q,{\sim})$.
In particular, we have 
$$
(Q^0_{\phi},Q^{1}_{\phi},{<_{\phi}}) =  (Q^0_{\phi'},Q^{1}_{\phi'},{<_{\phi'}}),
$$
where $Q^0_{\phi'}$, $Q^1_{\phi'}$, and ${<_{\phi'}}$ for the model $\phi'$ 
are defined analogously as for the model $\phi$.

Given the above, let $\camodules(Q) = \{Q\}$, $(Q^0,Q^{1},{<_{Q}}) = (Q^0_{\phi},Q^{1}_{\phi},{<_{\phi}})$, let $\pi(Q)$ be the word obtained from $\phi[Q]$ by replacing $\tau^0_{\phi}$ by $Q^0$ and 
$\tau^1_{\phi}$ by $Q^1$, and let $\Pi(Q) = \{\pi(Q)\}$.
Note that the reflection of $\pi(Q)$ equals to $\pi(Q)$.

\begin{definition}
We say a word $\phi^{adm}_Q$ on $Q^* \cup N_{\pqstree}(Q)$ is \emph{admissible by 
$\Pi(Q)$} if $\phi^{adm}_Q$ is obtained from the word $\pi(Q)$ by replacing 
$Q^0$ and $Q^1$ by $\tau^0$ and $\tau^1$, respectively, 
where $(\tau^0,\tau^1)$ is a model admissible by $\QQQ$.
\end{definition}

The next lemma characterizes the extended conformal models of $(Q,{\sim})$ for the case when $Q$ is a permutation component.
\begin{lemma}
Suppose $Q$ is a permutation component in $\mathcal{Q}$ and let $\phi_Q$ be a 
circular word on $Q^* \cup N_{\pqstree}(Q)$.
The word $\phi_Q$ is an extended conformal model of $(Q,{\sim})$ 
if and only if $\phi_Q$ is admissible by $\Pi(Q)$.
\end{lemma}

\input ./figures/proof_sketch/parallel/inside_notation.tex

\subsubsection{Prime non-permutation components}
Suppose $Q \in \mathcal{Q}$ is a prime non-permutation component.
The conformal models of $(Q,{\sim})$ are described in Subsection~\ref{subsec:conformal_models_prime}.
So, let $K(Q)$ be the equivalence classes of $K$-relation as introduced in Definition~\ref{def:prime-K-relation},
let $\mathbb{K} = (K^0,K^1,{<_K})$ be the metachord associated with the CA-module $K$ for $K \in K(Q)$, and
let $\Gamma(Q) = \{\gamma^0(Q),\gamma^1(Q)\}$ be the set 
containing the circular orders of the slots $\{K^0,K^1: K \in K(Q)\}$
such that the conformal models for $(Q,{\sim})$ coincide with the admissible models for $\Gamma(Q)$ - see Theorem~\ref{thm:conformal_models_prime_component}.

Suppose $R$ is a set containing a representant of every set in $K(M)$.
We assume $r^0 \in K^0$, where $r \in R$ is a vertex representing the set $K \in K(M)$.

First, we list some properties of the extended conformal models of $(Q,{\sim})$.
Lemma~\ref{lemma:prime_consistent_modules}.\eqref{item:prime_consistent_modules_skeleton} of Section~\ref{sec:prime_case_properties} asserts that: 
\begin{description}
 \item[\namedlabel{prop:parallel_prime_skeleton}{(PP1)}] There are two conformal models $\phi^0_R$ and $\phi^1_R$ of the graph $(R,{\sim})$, 
 one being the reflection of the other. 
 In particular, for every extended conformal model $\phi_Q$ of $(Q,{\sim})$
 we have either $\phi_Q \Vert R^* \equiv \phi^0_R$ or $\phi_Q \Vert R^* \equiv \phi^1_R$.
\end{description}
The next property follows from Lemma~~\ref{lemma:prime_consistent_modules}.\eqref{item:prime_consistent_modules_contiguous_subwords} of Section~\ref{sec:prime_case_properties} and from the fact that $\phi^Q \equiv \phi_Q \Vert Q^*$ is a conformal modal for $(Q,{\sim})$:
\begin{description} 
 \item[\namedlabel{prop:parallel_prime_contiguous_subwords}{(PP2)}] 
 For every extended conformal model $\phi_Q$ of $(Q,{\sim})$ 
 and every $K \in K(M)$ the sets $K^0$ and $K^1$ are contiguous
 in $\phi^Q$ and $(\phi^Q|K^0, \phi^Q|K^1)$ is a consistent permutation model of $(K,{\sim})$ in $\phi^Q$.
 \end {description}
Note that the above property does not assert that the set $K$ induces a consistent permutation model in $\phi_Q$.
Clearly, since $\phi^Q$ is conformal for $(Q,{\sim})$,
$(\phi^Q|K^0, \phi^Q|K^1)$ is admissible for $\KKK$.

For an extended conformal model $\phi_Q$ of $(Q,{\sim})$, $K \in K(Q)$, and $j \in \{0,1\}$, by $\tau(\phi_Q,K^j)$ we denote the shortest contiguous subword of $\phi_Q$ containing all the letters from $K^j$ and no letter from~$K^{1-j}$.
Properties \ref{prop:parallel_prime_contiguous_subwords} and \ref{prop:pqtree_Q_property} assert that $\tau(\phi_Q,K^j)$ is an extension
of the word $\phi^Q|K^j$ by some (possibly zero) letters from $N_{\pqstree}(Q)$ -- see Figure~\ref{fig:inside_notation} for an illustration.
Let $|\tau(\phi_Q,K^j)|$ denote the length of the word $\tau(\phi_Q,K^j)$.
Eventually, let $\inside(K)$ denote the set of the letters from $N_{\pqstree}(Q)$ that occur either in $\tau(\phi_Q,K^0)$ or in $\tau(\phi_Q,K^1)$.
The next claim proves some useful properties of the words $\tau(\phi_Q,K^j)$ and shows, in particular, that the set $\inside(K)$ does not depend on the choice of $\phi_Q$.
We refer to Figure~\ref{fig:tau_words} for an example illustrating the claim.
\begin{figure}[!h]
\begin{tikzpicture}[xscale=0.55,yscale=1,>=latex]
\coordinate (u1) at (0,3) {};
\coordinate (u2) at (1,3) {};
\coordinate (u3) at (2,3) {};
\coordinate (u4) at (3,3) {};
\coordinate (u5) at (4,3) {};
\coordinate (u6) at (5,3) {};
\coordinate (u7) at (6,3) {};
\coordinate (u8) at (7,3) {};
\coordinate (u9) at (8,3) {};

\coordinate (lu1) at (0,3.25) {};
\coordinate (lu2) at (1,3.25) {};
\coordinate (lu3) at (2,3.25) {};
\coordinate (lu4) at (3,3.25) {};
\coordinate (lu5) at (4,3.25) {};
\coordinate (lu6) at (5,3.25) {};
\coordinate (lu7) at (6,3.25) {};
\coordinate (lu8) at (7,3.25) {};
\coordinate (lu9) at (8,3.25) {};

\coordinate (lz0) at (-1.5,1.5) {};
\coordinate (lz1) at (9.5,1.5) {};

\coordinate (b1) at (0,0) {};
\coordinate (b2) at (1,0) {};
\coordinate (b3) at (2,0) {};
\coordinate (b4) at (3,0) {};
\coordinate (b5) at (4,0) {};
\coordinate (b6) at (5,0) {};
\coordinate (b7) at (6,0) {};
\coordinate (b8) at (7,0) {};
\coordinate (b9) at (8,0) {};

\coordinate (lb1) at (0,-0.2) {};
\coordinate (lb2) at (1,-0.2) {};
\coordinate (lb3) at (2,-0.2) {};
\coordinate (lb4) at (3,-0.2) {};
\coordinate (lb5) at (4,-0.2) {};
\coordinate (lb6) at (5,-0.2) {};
\coordinate (lb7) at (6,-0.2) {};
\coordinate (lb8) at (7,-0.2) {};
\coordinate (lb9) at (8,-0.2) {};

\coordinate (tau0) at (-2.5,3.25);
\coordinate (tau1) at (-2.5,-0.25);

\tikzstyle{every node}=[inner sep=2pt,fill=white]

\draw[->] (u1)--(b3);
\draw[<-] (u2)--(b5);
\draw[-,very thick] ($(u3)+(-0.4,0)$) -- ($(u3)+(0.4,0)$);
\draw[<-] (u4)--(b4);
\draw[<-] (u5)--(b7);
\draw[->, very thick, red] (u6)--(b9);
\draw[-,very thick] ($(u7)+(-0.4,0)$) -- ($(u7)+(0.4,0)$);
\draw[->] (u8)--(b1);
\draw[<-] (u9)--(b2);

\draw[very thick, -] ($(b6)+(-0.4,0)$) -- ($(b6)+(0.4,0)$);
\draw[very thick, -] ($(b8)+(-0.4,0)$) -- ($(b8)+(0.4,0)$);

\draw[->] (-1,3) -- (9,3);
\draw[dashed, red, thick, ->] (-1,1.5) -- (9,1.5);
\draw[<-] (-1,0) -- (9,0);

\tikzstyle{every node}=[inner sep=1pt]
\begin{tiny}
\node at (lz0) {$z^0$};
\node at (lz1) {$z^1$};

\node at (lu1) {$v^0_1$};
\node at (lu2) {$v^1_2$};
\node at (lu3) {$P_1$};
\node at (lu4) {$v^1_3$};
\node at (lu5) {$v^1_4$};
\node at (lu6) {$v^0_5$};
\node at (lu7) {$P_2$};
\node at (lu8) {$v^0_6$};
\node at (lu9) {$v^1_7$};

\node at (lb1) {$v^1_6$};
\node at (lb2) {$v^0_7$};
\node at (lb3) {$v^1_1$};
\node at (lb4) {$v^0_3$};
\node at (lb5) {$v^0_2$};
\node at (lb6) {$P_3$};
\node at (lb7) {$v^1_4$};
\node at (lb8) {$P_4$};
\node at (lb9) {$v^1_5$};

\node at (tau0) {$\tau(\phi_Q,K^0)$};
\node at (tau1) {$\tau(\phi_Q,K^1)$};

\end{tiny}
\end{tikzpicture}
\hspace{0.2cm}
\begin{tikzpicture}[xscale=0.60,yscale=1]
\draw[dashed] (0,-0.5) -- (0,3.5);
\end{tikzpicture}
\hspace{0.2cm}
\begin{tikzpicture}[xscale=0.55,yscale=1,>=latex]
\coordinate (u1) at (0,3) {};
\coordinate (u2) at (-1,3) {};
\coordinate (u3) at (-2,3) {};
\coordinate (u4) at (-3,3) {};
\coordinate (u5) at (-4,3) {};
\coordinate (u6) at (-5,3) {};
\coordinate (u7) at (-6,3) {};
\coordinate (u8) at (-7,3) {};
\coordinate (u9) at (-8,3) {};

\coordinate (lu1) at (0,3.25) {};
\coordinate (lu2) at (-1,3.25) {};
\coordinate (lu3) at (-2,3.25) {};
\coordinate (lu4) at (-3,3.25) {};
\coordinate (lu5) at (-4,3.25) {};
\coordinate (lu6) at (-5,3.25) {};
\coordinate (lu7) at (-6,3.25) {};
\coordinate (lu8) at (-7,3.25) {};
\coordinate (lu9) at (-8,3.25) {};

\coordinate (b1) at (0,0) {};
\coordinate (b2) at (-1,0) {};
\coordinate (b3) at (-2,0) {};
\coordinate (b4) at (-3,0) {};
\coordinate (b5) at (-4,0) {};
\coordinate (b6) at (-5,0) {};
\coordinate (b7) at (-6,0) {};
\coordinate (b8) at (-7,0) {};
\coordinate (b9) at (-8,0) {};

\coordinate (lb1) at (0,-0.2) {};
\coordinate (lb2) at (-1,-0.2) {};
\coordinate (lb3) at (-2,-0.2) {};
\coordinate (lb4) at (-3,-0.2) {};
\coordinate (lb5) at (-4,-0.2) {};
\coordinate (lb6) at (-5,-0.2) {};
\coordinate (lb7) at (-6,-0.2) {};
\coordinate (lb8) at (-7,-0.2) {};
\coordinate (lb9) at (-8,-0.2) {};

\coordinate (tau0) at (2.5,3.25);
\coordinate (tau1) at (2.5,-0.25);

\coordinate (lz0) at (1.5,1.5) {};
\coordinate (lz1) at (-9.5,1.5) {};

\tikzstyle{every node}=[inner sep=2pt,fill=white]

\draw[<-] (u1)--(b3);
\draw[->] (u2)--(b5);
\draw[-,very thick] ($(u3)+(-0.4,0)$) -- ($(u3)+(0.4,0)$);
\draw[->] (u4)--(b4);
\draw[->] (u5)--(b7);
\draw[<-, very thick, red] (u6)--(b9);
\draw[-,very thick] ($(u7)+(-0.4,0)$) -- ($(u7)+(0.4,0)$);
\draw[<-] (u8)--(b1);
\draw[->] (u9)--(b2);

\draw[very thick, -] ($(b6)+(-0.4,0)$) -- ($(b6)+(0.4,0)$);
\draw[very thick, -] ($(b8)+(-0.4,0)$) -- ($(b8)+(0.4,0)$);

\draw[<-] (1,3) -- (-9,3);
\draw[dashed, thick, red,<-] (1,1.5) -- (-9,1.5);
\draw[->] (1,0) -- (-9,0);

\tikzstyle{every node}=[inner sep=1pt]
\begin{tiny}
\node at (lz0) {$z^1$};
\node at (lz1) {$z^0$};

\node at (lu1) {$v^1_1$};
\node at (lu2) {$v^0_2$};
\node at (lu3) {$P_1$};
\node at (lu4) {$v^0_3$};
\node at (lu5) {$v^0_4$};
\node at (lu6) {$v^1_5$};
\node at (lu7) {$P_2$};
\node at (lu8) {$v^1_6$};
\node at (lu9) {$v^0_7$};

\node at (lb1) {$v^0_6$};
\node at (lb2) {$v^1_7$};
\node at (lb3) {$v^0_1$};
\node at (lb4) {$v^1_3$};
\node at (lb5) {$v^1_2$};
\node at (lb6) {$P_3$};
\node at (lb7) {$v^0_4$};
\node at (lb8) {$P_4$};
\node at (lb9) {$v^0_5$};

\node at (tau0) {$\tau(\phi^R_Q,K^1)$};
\node at (tau1) {$\tau(\phi^R_Q,K^0)$};

\end{tiny}
\end{tikzpicture}
\hspace{0.2cm}

\caption{\label{fig:tau_words} 
The words $\tau(\phi_Q,K^0)$, $\tau(\phi_Q,K^1)$ and the words $\tau(\phi^R_Q,K^0)$, $\tau(\phi^R_Q,K^1)$, 
where $\phi_Q$ and $\phi^R_Q$ are two extended models of $(Q,{\sim})$ and $\phi^R_Q$ is the reflection of $\phi_Q$.
We have $|\tau(\phi_Q,K^0)| = |\tau(\phi^R_Q,K^1)| = 9$.
}
\end{figure}
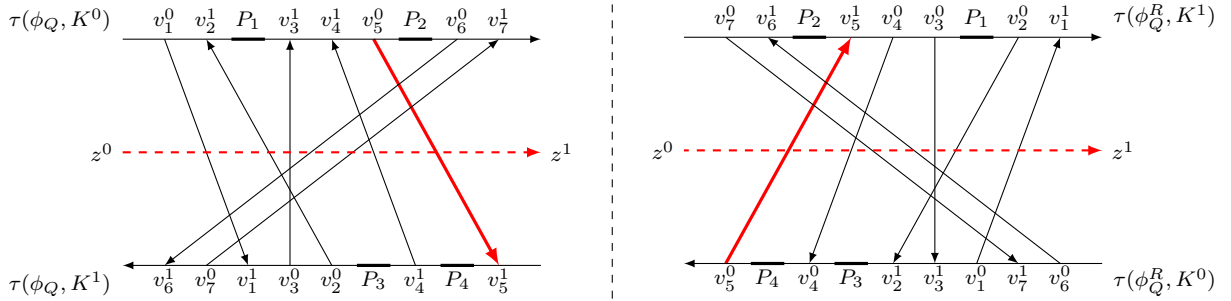
\begin{claim}
\label{claim:tau_phi_M_K_properties}
Suppose $\phi_Q, \phi'_Q$ are two conformal models of a Q-node $Q$ and suppose $a,b$ are two vertices in $K$.
\begin{enumerate}
 \item \label{item:inside_definition_1} If $\phi_Q \Vert Q^*$ and $\phi'_Q \Vert Q^*$ are admissible for $\gamma^{t}(Q)$ for some $t \in \{0,1\}$, 
 then for every $j \in \{0,1\}$, $a' \in \{a^0,a^1\}$, $b' \in \{b^0,b^1\}$, and $P \in N_{\pqstree}(Q)$:
 $$\text{$a' P b'$ is a subword of $\tau(\phi_Q,K^j)$ $\iff$  $a' P b'$ is a subword of $\tau(\phi'_Q,K^j)$}.$$
 \noindent In particular, $|\tau(\phi_Q,K^j)| = |\tau(\phi'_Q,K^j)|$ and $P$ occurs at position $k$ in $\tau(\phi_Q,K^j)$ if and only if $P$ occurs at position $k$ in $\tau(\phi'_Q,K^j)$.
 \item \label{item:inside_definition_2} If $\phi^Q \Vert Q^*$ is admissible for $\gamma^t(M)$ and $\phi'_Q \Vert Q^*$ is admissible for $\gamma^{1-t}(Q)$ for some $t \in \{0,1\}$, 
 then for every $j \in \{0,1\}$, $a' \in \{a^0,a^1\}$, $b' \in \{b^0,b^1\}$, 
 and $P \in N_{\pqstree}(Q)$:
 $$\text{$a' P b'$ is a subword of $\tau(\phi_Q,K^j)$ $\iff$ $b'' P a''$ is a subword of $\tau(\phi'_Q,K^{1-j})$},$$ 
 where $a'',b''$ are such that $\{a',a''\} = \{a^0,a^1\}$ and $\{b',b''\} = \{b^0,b^1\}$.
 \noindent In particular, $|\tau(\phi_Q,K^j)| = |\tau(\phi'_Q,K^{1-j})|$ and $P$ occurs  at position $k$ in $\tau(\phi_Q,K^j)$ 
 if and only if $P$ occurs at position $|\tau(\phi'_Q,K^{1-j})| - k +1$ in $\tau(\phi'_Q,K^{1-j})$.
\end{enumerate}
\end{claim}
\begin{proof}
Let $z$ be a vertex of $Q \setminus K$ such that $z \sim K$ -- such a vertex exists as $Q$ is prime.
Suppose $\phi_Q \Vert Q^*$ is admissible for $\gamma^0(Q)$
and suppose $\phi_Q(z)$ has the letters from $K^j$ on the left side.
Then, $\phi'_Q(z)$ has the letters from $K^j$ on the right side if $\phi'_Q \Vert Q$ is admissible to 
$\gamma^1(Q)$.

Suppose $a^1Pb^1$ is a subword of $\tau(\phi_Q,K^j)$ (the other cases are handled similarly).
It means that $P \in \leftside(z)$, $P \in \rightside(a)$, and $P \in \leftside(b)$.
Figure~\ref{fig:inside_property} shows possible relations between $\phi'_Q(a),\phi'_Q(b)$, and $P$, in conformal models
$\phi'_Q$ such that $\phi'_Q \Vert Q^*$ is admissible to $\gamma^0(Q)$ (to the left) and to $\gamma^1(Q)$ (to the right).

To show statement~\eqref{item:inside_definition_1} note that for every conformal model $\phi'_Q$ such that $\phi'_Q \Vert Q^*$ 
is admissible for $\gamma^0(Q)$,
the letters from $K^j$ occur on the left side of $\phi'_Q(z)$.
In particular, the letters $a^1,b^1,P$ occur on the left side of $\phi'_Q(z)$ as  $P \in \leftside(z)$.
Since $P \in \rightside(a)$ and $P \in \leftside(b)$, the letters $a^1,b^1,P$ occur in $\tau(\phi'_Q,K^j)$ in the order $a^1,P,b^1$.
The other cases are proved similarly.

To prove statement \eqref{item:inside_definition_2} suppose $\phi'_Q \Vert Q^*$ is admissible for $\gamma^1(Q)$.
In particular, it means that the letters $a^0,b^0$ are on the left side of $\phi'_Q(z)$.
Also, $P$ is on the left side of $\phi'_Q(z)$ as $P \in \leftside(z)$.
Now, $P \in \rightside(a)$ and $P \in \leftside(b)$ implies that the letters $a^0, b^0, P$ occur in $\tau(\phi'_Q,K^j)$ in the order $b^0,P,a^0$ -- see Figure~\ref{fig:inside_property} to the right.

\begin{figure}
\begin{tikzpicture}[xscale=0.6,yscale=0.6,>=latex]
\coordinate (center) at (0,0);

\coordinate (lmod) at ($(center)+(235:3.5cm)$) {};
\coordinate (lN) at ($(center)+(90:2.3cm)$) {};
\coordinate (lSju) at ($(center)+(90:3cm)$) {};
\coordinate (lSjb) at ($(center)+(270:3cm)$) {};

\coordinate (x0) at ($(center)+(180:2cm)$) {};
\coordinate (x1) at ($(center)+(0:2cm)$) {};
\coordinate (lx0) at ($(center)+(180:2.3cm)$) {};
\coordinate (lx1) at ($(center)+(0:2.3cm)$) {};

\coordinate (a0) at ($(center)+(250:2cm)$) {};
\coordinate (a1) at ($(center)+(110:2cm)$) {};
\coordinate (la0) at ($(center)+(250:2.3cm)$) {};
\coordinate (la1) at ($(center)+(110:2.3cm)$) {};

\coordinate (b0) at ($(center)+(290:2cm)$) {};
\coordinate (b1) at ($(center)+(70:2cm)$) {};
\coordinate (lb0) at ($(center)+(290:2.3cm)$) {};
\coordinate (lb1) at ($(center)+(70:2.3cm)$) {};

\draw[very thick] ([shift=(85:2cm)]0,0) arc (85:95:2cm);

\draw (0,0) circle (2cm);
\draw[dashed] (0,0) circle (2.7cm);
\draw[very thick] ([shift=(65:2.7cm)]0,0) arc (65:115:2.7cm);
\draw[very thick] ([shift=(245:2.7cm)]0,0) arc (245:295:2.7cm);

\draw[red,thick,dashed,->] (x0)--(x1);
\draw[thick,->] (a0)--(a1);
\draw[thick,->] (b0)--(b1);

\tikzstyle{every node}=[inner sep=1pt]
\begin{tiny}
\node at (lx0) {$z^0$};
\node at (lx1) {$z^1$};
\node at (la0) {$a^0$};
\node at (la1) {$a^1$};
\node at (lb0) {$b^0$};
\node at (lb1) {$b^1$};
\node at (lN) {$P$};
\node at (lSju) {$K^j$};
\node at (lSjb) {$K^{1-j}$};
\end{tiny}
\draw[white] (-2.5,-2.5)--(-2.5,-2.3);
\draw[white] (2.5,2.5)--(2.5,2.3);
\end{tikzpicture} 
\hspace{0.2cm}
\begin{tikzpicture}[xscale=0.6,yscale=0.6,>=latex]
\coordinate (center) at (0,0);

\coordinate (lmod) at ($(center)+(235:3.5cm)$) {};
\coordinate (lN) at ($(center)+(90:2.3cm)$) {};
\coordinate (lSju) at ($(center)+(90:3cm)$) {};
\coordinate (lSjb) at ($(center)+(270:3cm)$) {};

\coordinate (x0) at ($(center)+(180:2cm)$) {};
\coordinate (x1) at ($(center)+(0:2cm)$) {};
\coordinate (lx0) at ($(center)+(180:2.3cm)$) {};
\coordinate (lx1) at ($(center)+(0:2.3cm)$) {};

\coordinate (b0) at ($(center)+(250:2cm)$) {};
\coordinate (b1) at ($(center)+(70:2cm)$) {};
\coordinate (lb0) at ($(center)+(250:2.3cm)$) {};
\coordinate (lb1) at ($(center)+(70:2.3cm)$) {};

\coordinate (a0) at ($(center)+(290:2cm)$) {};
\coordinate (a1) at ($(center)+(110:2cm)$) {};
\coordinate (la0) at ($(center)+(290:2.3cm)$) {};
\coordinate (la1) at ($(center)+(110:2.3cm)$) {};

\draw[very thick] ([shift=(85:2cm)]0,0) arc (85:95:2cm);

\draw (0,0) circle (2cm);
\draw[dashed] (0,0) circle (2.7cm);
\draw[very thick] ([shift=(65:2.7cm)]0,0) arc (65:115:2.7cm);
\draw[very thick] ([shift=(245:2.7cm)]0,0) arc (245:295:2.7cm);

\draw[red,dashed,thick,->] (x0)--(x1);
\draw[thick,->] (a0)--(a1);
\draw[thick,->] (b0)--(b1);

\tikzstyle{every node}=[inner sep=1pt]
\begin{tiny}
\node at (lx0) {$z^0$};
\node at (lx1) {$z^1$};
\node at (la0) {$a^0$};
\node at (la1) {$a^1$};
\node at (lb0) {$b^0$};
\node at (lb1) {$b^1$};
\node at (lN) {$P$};
\node at (lSju) {$K^j$};
\node at (lSjb) {$K^{1-j}$};
\end{tiny}
\draw[white] (-2.5,-2.5)--(-2.5,-2.3);
\draw[white] (2.5,2.5)--(2.5,2.3);
\end{tikzpicture} 
\hspace{0.2cm}
\begin{tikzpicture}[xscale=0.6,yscale=0.6,>=latex]
\draw[dashed] (0,-3)--(0,3);
\end{tikzpicture} 
\hspace{0.2cm}
\begin{tikzpicture}[xscale=0.6,yscale=0.6,>=latex]
\coordinate (center) at (0,0);

\coordinate (lmod) at ($(center)+(235:3.5cm)$) {};
\coordinate (lN) at ($(center)+(90:2.3cm)$) {};
\coordinate (lSju) at ($(center)+(90:3cm)$) {};
\coordinate (lSjb) at ($(center)+(270:3cm)$) {};

\coordinate (x0) at ($(center)+(180:2cm)$) {};
\coordinate (x1) at ($(center)+(0:2cm)$) {};
\coordinate (lx0) at ($(center)+(180:2.3cm)$) {};
\coordinate (lx1) at ($(center)+(0:2.3cm)$) {};

\coordinate (a1) at ($(center)+(250:2cm)$) {};
\coordinate (a0) at ($(center)+(70:2cm)$) {};
\coordinate (la1) at ($(center)+(250:2.3cm)$) {};
\coordinate (la0) at ($(center)+(70:2.3cm)$) {};

\coordinate (b1) at ($(center)+(290:2cm)$) {};
\coordinate (b0) at ($(center)+(110:2cm)$) {};
\coordinate (lb1) at ($(center)+(290:2.3cm)$) {};
\coordinate (lb0) at ($(center)+(110:2.3cm)$) {};

\draw[very thick] ([shift=(85:2cm)]0,0) arc (85:95:2cm);

\draw (0,0) circle (2cm);
\draw[dashed] (0,0) circle (2.7cm);
\draw[very thick] ([shift=(65:2.7cm)]0,0) arc (65:115:2.7cm);
\draw[very thick] ([shift=(245:2.7cm)]0,0) arc (245:295:2.7cm);

\draw[red,dashed,thick,->] (x0)--(x1);
\draw[thick,->] (a0)--(a1);
\draw[thick,->] (b0)--(b1);

\tikzstyle{every node}=[inner sep=1pt]
\begin{tiny}
\node at (lx0) {$z^0$};
\node at (lx1) {$z^1$};
\node at (la0) {$a^0$};
\node at (la1) {$a^1$};
\node at (lb0) {$b^0$};
\node at (lb1) {$b^1$};
\node at (lN) {$P$};
\node at (lSju) {$K^{1-j}$};
\node at (lSjb) {$K^{j}$};
\end{tiny}
\draw[white] (-2.5,-2.5)--(-2.5,-2.3);
\draw[white] (2.5,2.5)--(2.5,2.3);
\end{tikzpicture} 
\hspace{0.2cm}
\begin{tikzpicture}[xscale=0.6,yscale=0.6,>=latex]
\coordinate (center) at (0,0);

\coordinate (lmod) at ($(center)+(235:3.5cm)$) {};
\coordinate (lN) at ($(center)+(90:2.3cm)$) {};
\coordinate (lSju) at ($(center)+(90:3cm)$) {};
\coordinate (lSjb) at ($(center)+(270:3cm)$) {};

\coordinate (x0) at ($(center)+(180:2cm)$) {};
\coordinate (x1) at ($(center)+(0:2cm)$) {};
\coordinate (lx0) at ($(center)+(180:2.3cm)$) {};
\coordinate (lx1) at ($(center)+(0:2.3cm)$) {};

\coordinate (b1) at ($(center)+(250:2cm)$) {};
\coordinate (b0) at ($(center)+(110:2cm)$) {};
\coordinate (lb1) at ($(center)+(250:2.3cm)$) {};
\coordinate (lb0) at ($(center)+(110:2.3cm)$) {};

\coordinate (a1) at ($(center)+(290:2cm)$) {};
\coordinate (a0) at ($(center)+(70:2cm)$) {};
\coordinate (la1) at ($(center)+(290:2.3cm)$) {};
\coordinate (la0) at ($(center)+(70:2.3cm)$) {};

\draw[very thick] ([shift=(85:2cm)]0,0) arc (85:95:2cm);

\draw (0,0) circle (2cm);
\draw[dashed] (0,0) circle (2.7cm);
\draw[very thick] ([shift=(65:2.7cm)]0,0) arc (65:115:2.7cm);
\draw[very thick] ([shift=(245:2.7cm)]0,0) arc (245:295:2.7cm);

\draw[red,dashed,thick,->] (x0)--(x1);
\draw[thick,->] (a0)--(a1);
\draw[thick,->] (b0)--(b1);

\tikzstyle{every node}=[inner sep=1pt]
\begin{tiny}
\node at (lx0) {$z^0$};
\node at (lx1) {$z^1$};
\node at (la0) {$a^0$};
\node at (la1) {$a^1$};
\node at (lb0) {$b^0$};
\node at (lb1) {$b^1$};
\node at (lN) {$P$};
\node at (lSju) {$K^{1-j}$};
\node at (lSjb) {$K^{j}$};
\end{tiny}
\draw[white] (-2.5,-2.5)--(-2.5,-2.3);
\draw[white] (2.5,2.5)--(2.5,2.3);
\end{tikzpicture} 

\caption{\label{fig:inside_property} To the left: models admissible for $\gamma^0(Q)$, to the right:
models admissible for $\gamma^1(Q)$.}
\end{figure}
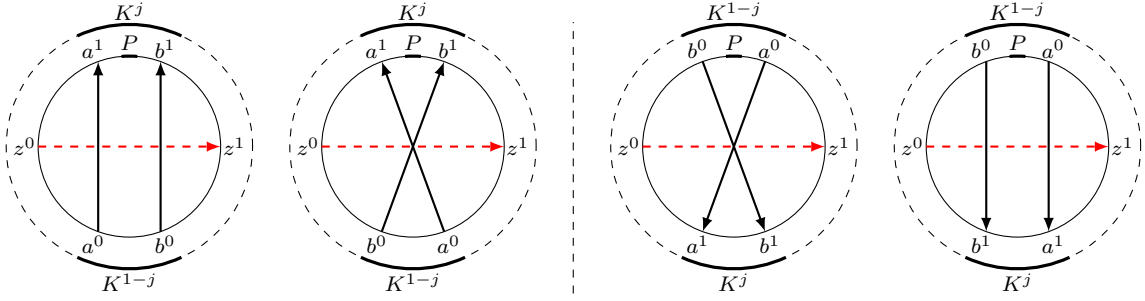

The remaining statements of the claim follow easily from the observations made above  -- see Figure~\ref{fig:tau_words}
for an illustration.
\end{proof}

Next, we show that the conformal models of $(Q,{\sim})$ extendable by the letters $N_{\pqstree}(Q)$ can be defined as the admissible models 
for appropriately ``refined'' circular orders of the slots $\gamma^0(Q)$ and $\gamma^1(Q)$
-- see Figure \ref{fig:circular_orders_of_the_slots_and_the_nodes} for an illustration.
To describe those ``refinements'', for every $K \in K(Q)$ we introduce first the set $\camodules(K)$ of \emph{CA-modules of $K$}; all the sets in $\camodules(K)$ for all $K \in K(Q)$ and all $Q \in \strongModules$ will constitute the set of all CA-modules of $V$.

\begin{definition}
Suppose $K$ is a member of $K(Q)$. 
For every $v \in K$, the \emph{left-right partition of the set $\inside(K)$} is a pair:
\begin{itemize}
 \item $(\leftside(v) \cap \inside(K), \rightside(v) \cap \inside(K))$ if $v^0 \in K^0$ and $v^1 \in K^1$,
 \item $(\rightside(v) \cap \inside(K), \leftside(v) \cap \inside(K))$ if $v^1 \in K^0$ and $v^0 \in K^1$.
\end{itemize}
A set $S \subseteq K$ is a \emph{CA-module of $K$} if $S$ is a maximal module (not necessary strong) in $(K,{\sim})$ such that all the vertices from $S$ admit the same left-right partition of the set $\inside(K)$.
We denote the set of all CA-modules of $K$ by $\camodules(K)$. 
\end{definition}
In Section~\ref{subsec:refinement_procedure} we describe so-called ``refinement procedure'' which computes the sets $\camodules(K)$ for every set $K \in K(M)$.
In particular, the procedure asserts that:  
\begin{description}
 \item[\namedlabel{prop:K_refinement_partition}{(R1)}] The set $\camodules(K)$ forms a partition of $K$.
\end{description}

Figure~\ref{fig:slot_refinement_example_no_reflection} shows an example of the set $\camodules(K)$ for some $K \in K(Q)$.
For this specific case the set $\camodules(K)$ consists of the modules $S_1,\ldots,S_6$; for example,
the left-right partition of $inside(K)$ for the vertices from $S_5$ equals to $(\{P_1,P_3\},\{P_2,P_4\})$.
We easily check that the vertices from every set~$S_i$ admit the same left-right partition of~$\inside(K)$.
In the modular decomposition $\strongModules(K,{\sim})$ of $(K,{\sim})$ the module
$K$ is prime and has four children:
$S_1$, $S_2 \cup S_3$, $S_4$, $S_5 \cup S_6$. 
We easily note that we can not extend any set $S_i$ into a larger module in $(K,{\sim})$ whose 
vertex set would admit the same left-right partition of $\inside(K)$.
\input ./figures/proof_sketch/parallel/slot_refinement_example_no_reflection.tex

Before we list the next properties of $\camodules(K)$, for every $S \in \camodules(K)$ we define the metachord $\mathbb{S} = (S^0,S^1,{<_S})$ for $S$ 
as the restriction of $\mathbb{K}$ to the set $S$. 
That is, we have $S^0 = K^0 \cap S^*$, $S^1 = K^1 \cap S^*$,
and ${<_{S}}$ equals to ${<_{K}}$ restricted to the set $S$.

Suppose $\phi_Q$ and $\phi'_Q$ are two extended conformal models of $(Q,{\sim})$ such that $\phi_Q \Vert Q^*$ and $\phi'_Q \Vert Q^*$ are admissible for $\gamma^t(M)$ for some $t \in \{0,1\}$.
Then, basing on Claim~\ref{claim:tau_phi_M_K_properties}, we show in Section~\ref{subsec:refinement_procedure} that $\phi_Q$ and $\phi'_Q$ satisfy the following properties with respect to the set~$\camodules(K)$:
\begin{description}
 \item[\namedlabel{prop:K_refinement_contiguous_subwords}{(R2)}] For every $S \in \camodules(K)$ the sets $S^0$ and $S^1$ are contiguous in $\tau(\phi_Q,K^0)$ and $\tau(\phi_Q,K^1)$, respectively, and the pair $(\phi_Q|S^0, \phi_Q|S^1)$ is a consistent permutation model of $(S,{\sim})$ admissible for $\mathbb{S}$. 
 \item[\namedlabel{prop:K_refinement_contiguous_subwords_order}{(R3)}] For every $S_1,S_2 \in S(K)$ and every $j \in \{0,1\}$, 
 the word $\phi_Q|S^{j}_{1}$ occurs before the word $\phi_Q|S^{j}_{2}$ 
 in $\tau(\phi_Q,K^j)$ if and only if the word $\phi'_Q|S^{j}_{1}$ occurs before the word $\phi'_Q|S^{j}_{2}$ in $\tau(\phi'_Q,K^j)$.
\end{description} 
See Figure~\ref{fig:slot_refinement_example_no_reflection} for an illustration.

Suppose $\phi,\phi^R$ are two conformal models of $G_{ov}$ such that $\phi^R$ is the reflection of $\phi$, 
$\phi \Vert Q^*$ is admissible for $\gamma^0(Q)$, 
and $\phi^R \Vert Q^*$ is admissible for $\gamma^1(Q)$.
Let $\camodules(Q) = \bigcup_{K \in K(Q)} \camodules(K)$ and suppose $S_1,\ldots,S_k$ are the members of $\camodules(Q)$.
Now:
\begin{itemize}
\item we replace the contiguous word $\phi_Q|S^j_i$ in $\phi_Q$ by the letter $S^j_i$, for $j \in \{0,1\}$ and $i \in [k]$, thus obtaining the circular word $\pi^0(Q)$ on $\{S^0_1,S^1_1,\ldots,S^0_k,S^1_k\} \cup N_{\pqstree}(Q)$,
\item we replace the contiguous word $\phi^R_Q|S^j_i$ in $\phi^R_Q$ by the letter $S^j_i$, for $j \in \{0,1\}$ and $i \in [k]$, thus obtaining the circular word $\pi^1(Q)$ on $\{S^0_1,S^1_1,\ldots,S^0_k,S^1_k\} \cup N_{\pqstree}(Q)$.
\end{itemize}
and we set $\Pi(Q) = \{\pi^0(Q),\pi^1(Q)\}$.
Note that $\pi^1(Q)$ is the reflection of $\pi^0(Q)$ as $\phi^R_Q$ is the reflection of $\phi_Q$ -- see Figure~\ref{fig:circular_orders_of_the_slots_and_the_nodes} for an illustration.

\input ./figures/proof_sketch/parallel/circular_orders_of_slots_and_nodes.tex

Equivalently, we can imagine $\pi^0(Q)$ as it arises from $\gamma^0(Q)$ 
by replacing every slot $K^0$ by the slots $\{S^0: S \in \camodules(K)\}$ and $K^1$ by the slots $\{S^1:S \in \camodules(K)\}$,
ordered as they occur in the words $\tau(\phi_Q,K^0)$ and $\tau(\phi_Q,K^1)$, and then appropriately extended by $N_{\pqstree}(Q)$.
Similarly for $\pi^1(Q)$.
Figure~\ref{fig:slot_refinement_example} illustrates that $\pi^1(Q)$ is indeed the reflection of $\pi^0(Q)$.
\input ./figures/proof_sketch/parallel/slot_refinement_example.tex

\begin{definition}
Let $j \in \{0,1\}$.
A circular word $\phi^{adm}_Q$ on the set $Q^{*} \cup N_{\pqstree}(Q)$ is \emph{admissible for $\pi^j(Q)$}
if $\phi^{adm}_Q$ arises from $\pi^j(Q)$ by replacing the slots $S^0_i,S^1_i$ for $i \in [k]$
by the words $\tau^0_i, \tau^1_i$, respectively, 
where $(\tau^0_i,\tau^1_i)$ is an admissible model for $\mathbb{S}_i$.
A circular word $\phi^{adm}_Q$ on the set $Q^{*} \cup N_{\pqstree}(Q)$ is \emph{admissible for $\Pi(Q)$} if $\phi^{adm}_Q$ is admissible for some $\pi^j(Q)$ in $\Pi(Q)$.
\end{definition}

We summarize the results of this part with the following lemma.
\begin{lemma}
\label{lem:parallel_prime_admissible_conformal}
Let $Q$ be a prime non-permutation component in $\mathcal{Q}$ and let $\phi_Q$ be a circular word on the set $Q^{*} \cup N_{\pqstree}(Q)$.
The word $\phi_Q$ is an extended conformal model for $(Q,{\sim})$ 
if and only if $\phi_Q$ is admissible for $\Pi(Q)$.
\end{lemma}
\begin{proof}
Suppose $\phi_Q$ is an extended conformal model for $(Q,{\sim})$.
We note that:
\begin{itemize}
 \item if $\phi_Q \Vert Q^*$ is admissible for $\gamma^0(Q)$, then $\phi_Q$ is admissible for $\pi^0(Q)$,
 \item if $\phi_Q \Vert Q^*$ is admissible for $\gamma^1(Q)$, then $\phi_Q$ is admissible for $\pi^1(Q)$,
\end{itemize}
which follows by Properties~\ref{prop:K_refinement_partition}--\ref{prop:K_refinement_contiguous_subwords_order}.

Suppose $\phi_Q$ is an admissible model for $\Pi(Q)$.
Assume that $\phi_Q$ is admissible for $\pi^0(Q)$. 
Then $\phi_Q \Vert Q^*$ is admissible for $\gamma^0(Q)$, and hence 
$\phi_Q \Vert Q^*$ is conformal for $(Q,{\sim})$.
Also, since $\phi_Q$ is admissible for $\pi^0(Q)$, $\phi_Q$
properly extends $\phi_Q \Vert Q^*$ by the letters from $N_\pqstree(Q)$. 
So, $\phi_Q$ is an extended conformal model of $(Q,{\sim})$.
\end{proof}

\subsubsection{Serial non-permutation components}
Let $Q$ be a serial non-permutation component in $\mathcal{Q}$ and let $M_1,\ldots,M_n$ be the children of $Q$ in $\strongModules(G_{ov})$.
Again, our goal is to describe the conformal models of $(Q,{\sim})$ 
that are extendable by the letters from $N_{\pqstree}(Q)$.
Similarly to the previous case, not every conformal model of $(Q,{\sim})$
can be extended.
However, we have much more freedom in generating the conformal models of $(Q,{\sim})$ 
in the case when $Q$ is serial compared to the case when $Q$ is prime.
It turns out, however, that in the serial case the conformal models of $(Q,{\sim})$ extendable by $N_{\pqstree}(Q)$ admit the same description as for the prime case.

As for prime non-permutation components,
we first partition the children of $Q$ into sets $K(Q)$.
The role of $K(Q)$ is analogous as for prime non-permutation children of~$V$.
In the definition given below we use the sets $\inside(M_i)$ for $i \in [n]$.
Note that they are correctly defined as the statements of Claim~\ref{claim:tau_phi_M_K_properties} 
are also valid for every set $M_i$ (as there is $z \in Q \setminus M_i$ such that $z \sim M_i$).

\begin{figure}[htp!]
\begin{tikzpicture}[scale=0.9,>=latex]
\coordinate (center) at (0.0,0.0) {};

\draw (center) circle (2cm);
\draw[dashed] (center) circle (2.55cm);

\coordinate (mod) at ($(center)+(225:3.5cm)$) {};


\coordinate (m1a1) at ($(center)+(97:2cm)$) {};
\coordinate (m1a2) at ($(center)+(83:2cm)$) {};
\coordinate (lm1a) at ($(center)+(90:2.25cm)$) {};
\coordinate (lm1a1) at ($(center)+(97:2.25cm)$) {};
\coordinate (lm1a2) at ($(center)+(83:2.25cm)$) {};

\coordinate (m1b1) at ($(center)+(277:2cm)$) {};
\coordinate (m1b2) at ($(center)+(263:2cm)$) {};
\coordinate (lm1b) at ($(center)+(270:2.25cm)$) {};
\coordinate (lm1b1) at ($(center)+(277:2.25cm)$) {};
\coordinate (lm1b2) at ($(center)+(263:2.25cm)$) {};

\coordinate (m2a1) at ($(center)+(52:2cm)$) {};
\coordinate (m2a2) at ($(center)+(38:2cm)$) {};
\coordinate (lm2a) at ($(center)+(45:2.25cm)$) {};
\coordinate (lm2a1) at ($(center)+(38:2.25cm)$) {};
\coordinate (lm2a2) at ($(center)+(52:2.25cm)$) {};

\coordinate (m2b1) at ($(center)+(232:2cm)$) {};
\coordinate (m2b2) at ($(center)+(218:2cm)$) {};
\coordinate (lm2b) at ($(center)+(225:2.25cm)$) {};
\coordinate (lm2b1) at ($(center)+(232:2.25cm)$) {};
\coordinate (lm2b2) at ($(center)+(218:2.25cm)$) {};

\draw[very thick,-] ([shift=(100:2.55cm)]0,0) arc (100:35:2.55cm);
\coordinate (lk11) at ($(center)+(67.5:2.8cm)$) {};

\draw[very thick,-] ([shift=(280:2.55cm)]0,0) arc (280:215:2.55cm);
\coordinate (lk01) at ($(center)+(247.5:2.8cm)$) {};

\coordinate (m3a1) at ($(center)+(7:2cm)$) {};
\coordinate (m3a2) at ($(center)+(-7:2cm)$) {};
\coordinate (lm3a) at ($(center)+(0:2.25cm)$) {};
\coordinate (lm3a1) at ($(center)+(7:2.25cm)$) {};
\coordinate (lm3a2) at ($(center)+(-7:2.25cm)$) {};

\coordinate (m3b1) at ($(center)+(187:2cm)$) {};
\coordinate (m3b2) at ($(center)+(173:2cm)$) {};
\coordinate (lm3b) at ($(center)+(180:2.25cm)$) {};
\coordinate (lm3b1) at ($(center)+(187:2.25cm)$) {};
\coordinate (lm3b2) at ($(center)+(173:2.25cm)$) {};

\draw[very thick,-] ([shift=(-10:2.55cm)]0,0) arc (-10:10:2.55cm);
\coordinate (lk02) at ($(center)+(0:2.8cm)$) {};

\draw[very thick,-] ([shift=(170:2.55cm)]0,0) arc (170:190:2.55cm);
\coordinate (lk12) at ($(center)+(180.5:2.8cm)$) {};

\coordinate (m4a1) at ($(center)+(-38:2cm)$) {};
\coordinate (m4a2) at ($(center)+(-52:2cm)$) {};
\coordinate (lm4a) at ($(center)+(-45:2.25cm)$) {};
\coordinate (lm4a1) at ($(center)+(-38:2.25cm)$) {};
\coordinate (lm4a2) at ($(center)+(-52:2.25cm)$) {};

\coordinate (m4b1) at ($(center)+(142:2cm)$) {};
\coordinate (m4b2) at ($(center)+(128:2cm)$) {};
\coordinate (lm4b) at ($(center)+(135:2.25cm)$) {};
\coordinate (lm4b1) at ($(center)+(142:2.25cm)$) {};
\coordinate (lm4b2) at ($(center)+(128:2.25cm)$) {};

\draw[very thick,-] ([shift=(-35:2.55cm)]0,0) arc (-35:-55:2.55cm);
\coordinate (lk13) at ($(center)+(-45:2.8cm)$) {};

\draw[very thick,-] ([shift=(125:2.55cm)]0,0) arc (125:145:2.55cm);
\coordinate (lk03) at ($(center)+(135.5:2.8cm)$) {};

\coordinate (n1) at ($(center)+(112.5:2.25cm)$) {};
\draw[very thick,-] ([shift=(107.5:2cm)]0,0) arc (107.5:117.5:2cm);

\coordinate (n2) at ($(center)+(22.5:2.25cm)$) {};
\draw[very thick,-] ([shift=(17.5:2cm)]0,0) arc (17.5:27.5:2cm);

\coordinate (n3) at ($(center)+(-45:1.75cm)$) {};
\draw[very thick,-] ([shift=(-40:2cm)]0,0) arc (-40:-50:2cm);

\draw[<-] (m1a1)--(m1b2);
\draw[->] (m1a2)--(m1b1);

\draw[<-] (m2a1)--(m2b2);
\draw[->] (m2a2)--(m2b1);

\draw[<-] (m3a1)--(m3b2);
\draw[->] (m3a2)--(m3b1);

\draw[<-] (m4a1)--(m4b2);
\draw[->] (m4a2)--(m4b1);

\tikzstyle{every node}=[inner sep=1pt]

\begin{tiny}

\node at (lm1a) {$M_1$};
\node at (lm1b) {$M_1$};

\node at (lm2a) {$M_2$};
\node at (lm2b) {$M_2$};

\node at (lm3a) {$M_3$};
\node at (lm3b) {$M_3$};

\node at (lm4a) {$M_4$};
\node at (lm4b) {$M_4$};

\node at (lk11) {$K_1$};
\node at (lk01) {$K_1$};

\node at (lk12) {$K_2$};
\node at (lk02) {$K_2$};

\node at (lk13) {$K_3$};
\node at (lk03) {$K_3$};

\node at (n1) {$P_1$};
\node at (n2) {$P_2$};
\node at (n3) {$P_3$};
\end{tiny}

\node at (mod) {$\phi_Q$};

\draw[white] (-3,-2.8)--(-3,-2.3);
\draw[white] (2.7,2.8)--(2.7,2.3);
\end{tikzpicture}

\caption{\label{fig:serial_K_sets} 
An extended conformal model $\phi_Q$ of $(Q,{\sim})$ for a serial non-permutation component $Q$ with four children $M_1,M_2,M_3,M_4$.
We have $\inside(M_4) = \{P_3\}$ and hence $M_4$ is a member of $K(Q)$.
Every chord representing a vertex from $M_1 \cup M_2$ has $P_1$ on one side and $P_2,P_3$ on the other side.
Every chord representing a vertex from $M_3$ has $P_1,P_2$ on one side and $P_3$ on the other side.
This shows that $K(Q)  = \{M_1 \cup M_2, M_3,M_4\}$.
The sets $K_i$ are represented by arcs on the outer circles.
}
\end{figure}
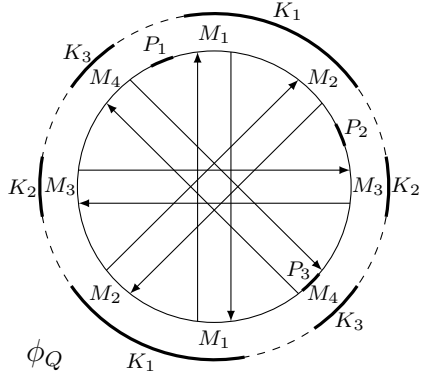

\begin{definition}
\label{def:serial_K_sets}
A partition $K(Q)$ of the module $Q$ is defined as follows:
\begin{itemize}
\item If $\inside(M_i) \neq \emptyset$, then $M_i$ is a member of $K(Q)$.
\item The remaining members of $K(Q)$ are the equivalence classes of the $K$-relation 
defined on the set $$\bigcup \{M_i: i \in [n] \text{ and } \inside(M_i) = \emptyset\}$$
such that for every $u,v \in \bigcup \{M_i: i \in [n] \text{ and } \inside(M_i) = \emptyset\}$:
$$
\begin{array}{ccc}
u K v\ & \iff 
\begin{array}{l}
\{\leftside(u) \cap N_{\pqstree}(Q), \rightside(u) \cap N_{\pqstree}(Q) \} = \\
\{\leftside(v) \cap N_{\pqstree}(Q), \rightside(v) \cap N_{\pqstree}(Q) \}.
\end{array}
\end{array}
$$
\end{itemize}
\end{definition}
Observe that every member of $K(Q)$ is the union of some children of $Q$ -- see Figure~\ref{fig:serial_K_sets} for an illustration.
Finally, observe that $|K(Q)| \geq 2$ as otherwise $Q$ would be a permutation component.

Next, we show that the sets in $K(Q)$ satisfy analogous properties as the corresponding sets for prime children of $V$.
Suppose $R$ is the set that contains an element from every set in $K(Q)$ (note that $(R,{\sim})$ is a clique and thus every oriented chord model of $(R,{\sim})$ is conformal).
Lemma~\ref{lemma:serial_consistent_modules} of Section~\ref{subsec:parallel_M_serial} proves that:
\begin{description}
 \item[\namedlabel{prop:parallel_serial_skeleton}{(PS1)}] There are two conformal models of $(R,{\sim})$, $\phi^0_R$ and its reflection $\phi^1_R$, 
 such that for every extended conformal model $\phi_Q$ of $(Q,{\sim})$ 
 we have either $\phi_Q \Vert R^* \equiv \phi^0_R$ or $\phi_Q \Vert R^* \equiv \phi^1_R$.
 \item[\namedlabel{prop:parallel_serial_contiguous_subwords}{(PS2)}]
 For every extended conformal model $\phi_Q$ of $(Q,{\sim})$ and every $K \in K(Q)$ the set $K$ induces a consistent permutation model in $\phi^Q \equiv \phi_Q \Vert Q^*$.
 Moreover, if $K$ is the union of at least two children of $Q$, the set $K$
 induces a consistent permutation model also in~$\phi_Q$.
\end{description}
See Figure~\ref{fig:serial_K_sets} for an illustration.

Eventually, for every $K \in K(Q)$ we can define the metachord $\mathbb{K} = (K^0,K^1,{<_K})$ so as for every extended conformal model $\phi_Q$ of $Q$ the consistent permutation model induced by the set $K$ in $\phi_Q \Vert Q^*$ is admissible for $\mathbb{K}$. 
Indeed, if $K=M_i$ for some $i \in [n]$, we can define $\mathbb{K}$ in the same way as in Claim~\ref{claim:metachords_for_children_of_serial} as the graph $(K,{\parallel})$ is connected.
Otherwise, we define $\mathbb{K}$ in the same way as in Claim~\ref{claim:metachords_for_children_of_prime}.
Indeed, since the set $K$ induces a consistent permutation model in every extended conformal model $\phi_Q$ of $(Q,{\sim})$, we may decide the orientations of the chords in $\mathbb{K}$ with respect to some P-node $P$ adjacent to $Q$ in $\pqstree$.

Now, we proceed the same way as for prime modules in $\mathcal{Q}$: we define
the CA-modules $\camodules(K)$ for every $K \in K(Q)$, the set $\Pi(Q) = \{\pi^0(Q), \pi^1(Q)\}$ of possible orderings of $Q$,  and the admissible models for $\Pi(Q)$.
Eventually, we get the following lemma:
\begin{lemma}
\label{lem:parallel_serial_admissible_conformal}
Let $Q$ be a serial non-permutation component in $\mathcal{Q}$ and let $\phi_Q$ be a circular word on the set $Q^{*} \cup N_{\pqstree}(Q)$.
The word $\phi_Q$ is an extended conformal of $(Q,{\sim})$ 
if and only if $\phi_Q$ is admissible for $\Pi(Q)$.
\end{lemma}

Finally, for every P-node $P$ in $\pqstree$ we set $\Pi(P)$ 
such that it contains all circular orders of the set $N_{\pqstree}(P)$.
Now, since the sets $\Pi(N)$ are defined for every node $N$ in the tree $\pqstree$, 
we may extend $\pqstree$ to the PQS-tree $\pqstree$.
Finally, we are ready to prove Theorem \ref{thm:main_theorem} for the case where $V$
is parallel module in $\strongModules(G_{ov})$.
\begin{proof}[Proof of Theorem~\ref{thm:main_theorem} for the case when $V$ is parallel in $\strongModules(G_{ov})$]
Let $\phi$ be a conformal model of $G_{ov}$. 
The results of this subsection show that:
\begin{itemize}
 \item For every Q-node $Q$ in $\pqstree$, $\phi[Q]$ is admissible for an ordering in $\Pi(Q)$.
 \item For every P-node $P$ in $\pqstree$, $\phi[P]$ is admissible for an ordering in $\Pi(P)$. 
\end{itemize}
In particular, it means that $\pi(\phi)$ is a member of $\Pi$ and $\phi$ is admissible for $\Pi$.

We recall that we have already shown that every admissible model for $\Pi$ is 
conformal for $G_{ov}$.
\end{proof}

This, together with Theorems~\ref{thm:conformal_models_serial_component} and~\ref{thm:conformal_models_prime_component}, completes the proof of Theorem~\ref{thm:main_theorem}.

\section{Modular decomposition $\mathcal{M}(G_{ov})$ and chord models of $G_{ov}$}
\label{sec:modular_decomposition_and_chord_models}
Let $G$ be a circular-arc graph with no universal vertices and no twins and let 
$G_{ov}$ be the overlap graph of $G$.
In this section we prove Properties~\ref{prop_serial:contiguous_subwords} -- \ref{prop_prime:contiguous_subwords_parallel} 
of the conformal models of $(Q,{\sim})$, where $Q$ is a component of $G_{ov}$
(that is, $Q$ is serial/prime and $Q=V$ or $Q$ is a child of parallel $V$ in $\strongModules(G_{ov})$).
As earlier, we assume $M_1,\ldots,M_n$ are the children of $Q$ in $\mathcal{M}(G_{ov})$.
For every $i \in [k]$ we denote by $N(M_i)$ the set
\begin{equation*}
\begin{array}{ccl}
N(M_i) &=& \{x \in Q \setminus M_i: x \sim M_i\}. \\
\end{array}
\end{equation*}

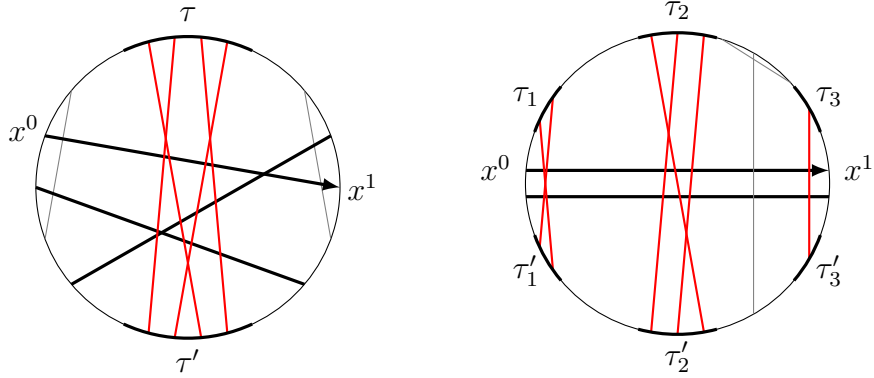
\begin{figure}[htp!]
\centering
\begin{tikzpicture}[scale=1,>=latex]
\coordinate (center) at (0,0) {};
\coordinate (ltau) at ($(center)+(90:2.3cm)$) {};
\coordinate (ltau') at ($(center)+(270:2.3cm)$) {};

\coordinate (u4) at ($(center)+(75:2cm)$) {};
\coordinate (u3) at ($(center)+(85:2cm)$) {};
\coordinate (u2) at ($(center)+(95:2cm)$) {};
\coordinate (u1) at ($(center)+(105:2cm)$) {};

\coordinate (b4) at ($(center)+(255:2cm)$) {};
\coordinate (b3) at ($(center)+(265:2cm)$) {};
\coordinate (b2) at ($(center)+(275:2cm)$) {};
\coordinate (b1) at ($(center)+(285:2cm)$) {};

\coordinate (l5) at ($(center)+(140:2cm)$) {};
\coordinate (l4) at ($(center)+(160:2cm)$) {};
\coordinate (l3) at ($(center)+(180:2cm)$) {};
\coordinate (l2) at ($(center)+(200:2cm)$) {};
\coordinate (l1) at ($(center)+(220:2cm)$) {};

\coordinate (ll4) at ($(center)+(160:2.3cm)$) {};

\coordinate (r5) at ($(center)+(-40:2cm)$) {};
\coordinate (r4) at ($(center)+(-20:2cm)$) {};
\coordinate (r3) at ($(center)+(0:2cm)$) {};
\coordinate (r2) at ($(center)+(20:2cm)$) {};
\coordinate (r1) at ($(center)+(40:2cm)$) {};

\coordinate (lr3) at ($(center)+(0:2.3cm)$) {};

\draw ($(center)$) circle (2cm);

\draw[gray] (l2)-- (l5);
\draw[gray] (r1)-- (r4);

\draw[black,very thick] (l1)-- (r2);
\draw[black,very thick] (l3)-- (r5);
\draw[black,very thick,->] (l4)-- (r3);

\draw[thick,red] (u1)-- (b2);
\draw[thick,red] (u2)-- (b4);
\draw[thick,red] (u3)-- (b1);
\draw[thick,red] (u4)-- (b3);

\draw[very thick] ([shift=(245:2cm)]0,0) arc (245:295:2cm);
\draw[very thick] ([shift=(65:2cm)]0,0) arc (65:115:2cm);

\tikzstyle{every node}=[inner sep=1pt]
\node at (ltau) {$\tau$};
\node at (ltau') {$\tau'$};
\node at (ll4) {$x^0$};
\node at (lr3) {$x^1$};

\end{tikzpicture} 
\hspace{1cm}
\begin{tikzpicture}[scale=1,>=latex]
\coordinate (center) at (0,0) {};
\coordinate (ltau1) at ($(center)+(150:2.3cm)$) {};
\coordinate (ltau'1) at ($(center)+(210:2.3cm)$) {};
\coordinate (ltau2) at ($(center)+(90:2.3cm)$) {};
\coordinate (ltau'2) at ($(center)+(270:2.3cm)$) {};
\coordinate (ltau3) at ($(center)+(30:2.3cm)$) {};
\coordinate (ltau'3) at ($(center)+(330:2.3cm)$) {};

\coordinate (tau1u1) at ($(center)+(145:2cm)$) {};
\coordinate (tau1u2) at ($(center)+(155:2cm)$) {};

\coordinate (tau1b1) at ($(center)+(205:2cm)$) {};
\coordinate (tau1b2) at ($(center)+(215:2cm)$) {};

\coordinate (tau2u1) at ($(center)+(80:2cm)$) {};
\coordinate (tau2u2) at ($(center)+(90:2cm)$) {};
\coordinate (tau2u3) at ($(center)+(100:2cm)$) {};

\coordinate (tau2b1) at ($(center)+(260:2cm)$) {};
\coordinate (tau2b2) at ($(center)+(270:2cm)$) {};
\coordinate (tau2b3) at ($(center)+(280:2cm)$) {};

\coordinate (tau3u1) at ($(center)+(30:2cm)$) {};
\coordinate (tau3b1) at ($(center)+(330:2cm)$) {};

\draw ($(center)$) circle (2cm);

\draw[black,very thick,<-] ($(center)+(5:2cm)$)--($(center)+(175:2cm)$);

\coordinate (lx0) at ($(center)+(175:2.4cm)$) {};
\coordinate (lx1) at ($(center)+(5:2.4cm)$) {};

\draw[black,very thick] ($(center)+(-5:2cm)$)--($(center)+(185:2cm)$);
\draw[gray] ($(center)+(60:2cm)$)--($(center)+(300:2cm)$);
\draw[gray] ($(center)+(42:2cm)$)--($(center)+(73:2cm)$);

\draw[red,thick] (tau1u1)-- (tau1b1);
\draw[red,thick] (tau1u2)-- (tau1b2);

\draw[red,thick] (tau2u1)-- (tau2b2);
\draw[red,thick] (tau2u2)-- (tau2b1);
\draw[red,thick] (tau2u3)-- (tau2b3);

\draw[red,thick] (tau3u1)-- (tau3b1);

\draw[very thick] ([shift=(20:2cm)]0,0) arc (20:40:2cm);
\draw[very thick] ([shift=(75:2cm)]0,0) arc (75:105:2cm);
\draw[very thick] ([shift=(140:2cm)]0,0) arc (140:160:2cm);

\draw[very thick] ([shift=(200:2cm)]0,0) arc (200:220:2cm);
\draw[very thick] ([shift=(255:2cm)]0,0) arc (255:285:2cm);
\draw[very thick] ([shift=(320:2cm)]0,0) arc (320:340:2cm);

\tikzstyle{every node}=[inner sep=1pt]
\node at (ltau1) {$\tau_1$};
\node at (ltau'1) {$\tau'_1$};
\node at (ltau2) {$\tau_2$};
\node at (ltau'2) {$\tau'_2$};
\node at (ltau3) {$\tau_3$};
\node at (ltau'3) {$\tau'_3$};

\node at (lx0) {$x^0$};
\node at (lx1) {$x^1$};

\end{tikzpicture} 

\caption{\label{fig:chord_models_for_proper_modules} Chord models of $(Q,{\sim})$ for the cases when
$Q$ is prime and $M_i$ is prime (to the left) and when
$Q$ is prime and $M_i$ is parallel. 
Chords of the module $M_i$ are in red, chords of the set $N(M_i)$ are in bold. 
When $M_i$ is parallel, we have $(\tau, \tau') = (\tau_1\tau_2\tau_3, \tau'_3\tau'_2\tau'_1)$, where
each $(\tau_i,\tau'_i)$ is a chord model of some child of $M_i$ in $\strongModules(G_{ov})$.
}
\end{figure}

\begin{lemma}
\label{lemma:circle_models_of_proper_modules_N_M}
Suppose $M_i$ is a prime/parallel child of $Q$ ($Q$ is as above, serial or prime).
For any chord model $\psi$ of $(Q,{\sim})$ we have
$$\psi \Vert (M_i \cup N(M_i)) \equiv \pi \tau \pi' \tau',$$
where $(\tau,\tau')$ is a permutation model of $(M_i,{\sim})$ and $\pi, \pi'$ are permutations of $N(M_i)$.
\end{lemma}
Figure~\ref{fig:chord_models_for_proper_modules} illustrates the claim for the case where $Q$ is prime and $M_i$ is prime (to the left) and when $Q$ is prime and $M_i$ 
is parallel (to the right).

\begin{proof}
Since $M_i$ is a prime/parallel child of $Q$ and $Q$ is serial/prime, 
there is $x \in Q \setminus M_i$ such that $x \sim M_i$.
We orient the chord $\psi(x)$ arbitrarily.
By Claim \ref{claim:permutation_graphs_in_G_ov} we have
$$\psi \Vert (M_i \cup \{x\})  \equiv x^{0} \tau x^{1} \tau',$$
where $(\tau,\tau')$ is a permutation model of $(M_i,{\sim})$.
We need to show that
$$\psi \Vert (M_i \cup N(M_i)) \equiv \pi \tau \pi' \tau',$$
where $\pi, \pi'$ are some permutations of $N(M_i)$.

Fix $z \in N(M_i)$ such that $z \neq x$.
Suppose for a contradiction that the chord $\psi(z)$ has one of its ends between
the ends of the chords corresponding to the letters of $\tau$.
That is, suppose that $x^{0}\tau_1 z \tau_2 x^{1}$ is a subword of $\psi \Vert (M_i \cup \{x,z\})$,
where $\tau_1$ and $\tau_2$ are non-empty words such that $\tau_1\tau_2 = \tau$.
Now, consider a partition of $M_i$ into two non-empty sets, $M^1_i$ and $M^2_i$:
$$
M^1_i = \{u \in M_i: u \in \tau_1\}  \quad \text{and} \quad  M^2_i = \{u \in M_i: u \in \tau_2\}.
$$
Since for every $u \in M_i$ the chord $\psi(u)$ intersects the chord $\psi(z)$, 
we have that $x^{1} \tau'_1 z \tau'_2 x^{0}$ is a subword of $\psi \Vert (M_i \cup \{x,z\})$, where
$\tau'_i$ is a permutation of the set of the letters in $\tau_i$ for every $i \in [2]$.
It means, in particular, that for every $u_1 \in M^1_i$ and every $u_2 \in M^2_i$ the chords $\psi(u_1)$ and $\psi(u_2)$ intersect.
So, we have $M^1_i \sim M^2_i$, which contradicts that $M_i$ is a prime or a parallel module in $\mathcal{M}(G_{ov})$.
\end{proof}
Since every child of a serial module is either prime or parallel, 
Lemma~\ref{lemma:circle_models_of_proper_modules_N_M} proves Properties~\ref{prop_serial:contiguous_subwords} and 
\ref{prop_serial:contiguous_subwords_overlap} of the conformal models of $(Q,{\sim})$ in the case when $Q$ is a serial component of $G_{ov}$.

The next lemma shows property \ref{prop_prime:contiguous_subwords_prime} of the conformal models $(Q,{\sim})$ when $Q$ is prime.
\begin{lemma}
\label{lemma:circle_models_of_a_proper_prime_module}
Let $M_i$ be a prime child of a prime component $Q$ and 
let $\psi$ be a chord model of $(Q,{\sim})$.
Then, the set $M_i$ induces a consistent permutation model $(\tau,\tau')$ of $(M_i,{\sim})$ in~$\psi$ 
(see Figure~\ref{fig:chord_models_for_proper_modules} to the left).
\end{lemma}
\begin{proof}
By Lemma \ref{lemma:circle_models_of_proper_modules_N_M},
$\psi \Vert (M_i \cup N(M_i)) \equiv \pi \tau \pi' \tau'$, where $\pi, \pi'$ are permutations of $N(M_i)$.
To complete the proof it suffices to show that for every $v \in Q \smallsetminus N(M_i)$ $$\text{either} \quad \psi \Vert (M_i \cup \{v\}) \equiv v v\tau \tau' \quad \text{or} \quad \psi \Vert (M_i \cup \{v\}) \equiv \tau v v \tau'.$$
Assume otherwise. 
Since $(Q,{\sim})$ is connected and since $M_i$ is a module in $(Q,{\sim})$, 
there is $u \in Q \smallsetminus N(M_i)$ such that
$$\psi \Vert (M_i \cup \{u\}) \equiv \tau_1 u \tau_2 \tau'_2 u \tau'_1,$$
where $\tau_1, \tau_2$ and $\tau'_1, \tau'_2$ are such that $\tau_1\tau_2 = \tau$, $\tau'_2\tau'_1=\tau'$,
$\tau_1, \tau_2, \tau'_1, \tau'_2$ are non-empty, 
and $\tau_i$ is a permutation of $\tau'_i$ for $i \in [2]$ as $u \parallel M_i$.
Hence, the sets
$$M^1_i = \{w \in M_i: w \in \tau_1\} \text{ and } M^2_i = \{w \in M_i: w \in \tau_2\}$$
partition $M_i$, and we have $M^1_i \neq \emptyset$, $M^2_i \neq \emptyset$, and $M^1_i \parallel M^2_i$.
So, $(M_i, {\sim})$ is not connected, which contradicts the fact that $M_i$ is a prime module in $\mathcal{M}(G_{ov})$.
\end{proof}

\begin{figure}[htp!]
\centering
\begin{tikzpicture}[scale=1]
\coordinate (center) at (0,0) {};
\coordinate (ltau1) at ($(center)+(150:2.3cm)$) {};
\coordinate (ltau'1) at ($(center)+(210:2.3cm)$) {};
\coordinate (ltau2) at ($(center)+(90:2.3cm)$) {};
\coordinate (ltau'2) at ($(center)+(270:2.3cm)$) {};
\coordinate (ltau3) at ($(center)+(30:2.3cm)$) {};
\coordinate (ltau'3) at ($(center)+(330:2.3cm)$) {};

\coordinate (tau1u1) at ($(center)+(145:2cm)$) {};
\coordinate (tau1u2) at ($(center)+(155:2cm)$) {};

\coordinate (tau1b1) at ($(center)+(205:2cm)$) {};
\coordinate (tau1b2) at ($(center)+(215:2cm)$) {};

\coordinate (tau2u1) at ($(center)+(80:2cm)$) {};
\coordinate (tau2u2) at ($(center)+(90:2cm)$) {};
\coordinate (tau2u3) at ($(center)+(100:2cm)$) {};

\coordinate (tau2b1) at ($(center)+(260:2cm)$) {};
\coordinate (tau2b2) at ($(center)+(270:2cm)$) {};
\coordinate (tau2b3) at ($(center)+(280:2cm)$) {};

\coordinate (tau3u1) at ($(center)+(30:2cm)$) {};
\coordinate (tau3b1) at ($(center)+(330:2cm)$) {};

\draw ($(center)$) circle (2cm);

\draw[black,very thick] ($(center)+(5:2cm)$)--($(center)+(175:2cm)$);
\draw[black,very thick] ($(center)+(-5:2cm)$)--($(center)+(185:2cm)$);
\draw[black] ($(center)+(60:2cm)$)--($(center)+(300:2cm)$);
\draw[black] ($(center)+(42:2cm)$)--($(center)+(73:2cm)$);

\draw[red,thick] (tau1u1)-- (tau1b1);
\draw[red,thick] (tau1u2)-- (tau1b2);

\draw[red,thick] (tau2u1)-- (tau2b2);
\draw[red,thick] (tau2u2)-- (tau2b1);
\draw[red,thick] (tau2u3)-- (tau2b3);

\draw[red,thick] (tau3u1)-- (tau3b1);

\draw[very thick] ([shift=(20:2cm)]0,0) arc (20:40:2cm);
\draw[very thick] ([shift=(75:2cm)]0,0) arc (75:105:2cm);
\draw[very thick] ([shift=(140:2cm)]0,0) arc (140:160:2cm);

\draw[very thick] ([shift=(200:2cm)]0,0) arc (200:220:2cm);
\draw[very thick] ([shift=(255:2cm)]0,0) arc (255:285:2cm);
\draw[very thick] ([shift=(320:2cm)]0,0) arc (320:340:2cm);

\tikzstyle{every node}=[inner sep=1pt]
\node at (ltau1) {$\tau_1$};
\node at (ltau'1) {$\tau'_1$};
\node at (ltau2) {$\tau_2$};
\node at (ltau'2) {$\tau'_2$};
\node at (ltau3) {$\tau_3$};
\node at (ltau'3) {$\tau'_3$};

\end{tikzpicture} 
\hspace{1cm}
\begin{tikzpicture}[scale=1]
\coordinate (center) at (0,0) {};
\coordinate (ltau1) at ($(center)+(165:2.3cm)$) {};
\coordinate (ltau'1) at ($(center)+(345:2.3cm)$) {};
\coordinate (ltau3) at ($(center)+(195:2.3cm)$) {};
\coordinate (ltau'3) at ($(center)+(15:2.3cm)$) {};
\coordinate (ltau2) at ($(center)+(90:2.3cm)$) {};
\coordinate (ltau'2) at ($(center)+(270:2.3cm)$) {};

\coordinate (u4) at ($(center)+(75:2cm)$) {};
\coordinate (u3) at ($(center)+(85:2cm)$) {};
\coordinate (u2) at ($(center)+(95:2cm)$) {};
\coordinate (u1) at ($(center)+(105:2cm)$) {};

\coordinate (b4) at ($(center)+(255:2cm)$) {};
\coordinate (b3) at ($(center)+(265:2cm)$) {};
\coordinate (b2) at ($(center)+(275:2cm)$) {};
\coordinate (b1) at ($(center)+(285:2cm)$) {};

\coordinate (tau1u1) at ($(center)+(160:2cm)$) {};
\coordinate (tau1u2) at ($(center)+(170:2cm)$) {};
\coordinate (tau1b1) at ($(center)+(340:2cm)$) {};
\coordinate (tau1b2) at ($(center)+(350:2cm)$) {};

\coordinate (tau3u1) at ($(center)+(10:2cm)$) {};
\coordinate (tau3u2) at ($(center)+(20:2cm)$) {};
\coordinate (tau3b1) at ($(center)+(190:2cm)$) {};
\coordinate (tau3b2) at ($(center)+(200:2cm)$) {};

\draw ($(center)$) circle (2cm);

\draw[very thick] ($(center)+(47.5:2cm)$)-- ($(center)+(225.5:2cm)$);
\draw ($(center)+(30:2cm)$)-- ($(center)+(65:2cm)$);

\draw[very thick] ($(center)+(132.5:2cm)$)-- ($(center)+(312.5:2cm)$);
\draw ($(center)+(295:2cm)$)-- ($(center)+(330:2cm)$);

\draw[thick,red] (tau1u1)-- (tau1b2);
\draw[thick,red] (tau1u2)-- (tau1b1);

\draw[thick,red] (tau3u1)-- (tau3b2);
\draw[thick,red] (tau3u2)-- (tau3b1);

\draw[thick,red] (u1)-- (b2);
\draw[thick,red] (u2)-- (b4);
\draw[thick,red] (u3)-- (b1);
\draw[thick,red] (u4)-- (b3);

\draw[very thick] ([shift=(5:2cm)]0,0) arc (5:25:2cm);
\draw[very thick] ([shift=(70:2cm)]0,0) arc (70:110:2cm);
\draw[very thick] ([shift=(155:2cm)]0,0) arc (155:175:2cm);

\draw[very thick] ([shift=(185:2cm)]0,0) arc (185:205:2cm);
\draw[very thick] ([shift=(250:2cm)]0,0) arc (250:290:2cm);
\draw[very thick] ([shift=(335:2cm)]0,0) arc (335:355:2cm);

\tikzstyle{every node}=[inner sep=1pt]
\node at (ltau1) {$\tau_1$};
\node at (ltau'1) {$\tau'_1$};
\node at (ltau2) {$\tau_2$};
\node at (ltau'2) {$\tau'_2$};
\node at (ltau3) {$\tau'_3$};
\node at (ltau'3) {$\tau_3$};

\end{tikzpicture} 

\caption{\label{fig:chord_models_prime_parallel_prime_serial} Chord models of $(Q,{\sim})$ for the cases when
$Q$ is prime and $M_i$ is parallel (to the left) and when $Q$ is prime and $M_i$ is serial (to the right). 
Chords of the module $M_i$ are in red, chords of the set $N(M_i)$ are in bold.}
\end{figure}
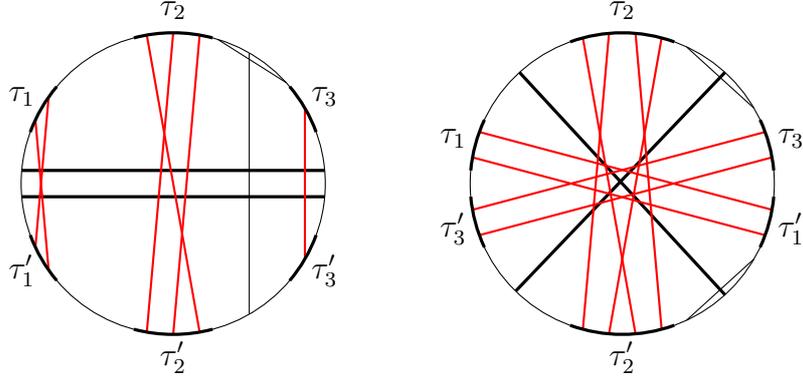

The next two lemmas show Property~\ref{prop_prime:contiguous_subwords_parallel} of the conformal models $(Q,{\sim})$ when $Q$ is prime.
See Figure \ref{fig:chord_models_prime_parallel_prime_serial} for an illustration.

\begin{lemma}
\label{lemma:circle_models_of_a_parallel_module}
Suppose $M_i$ is a parallel child of a prime component $Q$ and suppose $L_1,\ldots,L_k$ are the children of $M_i$.
Suppose $\psi$ is a chord model of $(Q,{\sim})$.
Then, 
$$\psi \Vert M_i \equiv \tau_{i_1} \ldots \tau_{i_k} \tau'_{i_k} \ldots \tau'_{i_1},$$
where $i_1,\ldots,i_k$ is a permutation of $[k]$ and $(\tau_{i_j}, \tau'_{i_j})$
is a permutation model of $(L_{i_j},{\sim})$ for every $j \in [k]$.
Moreover, for every $j \in [k]$ the words $\tau_{i_j}$ and $\tau'_{i_j}$ are contiguous~$\psi$ 
(see Figure~\ref{fig:chord_models_prime_parallel_prime_serial} to the left).
\end{lemma}
\begin{proof}
Since $Q$ is prime, we can pick $x \in Q \setminus M_i$ such that $x \sim M_i$.
Since $M_i$ is parallel, $L_j$ is either serial or prime for $j \in [k]$.
In particular, $(L_j,{\sim})$ is connected for every $j \in [k]$.
Since $x \sim M_i$ and since $L_j \parallel L_{j'}$ for every two distinct $j,j' \in [k]$,
$$\psi \Vert (M_i \cup \{x\}) \equiv x \tau_{i_1} \ldots \tau_{i_k} x \tau'_{i_k} \ldots \tau'_{i_1},$$
where $i_1,\ldots,i_k$ is a permutation of $[k]$ and $(\tau_{i_j}, \tau'_{i_j})$
is a permutation model of $(L_{i_j},{\sim})$ for every $j \in [k]$.
Since $M_i$ is parallel, by Lemma~\ref{lemma:circle_models_of_proper_modules_N_M} we have that 
$$\psi \Vert (M_i \cup N(M_i)) \equiv \pi \tau_{i_1} \ldots \tau_{i_k} \pi' \tau'_{i_1} \ldots \tau'_{i_k},$$
where $\pi$ and $\pi'$ are permutations of $N(M_i)$.
Finally, with an argument similar to the one used in the previous lemma,
we show that $\tau_{i_j}$ and $\tau'_{i_j}$ are contiguous subwords in $\psi$.
\end{proof}

Finally, the  last lemma of this section shows property \ref{prop_prime:contiguous_subwords_serial} of the conformal models $(Q,{\sim})$ when $Q$ is prime.
\begin{lemma}
\label{lemma:circle_models_of_a_serial_module}
Suppose $M_i$ is a serial child of a prime component $Q$ and suppose $L_1,\ldots,L_k$ are the children of $M_i$.
Suppose $\psi$ is a chord model of $(Q,{\sim})$.
Then
$$\psi \Vert M_i \equiv \tau_{i_1} \ldots \tau_{i_k} \tau'_{i_1} \ldots \tau'_{i_k},$$
where $i_1,\ldots,i_k$ is a permutation of $[k]$ 
and $(\tau_{i_j}, \tau'_{i_j})$ is a permutation model of $(L_{i_j},{\sim})$ for every $j \in [k]$.
Moreover, for every $j \in [k]$ the words $\tau_{i_j}$ and $\tau'_{i_j}$ are contiguous in~$\psi$ (see Figure \ref{fig:chord_models_prime_parallel_prime_serial} to the right).
\end{lemma}
\begin{proof}
Since $M_i$ is serial, $L_j$ is either prime or parallel for $j \in [k]$.
Using the same arguments as in Lemma~\ref{lemma:circle_models_of_proper_modules_N_M}, we deduce that
$$\psi \Vert M_i \equiv \tau_{i_1} \ldots \tau_{i_k} \tau'_{i_1} \ldots \tau'_{i_k},$$
where $i_1,\ldots,i_k$ is a permutation of $[k]$ 
and $(\tau_{i_j}, \tau'_{i_j})$ is a permutation model of $(L_{i_j},{\sim})$ for every $j \in [k]$.

Now, it remains to prove that $\tau_{i_j}$ and $\tau'_{i_j}$ are contiguous subwords in $\psi$.
Arguing as in the proof of Lemma~\ref{lemma:circle_models_of_proper_modules_N_M},
for every $x \in N(M_i)$ there is $j \in [k]$ such that
$$\psi \Vert (M_i \cup \{x\}) \equiv \tau_{i_1} \ldots \tau_{i_j} x \tau_{i_{j+1}} \ldots \tau_{i_k}\tau'_{i_1} \ldots \tau'_{i_j} x \tau'_{i_{j+1}} \ldots \tau'_{i_k}.$$
Assume that $\tau_{i_j}, \tau'_{i_j}$ for some $j \in [k]$ do not form two contiguous subwords in $\psi$.
Then, by the connectivity of $(Q,{\sim})$, there is $u \in Q \setminus N(M_i)$ such that
$u \parallel M_i$ and $\psi \Vert (L_{i_j} \cup \{u\}) \equiv \mu_1 u \mu_2 \mu'_2 u \mu'_1$, where
$\mu_1, \mu_2$ and $\mu'_1, \mu'_2$ are such that $\mu_1\mu_2 = \tau_{i_j}$,
$\mu'_{2}\mu'_1 = \tau'_{i_j}$, and $\mu_1, \mu_2, \mu'_1, \mu'_2$ are non-empty.
Since $u \parallel L_{i_j}$, $\mu_i$ is a permutation of $\mu'_i$ for $i \in [2]$.
Then, note that $\psi(u)$ intersects every chord from $\psi(M_i \setminus L_{i_j})$,
which is not possible as $u \parallel M_i$.
\end{proof}

\section{Conformal models of prime components -- appendix}
\label{sec:prime_case_properties}
The aim of this section is to prove Properties~\ref{prop:prime_skeleton} and~\ref{prop:prime_contiguous_subwords} used in Section~\ref{sec:proof_sketch}.
As in Section~\ref{sec:proof_sketch} we assume $Q$ is a prime component of $G_{ov}$, $M_1,\ldots,M_n$ are the children of $Q$ in $\mathcal{M}(G_{ov})$, and 
$K(Q) = \{K_1,\ldots,K_k\}$ are the equivalent classes of $K$-relation defined by Definition~\ref{def:prime-K-relation}. 

Property~\ref{prop:prime_skeleton} is proved in two steps.
First, we show that a subgraph of $G_{ov}$ induced by the set containing a vertex from every child $M_i$ of $Q$ has two conformal models, one being the reflection of the other.
Then, we prove that a subgraph of $G_{ov}$ induced by a set containing a vertex from every set $K_i$ in $K(Q)$ has two conformal models, one being the reflection of the other.

In order to accomplish the first step, we need to describe the structure of the chord models of a circle graph.
A description presented below is by Chaplick, Fulek, and Klav\'{i}k~\cite{CFK13}, who used the following concepts in their work:
the split decomposition due to Cunningham~\cite{Cun82}, 
Theorem~\ref{thm:cicle_representation_no_split} due to
Gabor, Supowit, and Hsu~\cite{GSH89}, $\diamond$~relation due to Chaplick, Fulek, and Klav\'{i}k~\cite{CFK13} (who were inspired by Naji~\cite{Naji85}), and maximal splits are due to Chaplick, Fulek, and Klav\'{i}k~\cite{CFK13}.
We refer to~\cite{CFK13} for more details.

\begin{figure}[!htp]

\begin{tikzpicture}[scale=0.4]
    \coordinate (A_center) at (-2,0) {};
    \coordinate (lA_center) at (-2,0.6) {};
    \coordinate (lA) at (-2,2.5) {};
    
    \coordinate (Ar1) at ($(A_center)+(-40:1 and 2)$) {};
    \coordinate (Ar2) at ($(A_center)+(-20:1 and 2)$) {};
    \coordinate (Ar3) at ($(A_center)+(0:1 and 2)$) {};
    \coordinate (Ar4) at ($(A_center)+(20:1 and 2)$) {};
    \coordinate (Ar5) at ($(A_center)+(40:1 and 2)$) {};
    \coordinate (Al1) at ($(A_center)+(220:1 and 2)$) {};
    \coordinate (Al2) at ($(A_center)+(200:1 and 2)$) {};
    \coordinate (Al3) at ($(A_center)+(180:1 and 2)$) {};
    \coordinate (Al4) at ($(A_center)+(160:1 and 2)$) {};
    \coordinate (Al5) at ($(A_center)+(140:1 and 2)$) {};

    \coordinate (B_center) at (2,0) {};    
    \coordinate (lB_center) at (2,0.7) {};    
    \coordinate (lB) at (2,2.5) {};

    \coordinate (Br1) at ($(B_center)+(-40:1 and 2)$) {};
    \coordinate (Br2) at ($(B_center)+(-20:1 and 2)$) {};
    \coordinate (Br3) at ($(B_center)+(0:1 and 2)$) {};
    \coordinate (Br4) at ($(B_center)+(20:1 and 2)$) {};
    \coordinate (Br5) at ($(B_center)+(40:1 and 2)$) {};
    \coordinate (Bl1) at ($(B_center)+(220:1 and 2)$) {};
    \coordinate (Bl2) at ($(B_center)+(200:1 and 2)$) {};
    \coordinate (Bl3) at ($(B_center)+(180:1 and 2)$) {};
    \coordinate (Bl4) at ($(B_center)+(160:1 and 2)$) {};
    \coordinate (Bl5) at ($(B_center)+(140:1 and 2)$) {};
    
    \coordinate (AA_center) at (-5.5,0) {};
    \coordinate (lAA) at (-5.5,2.2) {};
    \coordinate (AAr1) at ($(AA_center)+(-40:1 and 1.5)$) {};
    \coordinate (AAr2) at ($(AA_center)+(-20:1 and 1.5)$) {};
    \coordinate (AAr3) at ($(AA_center)+(0:1 and 1.5)$) {};
    \coordinate (AAr4) at ($(AA_center)+(20:1 and 1.5)$) {};
    \coordinate (AAr5) at ($(AA_center)+(40:1 and 1.5)$) {};

    \coordinate (AB_center) at (5.5,0) {};
    \coordinate (lAB) at (5.5,2.2) {};
    \coordinate (ABl1) at ($(AB_center)+(220:1 and 1.5)$) {};
    \coordinate (ABl2) at ($(AB_center)+(200:1 and 1.5)$) {};
    \coordinate (ABl3) at ($(AB_center)+(180:1 and 1.5)$) {};
    \coordinate (ABl4) at ($(AB_center)+(160:1 and 1.5)$) {};
    \coordinate (ABl5) at ($(AB_center)+(140:1 and 1.5)$) {};

    \draw[fill=gray!60] (A_center) ellipse (1 and 2);
    \draw[fill=gray!60] (B_center) ellipse (1 and 2);
    \draw[fill=gray!30] (AA_center) ellipse (1 and 1.5);
    \draw[fill=gray!30] (AB_center) ellipse (1 and 1.5);
    
    \tikzstyle{every node}=[circle,minimum size=5pt,inner sep=0pt,draw,fill]
    \node at (A_center) {};
    \node at (B_center) {};

    \tikzstyle{every node}=[inner sep=1pt]
    \node at (lA_center) {$a$};
    \node at (lB_center) {$b$};
    \node at (lA) {$A$};
    \node at (lB) {$B$};
    \node at (lAA) {$\alpha(A)$};
    \node at (lAB) {$\alpha(B)$};

    \draw[-] (Ar1)--(Bl1);
    \draw[-] (Ar1)--(Bl2);
    \draw[-] (Ar1)--(Bl3);
    \draw[-] (Ar1)--(Bl4);
    \draw[-] (Ar1)--(Bl5);

    \draw[-] (Ar2)--(Bl1);
    \draw[-] (Ar2)--(Bl2);
    \draw[-] (Ar2)--(Bl3);
    \draw[-] (Ar2)--(Bl4);
    \draw[-] (Ar2)--(Bl5);
    
    \draw[-] (Ar3)--(Bl1);
    \draw[-] (Ar3)--(Bl2);
    \draw[-] (Ar3)--(Bl3);
    \draw[-] (Ar3)--(Bl4);
    \draw[-] (Ar3)--(Bl5);
    
    \draw[-] (Ar4)--(Bl1);
    \draw[-] (Ar4)--(Bl2);
    \draw[-] (Ar4)--(Bl3);
    \draw[-] (Ar4)--(Bl4);
    \draw[-] (Ar4)--(Bl5);
    
    \draw[-] (Ar5)--(Bl1);
    \draw[-] (Ar5)--(Bl2);
    \draw[-] (Ar5)--(Bl3);
    \draw[-] (Ar5)--(Bl4);
    \draw[-] (Ar5)--(Bl5);

    \draw[-] (AAr1)--(Al1);
    \draw[-] (AAr2)--(Al2);
    \draw[-] (AAr3)--(Al3);
    \draw[-] (AAr4)--(Al4);
    \draw[-] (AAr5)--(Al5);

    \draw[-] (Br1)--(ABl1);
    \draw[-] (Br2)--(ABl2);
    \draw[-] (Br3)--(ABl3);
    \draw[-] (Br4)--(ABl4);
    \draw[-] (Br5)--(ABl5);
    
    \draw[white] (-8,-4)--(-7.5,-4);
\end{tikzpicture}
\hspace{0.5cm}
\begin{tikzpicture}[scale=0.6,>=latex]
\coordinate (center) at (0,0) {};

\coordinate (tauA) at ($(center)+(90:2.4)$) {};
\coordinate (tau'A) at ($(center)+(270:2.4)$) {};
\coordinate (tauB) at ($(center)+(0:2.4)$) {};
\coordinate (tau'B) at ($(center)+(180:2.4)$) {};

\coordinate (A) at ($(center)+(70:1.3)$) {};
\coordinate (B) at ($(center)+(200:1.3)$) {};

\coordinate (u1) at ($(center)+(120:2)$) {};
\coordinate (u2) at ($(center)+(110:2)$) {};
\coordinate (u3) at ($(center)+(93:2)$) {};
\coordinate (u4) at ($(center)+(87:2)$) {};
\coordinate (u5) at ($(center)+(70:2)$) {};
\coordinate (u6) at ($(center)+(60:2)$) {};

\coordinate (b1) at ($(center)+(240:2)$) {};
\coordinate (b2) at ($(center)+(250:2)$) {};
\coordinate (b3) at ($(center)+(267:2)$) {};
\coordinate (b4) at ($(center)+(273:2)$) {};
\coordinate (b5) at ($(center)+(290:2)$) {};
\coordinate (b6) at ($(center)+(300:2)$) {};

\coordinate (l1) at ($(center)+(160:2)$) {};
\coordinate (l2) at ($(center)+(177:2)$) {};
\coordinate (l3) at ($(center)+(183:2)$) {};
\coordinate (l4) at ($(center)+(200:2)$) {};

\coordinate (r1) at ($(center)+(3:2)$) {};
\coordinate (r2) at ($(center)+(-3:2)$) {};

\draw (center) circle (2cm);

\draw[thick,-] (u1)--(u6);
\draw[thick,-] (u2)--(u5);
\draw[thick,-] (u3)--(b3);
\draw[thick,-] (u4)--(b4);
\draw[thick,-] (b2)--(b5);

\draw[thick,-] (l2)--(r1);
\draw[thick,-] (l3)--(r2);
\draw[thick,-] (l1)--(l4);

\draw[red, very thick,-] ([shift=(50:2cm)]0,0) arc (50:130:2cm);
\draw[blue, very thick,-] ([shift=(230:2cm)]0,0) arc (230:310:2cm);

\draw[black, very thick,-] ([shift=(-30:2cm)]0,0) arc (-30:30:2cm);
\draw[black, very thick,-] ([shift=(150:2cm)]0,0) arc (150:210:2cm);

\tikzstyle{every node}=[inner sep=1pt]
\node[red] at (tauA) {$\tau_A$};
\node[blue] at (tau'A) {$\tau'_A$};;
\node[black] at (tauB) {$\tau_B$};
\node[black] at (tau'B) {$\tau'_B$};
\node at (A) {$A$};
\node at (B) {$B$};

\draw[white] (-2.5,-3.0)--(-2,-3.0);
\end{tikzpicture}
\hspace{0.25cm}
\begin{tikzpicture}[scale=0.6,>=latex]
\coordinate (center) at (0,0) {};

\coordinate (tauA) at ($(center)+(90:2.4)$) {};
\coordinate (tau'A) at ($(center)+(270:2.4)$) {};
\coordinate (tauB) at ($(center)+(0:2.4)$) {};
\coordinate (tau'B) at ($(center)+(180:2.4)$) {};

\coordinate (A) at ($(center)+(70:1.3)$) {};
\coordinate (B) at ($(center)+(200:1.3)$) {};

\coordinate (u1) at ($(center)+(120:2)$) {};
\coordinate (u2) at ($(center)+(110:2)$) {};
\coordinate (u3) at ($(center)+(93:2)$) {};
\coordinate (u4) at ($(center)+(87:2)$) {};
\coordinate (u5) at ($(center)+(70:2)$) {};
\coordinate (u6) at ($(center)+(60:2)$) {};

\coordinate (b1) at ($(center)+(240:2)$) {};
\coordinate (b2) at ($(center)+(250:2)$) {};
\coordinate (b3) at ($(center)+(267:2)$) {};
\coordinate (b4) at ($(center)+(273:2)$) {};
\coordinate (b5) at ($(center)+(290:2)$) {};
\coordinate (b6) at ($(center)+(300:2)$) {};

\coordinate (l1) at ($(center)+(160:2)$) {};
\coordinate (l2) at ($(center)+(177:2)$) {};
\coordinate (l3) at ($(center)+(183:2)$) {};
\coordinate (l4) at ($(center)+(200:2)$) {};

\coordinate (r1) at ($(center)+(3:2)$) {};
\coordinate (r2) at ($(center)+(-3:2)$) {};

\draw (center) circle (2cm);

\draw[thick,-] (u2)--(u5);
\draw[thick,-] (u3)--(b3);
\draw[thick,-] (u4)--(b4);
\draw[thick,-] (b2)--(b5);
\draw[thick,-] (b1)--(b6);

\draw[thick,-] (l2)--(r1);
\draw[thick,-] (l3)--(r2);
\draw[thick,-] (l1)--(l4);

\draw[blue, very thick,-] ([shift=(50:2cm)]0,0) arc (50:130:2cm);
\draw[red, very thick,-] ([shift=(230:2cm)]0,0) arc (230:310:2cm);

\draw[black, very thick,-] ([shift=(-30:2cm)]0,0) arc (-30:30:2cm);
\draw[black, very thick,-] ([shift=(150:2cm)]0,0) arc (150:210:2cm);

\tikzstyle{every node}=[inner sep=1pt]
\node[blue] at (tauA) {$\tau'_A$};
\node[red] at (tau'A) {$\tau_A$};;
\node[black] at (tauB) {$\tau_B$};
\node[black] at (tau'B) {$\tau'_B$};
\node at (A) {$A$};
\node at (B) {$B$};

\draw[white] (-2.5,-3.0)--(-2,-3.0);

\end{tikzpicture}
\caption{\label{fig:split} Split $(\alpha(A),A,\alpha(B),B)$ in $G_U$ and two possible chord models of $G_{ov}$: $\tau_A\tau_B\tau'_A\tau'_B$ and $\tau'_A\tau_B\tau_A\tau'_B$,
where $b\tau_Ab\tau'_A$ and $a\tau_Ba\tau'_B$ are chord models of $(A \cup \alpha(A) \cup \{b\}, {\sim})$ and $(B \cup \alpha(B) \cup \{a\}, {\sim})$ for some $a \in A$ and some $b \in B$, respectively.}
\end{figure}
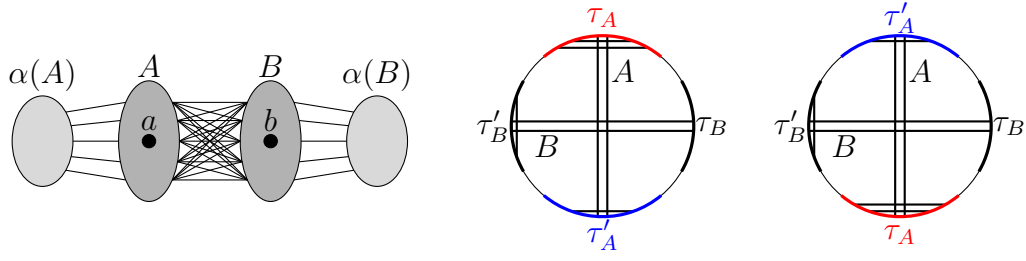

Suppose $G_U = (U,{\sim})$ is a connected circle graph.
A quadruple $(A, \alpha(A), B, \alpha(B))$ is a \emph{split} in $G_U$ if: 
\begin{itemize}
\item The sets $A$, $B$, $\alpha(A)$, $\alpha(B)$ form a partition of $U$,
\item $A \neq \emptyset$ and $B \neq \emptyset$, but possibly $\alpha(A) = \emptyset$ or $\alpha(B) = \emptyset$,
\item $A \sim B$.
\item $\alpha(A) \parallel (B \cup \alpha(B))$ and $\alpha(B) \parallel A \cup \alpha(A)$.
\end{itemize}
See Figure~\ref{fig:split} to the left.
Since $G_U$ is connected, $(A,\alpha(A),B,\alpha(B))$ can be uniquely recovered from the sets $A$ and $B$.
Hence, without losing any information, we say that $(A,\alpha(A),B,\alpha(B))$ is the \emph{split between $A$ and $B$},
and we denote $(A,\alpha(A),B,\alpha(B))$ simply by $(A,B)$.

A split $(A,B)$ is \emph{non-trivial} if $|A \cup \alpha(A)| \geq 2$ and $|B \cup \alpha(B)| \geq 2$;
otherwise $(A,B)$ is \emph{trivial}.
\begin{theorem}[\cite{GSH89}]
\label{thm:cicle_representation_no_split}
If $G_U$ has no non-trivial split, $G_U$ has only two chord models, one being the reflection of the other.
\end{theorem}
On the other hand, if $G_U$ has non-trivial splits, 
$G_U$ has more then two non-equivalent chord models -- see Figure~\ref{fig:split} to the right.

A split in $G_U$ between $A$ and $B$ is \emph{maximal}
if there is no split in $G_U$ between $A'$ and $B'$, where $A'$ and $B'$ are such that $A \subseteq A'$, $B \subseteq B'$, and $|A| < |A'|$ or $|B| < |B|'$.
Lemma~1 in \cite{CFK13} provides the following characterization of maximal splits in $G_U$: 
a split between $A$ and $B$ is maximal if and only if
there exists no $C \subseteq \alpha(A)$ such that $C$ induces a connected component in $(\alpha(A),{\sim})$ and for every vertex $u  \in C$ either $u \sim A$ or $u \parallel A$, 
and similarly for $\alpha(B)$ and $B$.
This observation allows to present the algorithm for computing a maximal split in $G_U$ (see \cite{CFK13} for more details):
\begin{itemize}
 \item start with any non-trivial split between $A$ and $B$,
 \item while there exists $C$ witnessing that $(A,B)$ is not maximal in~$G_{U}$: 
 if $C \subseteq \alpha(A)$ set $A = A$ and $B = B \cup C'$, 
 where $C'$ is the set of all vertices from $C$ adjacent to $A$,
 and if $C \subseteq \alpha(B)$, set $B = B$ and $A = A \cup C'$,
 where $C'$ is the set of all vertices from $C$ adjacent to $B$,
 \item return $(A,B)$.
\end{itemize}

Suppose $(A,B)$ is a maximal split in $G_U$ produced by the above algorithm.
Note that $(A,B)$ might be trivial. 
In this case we have either $|A|=1$ and $|\alpha(A)|=0$
or $|B|=1$ and $|\alpha(B)|=0$.
Lemma~2 of~\cite{CFK13} proves the following:
if $(A,B)$ is trivial with $|A|= \{a\}$ and $\alpha(A) = \emptyset$,
then $a$ is an \emph{articulation point} in $G_U$ (i.e. $(U \setminus \{a\}, {\sim})$ is disconnected).
In particular, if at some point the split $(A,B)$ maintained by the algorithm is trivial,
$(A,B)$ can not be further extended, the algorithm stops and returns $(A,B)$.
For a purpose mentioned later, 
we restrict choices made by the algorithm: in each step, having the choice of extending $(A,B)$ into a non-trivial or a trivial split,
the algorithm always takes the first option.
\subsection{The structure of chord models with respect to a non-trivial maximal split}
\label{subsec:structure_non_trivial_split}
Let $(A,B)$ be a non-trivial maximal split in $G_U$ returned by the algorithm computing a maximal split.
Let $C = A\cup B$. 
Following~\cite{CFK13}, let $\diamond$ be the smallest equivalence relation on $C$ containing the pairs $(u,v) \in C \times C$ such that:
\begin{itemize}
\item $u \parallel v$,
\item $u,v$ are connected by a path in $(U,{\sim})$ with 
all the inner vertices in $\alpha(A) \cup \alpha(B)$.
\end{itemize}
Suppose $C_1,\ldots,C_k$ are the equivalence classes of $\diamond$~relation.
Note that $C_i \subseteq A$ or $C_i \subseteq B$ for every $i \in [k]$ and
thus $k \geq 2$.
Observe that: 
\begin{itemize}
\item $C_i \sim C_j$ for every distinct $i,j \in [k]$ (if $u \parallel v$ for some $u \in C_i$ and $v \in C_j$, then $u \diamond v$, and $u,v$ can not be in different classes of $\diamond$~relation).
\end{itemize}
Next, since the vertices of every component of $(U \setminus C, {\sim})$ can be adjacent to exactly one set $C_i$ among $C_1,\ldots,C_k$, we can partition the set $U \setminus C$ into 
the sets $\alpha(C_1),\ldots,\alpha(C_k)$ such that:
\begin{itemize}
 \item $\alpha(C_i) \parallel (\alpha(C_j) \cup C_j)$ for every two distinct $i,j \in [k]$.
\end{itemize}
Note that $\alpha(C_i)$ might be empty. 
See Figure~\ref{fig:non_trivial_maximal_split} for an illustration.

\input ./figures/prime_properties/non_trivial_maximal_split.tex
Further, let $G_i$ by a graph obtained from $G_U$ by contracting the vertices
from $U \setminus (C_i \cup \alpha(C_i))$ into a single vertex $v_i$ adjacent to every vertex in $C_i$.
That is, $G_i$ is such that $V(G_i) = C_i \cup \alpha(C_i) \cup \{v_i\}$, 
$v_iv \in E(G_i)$ for every $v \in C_i$, $v_iv \notin E(G_i)$ for every $v \in \alpha(C_i)$, 
and $uv \in E(G_i)$ if and only if $u \sim v$ for $u,v \in C_i \cup \alpha(C_i)$.
Note that every chord model of $G_i$ has the form $v_i \tau_i v_i \tau'_i$, where
every $v \in C_i$ occurs once in both words $\tau_i$ and $\tau'_i$ and 
every $v \in \alpha(C_i)$ occurs twice either in $\tau_i$ or in $\tau'_i$.
The next theorem describes the relationship between the set of all chord models of $G_U$ and 
the set of all chord models of~$G_i$.

\begin{theorem}[Proposition~1 from \cite{CFK13}] 
\label{thm:non_trivial_split_representations}
The following statements hold:
\begin{enumerate}
\item If $v_i \tau_i v_i \tau'_i$ is a chord model of $G_i$ for $i \in [k]$,
$i_1, \ldots, i_k$ is a permutation of $[k]$, and the words $\mu_i,\mu'_i$ are such that $\{\mu_i,\mu'_i\} = \{\tau_i,\tau'_i\}$ for $i \in [k]$, then 
$$\tau \equiv \mu_{i_1}\ldots \mu_{i_k}\mu'_{i_1}\ldots \mu'_{i_k},$$
is a chord model of $G_U$.
\item If $\tau$ is a chord model of $G_U$, then 
$$\tau \equiv \mu_{i_1}\ldots \mu_{i_k}\mu'_{i_1}\ldots \mu'_{i_k},$$
where $i_1, \ldots, i_k$ is a permutation of $[k]$  and $v_{i_j} \mu_{i_j} v_{i_j} \mu'_{i_j}$ is a chord model of $G_{i_j}$ for $j \in [k]$.
\end{enumerate}
\end{theorem}
See Figure \ref{fig:non_trivial_maximal_split} for an illustration.

\subsection{The structure of chord models with respect to a trivial maximal split}
\label{subsec:structure_trivial_split}
\begin{figure}[!htp]
\begin{tikzpicture}[scale=0.35]
    \coordinate (a) at (0,-2) {};
    \coordinate (la) at (0,-3) {};
    \coordinate (C1_center) at (-6,2) {};
    \coordinate (C2_center) at (0,2) {};
    \coordinate (C3_center) at (6,2) {};
    
    \coordinate (AC1_center) at (-6,5) {};
    \coordinate (AC2_center) at (0,5) {};
    \coordinate (AC3_center) at (6,5) {};

    \coordinate (lC1) at ($(C1_center)+(180:2.8 and 1)$) {};
    \coordinate (lC2) at ($(C2_center)+(180:2.8 and 1)$) {};
    \coordinate (lC3) at ($(C3_center)+(180:2.8 and 1)$) {};

    \coordinate (lAC1) at ($(AC1_center)+(180:2.6 and 1)$) {};
    \coordinate (lAC2) at ($(AC2_center)+(180:2.6 and 1)$) {};
    \coordinate (lAC3) at ($(AC3_center)+(180:2.6 and 1)$) {};

    \coordinate (C1l1) at ($(C1_center)+(240:2 and 1)$) {};
    \coordinate (C1l2) at ($(C1_center)+(270:2 and 1)$) {};
    \coordinate (C1l3) at ($(C1_center)+(290:2 and 1)$) {};
    \coordinate (C1l4) at ($(C1_center)+(310:2 and 1)$) {};

    \coordinate (C1u1) at ($(C1_center)+(120:2 and 1)$) {};
    \coordinate (C1u2) at ($(C1_center)+(90:2 and 1)$) {};
    \coordinate (C1u3) at ($(C1_center)+(60:2 and 1)$) {};

    \coordinate (AC1l1) at ($(AC1_center)+(240:1 and 1)$) {};
    \coordinate (AC1l2) at ($(AC1_center)+(270:1 and 1)$) {};
    \coordinate (AC1l3) at ($(AC1_center)+(300:1 and 1)$) {};

    \coordinate (C2l1) at ($(C2_center)+(300:2 and 1)$) {};
    \coordinate (C2l2) at ($(C2_center)+(285:2 and 1)$) {};
    \coordinate (C2l3) at ($(C2_center)+(270:2 and 1)$) {};
    \coordinate (C2l4) at ($(C2_center)+(255:2 and 1)$) {};
    \coordinate (C2l5) at ($(C2_center)+(240:2 and 1)$) {};

    \coordinate (C2u1) at ($(C2_center)+(120:2 and 1)$) {};
    \coordinate (C2u2) at ($(C2_center)+(90:2 and 1)$) {};
    \coordinate (C2u3) at ($(C2_center)+(60:2 and 1)$) {};

    \coordinate (AC2l1) at ($(AC2_center)+(240:1 and 1)$) {};
    \coordinate (AC2l2) at ($(AC2_center)+(270:1 and 1)$) {};
    \coordinate (AC2l3) at ($(AC2_center)+(300:1 and 1)$) {};

    \coordinate (C3l1) at ($(C3_center)+(300:2 and 1)$) {};
    \coordinate (C3l2) at ($(C3_center)+(270:2 and 1)$) {};
    \coordinate (C3l3) at ($(C3_center)+(250:2 and 1)$) {};
    \coordinate (C3l4) at ($(C3_center)+(230:2 and 1)$) {};
    
    \coordinate (C3u1) at ($(C3_center)+(120:2 and 1)$) {};
    \coordinate (C3u2) at ($(C3_center)+(90:2 and 1)$) {};
    \coordinate (C3u3) at ($(C3_center)+(60:2 and 1)$) {};

    \coordinate (AC3l1) at ($(AC3_center)+(240:1 and 1)$) {};
    \coordinate (AC3l2) at ($(AC3_center)+(270:1 and 1)$) {};
    \coordinate (AC3l3) at ($(AC3_center)+(300:1 and 1)$) {};

    \draw[fill=gray!60] (C1_center) ellipse (2 and 1);
    \draw[fill=gray!60] (C2_center) ellipse (2 and 1);
    \draw[fill=gray!60] (C3_center) ellipse (2 and 1);

    \draw[fill=gray!30] (AC1_center) ellipse (1 and 1);
    \draw[fill=gray!30] (AC2_center) ellipse (1 and 1);
    \draw[fill=gray!30] (AC3_center) ellipse (1 and 1);

    \tikzstyle{every node}=[inner sep=1pt]
    \node at (lC1) {$C_1$};
    \node at (lC2) {$C_2$};
    \node at (lC3) {$C_3$};
    \node at (lAC1) {$\alpha(C_1)$};
    \node at (lAC2) {$\alpha(C_2)$};
    \node at (lAC3) {$\alpha(C_3)$};
    \node at (la) {$a$};
    
    \tikzstyle{every node}=[circle,minimum size=5pt,inner sep=0pt,draw,fill]
    \node at (a) {};

    \draw[-] (a)--(C1l1);
    \draw[-] (a)--(C1l2);
    \draw[-] (a)--(C1l3);
    \draw[-] (a)--(C1l4);
    \draw[-] (a)--(C2l1);
    \draw[-] (a)--(C2l2);
    \draw[-] (a)--(C2l3);
    \draw[-] (a)--(C2l4);
    \draw[-] (a)--(C2l5);
    \draw[-] (a)--(C3l1);
    \draw[-] (a)--(C3l2);
    \draw[-] (a)--(C3l3);
    \draw[-] (a)--(C3l4);
    
    \draw[-] (C1u1)--(AC1l1);
    \draw[-] (C1u2)--(AC1l2);
    \draw[-] (C1u3)--(AC1l3);
    
    \draw[-] (C2u1)--(AC2l1);
    \draw[-] (C2u2)--(AC2l2);
    \draw[-] (C2u3)--(AC2l3);

    \draw[-] (C3u1)--(AC3l1);
    \draw[-] (C3u2)--(AC3l2);
    \draw[-] (C3u3)--(AC3l3);
\draw[white] (-2.5,-4)--(-2,-4);
\end{tikzpicture}
\hspace{0.5cm}
\begin{tikzpicture}[scale=0.7,>=latex]
\coordinate (center) at (0,0) {};

\coordinate (tau2) at ($(center)+(45:2.4)$) {};
\coordinate (tau'2) at ($(center)+(315:2.4)$) {};

\coordinate (tau3) at ($(center)+(90:2.4)$) {};
\coordinate (tau'3) at ($(center)+(270:2.4)$) {};

\coordinate (tau4) at ($(center)+(135:2.4)$) {};
\coordinate (tau'4) at ($(center)+(225:2.4)$) {};

\coordinate (r) at ($(center)+(0:2)$) {};
\coordinate (ur) at ($(center)+(45:2)$) {};
\coordinate (u) at ($(center)+(90:2)$) {};
\coordinate (ul) at ($(center)+(135:2)$) {};
\coordinate (l) at ($(center)+(180:2)$) {};
\coordinate (bl) at ($(center)+(225:2)$) {};
\coordinate (b) at ($(center)+(270:2)$) {};
\coordinate (br) at ($(center)+(315:2)$) {};

\draw (center) circle (2cm);

\draw[very thick,-] (r)--(l);
\draw[very thick,-] (ur)--(br);
\draw[very thick,-] (u)--(b);
\draw[very thick,-] (ul)--(bl);

\draw[very thick] ([shift=(30:2cm)]0,0) arc (30:60:2cm);
\draw[very thick] ([shift=(75:2cm)]0,0) arc (75:105:2cm);
\draw[very thick] ([shift=(120:2cm)]0,0) arc (120:150:2cm);
\draw[very thick] ([shift=(210:2cm)]0,0) arc (210:240:2cm);
\draw[very thick] ([shift=(255:2cm)]0,0) arc (255:285:2cm);
\draw[very thick] ([shift=(300:2cm)]0,0) arc (300:330:2cm);

\tikzstyle{every node}=[inner sep=1pt]
\node at (tau1) {$a$};
\node at (tau'1) {$a$};
\node at (tau2) {$\tau_1$};
\node at (tau'2) {$\tau'_1$};
\node at (tau3) {$\tau'_2$};
\node at (tau'3) {$\tau_2$};
\node at (tau4) {$\tau_3$};
\node at (tau'4) {$\tau'_3$};

\draw[white] (-2.5,-3)--(-2,-3);
\end{tikzpicture}
\begin{tikzpicture}[scale=0.7,>=latex]
\coordinate (center) at (0,0) {};

\coordinate (tau2) at ($(center)+(45:2.4)$) {};
\coordinate (tau'2) at ($(center)+(315:2.4)$) {};

\coordinate (tau3) at ($(center)+(90:2.4)$) {};
\coordinate (tau'3) at ($(center)+(270:2.4)$) {};

\coordinate (tau4) at ($(center)+(135:2.4)$) {};
\coordinate (tau'4) at ($(center)+(225:2.4)$) {};

\coordinate (r) at ($(center)+(0:2)$) {};
\coordinate (ur) at ($(center)+(45:2)$) {};
\coordinate (u) at ($(center)+(90:2)$) {};
\coordinate (ul) at ($(center)+(135:2)$) {};
\coordinate (l) at ($(center)+(180:2)$) {};
\coordinate (bl) at ($(center)+(225:2)$) {};
\coordinate (b) at ($(center)+(270:2)$) {};
\coordinate (br) at ($(center)+(315:2)$) {};

\draw (center) circle (2cm);

\draw[very thick,-] (r)--(l);
\draw[very thick,-] (ur)--(br);
\draw[very thick,-] (u)--(b);
\draw[very thick,-] (ul)--(bl);

\draw[very thick] ([shift=(30:2cm)]0,0) arc (30:60:2cm);
\draw[very thick] ([shift=(75:2cm)]0,0) arc (75:105:2cm);
\draw[very thick] ([shift=(120:2cm)]0,0) arc (120:150:2cm);
\draw[very thick] ([shift=(210:2cm)]0,0) arc (210:240:2cm);
\draw[very thick] ([shift=(255:2cm)]0,0) arc (255:285:2cm);
\draw[very thick] ([shift=(300:2cm)]0,0) arc (300:330:2cm);

\tikzstyle{every node}=[inner sep=1pt]
\node at (tau1) {$a$};
\node at (tau'1) {$a$};
\node at (tau2) {$\tau'_2$};
\node at (tau'2) {$\tau_2$};
\node at (tau3) {$\tau_1$};
\node at (tau'3) {$\tau'_1$};
\node at (tau4) {$\tau_3$};
\node at (tau'4) {$\tau'_3$};

\draw[white] (-2.5,-3)--(-2,-3);
\end{tikzpicture}
\caption{\label{fig:trivial_maximal_split} Maximal trivial split. 
Given chord models $a\tau_ia\tau'_i$ of $G_i$ for $i \in [3]$,
two non-equivalent chord models of $G_U$, namely $a\tau_3\tau'_2\tau_1a\tau'_1\tau_2\tau'_3$ and $a\tau_3\tau_1\tau'_2a\tau_2\tau'_1\tau'_3$, are shown to the right.}
\end{figure}
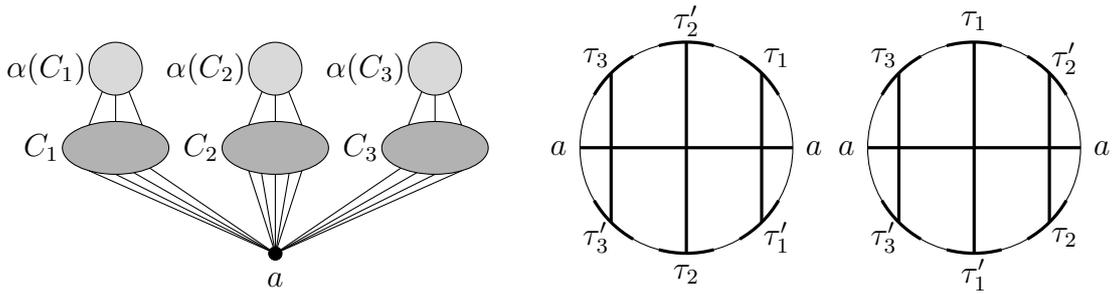

Let $(A,B)$ be a trivial maximal split in $G_U$
returned by the algorithm computing a maximal split.
Without loss of generality we assume that $A = \{a\}$ and $\alpha(A) = \emptyset$.
Lemma~2 in~\cite{CFK13} proves that $a$ is the articulation point of $G_U$.
Let $D_1,\dots,D_k$ for some $k \geq 2$ be the connected components of $(U \setminus \{a\},{\sim})$.
Let $C_i = \{v \in D_i: v \sim a\}$ and $\alpha(C_i) = \{v \in D_i: v \parallel a\}$
-- see Figure~\ref{fig:trivial_maximal_split} to the left.
Let $G_i$ be the restriction of $G_U$ to the set $\{a\} \cup C_i \cup \alpha(C_i)$, 
i.e. $G_i = (\{a\} \cup C_i \cup \alpha(C_i), {\sim})$.
Note that every chord model of $G_i$ has the form $a \tau_i a \tau'_i$, where
every $v \in C_i$ occurs once in both $\tau_i$ and $\tau'_i$ and
every $v \in \alpha(C_i)$ occurs twice either in $\tau_i$ or in $\tau'_i$.
The next theorem describes the relation between the set of all chord models of $G_U$ and the set of all chord models of $G_i$.

\begin{theorem}[Proposition 2 in \cite{CFK13}] 
\label{thm:trivial_split_representations}
The following statements hold:
\begin{enumerate}
\item If $a \tau_i a \tau'_i$ is a chord model of $G_i$ for $i \in [k]$,
$i_1, \ldots, i_k$ is a permutation of $[k]$, and the words $\mu_i,\mu'_i$ are 
such that $\{\mu_i,\mu'_i\} = \{\tau_i,\tau'_i\}$ for $i \in [k]$, then 
$$\tau \equiv a\mu_{i_1}\ldots \mu_{i_k} a\mu'_{i_k}\ldots \mu'_{i_1}$$
is a chord model of $G_U$.
\item If $\tau$ is a chord model of $G_U$, then 
$$\tau \equiv a\mu_{i_1} \ldots \mu_{i_k}a\mu'_{i_k} \ldots \mu'_{i_1},$$
where $i_1, \ldots, i_k$ is a permutation of $[k]$ and $a \mu_{i_j} a \mu'_{i_j}$ is a chord model of $G_{i_j}$ for $j \in [k]$.
\end{enumerate}
\end{theorem}
See Figure~\ref{fig:trivial_maximal_split} for an illustration.
Note that we can use the above theorem to describe the chord models of $G_U$ 
whenever $G_U$ contains some articulation point.

This completes the description of the chord models of $G_{U}$.

Now we are ready to prove the first step of Property~\ref{prop:prime_skeleton}.
Let $U$ be the set containing a vertex from every child $M_i$ of $Q$.
In particular, $(U,{\sim})$ is a prime graph, that is, $(U,{\sim})$ contains no modules other than the trivial ones.
\begin{lemma}
\label{lemma:two_models_of_a_prime_graph}
The prime graph $(U,{\sim})$ has exactly two conformal models, one being the reflection of the other.
\end{lemma}
\begin{proof}
Since $G$ is a circular-arc graph, $(U,{\sim})$ has at least one conformal model.
Our goal is to prove that this model, up to reflection, is unique.

We prove the lemma by induction on the number of vertices in $(U,{\sim})$.
The smallest prime graph is the path $P_4$ on four vertices.
Suppose $(U,{\sim})$ is isomorphic to $P_4$, that is, suppose $U = \{u_1,u_2,u_3,u_4\}$ and ${\sim}$ contains the edges $u_1 \sim u_2$,  $u_2\sim u_3$, and $u_3 \sim u_4$.
Clearly, for every conformal model $\phi$ of $(U,{\sim})$ we have 
either $\phi \Vert \{u_2,u_3\} = u^0_2u^1_3u^1_2u_3^0$ or 
$\phi \Vert \{u_2,u_3\} = u^0_2u^0_3u^1_2u_3^1$ -- see Figure \ref{fig:path_p4}.
Note that there is a unique conformal model $\phi^0$ of $(U,{\sim})$
such that $\phi^0 \Vert \{u_2,u_3\} = u^0_2u^1_3u^1_2u_3^0$.
Indeed, the placement and the orientation of the chord $\phi^0(u_1)$ in $\phi^0$ is uniquely determined and depends only on whether $u_1 \in \leftside(u_3)$ and on whether $u_3 \in \leftside(u_1)$ -- see Figure \ref{fig:path_p4}.
Similarly, the placement and the orientation of $\phi^0(u_4)$ is uniquely determined
and depends only on whether $u_4 \in \leftside(u_2)$ and on whether $u_2 \in \leftside(u_4)$.
Thus, $\phi^0$ is unique.
We prove the same way that there is a unique conformal model $\phi^1$ of $(U,{\sim})$ extending $u^0_2u^0_3u^1_2u_3^1$.
Finally, we note that $\phi^1$ is the reflection of $\phi^0$ -- see Figure~\ref{fig:path_p4} for an illustration. 
This completes the proof of the base of the induction.
\begin{figure}
\begin{tikzpicture}[xscale=0.80,yscale=0.80,>=latex]
\coordinate (center) at (0,0) {};
\coordinate (u2l) at ($(center)+(180:2cm)$) {};
\coordinate (u2r) at ($(center)+(0:2cm)$) {};
\coordinate (u3u) at ($(center)+(90:2cm)$) {};
\coordinate (u3b) at ($(center)+(270:2cm)$) {};
\coordinate (u1u) at ($(center)+(150:2cm)$) {};
\coordinate (u1b) at ($(center)+(210:2cm)$) {};
\coordinate (u4l) at ($(center)+(120:2cm)$) {};
\coordinate (u4r) at ($(center)+(60:2cm)$) {};

\coordinate (lu2l) at ($(center)+(180:2.3cm)$) {};
\coordinate (lu2r) at ($(center)+(0:2.3cm)$) {};
\coordinate (lu3u) at ($(center)+(90:2.3cm)$) {};
\coordinate (lu3b) at ($(center)+(270:2.3cm)$) {};
\coordinate (lu1u) at ($(center)+(150:2.3cm)$) {};
\coordinate (lu1b) at ($(center)+(210:2.3cm)$) {};
\coordinate (lu4l) at ($(center)+(120:2.3cm)$) {};
\coordinate (lu4r) at ($(center)+(60:2.3cm)$) {};

\draw (0,0) circle (2cm);
\draw[->,thick] (u2l)--(u2r);
\draw[->,thick] (u3b)--(u3u);
\draw[->,red] (u1b)--(u1u);
\draw[<-,red] (u4l)--(u4r);

\tikzstyle{every node}=[inner sep=1pt]
\begin{tiny}
\node at (lu1u) {$u^1_1$};
\node at (lu1b) {$u^0_1$};
\node at (lu2l) {$u^0_2$};
\node at (lu2r) {$u^1_2$};
\node at (lu3u) {$u^1_3$};
\node at (lu3b) {$u^0_3$};
\node at (lu4l) {$u^1_4$};
\node at (lu4r) {$u^0_4$};
\end{tiny}
\end{tikzpicture} 
\hspace{1cm}
\begin{tikzpicture}[xscale=0.80,yscale=0.80,>=latex,shorten >=-0.4pt,shorten <=-0.4pt]

\coordinate (u2l) at ($(center)+(180:2cm)$) {};
\coordinate (u2r) at ($(center)+(0:2cm)$) {};
\coordinate (u3b) at ($(center)+(90:2cm)$) {};
\coordinate (u3u) at ($(center)+(270:2cm)$) {};
\coordinate (u1b) at ($(center)+(30:2cm)$) {};
\coordinate (u1u) at ($(center)+(330:2cm)$) {};
\coordinate (u4l) at ($(center)+(120:2cm)$) {};
\coordinate (u4r) at ($(center)+(60:2cm)$) {};

\coordinate (lu2l) at ($(center)+(180:2.3cm)$) {};
\coordinate (lu2r) at ($(center)+(0:2.3cm)$) {};
\coordinate (lu3b) at ($(center)+(90:2.3cm)$) {};
\coordinate (lu3u) at ($(center)+(270:2.3cm)$) {};
\coordinate (lu1b) at ($(center)+(30:2.3cm)$) {};
\coordinate (lu1u) at ($(center)+(330:2.3cm)$) {};
\coordinate (lu4l) at ($(center)+(120:2.3cm)$) {};
\coordinate (lu4r) at ($(center)+(60:2.3cm)$) {};

\draw (0,0) circle (2cm);
\draw[->,thick] (u2l)--(u2r);
\draw[->,thick] (u3b)--(u3u);
\draw[->,red] (u1b)--(u1u);
\draw[<-,red] (u4l)--(u4r);

\tikzstyle{every node}=[inner sep=1pt]
\begin{tiny}
\node at (lu1u) {$u^1_1$};
\node at (lu1b) {$u^0_1$};
\node at (lu2l) {$u^0_2$};
\node at (lu2r) {$u^1_2$};
\node at (lu3u) {$u^1_3$};
\node at (lu3b) {$u^0_3$};
\node at (lu4l) {$u^1_4$};
\node at (lu4r) {$u^0_4$};
\end{tiny}
\end{tikzpicture} 

\caption{\label{fig:path_p4} The placements and the orientations of $\phi(u_1)$ and $\phi(u_4)$ for the case:
$u_1 \in \leftside(u_3)$, $u_3 \in \rightside(u_1)$, $u_4 \in \leftside(u_2)$ and $u_2 \in \leftside(u_4)$.
}

\end{figure}

Suppose $(U,{\sim})$ has at least $5$ vertices.
If $(U,{\sim})$ has no non-trivial splits, 
Theorem~\ref{thm:cicle_representation_no_split} asserts that $(U,{\sim})$ has two chord models, say $\psi$ and $\psi^R$, where $\psi^R$ is the reflection of~$\psi$.
Since $(U,{\sim})$ is prime, there is a unique orientation of the chords in $\psi$ and $\psi^R$, which leads to conformal models $\phi$ and $\phi^R$ of $(U,{\sim})$, respectively.
Since $\phi$ and $\phi^R$ are the only two conformal models of $(U,{\sim})$, $\phi^R$ must be the reflection of $\phi$.

Suppose $(U,{\sim})$ has a non-trivial split.
In this case the proof goes as follows.
We take a maximal split in $(U,{\sim})$ produced by the algorithm introduced above, and then, using a structure induced by this split,
we divide the graph $(U,{\sim})$ into so-called \emph{probes}.
A probe is a connected induced subgraph of $(U,{\sim})$ which, as we will prove, 
has a unique, up to reflection, conformal model (a probe can be seen as an \emph{almost prime} graph).
Finally, we show that there is a unique way to fit the models of the probes together to get a conformal model of $(U,{\sim})$.

\begin{definition}
A \emph{probe} in $(U,{\sim})$ is a quadruple $(y,x,X,\alpha(X))$, where $x,y \in U$, $X,\alpha(X) \subsetneq U$, that satisfies the following properties:
\begin{enumerate}
 \item \label{item:probe_P_subset} the sets $\{x\}$, $\{y\}$, $X$, $\alpha(X)$ are non-empty and pairwise disjoint,  
 and the set $P = \{x,y\} \cup X \cup \alpha(X)$ is a proper subset of $U$,
 \item \label{item:probe_yx_relation} $y \sim x$, $y \parallel X \cup \alpha(X)$,
 $x \sim  X$, $x \parallel \alpha(X)$, and the graph $(P,{\sim})$ is connected,
 \item \label{item:probe_neighborood_outside} for every $t \in U \setminus P$, 
 either $t \parallel (X \cup \alpha(X))$, or $t \sim X  \text{ and } t \parallel \alpha(X)$, 
 or $t \sim (X \cup \alpha(X))$.
\end{enumerate}
\end{definition}
See Figure~\ref{fig:probe} for an illustration.
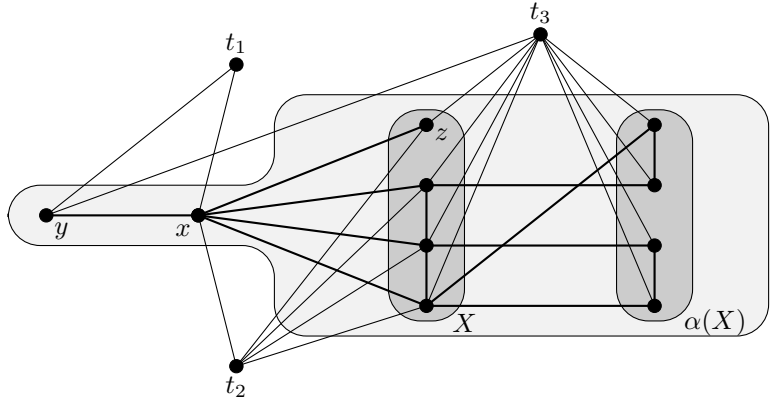
\begin{figure}[htp!]
\centering
\begin{tikzpicture}[xscale=1,yscale=0.4]
\coordinate (center) at (0,0) {};
\coordinate (x) at (0,0) {};
\coordinate (lx) at (-0.2,-0.5) {};

\coordinate (y) at (-2,0) {};
\coordinate (ly) at (-1.8,-0.5) {};

\coordinate (t1) at (0.5,5) {};
\coordinate (lt1) at (0.5,5.7) {};

\coordinate (t2) at (0.5,-5) {};
\coordinate (lt2) at (0.5,-5.7) {};

\coordinate (t3) at (4.5,6) {};
\coordinate (lt3) at (4.5,6.7) {};

\coordinate (x1) at (3,3) {};
\coordinate (lx1) at (3.2,2.7) {};
\coordinate (x2) at (3,1) {};
\coordinate (x3) at (3,-1) {};
\coordinate (x4) at (3,-3) {};
\coordinate (lX) at (3.5,-3.55) {};

\coordinate (ax1) at (6,3) {};
\coordinate (ax2) at (6,1) {};
\coordinate (ax3) at (6,-1) {};
\coordinate (ax4) at (6,-3) {};
\coordinate (lAX) at (6.8,-3.55) {};

\draw[rounded corners=12, fill=gray!10] (-2.5,1)--(1,1) -- (1,4) -- (7.5,4)--(7.5,-4)--(1,-4)--(1,-1)--(-2.5,-1)--cycle;
\draw[rounded corners=12, fill=gray!40] (2.5,-3.5)--(2.5,3.5) -- (3.5,3.5) -- (3.5,-3.5)--cycle;
\draw[rounded corners=12, fill=gray!40] (5.5,-3.5)--(5.5,3.5) -- (6.5,3.5) -- (6.5,-3.5)--cycle;

\tikzstyle{every node}=[circle,minimum size=5pt,inner sep=0pt,draw,fill]
\node at (center) {};
\node at (y) {};
\node at (x) {};

\node at (x1) {};
\node at (x2) {};
\node at (x3) {};
\node at (x4) {};

\node at (ax1) {};
\node at (ax2) {};
\node at (ax3) {};
\node at (ax4) {};

\node at (t1) {};
\node at (t2) {};
\node at (t3) {};

\tikzstyle{every node}=[inner sep=1pt]

\path (y) edge[thick] (x);

\path (x) edge[thick] (x1);
\path (x) edge[thick] (x2);
\path (x) edge[thick] (x3);
\path (x) edge[thick] (x4);

\path (x2) edge[thick] (ax2);
\path (x3) edge[thick] (ax3);
\path (x4) edge[thick] (ax4);
\path (x4) edge[thick] (ax1);

\path (x2) edge[thick] (x3);
\path (x3) edge[thick] (x4);

\path (ax1) edge[thick] (ax2);
\path (ax3) edge[thick] (ax4);

\path (t1) edge (x);
\path (t1) edge (y);

\path (t2) edge (x);
\path (t2) edge (x1);
\path (t2) edge (x2);
\path (t2) edge (x3);
\path (t2) edge (x4);

\path (t3) edge (y);
\path (t3) edge (x1);
\path (t3) edge (x2);
\path (t3) edge (x3);
\path (t3) edge (x4);
\path (t3) edge (ax1);
\path (t3) edge (ax2);
\path (t3) edge (ax3);
\path (t3) edge (ax4);

\begin{footnotesize}
\tikzstyle{every node}=[inner sep=2pt]
\node at (lX) {$X$};
\node at (lAX) {$\alpha(X)$};
\node at (lx) {$x$};
\node at (lx1) {$z$};
\node at (ly) {$y$};

\node at (lt1) {$t_1$};
\node at (lt2) {$t_2$};
\node at (lt3) {$t_3$};
\end{footnotesize}
\end{tikzpicture}
\caption{\label{fig:probe} A probe $(y,x,X,\alpha(X))$ in $(U,{\sim})$ and the vertices $t_1,t_2,t_3$ from $U \setminus P$.}
\end{figure}

\begin{claim}
\label{claim:probes_have_unique_model}
Let $(y,x,X,\alpha(X))$ be a probe in $(U,{\sim})$ and let $P = \{x,y\} \cup X \cup \alpha(X)$.
Then, the graph $(P,{\sim})$ has a unique, up to reflection, conformal model.
\end{claim}
\begin{proof}
Let $Z = \{z \in X: \text{$z \parallel (P \setminus \{x,z\})$}\}$.
Note that $|Z| \leq 1$ as otherwise, by Property~\eqref{item:probe_neighborood_outside} of the probe~$P$, $Z$~would be a non-trivial module in $(U,{\sim})$ -- see Figure \ref{fig:probe}.
We claim that:
\begin{itemize}
\item If $|Z| = 1$, then $\{y\} \cup Z$ is the only non-trivial module in $(P,{\sim})$.
\item If $Z = \emptyset$, then $(P,{\sim})$ has no non-trivial modules.
\end{itemize}

Let $M$ be a non-trivial module in $(P,{\sim})$.
We consider four cases depending on the intersection of $M$ with the set $\{y,x\}$.

Suppose $M \cap \{y,x\} = \emptyset$. 
Since $x \notin M$ and since $x \sim X$ and $x \parallel \alpha(X)$, 
we must have either $M \subseteq X$ or $M \subseteq \alpha(X)$.
Then, by Property~\eqref{item:probe_neighborood_outside} of $P$, 
every $t \in U \setminus P$ satisfies either $t \sim M$ or $t \parallel M$.
So, $M$ is a non-trivial module in $(U,{\sim})$, which can not be the case as
$(U,{\sim})$ is prime.

Suppose $M \cap \{y,x\}  = \{y,x\}$.
Since $X \sim x$ and $X \parallel y$, we must have $X \subseteq M$.
Since $(P,{\sim})$ is connected, we need to have $\alpha(X) \subset M$
as otherwise we would find a vertex $u \in P \setminus M$ such that $u$
is adjacent to a vertex in $M$ and non-adjacent to the vertex $y$ from $M$.
So, $M =P$, which contradicts that $M$ is a non-trivial module in $(P,{\sim})$.

Suppose $M \cap \{y,x\} = \{x\}$.
Since $y \sim x$ and $ y \parallel X \cup \alpha(X)$, 
we must have $M \cap (X \cup \alpha(X))= \emptyset$.
Hence, $M$ is trivial in $(P,{\sim})$, a contradiction.

Suppose $M \cap \{y,x\} = \{y\}$.
Note that $M \cap \alpha(X) = \emptyset$.
Otherwise, the vertex $x$ from outside $M$ is adjacent to the vertex $y$ in $M$ and
non-adjacent to a vertex from $M \cap \alpha(X)$, which can not be the case.
Let $M_X = M \cap X$.
If $M_X = \emptyset$, then $M = \{y\}$ and $M$ is trivial.
So, we must have $M_X \neq \emptyset$.
Note that for every vertex $u \in (X \cup \alpha(X)) \setminus M_X$ we have $u \parallel M_x$.
Otherwise, the vertex $u$ from outside $M$ would have a neighbour in $M$ and the non-neighbour $y$ in $M$.
If $|M_X| \geq 2$, then by Property~\eqref{item:probe_neighborood_outside} of~$P$, 
$M_X$ would be a non-trivial module in $(U,{\sim})$, which can not be the case.
So, $M$ is a non-trivial module of $(P,{\sim})$ if and only if $|M_X| = 1$, i.e., when $M_X = \{z\}$ for some $z \in X$.
In this case, $z$ is adjacent only to the vertex $x$ in $(P,{\sim})$, which shows $Z = \{z\}$.
So, we have $M = \{y,z\}$, which completes the proof of our claim.

Next we show that $(P,{\sim})$ has a unique, up to reflection, conformal model.

Suppose $Z = \emptyset$.
Then, the graph $(P,{\sim})$ contains no non-trivial modules.
One can easily check that $(P,{\sim})$ and $(P,{\parallel})$ are connected and hence 
$(P,{\sim})$ is prime.
Since $P$ has strictly fewer vertices than $(U,{\sim})$, the inductive hypothesis implies that $(P,{\sim})$ has a unique, up to reflection, conformal model.

Suppose $Z = \{z\}$. 
Then $\{y,z\}$ is the only non-trivial module in $(P,{\sim})$.
Since $(P \setminus \{z\},{\sim})$ and $(P \setminus \{z\},{\parallel})$ are connected and $(P \setminus \{z\}, {\sim})$ contains no non-trivial modules,
$(P \setminus \{z\}, {\sim})$ is prime.
By the inductive hypothesis, $(P \setminus \{z\},{\sim})$ has exactly two conformal models, $\phi$ and $\phi^R$, where $\phi^R$ is the reflection of~$\phi$.
Note that the vertex $x$ is the articulation point in the graph $(P \setminus \{z\},{\sim})$.
Suppose that $(P \setminus \{z,x\},{\sim})$ has exactly $k$ connected components, 
say $D_1,\ldots,D_k$, for some $k \geq 2$.
Note that $D_i  = \{y\}$ for some $i \in [k]$.
By Theorem~\ref{thm:trivial_split_representations}, $\phi \equiv x^0\tau_{i_1}\ldots \tau_{i_{k}}x^1 \tau'_{i_k}\ldots\tau'_{i_1}$, where $i_1,\ldots,i_k$ is a permutation of~$[k]$ and $x^0 \tau_{i_j} x^1 \tau'_{i_j}$ is a conformal model of $(\{x\} \cup D_{i_j},{\sim})$ for $j \in [k]$.
We show that there is a unique extension of $\phi$ by the oriented chord $\phi(z)$ 
such that the extended $\phi$ is conformal for $(P,{\sim})$.
Clearly, the extended $\phi$ must be of the form:
$$\phi \equiv x^0\tau_{i_1}\ldots \tau_{i_l} z' \tau_{i_{l+1}} \ldots \tau_{i_{k}}x^1 \tau'_{i_k} \ldots \tau'_{i_{l+1}} z'' \tau'_{i_{l}} \ldots \tau'_{i_1} \text{ for some } l \in \{0,\ldots,k\},$$
where  $z'$ and $z''$ are such that $\{z',z''\} = \{z^{0},z^{1}\}.$
Indeed, for every $i \in [k]$ we pick a vertex $a_i$ in $D_i$ such that $x \sim a_i$.
Then, the chord $\phi(z)$ must be on the left (right) side of $\phi(a_i)$ if $z \in \leftside(a_i)$ ($z \in \rightside(a_i)$, respectively).
Hence, the placement of the chord $\phi(z)$ in $\phi$ is uniquely determined.
The orientation of $\phi(z)$ can be set based on whether $y \in \leftside(z)$ or $y \in \rightside(z)$.
\end{proof}

Suppose $(U,{\sim})$ has a non-trivial split.
We use the algorithm introduced above to compute a maximal split $(A,B)$ in $(U,{\sim})$. 
\begin{itemize}
 \item If $(A,B)$ is non-trivial, we assume $C_1,\ldots,C_k$ and $\alpha(C_1),\ldots,\alpha(C_k)$ are such as defined in Subsection~\ref{subsec:structure_non_trivial_split}.
 In this case $C_i \neq \emptyset$ for every $i \in [k]$ and $C_1 \cup \ldots \cup C_k = A \cup B$.
 \item If $(A,B)$ is trivial, we assume $A=\{a\}$, $\alpha(A)=\emptyset$, and $C_1,\ldots,C_k$ and $\alpha(C_1),\ldots,\alpha(C_k)$ are such as defined in Subsection~\ref{subsec:structure_trivial_split}.
 In this case $C_i \neq \emptyset$ for every $i \in [k]$ and $(A,B) = (\{a\}, C_1 \cup \ldots \cup C_k)$.
\end{itemize}
See Figure~\ref{fig:probes_in_splits} for an illustration.
\input ./figures/prime_properties/probes_in_splits.tex

\noindent Next, we partition the set $[k]$ into two subsets, $I_1$ and $I_2$, as follows:
\begin{itemize}
 \item if $|C_i \cup \alpha(C_i)| = 1$, then $i \in I_1$,
 \item if $|C_i \cup \alpha(C_i)| \geq 2$, then $i \in I_2$.
\end{itemize}
Note that $|I_1| \leq 1$ as otherwise $\bigcup_{i \in I_1} C_i$ would be a non-trivial module in $(U,{\sim})$.
Without loss of generality we assume $C_1,\ldots,C_k$ are enumerated such that 
$I_1 = \{k\}$ if $I_1 \neq \emptyset$.
Observe that for every  $i \in I_2$ we have $\alpha(C_i) \neq \emptyset$ as otherwise
$C_i$ would be a non-trivial module in $(U,{\sim})$ and, since $(U,{\sim})$ is connected, some vertex in $C_i$ is adjacent to some vertex in $\alpha(C_i)$.
Hence, for every $i \in [k]$ we may pick vertices $a_i,b_i \in C_i \cup \alpha(C_i)$ such that:
\begin{itemize}
\item if $i \in I_2$, then $a_i \in C_i$, $b_i \in \alpha(C_i)$, and $a_i \sim b_i$,
\item if $i \in I_1$, then $a_i = b_i$, where $a_i$ is the only vertex in $C_i$,
\end{itemize}
see Figure~\ref{fig:probes_in_splits}.
We split the proof into two cases, depending on whether the following condition is satisfied:
\begin{equation}
\label{eq:probe_condition}
\begin{array}{c}
\text{For every $i \in I_2$ there exist $x,y \in U \setminus (C_i \cup \alpha(C_i))$ such that} \\
\text{$(y,x,C_i,\alpha(C_i))$ is a probe in $(U,{\sim})$}.
\end{array}
\tag{*}
\end{equation}
We claim that Condition~\eqref{eq:probe_condition} is not satisfied only when $(A,B)$ is a trivial split,
$k=2$, and $|C_2 \cup \alpha(C_2)|=1$ -- this case is shown in Figure \ref{fig:probes_in_splits} to the right.
Suppose $k\geq 3$ and suppose $i \in I_2$.
If $(A,B)$ is non-trivial, 
the set $C_i \cup \alpha(C_i)$ can be extended to a probe 
by the vertices $a_j,b_j$, where $j$ is any index in $I_2$ different from~$i$.
If $(A,B)$ is trivial, the set $C_i \cup \alpha(C_i)$ can be extended to a probe
by the vertices $a, a_j$, where $j$ is any index in $[k]$ different from $i$.
Suppose $k=2$.
Suppose $(A,B)$ is non-trivial.
Note that $|C_i \cup \alpha(C_i)|\geq 3$ for every $i \in [2]$.
Otherwise, the only vertex $a_i$ in $C_i$ is adjacent to the only vertex $b_i$ in $\alpha(C_i)$, 
and then the split $(A,B)$ is not maximal.
Hence, for every $i \in [2]$ the set $C_i \cup \alpha(C_i)$ can be extended to a probe by the vertices $a_j,b_j$, 
where $j \in [2]$ is such that $j \neq i$.
If $(A,B)$ is trivial and $|C_2 \cup \alpha(C_2)| \geq 2$, then the set $C_i \cup \alpha(C_i)$ can be extended to a probe by the vertices $a,a_j$, 
where $j \in [2]$ is such that $i \neq j$.
So, Condition~\eqref{eq:probe_condition} might be not satisfied only when 
$(A,B) = (\{a\}, C_1 \cup C_2)$ is trivial,
$k=2$, $|C_2| = 1$, and $|\alpha(C_2)|=0$.
Indeed, in this case~\eqref{eq:probe_condition} does not hold.

In the remaining part of the proof we consider two cases, depending on whether Condition~\eqref{eq:probe_condition} is satisfied. 

Suppose Condition~\eqref{eq:probe_condition} is satisfied.
Let 
$$ R_i = \left\{
\begin{array}{lll}
 \{a_1, b_1, \ldots, a_i,b_i\}& \text{if} & (A,B) \text{ is non-trivial},\\
 \{a, a_1,b_1, \ldots, a_i,b_i\} & \text{if} & (A,B) \text{ is trivial},
\end{array}
\right.
$$
let
$$
S = \left\{
\begin{array}{lll}
 \{a_1, a_2\}& \text{if} & (A,B) \text{ is non-trivial},\\
 \{a,a_1\} & \text{if} & (A,B) \text{ is trivial},
\end{array}
\right.
$$
and let $R = R_k$.
Eventually, let
$$
\begin{array}{lll}
 \phi^{0}_S \equiv a_1^0a^0_2a_{1}^1a_2^1 \quad \text{and} \quad \phi^{1}_S \equiv a_1^0a^1_2a_{1}^1a_2^0 &\text{if} & (A,B) \text{ is non-trivial},\\
 \phi^{0}_S \equiv a^0a^0_1a^1a_1^1 \quad \text{and} \quad \phi^{1}_S \equiv a^0a^1_1a^1a^0_1 & \text{if} & (A,B) \text{ is trivial}.\\
\end{array}
$$
In both cases, $\phi^{0}_S$ is the reflection of $\phi^{1}_S$ and any
conformal model $\phi$ of $(U,{\sim})$ extends either $\phi^{0}_S$ or $\phi^{1}_S$.
We claim that:
\begin{equation}
\label{eq:unique_models_of_skeletons}
\begin{array}{c}
\text{For every $m \in \{0,1\}$ there is a unique conformal model $\phi^{m}_R$ of $(R,{\sim})$} \\ 
\text{such that $\phi^m_R \Vert S^* \equiv \phi^m_S$.}
\end{array}
\end{equation}
\begin{equation}
\begin{array}{c}
\label{eq:unique_extension_of__skeletons}
\text{For every $m \in \{0,1\}$ there is a unique conformal model $\phi^m$ of $(U,{\sim})$} \\
\text{such that $\phi^m \Vert R^* \equiv \phi^m_R$.}
\end{array}
\end{equation}
Then, $\phi^{1}_R$ must be the reflection of $\phi^{0}_R$, 
and $\phi^1$ must be the reflection of $\phi^0$.
This will complete the lemma in the case when Condition~\eqref{eq:probe_condition} is satisfied.

First we prove \eqref{eq:unique_models_of_skeletons}.
Let $m=0$.
Suppose $(A,B)$ is non-trivial.
We claim that for every $i \in [2,k]$ there is a unique conformal model $\phi$ of $(R_i,{\sim})$ extending $\phi^{0}_S$.
To prove our claim for $i=2$ we need to argue that
there is a unique extension $\phi$ of $\phi^{0}_S$ by the chords $\phi(b_1)$ and $\phi(b_2)$.
However, this can be shown using the same arguments as in the proof that $P_4$ has a unique, up to reflection, conformal model.
Suppose $\phi$ is a unique conformal model of $(R_{i-1}, {\sim})$ extending $\phi^0_S$.
We show that there is a unique conformal extension of $\phi$ by the chords $\phi(a_i)$ and $\phi(b_i)$.
Note that $(\{a_1,\ldots,a_{i-1}\}, {\sim})$ is a clique in $(R_{i-1},{\sim})$, and
hence the chords $\{\phi(a_1),\ldots,\phi(a_{i-1})\}$ are pairwise intersecting.
There are $2(i-1)$ possible placements for the oriented chord $\phi(a_i)$.
Any such placement of $\phi(a)$ determines the pair $(X,Y)$, where $X,Y$ is a partition of $\{b_1,\ldots,b_{i-1}\}$ such that the chords from $\phi(X)$ are on the left side of $\phi(a_i)$ and the chords from $\phi(Y)$ are on the right side of $\phi(a_i)$ -- see Figure \ref{fig:R_models}.
Note that the pairs $(X,Y)$ corresponding to different placements of $\phi(a_i)$ are different.
Hence, to keep $\phi$ conformal, only one choice for $\phi(a_i)$ coincides with the pair
$$(\{b_1,\ldots,b_{i-1}\} \cap \leftside(a_i), \{b_1,\ldots,b_{i-1}\} \cap \rightside(a_i)).$$
In particular, it means that the placement of $\phi(a_i)$ is uniquely determined.
One can easily observe that the extension of $\phi$ by the chord $\phi(b_i)$ is also uniquely determined.

\begin{figure}
\begin{tikzpicture}[xscale=1.0,yscale=1.0,>=latex]
\coordinate (center) at (0,0);
\draw (0,0) circle (2cm);

\coordinate (a10) at ($(center)+(180:2cm)$) {};
\coordinate (a11) at ($(center)+(0:2cm)$) {};
\coordinate (b10) at ($(center)+(165:2cm)$) {};
\coordinate (b11) at ($(center)+(195:2cm)$) {};

\coordinate (la10) at ($(center)+(180:2.3cm)$) {};
\coordinate (la11) at ($(center)+(0:2.3cm)$) {};
\coordinate (lb10) at ($(center)+(165:2.3cm)$) {};
\coordinate (lb11) at ($(center)+(195:2.3cm)$) {};

\coordinate (a20) at ($(center)+(90:2cm)$) {};
\coordinate (a21) at ($(center)+(270:2cm)$) {};
\coordinate (b20) at ($(center)+(75:2cm)$) {};
\coordinate (b21) at ($(center)+(105:2cm)$) {};

\coordinate (la20) at ($(center)+(90:2.3cm)$) {};
\coordinate (la21) at ($(center)+(270:2.3cm)$) {};
\coordinate (lb20) at ($(center)+(75:2.3cm)$) {};
\coordinate (lb21) at ($(center)+(105:2.3cm)$) {};

\coordinate (a30) at ($(center)+(315:2cm)$) {};
\coordinate (a31) at ($(center)+(135:2cm)$) {};
\coordinate (b30) at ($(center)+(300:2cm)$) {};
\coordinate (b31) at ($(center)+(330:2cm)$) {};

\coordinate (la30) at ($(center)+(315:2.3cm)$) {};
\coordinate (la31) at ($(center)+(135:2.3cm)$) {};
\coordinate (lb30) at ($(center)+(300:2.3cm)$) {};
\coordinate (lb31) at ($(center)+(330:2.3cm)$) {};

\coordinate (ra40) at ($(center)+(45:2cm)$) {};
\coordinate (ra41) at ($(center)+(225:2cm)$) {};

\coordinate (lra40) at ($(center)+(45:2.3cm)$) {};
\coordinate (lra41) at ($(center)+(225:2.3cm)$) {};

\draw[very thick,red,<-] ([shift=(65:2cm)]0,0) arc (65:112:2cm);
\coordinate (pi2) at ($(center)+(70:1.75cm)$) {};
\draw[very thick,red,<-] ([shift=(255:2cm)]0,0) arc (255:282:2cm);
\coordinate (pi'2) at ($(center)+(260:1.75cm)$) {};

\draw[very thick,red,<-] ([shift=(155:2cm)]0,0) arc (155:202:2cm);
\coordinate (pi1) at ($(center)+(160:1.75cm)$) {};
\coordinate (pi'1) at ($(center)+(-10:1.75cm)$) {};
\draw[very thick,red,<-] ([shift=(-15:2cm)]0,0) arc (-15:12:2cm);

\draw[very thick,red,<-] ([shift=(290:2cm)]0,0) arc (290:337:2cm);
\coordinate (pi3) at ($(center)+(295:1.75cm)$) {};
\draw[very thick,red,<-] ([shift=(120:2cm)]0,0) arc (120:147:2cm);
\coordinate (pi'3) at ($(center)+(125:1.75cm)$) {};

\draw[thick,->] (a10)--(a11);
\draw[thick,->] (b10)--(b11);
\draw[thick,->] (a20)--(a21);
\draw[thick,->] (b20)--(b21);
\draw[thick,->] (a30)--(a31);
\draw[thick,->] (b30)--(b31);
\draw[thick,red,->] (ra40)--(ra41);

\tikzstyle{every node}=[inner sep=1pt]
\begin{tiny}
\node at (la10) {$a_1^0$};
\node at (la11) {$a_1^1$};
\node at (lb10) {$b_1^0$};
\node at (lb11) {$b_1^1$};
\node at (la20) {$a_2^0$};
\node at (la21) {$a_2^1$};
\node at (lb20) {$b_2^0$};
\node at (lb21) {$b_2^1$};
\node at (la30) {$a_3^0$};
\node at (la31) {$a_3^1$};
\node at (lb30) {$b_3^0$};
\node at (lb31) {$b_3^1$};
\node at (lra40) {$a_4^0$};
\node at (lra41) {$a_4^1$};
\node at (pi1) {$\pi_1$};
\node at (pi'1) {$\pi'_1$};
\node at (pi2) {$\pi_2$};
\node at (pi'2) {$\pi'_2$};
\node at (pi3) {$\pi_3$};
\node at (pi'3) {$\pi'_3$};
\end{tiny}
\end{tikzpicture} 
\hspace{0.5cm}
\begin{tikzpicture}[xscale=1.0,yscale=1.0,>=latex]

\draw (0,0) circle (2cm);

\coordinate (a0) at ($(center)+(180:2cm)$) {};
\coordinate (a1) at ($(center)+(0:2cm)$) {};

\coordinate (la0) at ($(center)+(180:2.3cm)$) {};
\coordinate (la1) at ($(center)+(0:2.3cm)$) {};

\coordinate (a10) at ($(center)+(90:2cm)$) {};
\coordinate (a11) at ($(center)+(270:2cm)$) {};
\coordinate (b10) at ($(center)+(105:2cm)$) {};
\coordinate (b11) at ($(center)+(75:2cm)$) {};

\coordinate (la10) at ($(center)+(90:2.3cm)$) {};
\coordinate (la11) at ($(center)+(270:2.3cm)$) {};
\coordinate (lb10) at ($(center)+(105:2.3cm)$) {};
\coordinate (lb11) at ($(center)+(75:2.3cm)$) {};

\draw[very thick,red,<-] ([shift=(65:2cm)]0,0) arc (65:112:2cm);
\coordinate (pi1) at ($(center)+(70:1.75cm)$) {};
\draw[very thick,red,<-] ([shift=(255:2cm)]0,0) arc (255:282:2cm);
\coordinate (pi'1) at ($(center)+(260:1.75cm)$) {};

\coordinate (a20) at ($(center)+(220:2cm)$) {};
\coordinate (a21) at ($(center)+(140:2cm)$) {};
\coordinate (b20) at ($(center)+(205:2cm)$) {};
\coordinate (b21) at ($(center)+(235:2cm)$) {};

\coordinate (la20) at ($(center)+(220:2.3cm)$) {};
\coordinate (la21) at ($(center)+(140:2.3cm)$) {};
\coordinate (lb20) at ($(center)+(205:2.3cm)$) {};
\coordinate (lb21) at ($(center)+(235:2.3cm)$) {};

\draw[very thick,red,<-] ([shift=(125:2cm)]0,0) arc (125:152:2cm);
\coordinate (pi'2) at ($(center)+(135:1.75cm)$) {};
\draw[very thick,red,<-] ([shift=(195:2cm)]0,0) arc (195:242:2cm);
\coordinate (pi2) at ($(center)+(237:1.8cm)$) {};

\coordinate (a31) at ($(center)+(40:2cm)$) {};
\coordinate (a30) at ($(center)+(320:2cm)$) {};
\coordinate (b30) at ($(center)+(305:2cm)$) {};
\coordinate (b31) at ($(center)+(335:2cm)$) {};

\coordinate (la31) at ($(center)+(40:2.3cm)$) {};
\coordinate (la30) at ($(center)+(320:2.3cm)$) {};
\coordinate (lb30) at ($(center)+(305:2.3cm)$) {};
\coordinate (lb31) at ($(center)+(335:2.3cm)$) {};

\draw[very thick,red,<-] ([shift=(25:2cm)]0,0) arc (25:52:2cm);
\coordinate (pi'3) at ($(center)+(45:1.75cm)$) {};
\draw[very thick,red,<-] ([shift=(295:2cm)]0,0) arc (295:342:2cm);
\coordinate (pi3) at ($(center)+(300:1.75cm)$) {};

\coordinate (ra41) at ($(center)+(119:2cm)$) {};
\coordinate (ra40) at ($(center)+(250:2cm)$) {};

\coordinate (lra41) at ($(center)+(119:2.3cm)$) {};
\coordinate (lra40) at ($(center)+(250:2.3cm)$) {};

\draw[thick,->] (a0)--(a1);
\draw[thick,->] (a10)--(a11);
\draw[thick,->] (b10)--(b11);
\draw[thick,->] (a20)--(a21);
\draw[thick,->] (b20)--(b21);
\draw[thick,->] (a30)--(a31);
\draw[thick,->] (b30)--(b31);
\draw[thick,red,->] (ra40)--(ra41);

\tikzstyle{every node}=[inner sep=1pt]
\begin{tiny}
\node at (la0) {$a^0$};
\node at (la1) {$a^1$};
\node at (la10) {$a_1^0$};
\node at (la11) {$a_1^1$};
\node at (lb10) {$b_1^0$};
\node at (lb11) {$b_1^1$};
\node at (la20) {$a_2^0$};
\node at (la21) {$a_2^1$};
\node at (lb20) {$b_2^0$};
\node at (lb21) {$b_2^1$};
\node at (la30) {$a_3^0$};
\node at (la31) {$a_3^1$};
\node at (lb30) {$b_3^0$};
\node at (lb31) {$b_3^1$};
\node at (lra40) {$a_4^0$};
\node at (lra41) {$a_4^1$};
\node at (pi1) {$\pi_1$};
\node at (pi'1) {$\pi'_1$};
\node at (pi2) {$\pi_2$};
\node at (pi'2) {$\pi'_2$};
\node at (pi3) {$\pi_3$};
\node at (pi'3) {$\pi'_3$};
\end{tiny}
\end{tikzpicture} 

\caption{\label{fig:R_models}
Extending $\phi$ by the chord $\phi(a_4)$.
To the left: $(A,B)$ is non-trivial, $A \cup B = C_1 \cup \ldots \cup C_4$,
$I_2 = \{1,2,3\}$, and $I_1 = \{4\}$.
The chord $\phi(a_4)$ induces a partition $(X,Y) = (\{b_3\}, \{b_1,b_2\})$.
The only conformal model $\phi^0$ of $(U,{\sim})$ extending $\phi^0_R$ is $\pi_1\pi'_3\pi_2a^0_4\pi'_1\pi_3\pi'_2a_4^1$.
To the right: $(A,B)$ is trivial, $(A,B) = (\{a\},C_1 \cup \ldots \cup C_4)$,
$I_2 = \{1,2,3\}$, and $I_1 = \{4\}$.
The chord $\phi(a_4)$ induces a partition $(X,Y) = (\{b_2\}, \{b_1,b_3\})$.
The only conformal model $\phi^0$ of $(U,{\sim})$ extending $\phi^0_R$ is $a^0 \pi'_2 a^1_4 \pi_1 \pi'_3 a^1 \pi_3 \pi'_1 a^0_4 \pi_2$.
}

\end{figure}
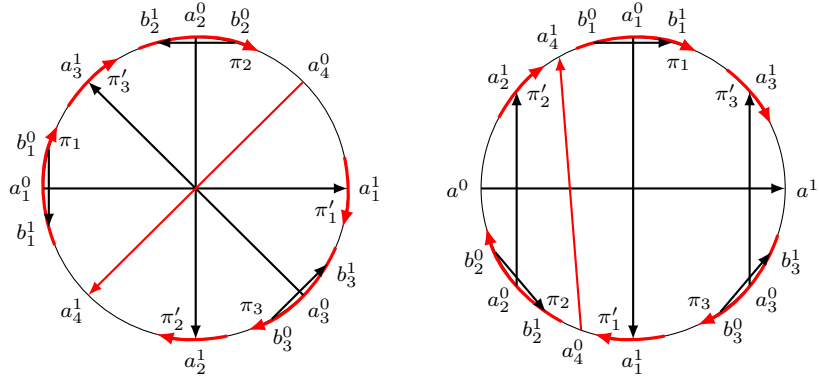

The case when the split $(A,B)$ is trivial is handled similarly.
This time, however, we need to place $\phi(a_i)$ so that it intersects $\phi(a)$ and does not intersect $\phi(a_1),\ldots,\phi(a_{i-1})$.
Again, with every possible placement of $\phi(a_i)$ we associate
the pair $(X,Y)$, defined in the same way as in the previous case, 
and we note that for different placements of $\phi(a_i)$ the pairs $(X,Y)$ are distinct -- see Figure~\ref{fig:R_models} for an illustration.

Next, we prove \eqref{eq:unique_extension_of__skeletons}.
Let $i \in I_2$, let $P_i = \{y,x\} \cup C_i \cup \alpha(C_i)$ be a probe in $(U,{\sim})$ for some $x,y \in U$, and let $\phi^{0}_i$ be the unique conformal model of $(P_i,{\sim})$ 
such that $\phi^0_i \Vert \{a_i,b_i\}^* \equiv \phi^0_R \Vert \{a_i,b_i\}^*$.
Assume that $\phi^{0}_i \Vert (\{x\} \cup C_i \cup \alpha(C_i))^* \equiv x' \pi_i x'' \pi'_i$, 
where $\{x',x''\} = \{x^{0},x^{1}\}$ and $\pi_i, \pi'_i$ are chosen such that both the letters $b^0_i,b^1_i$ are contained in $\pi_i$.
Eventually, assume that for every $i \in I_2$ the letters $a'_i,a''_i,b'_i,b''_i$
are chosen such that $b'_ia'_ib''_i$ and $a''_i$ are contiguous subwords of $\phi^0_R$, 
where $\{a'_i,a''_i\} = \{a^0_i,a^1_i\}$ and $\{b'_i,b''_i\} = \{b^0_i,b^1_i\}$.
Now, having in mind Theorems~\ref{thm:non_trivial_split_representations} and~\ref{thm:trivial_split_representations}, 
we conclude there is a unique conformal model $\phi^0$ of $(U,{\sim})$ extending $\phi^0_R$ and $\phi^0_i$ for every $i \in I_2$:
$\phi^0$ is obtained from $\phi^0_R$ by substituting $b'_ia'_ib''_i$ by $\pi_i$ and $a''_i$ by $\pi'_i$ for every $i \in I_2$ -- see Figure \ref{fig:R_models} for an illustration.

This completes the proof of the lemma for the case when Condition~\eqref{eq:probe_condition} is satisfied.

Suppose now that Condition \eqref{eq:probe_condition} is not satisfied. 
So, we have $k=2$, the split $(A,B) = (\{a\}, C_1 \cup C_2)$ is trivial, $C_2 = \{a_2\}$, and $\alpha(C_2)= \emptyset$.
If $|C_1 \cup \alpha(C_1)| =2$, then $C_1 = \{a_1\}$, $\alpha(C_1) = \{b_1\}$, $U = \{a_1,b_1,a,a_2\}$, which contradicts our assumption that $|U| \geq 5$.
So, in the remaining we assume $|C_1 \cup \alpha(C_1)| \geq 3$. 
We consider two cases depending on whether the graph $(\{a\} \cup C_1 \cup \alpha(C_1),{\sim})$ is prime.

Suppose $(\{a\} \cup C_1 \cup \alpha(C_1),{\sim})$ is prime. 
By the inductive hypothesis, $(\{a\} \cup C_1 \cup \alpha(C_1),{\sim})$ has two conformal models, 
$\phi$ and its reflection $\phi^R$.
Now, since $a_2 \parallel C_1 \cup \alpha(C_1)$, $a \sim a_2$, $a \sim C_1$,
we easily deduce that $\phi$ and $\phi^R$ can be uniquely extended by the chord of $a_2$ to
the conformal models of $(U,{\sim})$, one of each being the reflection of the other.

Suppose $(\{a\} \cup C_1 \cup \alpha(C_1), {\sim})$ has a non-trivial module~$M$ -- see Figure \ref{fig:probe_splitting}.
Observe that $a \in M$.
Otherwise, we would have either $M \subseteq C_1$ or $M \subseteq \alpha(C_1)$,
and $M$ would also be a non-trivial module of $(U,{\sim})$, which can not be the case.
For the remaining part of the proof, let $M_1 = M \cap C_1$ and 
$M_2 = C_1 \setminus M$.
We claim that $M_1 \neq \emptyset$ and $M_2 \neq \emptyset$.
Suppose that $M_1 = \emptyset$.
Then, $M \cap \alpha(C_1) \neq \emptyset$ as $M$ is a non-trivial module in $(\{a\} \cup C_1 \cup \alpha(C_1),{\sim})$.
Since $M_1 = C_1 \cap M = \emptyset$ and $C_1 \sim a \in M$, we deduce that $C_1 \sim (M \cap \alpha(C_1))$.
Furthermore, since $\alpha(C_1) \parallel a$, we have $(\alpha(C_1) \setminus M) \parallel (M \cap \alpha(C_1))$.
Note that the trivial split $(\{a\},C_1 \cup \{a_2\})$ returned by the algorithm had to arise as a result
of extending a non-trivial split $(\{a\},C_1)$.
However, at this point the algorithm could extend $(\{a\},C_1)$ 
into a trivial split $(\{a\},C_1 \cup \{a_2\})$ or
into a non-trivial split $(\{a\} \cup (M \cap \alpha(C_1)), C_1)$.
The restriction made on the algorithm choices asserts that the second option would be taken, which contradicts that the algorithm returns $(\{a\},C_1 \cup \{a_2\})$.
This proves $M_1 \neq \emptyset$.
Now, we prove $M_2 \neq \emptyset$.
Assuming otherwise, $(\{a\} \cup C_1) \subseteq M$ and $a \parallel \alpha(C_1)$ yields $\alpha(C_1) \subseteq M$ by connectivity of $(U,{\sim})$.
Then, $M$ is trivial in $(\{a\} \cup C_1 \cup \alpha(C_1), {\sim})$, which is not the case.
This shows $M_2 \neq \emptyset$.

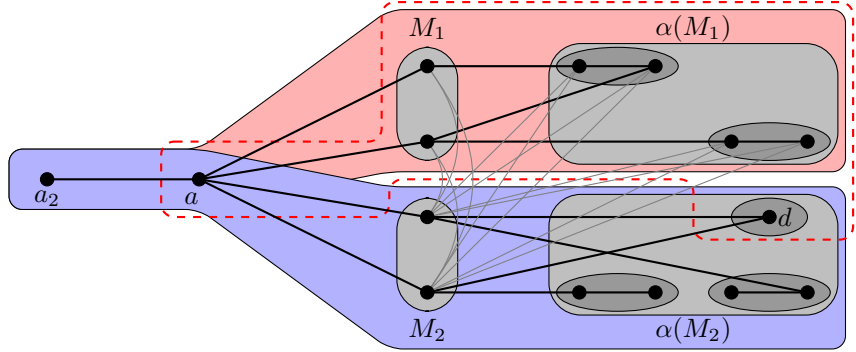
\begin{figure}[htp!]
\centering
\begin{tikzpicture}[xscale=1,yscale=0.5]
\coordinate (a) at (0,0) {};
\coordinate (la) at (-0.1,-0.5) {};

\coordinate (a2) at (-2,0) {};
\coordinate (la2) at (-2,-0.5) {};

\coordinate (M1) at (3,4) {};
\coordinate (x1) at (3,3) {};
\coordinate (x2) at (3,1) {};
\coordinate (x3) at (3,-1) {};
\coordinate (x4) at (3,-3) {};
\coordinate (M2) at (3,-4) {};

\coordinate (AM1) at (6.5,4) {};
\coordinate (AM2) at (6.5,-4) {};

\coordinate (d11) at (5,3.0) {};
\coordinate (d12) at (6,3.0) {};

\coordinate (d21) at (7,1) {};
\coordinate (d22) at (8,1) {};

\coordinate (d31) at (7.5,-1) {};
\coordinate (ld31) at (7.7,-1) {};

\coordinate (d41) at (5,-3) {};
\coordinate (d42) at (6,-3) {};

\coordinate (d51) at (7,-3) {};
\coordinate (d52) at (8,-3) {};

\begin{scope}[fill opacity=0.5]
\draw[rounded corners=5, fill=red!30] (-2.5,0.8)--(0,0.8)--(2.5,4.5) -- (8.5,4.5) -- (8.5,0.2)--(2.5,0.2)--(0,-0.8) -- (-2.5,-0.8)--cycle;
\draw[rounded corners=5, fill=blue!30] (-2.5,-0.8)--(0,-0.8)--(2.5,-4.5) -- (8.5,-4.5) -- (8.5,-0.2)--(2.5,-0.2)--(0,0.8) -- (-2.5,0.8)--cycle;
\end{scope}

\draw[rounded corners=12, fill=gray!50] (2.6,0.5)--(2.6,3.5) -- (3.4,3.5) -- (3.4,0.5)--cycle;
\draw[rounded corners=12, fill=gray!50] (2.6,-0.5)--(2.6,-3.5) -- (3.4,-3.5) -- (3.4,-0.5)--cycle;

\draw[rounded corners=12, fill=gray!50] (4.6,0.4)--(4.6,3.6) -- (8.4,3.6) -- (8.4,0.4)--cycle;
\draw[rounded corners=12, fill=gray!50] (4.6,-0.4)--(4.6,-3.6) -- (8.4,-3.6) -- (8.4,-0.4)--cycle;

\draw[fill=gray!80] (5.5,3.0) ellipse (0.8 and 0.5);
\draw[fill=gray!80] (7.5,1) ellipse (0.8 and 0.5);
\draw[fill=gray!80] (7.5,-1) ellipse (0.5 and 0.5);
\draw[fill=gray!80] (5.5,-3) ellipse (0.8 and 0.5);
\draw[fill=gray!80] (7.5,-3) ellipse (0.8 and 0.5);

\draw[rounded corners=5, thick, red, dashed] (-0.5,1)--(2.4,1)--(2.4,4.7) -- (8.6,4.7) -- (8.6,-1.6)--(6.5,-1.6)--(6.5,0)--(2.5,0)--(2.5,-1) -- (-0.5,-1)--cycle;


\tikzstyle{every node}=[inner sep=1pt]

\path (a2) edge[thick] (a);

\path (a) edge[thick] (x1);
\path (a) edge[thick] (x2);
\path (a) edge[thick] (x3);
\path (a) edge[thick] (x4);

\path (x1) edge[thick] (d11);
\path (x2) edge[thick] (d12);
\path (x2) edge[thick] (d21);
\path (x3) edge[thick] (d31);
\path (x4) edge[thick] (d31);
\path (x4) edge[thick] (d41);
\path (x3) edge[thick] (d52);

\path (d11) edge[thick] (d12);
\path (d21) edge[thick] (d22);
\path (d41) edge[thick] (d42);
\path (d51) edge[thick] (d52);

\path (x1) edge[bend left=20,gray] (x3);
\path (x2) edge[bend left=20,gray] (x4);
\path (x1) edge[bend left=20,gray] (x4);
\path (x2) edge[bend left=20,gray] (x3);

\path (x3) edge[gray] (d11);
\path (x3) edge[gray] (d12);
\path (x3) edge[gray] (d21);
\path (x3) edge[gray] (d22);

\path (x4) edge[gray] (d11);
\path (x4) edge[gray] (d12);
\path (x4) edge[gray] (d21);
\path (x4) edge[gray] (d22);

\tikzstyle{every node}=[circle,minimum size=5pt,inner sep=0pt,draw,fill]
\node at (a) {};
\node at (a2) {};

\node at (x1) {};
\node at (x2) {};
\node at (x3) {};
\node at (x4) {};

\node at (d11) {};
\node at (d12) {};
\node at (d21) {};
\node at (d22) {};
\node at (d31) {};
\node at (d41) {};
\node at (d42) {};
\node at (d51) {};
\node at (d52) {};

\begin{footnotesize}
\tikzstyle{every node}=[inner sep=2pt]
\node at (la) {$a$};
\node at (la2) {$a_2$};
\node at (M1) {$M_1$};
\node at (M2) {$M_2$};
\node at (AM1) {$\alpha(M_1)$};
\node at (AM2) {$\alpha(M_2)$};
\node at (ld31) {$d$};
\end{footnotesize}
\end{tikzpicture}
\caption{\label{fig:probe_splitting} Probes $(a_2,a,M_1,\alpha(M_1))$ and $(a_2,a,M_2,\alpha(M_2))$ in $(\{a_2,a\} \cup C_1 \cup \alpha(C_1),{\sim})$. 
Module $M$ in $(\{a\} \cup C_1 \cup \alpha(C_1))$ is surrounded by a red dashed line.
The components of $(\alpha(C_1),{\sim})$ are are marked in dark gray.}
\end{figure}

Now, note that for every component $D$ of the graph $(\alpha(C_1), {\sim})$
we have either $D \subseteq M$ or $D \cap M = \emptyset$.
Indeed, if some vertex of $D$ is in $M$, then $D \subseteq M$ as $(D,{\sim})$
is connected, $a \in M$, and $a \parallel D$.
We partition the vertices of $\alpha(C_1)$ into two sets, $\alpha(M_1)$ and $\alpha(M_2)$: 
we put the vertices from a component $D$ of $(\alpha(C_1),{\sim})$ to the set $\alpha(M_1)$ if there is an edge between a vertex in $D$ and a vertex in $M_1$;
otherwise, we put the vertices from $D$ to $\alpha(M_2)$.
In particular, note that for every component $D$ of $(\alpha(C_1),{\sim})$:
\begin{itemize}
\item if $D \subseteq \alpha(M_1)$, then $D \subseteq M$ (as $(\{a\} \cup M_1) \subseteq M$, some vertex in $D$ is adjacent to a vertex in $M_1$, and $a \parallel D$),
\item if $D \subseteq \alpha(M_2)$, then $D \parallel M_1$ and some vertex in $D$ is adjacent to a vertex in $M_2$ (as $(U,{\sim})$ is connected).
\end{itemize}
Note that there might be a component $D$ of $(\alpha(C_1),{\sim})$ such that $D \subseteq \alpha(M_2)$ and $D \subseteq M$.
In this case we have $D \parallel M_1$ and $D \sim M_2$, which shows that $D$ is a module in $(U,{\sim})$. 
Hence, $D = \{d\}$ for some $d \in \alpha(C_1)$ as otherwise $D$ would be a non-trivial module in $(U,{\sim})$ --  see Figure \ref{fig:probe_splitting} for an illustration.
Observe that $M_2 \sim (M_1 \cup \alpha(M_1))$ as otherwise
there would be a vertex in $M_2 \setminus M$ adjacent to $a$ in $M$ and non-adjacent to some vertex in $M$.
Summing up, we have (see Figure~\ref{fig:probe_splitting} for an illustration):
\begin{itemize}
 \item $M_1 \neq \emptyset$, $M_2 \neq \emptyset$, and ($\alpha(M_1) \neq \emptyset$ or $\alpha(M_2) \neq \emptyset$),
 \item $M_2 \sim (M_1 \cup \alpha(M_1))$,
 \item $\alpha(M_2) \parallel (M_1 \cup \alpha(M_1))$.
\end{itemize}
In particular, $(M_1 \cup \alpha(M_1),{\sim})$ is a permutation graph, and
for $i \in [2]$, if $\alpha(M_i) \neq \emptyset$, then the quadruple $(a_2,a,M_i,\alpha(M_i))$ is a probe in $(U,{\sim})$.
On the other hand, if $\alpha(M_i) = \emptyset$ then $|M_i|=1$ as otherwise $M_i$ would be a non-trivial module in $(U,{\sim})$.

Let $S = \{a,a_2\}$. 
We show that there is a unique conformal model $\phi$ of $(U,{\sim})$ that extends 
$\phi_S \equiv a^0a^1_2a^1a^0_2$.
Since $(C_1 \cup \alpha(C_1),{\sim})$ is connected and $a_2 \parallel (C_1 \cup \alpha(C_1))$,
we have either $(C_1 \cup \alpha(C_1)) \subseteq \rightside(a_2)$ or $(C_1 \cup \alpha(C_1)) \subseteq \leftside(a_2)$.
Suppose the first case holds.
Then, any conformal model $\phi$ of $(U,{\sim})$ extending $\phi_S$ must be of the form:
\begin{equation}
\label{eq:general_form_of_phi_U_prime}
\phi \equiv a^0a^1_2\tau'_{\phi} a^1 \tau''_{\phi} a^0_2,
\end{equation}
see Figure \ref{fig:P1P2_model}.
Let $P_i = \{a_2,a\} \cup M_i \cup \alpha(M_i)$ for $i \in [2]$.
Note that $(P_i,{\sim})$ admits a unique conformal model $\phi_{P_i}$ extending $\phi_S$, 
which follows from Claim~\ref{claim:probes_have_unique_model} if $\alpha(M_i) \neq \emptyset$ and from $|M_i|=1$ if $\alpha(M_i) = \emptyset$.
Suppose 
$$\phi_{P_i} \equiv a^0a^1_2\pi'_i a^1 \pi''_ia^0_2$$
for some words $\pi'_i, \pi''_i$ such that $\pi'_i \pi''_i$ is a permutation of $(M_i \cup \alpha(M_i))^*$.
\begin{claim}
\label{claim:properties_of_conformal_models_of_U}
Let $\phi$ be a conformal model of $(U,{\sim})$ of the form \eqref{eq:general_form_of_phi_U_prime}.
Then:
\begin{enumerate}
\item $\pi'_1$ and $\pi'_2$ are subwords of $\tau'_{\phi}$ and $|\pi'_1| + |\pi'_2| = |\tau'_{\phi}|$,
\item $\pi''_1$ and $\pi''_2$ are subwords of $\tau''_{\phi}$ and $|\pi''_1| + |\pi''_2| = |\tau''_{\phi}|$.
\item \label{item:prop_v_separates_u_and_a} For every $u \in \alpha(M_1)$ and every $v \in \alpha(M_2)$, either
 $\phi(u)$ and $\phi(v)$ are on different sides of $\phi(a)$, or
 there are on the same side of $\phi(a)$ and then $\phi(u)$ has the chords $\phi(v)$ and $\phi(a)$ on different sides.
\end{enumerate}
\end{claim}
\begin{proof}
See Figure~\ref{fig:P1P2_model} for an illustration.
The first two statements follow from the fact that $\phi \Vert P^*_i \equiv \phi_{P_i}$ for $i \in [2]$.

To show \eqref{item:prop_v_separates_u_and_a}, suppose $\phi(u)$ and $\phi(v)$ are on the same side of $\phi(a)$, 
but the chord $\phi(v)$ has the chords $\phi(u)$ and $\phi(a)$ on different sides.
Then, $\phi(v)$ intersects some chord from $\phi(P_1)$ as $(P_1,{\sim})$ is connected.
However, this is not possible as $v \parallel P_1$.
So, suppose $\phi(u)$, $\phi(v)$, and $\phi(a)$ are in series, that is, no chord from $\phi(a)$, $\phi(u)$, and $\phi(v)$
has the remaining two chords on its different sides.
Note that there is $v' \in \alpha(M_2)$ such that 
$v$ and $v'$ are in the same component $(D,{\sim})$ of $(\alpha(M_2),{\sim})$
and $v'$ is adjacent to some vertex $y \in M_2$.
Since $\phi(y)$ intersects both the chords $\phi(u)$ and $\phi(v')$ and $u \parallel D$,
we deduce that the chord $\phi(v')$ has the chords $\phi(a)$ and $\phi(u)$ on its different sides.
However, we already have shown that such a case is not possible.
\end{proof}

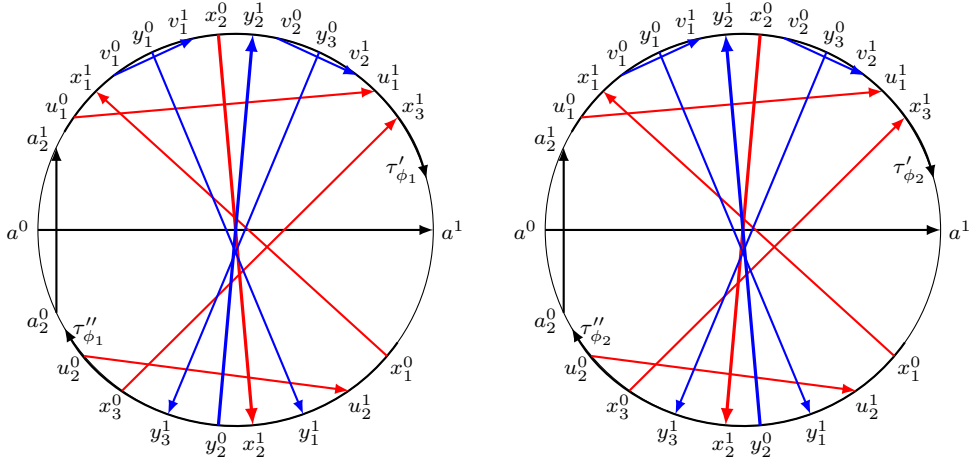
\begin{figure}
\begin{tikzpicture}[xscale=1.3,yscale=1.3,>=latex]
\coordinate (center) at (0,0);

\coordinate (a0) at ($(center)+(180:2cm)$) {};
\coordinate (a1) at ($(center)+(0:2cm)$) {};
\coordinate (la0) at ($(center)+(180:2.2cm)$) {};
\coordinate (la1) at ($(center)+(0:2.2cm)$) {};

\coordinate (a20) at ($(center)+(205:2cm)$) {};
\coordinate (a21) at ($(center)+(155:2cm)$) {};
\coordinate (la20) at ($(center)+(205:2.2cm)$) {};
\coordinate (la21) at ($(center)+(155:2.2cm)$) {};

\coordinate (x11) at ($(center)+(135:2cm)$) {};
\coordinate (x10) at ($(center)+(320:2cm)$) {};
\coordinate (lx11) at ($(center)+(135:2.2cm)$) {};
\coordinate (lx10) at ($(center)+(320:2.2cm)$) {};

\coordinate (x20) at ($(center)+(95:2cm)$) {};
\coordinate (x21) at ($(center)+(275:2cm)$) {};
\coordinate (lx20) at ($(center)+(95:2.2cm)$) {};
\coordinate (lx21) at ($(center)+(275:2.2cm)$) {};

\coordinate (x31) at ($(center)+(35:2cm)$) {};
\coordinate (x30) at ($(center)+(235:2cm)$) {};
\coordinate (lx31) at ($(center)+(35:2.2cm)$) {};
\coordinate (lx30) at ($(center)+(235:2.2cm)$) {};

\coordinate (u10) at ($(center)+(145:2cm)$) {};
\coordinate (u11) at ($(center)+(45:2cm)$) {};
\coordinate (lu10) at ($(center)+(145:2.2cm)$) {};
\coordinate (lu11) at ($(center)+(45:2.2cm)$) {};

\coordinate (u20) at ($(center)+(220:2cm)$) {};
\coordinate (u21) at ($(center)+(305:2cm)$) {};
\coordinate (lu20) at ($(center)+(220:2.2cm)$) {};
\coordinate (lu21) at ($(center)+(305:2.2cm)$) {};

\coordinate (y10) at ($(center)+(115:2cm)$) {};
\coordinate (y11) at ($(center)+(290:2cm)$) {};
\coordinate (ly10) at ($(center)+(115:2.2cm)$) {};
\coordinate (ly11) at ($(center)+(290:2.2cm)$) {};

\coordinate (y21) at ($(center)+(85:2cm)$) {};
\coordinate (y20) at ($(center)+(265:2cm)$) {};
\coordinate (ly21) at ($(center)+(85:2.2cm)$) {};
\coordinate (ly20) at ($(center)+(265:2.2cm)$) {};

\coordinate (y30) at ($(center)+(65:2cm)$) {};
\coordinate (y31) at ($(center)+(250:2cm)$) {};
\coordinate (ly30) at ($(center)+(65:2.2cm)$) {};
\coordinate (ly31) at ($(center)+(250:2.2cm)$) {};

\coordinate (v10) at ($(center)+(128:2cm)$) {};
\coordinate (v11) at ($(center)+(102:2cm)$) {};
\coordinate (lv10) at ($(center)+(125:2.2cm)$) {};
\coordinate (lv11) at ($(center)+(105:2.2cm)$) {};

\coordinate (v20) at ($(center)+(78:2cm)$) {};
\coordinate (v21) at ($(center)+(52:2cm)$) {};
\coordinate (lv20) at ($(center)+(75:2.2cm)$) {};
\coordinate (lv21) at ($(center)+(55:2.2cm)$) {};

\draw (0,0) circle (2cm);

\draw[thick,->] ([shift=(150:2cm)]0,0) arc (150:15:2cm);
\coordinate (tauprimphi) at ($(center)+(20:1.8cm)$) {};

\coordinate (tauprimphi) at ($(center)+(20:1.8cm)$) {};

\draw[thick,->] ([shift=(325:2cm)]0,0) arc (325:210:2cm);
\coordinate (taubisphi) at ($(center)+(215:1.8cm)$) {};

\draw[thick,->] (a0)--(a1);
\draw[thick,->] (a20)--(a21);
\draw[thick,red,->] (x10)--(x11);
\draw[very thick,red,->] (x20)--(x21);
\draw[thick,red,->] (x30)--(x31);
\draw[thick,red,->] (u10)--(u11);
\draw[thick,red,->] (u20)--(u21);

\draw[thick,blue,->] (y10)--(y11);
\draw[very thick,blue,->] (y20)--(y21);
\draw[thick,blue,->] (y30)--(y31);

\draw[thick,blue,->] (v10)--(v11);
\draw[thick,blue,->] (v20)--(v21);

\tikzstyle{every node}=[inner sep=1pt]
\begin{tiny}
\node at (la0) {$a^0$};
\node at (la1) {$a^1$};
\node at (la20) {$a_2^0$};
\node at (la21) {$a_2^1$};

\node at (lx10) {$x_1^0$};
\node at (lx11) {$x_1^1$};
\node at (lx20) {$x_2^0$};
\node at (lx21) {$x_2^1$};
\node at (lx30) {$x_3^0$};
\node at (lx31) {$x_3^1$};
\node at (lu10) {$u_1^0$};
\node at (lu11) {$u_1^1$};
\node at (lu20) {$u_2^0$};
\node at (lu21) {$u_2^1$};

\node at (ly10) {$y_1^0$};
\node at (ly11) {$y_1^1$};
\node at (ly20) {$y_2^0$};
\node at (ly21) {$y_2^1$};
\node at (ly30) {$y_3^0$};
\node at (ly31) {$y_3^1$};
\node at (lv10) {$v_1^0$};
\node at (lv11) {$v_1^1$};
\node at (lv20) {$v_2^0$};
\node at (lv21) {$v_2^1$};

\node at (tauprimphi) {$\tau'_{\phi_1}$};
\node at (taubisphi) {$\tau''_{\phi_1}$};
\end{tiny}
\end{tikzpicture} 
\hspace{0.3cm}
\begin{tikzpicture}[xscale=1.3,yscale=1.3,>=latex]
\coordinate (center) at (0,0);

\coordinate (a0) at ($(center)+(180:2cm)$) {};
\coordinate (a1) at ($(center)+(0:2cm)$) {};
\coordinate (la0) at ($(center)+(180:2.2cm)$) {};
\coordinate (la1) at ($(center)+(0:2.2cm)$) {};

\coordinate (a20) at ($(center)+(205:2cm)$) {};
\coordinate (a21) at ($(center)+(155:2cm)$) {};
\coordinate (la20) at ($(center)+(205:2.2cm)$) {};
\coordinate (la21) at ($(center)+(155:2.2cm)$) {};

\coordinate (x11) at ($(center)+(135:2cm)$) {};
\coordinate (x10) at ($(center)+(320:2cm)$) {};
\coordinate (lx11) at ($(center)+(135:2.2cm)$) {};
\coordinate (lx10) at ($(center)+(320:2.2cm)$) {};

\coordinate (x20) at ($(center)+(85:2cm)$) {};
\coordinate (x21) at ($(center)+(265:2cm)$) {};
\coordinate (lx20) at ($(center)+(85:2.2cm)$) {};
\coordinate (lx21) at ($(center)+(265:2.2cm)$) {};

\coordinate (x31) at ($(center)+(35:2cm)$) {};
\coordinate (x30) at ($(center)+(235:2cm)$) {};
\coordinate (lx31) at ($(center)+(35:2.2cm)$) {};
\coordinate (lx30) at ($(center)+(235:2.2cm)$) {};

\coordinate (u10) at ($(center)+(145:2cm)$) {};
\coordinate (u11) at ($(center)+(45:2cm)$) {};
\coordinate (lu10) at ($(center)+(145:2.2cm)$) {};
\coordinate (lu11) at ($(center)+(45:2.2cm)$) {};

\coordinate (u20) at ($(center)+(220:2cm)$) {};
\coordinate (u21) at ($(center)+(305:2cm)$) {};
\coordinate (lu20) at ($(center)+(220:2.2cm)$) {};
\coordinate (lu21) at ($(center)+(305:2.2cm)$) {};

\coordinate (y10) at ($(center)+(115:2cm)$) {};
\coordinate (y11) at ($(center)+(290:2cm)$) {};
\coordinate (ly10) at ($(center)+(115:2.2cm)$) {};
\coordinate (ly11) at ($(center)+(290:2.2cm)$) {};

\coordinate (y21) at ($(center)+(95:2cm)$) {};
\coordinate (y20) at ($(center)+(275:2cm)$) {};
\coordinate (ly21) at ($(center)+(95:2.2cm)$) {};
\coordinate (ly20) at ($(center)+(275:2.2cm)$) {};

\coordinate (y30) at ($(center)+(65:2cm)$) {};
\coordinate (y31) at ($(center)+(250:2cm)$) {};
\coordinate (ly30) at ($(center)+(65:2.2cm)$) {};
\coordinate (ly31) at ($(center)+(250:2.2cm)$) {};

\coordinate (v10) at ($(center)+(128:2cm)$) {};
\coordinate (v11) at ($(center)+(102:2cm)$) {};
\coordinate (lv10) at ($(center)+(125:2.2cm)$) {};
\coordinate (lv11) at ($(center)+(105:2.2cm)$) {};

\coordinate (v20) at ($(center)+(78:2cm)$) {};
\coordinate (v21) at ($(center)+(52:2cm)$) {};
\coordinate (lv20) at ($(center)+(75:2.2cm)$) {};
\coordinate (lv21) at ($(center)+(55:2.2cm)$) {};

\draw (0,0) circle (2cm);

\draw[thick,->] ([shift=(150:2cm)]0,0) arc (150:15:2cm);
\coordinate (tauprimphi) at ($(center)+(20:1.8cm)$) {};

\coordinate (tauprimphi) at ($(center)+(20:1.8cm)$) {};

\draw[thick,->] ([shift=(325:2cm)]0,0) arc (325:210:2cm);
\coordinate (taubisphi) at ($(center)+(215:1.8cm)$) {};

\draw[thick,->] (a0)--(a1);
\draw[thick,->] (a20)--(a21);
\draw[thick,red,->] (x10)--(x11);
\draw[very thick,red,->] (x20)--(x21);
\draw[thick,red,->] (x30)--(x31);
\draw[thick,red,->] (u10)--(u11);
\draw[thick,red,->] (u20)--(u21);

\draw[thick,blue,->] (y10)--(y11);
\draw[very thick,blue,->] (y20)--(y21);
\draw[thick,blue,->] (y30)--(y31);

\draw[thick,blue,->] (v10)--(v11);
\draw[thick,blue,->] (v20)--(v21);

\tikzstyle{every node}=[inner sep=1pt]
\begin{tiny}
\node at (la0) {$a^0$};
\node at (la1) {$a^1$};
\node at (la20) {$a_2^0$};
\node at (la21) {$a_2^1$};

\node at (lx10) {$x_1^0$};
\node at (lx11) {$x_1^1$};
\node at (lx20) {$x_2^0$};
\node at (lx21) {$x_2^1$};
\node at (lx30) {$x_3^0$};
\node at (lx31) {$x_3^1$};
\node at (lu10) {$u_1^0$};
\node at (lu11) {$u_1^1$};
\node at (lu20) {$u_2^0$};
\node at (lu21) {$u_2^1$};

\node at (ly10) {$y_1^0$};
\node at (ly11) {$y_1^1$};
\node at (ly20) {$y_2^0$};
\node at (ly21) {$y_2^1$};
\node at (ly30) {$y_3^0$};
\node at (ly31) {$y_3^1$};
\node at (lv10) {$v_1^0$};
\node at (lv11) {$v_1^1$};
\node at (lv20) {$v_2^0$};
\node at (lv21) {$v_2^1$};

\node at (tauprimphi) {$\tau'_{\phi_2}$};
\node at (taubisphi) {$\tau''_{\phi_2}$};
\end{tiny}
\end{tikzpicture} 
\caption{\label{fig:P1P2_model}
Examples of conformal models $\phi_1,\phi_2$ of $(U,{\sim})$ extending $a^0a^1_2a^1a^0_2$:
$M_1 = \{x_1,x_2,x_3\}$, $\alpha(M_1) = \{u_1,u_2\}$, $M_2 = \{y_1,y_2,y_3\}$, $\alpha(M_2) = \{v_1,v_2\}$.
The chords representing the vertices from $M_1 \cup \alpha(M_1)$ are in red, 
the chords representing the vertices from $M_2 \cup \alpha(M_2)$ are in blue. 
We have $\pi'_1 = u^0_1x^1_1x^0_2u^1_1x^1_3$, $\pi'_2 = v^0_1y^0_1v^1_1y_2^1v^0_2y^0_3v_2^1$,
$\pi''_1 = x^0_1u^1_2x^1_2x^0_3u^0_2$, and $\pi''_2 = y^1_1y^0_2y^1_3$.
The vertices $x_2$ and $y_2$ are mixed in $\tau'_{\phi_1}$ and $\tau'_{\phi_2}$ and in $\tau''_{\phi_1}$ and $\tau''_{\phi_2}$.
}

\end{figure}

Our goal is to show that there is unique way to compose the words $\pi'_1$ and $\pi'_2$ and the words
$\pi''_1$ and $\pi''_2$ to get a conformal model of $(U,{\sim})$ of the form \eqref{eq:general_form_of_phi_U_prime}.
Suppose that there are two such models, say $\phi_1$ and $\phi_2$. 
Let 
$$\phi_1 \equiv a^0a^1_2\tau'_{\phi_1} a^1 \tau''_{\phi_1} a^0_2 \quad \text{and} \quad \phi_2 \equiv a^0a^1_2\tau'_{\phi_2} a^1 \tau''_{\phi_2} a^0_2.$$
We say that $x \in M_1 \cup \alpha(M_1)$ and $y \in M_2 \cup \alpha(M_2)$ are \emph{mixed in $\tau'_{\phi_1}$ and $\tau'_{\phi_2}$}
if there are $x' \in \{x^{0},x^{1}\}$ and $y' \in \{y^0,y^1\}$ such that 
$x'$ and $y'$ occur in different orders in the words $\tau'_{\phi_1}$ and $\tau'_{\phi_2}$
-- see Figure \ref{fig:P1P2_model}.
We introduce the notion of being mixed in $\tau''_{\phi_1}$ and $\tau''_{\phi_2}$ similarly.
Clearly, if $\phi_1$ and $\phi_2$ are non-equivalent,
there are vertices $x \in M_1 \cup \alpha(M_1)$ and $y \in M_2 \cup \alpha(M_2)$ such that $x$ and $y$ are mixed either 
in $\tau'_{\phi_1}$ and $\tau'_{\phi_2}$ or in $\tau''_{\phi_1}$ and $\tau''_{\phi_2}$.
Suppose $x$ and $y$ are mixed in $\tau'_{\phi_1}$ and $\tau'_{\phi_2}$.
We claim that $x \in M_1$ and $y \in M_2$.
We can not have $x \in M_1$ and $y \in \alpha(M_2)$ 
as we have $x \parallel y$. 
We can not have $x \in \alpha(M_1)$ and $y \in M_2$ as 
in every conformal model $\phi$ of $(U,{\sim})$ of the form~\eqref{eq:general_form_of_phi_U_prime} the chord $\phi(x)$ intersects $\phi(y)$ and the orientation of $\phi(x)$ with respect to $\phi(y)$ is the same ($\phi(x)$ has $\phi(a)$ on the same side).
Finally, by Claim~\ref{claim:properties_of_conformal_models_of_U}.\eqref{item:prop_v_separates_u_and_a} we can not have $x \in \alpha(M_1)$ and $y \in \alpha(M_2)$.
So, we must have $x \in M_1$ and $y \in M_2$.
Thus, we have $x \sim y$, which means that $x$ and $y$ are also mixed in $\tau''_{\phi_1}$ and $\tau''_{\phi_2}$.
So, from now we abbreviate and we say that $x$ and $y$ \emph{are mixed} if $x$ and $y$ are mixed in $\tau'_{\phi_1}$ and $\tau'_{\phi_2}$ and in $\tau''_{\phi_1}$ and $\tau''_{\phi_2}$.
We claim that:
\begin{equation}
\label{eqref:mixed_properties}
\begin{array}{c}
\text{If $x \in M_1$ and $y \in M_2$ are mixed, then } \\
\{x,y\} \sim \alpha(M_1) \text{ and } \{x, y\} \parallel \alpha(M_2).
\end{array}
\end{equation}
We prove $\{x,y\} \sim \alpha(M_1)$. 
Clearly, $y \sim \alpha(M_1)$ and $x \parallel \alpha(M_2)$ by the properties of the probes~$P_1$ and~$P_2$.
Suppose there is $u \in \alpha(M_1)$ such that $x \parallel u$.
If this is the case, the relative position of $\phi(x)$ and $\phi(u)$ is the same
in any conformal model $\phi$ of $(U,{\sim})$ of the form~\eqref{eq:general_form_of_phi_U_prime}.
Since $\phi_i(y)$ intersects $\phi_i(u)$ for every $i \in [2]$, $x$ and $y$ can not be mixed.
The second statement of \eqref{eqref:mixed_properties} is proved similarly.
Now, note that $(C_1,\sim)$ is a permutation subgraph of $(U,{\sim})$ as $a \sim C_1$.
Hence, if $x \in M_1$ is mixed with $y \in M_2$ and $x \parallel x_1$ for some $x_1 \in M_1$, 
then $x_1$ is also mixed with $y$.
Similarly, if $x \in M_1$ is mixed with $y \in M_2$ and $y \parallel y_1$ for some $y_1 \in M_2$, 
then $y_1$ is also mixed with $x$.
Now, let
$$W = \bigcup \{ z,t : \text{ $z$ and $t$ are mixed}\},$$
that is, $W$ contains all the elements in $C_1$ that are mixed with some other element in~$C_1$.
Note that $|W| \geq 2$ as there are at least two elements that are mixed.  
Moreover, $W \subseteq C_1 \subsetneq U$ and 
$W \sim (C_1 \setminus W)$ by the observations given above.
Since $W \sim \alpha(M_1)$ and $W \parallel \alpha(M_2)$ by~\eqref{eqref:mixed_properties}, $W$ is a non-trivial module in $(U,{\sim})$, which yields a contradiction.
\end{proof}

Before we prove Properties~\ref{prop:prime_skeleton} and \ref{prop:prime_contiguous_subwords} of $K(Q)$ we show a claim that allows to define $K$-relation in a different way, by means of the set $U$.
\begin{claim} Suppose $K$ is as defined in Definition \ref{def:prime-K-relation}.
\label{claim:prime-K-relation-U}
\begin{enumerate}
\item \label{item:prime-K-relation-U-parallel} If $M_i$ is parallel, then for every $v,w \in M_i$:
$$
v K w  \quad \iff \quad
\begin{array}{c}
\vspace{1pt}
\text{either $\{v,w\} \subseteq \leftside(u)$ or $\{v,w\} \subseteq \rightside(u)$,} \\
\text{for every $u \in U \setminus M_i$ such that $u \parallel M_i$.}
\end{array}
$$
\item \label{item:prime-K-relation-U-serial} If $M_i$ is serial, then for every $v,w \in M_i$:
$$
v K w  \quad \iff \quad 
\begin{array}{c}
\vspace{1pt}
\{ \leftside(v) \cap (U \setminus M_i), \rightside(v) \cap (U \setminus M_i) \} = \\
\{ \leftside(w) \cap (U \setminus M_i), \rightside(w) \cap (U \setminus M_i)\}.
\end{array}
$$
\end{enumerate}
\end{claim}
\begin{proof}
To prove \eqref{item:prime-K-relation-U-parallel} we need to show, for every $v,w \in M_i$,
the equivalence between the following two statements:
\begin{equation}
\label{eq:parallel_U}
\{v,w\} \subseteq \leftside(u) \text{ or } \{v,w\} \subseteq \rightside(u) \text{ for every $u \in U \setminus M_i$ such that $u \parallel M_i$}.
\end{equation}
\begin{equation}
\label{eq:parallel_M}
\{v,w\} \subseteq \leftside(u) \text{ or } \{v,w\} \subseteq \rightside(u) \text{ for every $u \in Q \setminus M_i$ such that $u \parallel M_i$}.
\end{equation}
We need to show that~\eqref{eq:parallel_U} implies~\eqref{eq:parallel_M}, as the inverse implication is obvious.
Suppose~\eqref{eq:parallel_U} holds, 
but there is $m \in Q \setminus M_i$ such that $v,w$ are on different sides of $m$ -- see Figure \ref{fig:K-relation-U-set} to the left.
Suppose $m \in M_j$ for some $j \in [n]$ different from $i$.
Let $\phi$ be a conformal model of $(Q,{\sim})$
and let $x \in Q \setminus M_i$ be such that $x \sim M_i$.
In particular, $\phi(x)$ intersects $\phi(v)$ and $\phi(w)$, and hence $\phi(x)$ intersects also $\phi(m)$.
Thus, $x \in Q \setminus (M_i \cup M_j)$.
Now, let $u \in U$ be a vertex such that $u \in M_j$.
Note that $\phi(u)$ intersects $\phi(x)$.
Moreover, $\phi(u)$ has $\phi(v)$ and $\phi(w)$ on different sides. 
Otherwise, $M_j$ would be also parallel and $M_i \cup M_j$ would be a non-trivial module in $M$, which is not possible as both $M_i$ and $M_j$ are 
maximal non-trivial modules in $(Q,{\sim})$.

\begin{figure}[htp!]
\begin{tikzpicture}[xscale=1,yscale=1,>=latex,shorten >=-0.4pt,shorten <=-0.4pt]
\coordinate (center) at (0,0);

\coordinate (v0) at ($(center)+(240:2cm)$) {};
\coordinate (v1) at ($(center)+(120:2cm)$) {};
\coordinate (lv0) at ($(center)+(240:2.3cm)$) {};
\coordinate (lv1) at ($(center)+(120:2.3cm)$) {};

\coordinate (w0) at ($(center)+(300:2cm)$) {};
\coordinate (w1) at ($(center)+(60:2cm)$) {};
\coordinate (lw0) at ($(center)+(300:2.3cm)$) {};
\coordinate (lw1) at ($(center)+(60:2.3cm)$) {};

\coordinate (x0) at ($(center)+(180:2cm)$) {};
\coordinate (x1) at ($(center)+(0:2cm)$) {};
\coordinate (lx0) at ($(center)+(180:2.3cm)$) {};
\coordinate (lx1) at ($(center)+(0:2.3cm)$) {};

\coordinate (m0) at ($(center)+(280:2cm)$) {};
\coordinate (m1) at ($(center)+(100:2cm)$) {};
\coordinate (lm0) at ($(center)+(280:2.3cm)$) {};
\coordinate (lm1) at ($(center)+(100:2.3cm)$) {};

\coordinate (u0) at ($(center)+(260:2cm)$) {};
\coordinate (u1) at ($(center)+(80:2cm)$) {};
\coordinate (lu0) at ($(center)+(260:2.3cm)$) {};
\coordinate (lu1) at ($(center)+(80:2.3cm)$) {};

\coordinate (un0) at ($(center)+(320:2cm)$) {};
\coordinate (un1) at ($(center)+(40:2cm)$) {};
\coordinate (lun0) at ($(center)+(320:2.3cm)$) {};
\coordinate (lun1) at ($(center)+(40:2.3cm)$) {};

\draw (0,0) circle (2cm);

\draw[thick,->] (x0)--(x1);
\draw[thick,<-] (v0)--(v1);
\draw[thick,->] (w0)--(w1);
\draw[thick,->] (m0)--(m1);
\draw[red, thick,->] (u0)--(u1);
\draw[red, dashed, thick,->] (un0)--(un1);

\tikzstyle{every node}=[inner sep=1pt]
\begin{tiny}
\node at (lx0) {$x^0$};
\node at (lx1) {$x^1$};

\node at (lv0) {$v^1$};
\node at (lv1) {$v^0$};

\node at (lw0) {$w^0$};
\node at (lw1) {$w^1$};

\node at (lu0) {$u^0$};
\node at (lu1) {$u^1$};

\node at (lm0) {$m^0$};
\node at (lm1) {$m^1$};

\node at (lun0) {$u^0$};
\node at (lun1) {$u^1$};
\end{tiny}
\draw[white] (-2.5,-2.5)--(-2.5,-2.3);
\draw[white] (2.5,2.5)--(2.5,2.3);
\end{tikzpicture} 
\hspace{0.5cm}
\begin{tikzpicture}[xscale=1,yscale=1,>=latex,shorten >=-0.4pt,shorten <=-0.4pt]
\coordinate (center) at (0,0);

\coordinate (v0) at ($(center)+(135:2cm)$) {};
\coordinate (v1) at ($(center)+(315:2cm)$) {};
\coordinate (lv0) at ($(center)+(135:2.3cm)$) {};
\coordinate (lv1) at ($(center)+(315:2.3cm)$) {};

\coordinate (w0) at ($(center)+(225:2cm)$) {};
\coordinate (w1) at ($(center)+(45:2cm)$) {};
\coordinate (lw0) at ($(center)+(225:2.3cm)$) {};
\coordinate (lw1) at ($(center)+(45:2.3cm)$) {};

\coordinate (m10) at ($(center)+(340:2cm)$) {};
\coordinate (m11) at ($(center)+(20:2cm)$) {};
\coordinate (lm10) at ($(center)+(340:2.35cm)$) {};
\coordinate (lm11) at ($(center)+(20:2.35cm)$) {};

\coordinate (m20) at ($(center)+(110:2cm)$) {};
\coordinate (m21) at ($(center)+(70:2cm)$) {};
\coordinate (lm20) at ($(center)+(110:2.35cm)$) {};
\coordinate (lm21) at ($(center)+(70:2.35cm)$) {};

\coordinate (u10) at ($(center)+(325:2cm)$) {};
\coordinate (u11) at ($(center)+(35:2cm)$) {};
\coordinate (lu10) at ($(center)+(325:2.3cm)$) {};
\coordinate (lu11) at ($(center)+(35:2.3cm)$) {};

\coordinate (uu10) at ($(center)+(145:2cm)$) {};
\coordinate (uu11) at ($(center)+(215:2cm)$) {};
\coordinate (luu10) at ($(center)+(145:2.3cm)$) {};
\coordinate (luu11) at ($(center)+(215:2.3cm)$) {};

\coordinate (u20) at ($(center)+(55:2cm)$) {};
\coordinate (u21) at ($(center)+(125:2cm)$) {};
\coordinate (lu20) at ($(center)+(55:2.3cm)$) {};
\coordinate (lu21) at ($(center)+(125:2.3cm)$) {};

\coordinate (uu20) at ($(center)+(235:2cm)$) {};
\coordinate (uu21) at ($(center)+(305:2cm)$) {};
\coordinate (luu20) at ($(center)+(235:2.3cm)$) {};
\coordinate (luu21) at ($(center)+(305:2.3cm)$) {};

\coordinate (phi_u1) at ($(center)+(0:1.2cm)$) {};
\coordinate (phi_uu1) at ($(center)+(180:1.2cm)$) {};
\coordinate (phi_u2) at ($(center)+(90:1.4cm)$) {};
\coordinate (phi_uu2) at ($(center)+(270:1.4cm)$) {};

\draw (0,0) circle (2cm);

\draw[thick,<-] (v0)--(v1);
\draw[thick,->] (w0)--(w1);

\draw[thick,->] (m10)--(m11);

\draw[thick,->] (m20)--(m21);

\draw[red, thick,-] (u10)--(u11);
\draw[red, thick,-] (uu10)--(uu11);

\draw[blue, thick,-] (u20)--(u21);
\draw[blue, thick,-] (uu20)--(uu21);

\tikzstyle{every node}=[inner sep=1pt]
\begin{tiny}

\node at (phi_u1) {$\phi(u_1)$};
\node at (phi_uu1) {$\phi(u_1)$};

\node at (phi_u2) {$\phi(u_2)$};
\node at (phi_uu2) {$\phi(u_2)$};

\node at (lv0) {$v^1$};
\node at (lv1) {$v^0$};

\node at (lw0) {$w^0$};
\node at (lw1) {$w^1$};

\node at (lm10) {$m_1^0$};
\node at (lm11) {$m_1^1$};

\node at (lm20) {$m_2^0$};
\node at (lm21) {$m_2^1$};
\end{tiny}
\draw[white] (-2.5,-2.5)--(-2.5,-2.3);
\draw[white] (2.5,2.5)--(2.5,2.3);
\end{tikzpicture} 
\caption{\label{fig:K-relation-U-set}\ 
}
\end{figure}

To prove \eqref{item:prime-K-relation-U-serial} we need to show that for every $v,w \in M_i$ the statements \eqref{eq:serial_U} and \eqref{eq:serial_M} are equivalent, where:
\begin{equation}
\label{eq:serial_U}
\begin{array}{l}
\{ \leftside(v) \cap (U \setminus M_i), \rightside(v) \cap (U \setminus M_i) \} = \\
\{ \leftside(w) \cap (U \setminus M_i), \rightside(w) \cap (U \setminus M_i)\}.
\end{array}
\end{equation}
\begin{equation}
\label{eq:serial_M}
\begin{array}{l}
\{ \leftside(v) \cap (Q \setminus M_i), \rightside(v) \cap (Q \setminus M_i) \} = \\
\{ \leftside(w) \cap (Q \setminus M_i), \rightside(w) \cap (Q \setminus M_i)\}.
\end{array}
\end{equation}
We need to show that \eqref{eq:serial_U} implies \eqref{eq:serial_M}
as the inverse implication is obvious.
Suppose for a contrary that \eqref{eq:serial_U} holds but \eqref{eq:serial_M} is not satisfied.
In particular, $v$ and $w$ are from different children of $M_i$, as otherwise the equivalence between~\eqref{eq:serial_U} and~\eqref{eq:serial_M} follows by Property~\ref{prop_prime:contiguous_subwords_serial}.
Hence, $v \sim w$.
Now, suppose there are $m_1, m_2 \in Q \setminus M_i$ such that 
$m_1,m_2$ are on one side of $v$ and $m_1,m_2$ are on different sides of $w$.
Suppose first that $m_1,m_2$ are on the right side of $v$, $m_1$
is on the right side of $w$, and $m_2$ is on the left side of $w$ -- see Figure~\ref{fig:K-relation-U-set} to the right for an illustration.
Suppose $\phi$ is a conformal model of $(Q,{\sim})$.
Note that $m_1,m_2$ can not belong to the same child $M_j$ of $Q$. 
Otherwise, let $x \in M \setminus M_j$ be a vertex such that $x \sim M_j$.
Then, the chord $\phi(x)$ intersects $\phi(w)$ and does not intersect $\phi(v)$.
Then, $x \in M_i$ as $x \sim w$ and $x \parallel v$. 
On the other hand, $x \notin M_i$ as $x \sim \{m_1,m_2\}$ and $M_i \parallel \{m_1,m_2\}$.
Suppose $m_1 \in M_j$ and $m_2 \in M_{j'}$ for some $j \neq j'$ distinct from~$i$. 
Let $u_1, u_2 \in U$ be such that $u_1 \in M_j$ and $u_2 \in M_{j'}$.
Note that the chord $\phi(u_1)$ is either between $w^1$ and $v^0$ or between $w^0$ and $v^1$ in $\phi$ as there is $y \in Q \setminus (M_j \cup M_i)$ such that $y \sim \{u_1,m_1\}$ and either $y \sim \{v,w\}$ or $y \parallel \{v,w\}$.
Similarly, the chord $\phi(u_2)$ is either between $v^1$  and $w^1$ or between $v^0$ and $w^0$ -- see Figure~\ref{fig:K-relation-U-set} to the right.
In any case, $u_1$ and $u_2$ contradict \eqref{eq:serial_U}. 
The other cases are proved analogously.
\end{proof}

The next lemma proves Properties~\ref{prop:prime_skeleton} and~\ref{prop:prime_contiguous_subwords} of the set $K(Q)$.
\begin{lemma}
\label{lemma:prime_consistent_modules}
Let $K_1,\ldots,K_k$ be the sets in $K(Q)$ and let $R = \{r_1,\ldots,r_k\}$ be the set such that $r_i \in K_i$ for $i \in [k]$.
Then:
\begin{enumerate}
\item\label{item:prime_consistent_modules_skeleton}
 The graph $(R, {\sim})$ has exactly two conformal models, $\phi^{0}_R$ and $\phi^{1}_R$, one being the reflection of the other.
\item \label{item:prime_consistent_modules_contiguous_subwords}
For every conformal model $\phi$ of $(Q,{\sim})$ and every $j \in [k]$, 
the set $K_j$ induces a consistent permutation model of $(K_j,{\sim})$ in~$\phi$.
\end{enumerate}
\end{lemma}
\begin{proof}
Note that $n \leq k$.
Without loss of generality we assume that $U = \{r_1,\ldots,r_{n}\}$ and
$r_i \in M_i$ for $i \in [n]$. 
Let $R_j = \{r_1,\ldots,r_j\}$ for every $j \in [k]$.
By Lemma~\ref{lemma:two_models_of_a_prime_graph}, $(U,{\sim})$ has two conformal models,
$\phi^0_U$ and its reflection $\phi^1_U$.
Our proof is based on the following statement:
\begin{equation}
\label{eq:extending_conformal_models_of_U_on_R}
\begin{array}{c}
\text{For every $j = \{|U|,\ldots,k\}$ and every $m \in \{0,1\}$ there is a unique}\\
\text{conformal model $\phi^{m}_{j}$ of $(R_j,{\sim})$ such that $\phi^{m}_{j} \Vert U^* \equiv \phi^{m}_U$.}
\end{array}
\end{equation}
Clearly, statement \eqref{item:prime_consistent_modules_skeleton} follows from statement \eqref{eq:extending_conformal_models_of_U_on_R} for $j = k$.

We prove the statement by induction on $j$.
For $j=|U|=n$ statement~\eqref{eq:extending_conformal_models_of_U_on_R} is satisfied as $R_n = U$.
Suppose~\eqref{eq:extending_conformal_models_of_U_on_R} holds for $j = l-1$ for some $l > |U|$.
Our goal is to prove it for $j=l$.
Suppose $r_l$ is in the module $M_i$ for some $i \in [k]$.
From the inductive hypothesis, there is a unique extension $\phi^m_{l-1}$ of $\phi^{m}_U$ on the set $R_{l-1}$.
Suppose for a contradiction that there are two non-equivalent conformal models of $(R_l,{\sim})$ extending $\phi^{m}_{l-1}$ by the chord for $r_l$.
Equivalently,
there is a circular word $\phi$ extending $\phi^{m}_{l-1}$ by the letters $x^{0},x^{1}, y^{0}, y^{1}$ 
such that $\phi' \equiv \phi \Vert (R_{l-1} \cup \{x\})^*$ and $\phi'' \equiv \phi \Vert (R_{l-1} \cup \{y\})^*$ are two non-equivalent conformal models of $(R_l,{\sim})$
after replacing $x^0,x^1$ by $r^0_l,r^1_l$ in $\phi'$ and $y^0,y^1$ by $r^0_l,r^1_l$ in $\phi''$, respectively.
Note that for every $r \in R_{l-1}$ the circular word $\phi$ satisfies the following properties: 
\begin{itemize}
\item $r \in \leftside(r_l) \iff \text{$\phi(r)$ is on the left side of $\phi(x)$ and $\phi(y)$}$,
\item $r \in \rightside(r_l) \iff \text{$\phi(r)$ is on the right side of $\phi(x)$ and $\phi(y)$}$,
\item $r_l \in \leftside(r) \iff \text{$\phi(x)$ and $\phi(y)$ are on the left side of $\phi(r)$}$,
\item $r_l \in \rightside(r) \iff \text{$\phi(x)$ and $\phi(y)$ are on the right side of $\phi(r)$}$.
\end{itemize}
We consider two cases depending on whether the chords $\phi(x)$ and $\phi(y)$ intersect in~$\phi$.

Suppose $\phi(x)$ and $\phi(y)$ do not intersect.
Suppose we have $\phi \Vert \{x,y\}^* \equiv x^0x^1y^1y^0$ -- see Figure~\ref{fig:prime_skeleton_unique_model} to the left.
Since $\phi'$ and $\phi''$ are non-equivalent, there is $r \in R_{l-1}$
such that $\phi(r)$ has one of its ends between $x^1$ and $y^1$ or between $y^0$ and $x^0$. 
Since the chord $\phi(r)$ can not intersect both $\phi(x)$ and $\phi(y)$, 
$\phi(r)$ must be on the right side of $\phi(x)$ and the left side of $\phi(y)$.
However, this contradicts the properties of $\phi$ listed above.
So, suppose we have $\phi \Vert \{x,y\}^* \equiv x^0x^1y^0y^1$ -- see Figure~\ref{fig:prime_skeleton_unique_model} in the middle.
Let $r \in R_{l-1}$ be such that $r \parallel r_l$. 
The chord $\phi(r)$ must lie on the right side of $\phi(x)$ and the right side of $\phi(y)$ as
any other placement of $\phi(r)$ would contradict the properties of $\phi$.
For the same reason, $\phi(r)$ can not have $\phi(x)$ and $\phi(y)$ on its different sides.
Hence, $\phi(r)$ has both its ends either between $x^{1}$ and $y^{0}$ or
between $y^{1}$ and $x^{0}$.
Note that $r$ can not belong to $M_i$ as otherwise the chord $\phi(u)$, where $u \in U \setminus M_i$ is such that $u \sim M_i$, 
could not intersect $\phi(r)$, $\phi(x)$, and $\phi(y)$ at the same time.
Let $P$ be a path in $(R_l,{\sim})$ joining $r$ and $r_l$ with all inner vertices in $U$.
Note that there must be a vertex $u \in U$ in the path $P$ such that $\phi(u)$ has $\phi(x)$ and $\phi(y)$ on its different sides. 
This contradicts the properties of $\phi$.
The remaining cases are proved similarly.

\begin{figure}[htp!]
\begin{tikzpicture}[xscale=0.8,yscale=0.8,>=latex]
\coordinate (center) at (0,0);

\coordinate (x0) at ($(center)+(210:2cm)$) {};
\coordinate (x1) at ($(center)+(150:2cm)$) {};
\coordinate (lx0) at ($(center)+(210:2.3cm)$) {};
\coordinate (lx1) at ($(center)+(150:2.3cm)$) {};

\coordinate (y0) at ($(center)+(330:2cm)$) {};
\coordinate (y1) at ($(center)+(30:2cm)$) {};
\coordinate (ly0) at ($(center)+(330:2.3cm)$) {};
\coordinate (ly1) at ($(center)+(30:2.3cm)$) {};

\coordinate (s0) at ($(center)+(130:2cm)$) {};
\coordinate (s1) at ($(center)+(80:2cm)$) {};
\coordinate (ls0) at ($(center)+(130:2.3cm)$) {};
\coordinate (ls1) at ($(center)+(80:2.3cm)$) {};

\coordinate (s20) at ($(center)+(70:2cm)$) {};
\coordinate (s21) at ($(center)+(270:2cm)$) {};
\coordinate (ls20) at ($(center)+(70:2.3cm)$) {};
\coordinate (ls21) at ($(center)+(270:2.3cm)$) {};

\draw (0,0) circle (2cm);

\draw[red,thick,->] (x0)--(x1);
\draw[red,thick,->] (y0)--(y1);
\draw[thick,->] (s0)--(s1);
\draw[thick,->] (s20)--(s21);

\tikzstyle{every node}=[inner sep=1pt]
\begin{tiny}
\node at (lx0) {$x^0$};
\node at (lx1) {$x^1$};
\node at (ly0) {$y^0$};
\node at (ly1) {$y^1$};
\node at (ls0) {$r^0$};
\node at (ls1) {$r^1$};
\node at (ls20) {$r^0$};
\node at (ls21) {$r^1$};
\end{tiny}
\draw[white] (-2.5,-2.5)--(-2.5,-2.3);
\draw[white] (2.5,2.5)--(2.5,2.3);
\end{tikzpicture} 
\hspace{0.5cm}
\begin{tikzpicture}[xscale=0.8,yscale=0.8,>=latex]
\coordinate (center) at (0,0);

\coordinate (x0) at ($(center)+(210:2cm)$) {};
\coordinate (x1) at ($(center)+(150:2cm)$) {};
\coordinate (lx0) at ($(center)+(210:2.3cm)$) {};
\coordinate (lx1) at ($(center)+(150:2.3cm)$) {};

\coordinate (y1) at ($(center)+(330:2cm)$) {};
\coordinate (y0) at ($(center)+(30:2cm)$) {};
\coordinate (ly1) at ($(center)+(330:2.3cm)$) {};
\coordinate (ly0) at ($(center)+(30:2.3cm)$) {};

\coordinate (s0) at ($(center)+(130:2cm)$) {};
\coordinate (s1) at ($(center)+(80:2cm)$) {};
\coordinate (ls0) at ($(center)+(130:2.3cm)$) {};
\coordinate (ls1) at ($(center)+(80:2.3cm)$) {};

\coordinate (s10) at ($(center)+(110:2cm)$) {};
\coordinate (s11) at ($(center)+(50:2cm)$) {};
\coordinate (ls10) at ($(center)+(110:2.3cm)$) {};
\coordinate (ls11) at ($(center)+(50:2.3cm)$) {};

\coordinate (s20) at ($(center)+(70:2cm)$) {};
\coordinate (s21) at ($(center)+(270:2cm)$) {};
\coordinate (ls20) at ($(center)+(70:2.3cm)$) {};
\coordinate (ls21) at ($(center)+(270:2.3cm)$) {};

\coordinate (s30) at ($(center)+(-5:2cm)$) {};
\coordinate (s31) at ($(center)+(185:2cm)$) {};
\coordinate (ls30) at ($(center)+(-5:2.3cm)$) {};
\coordinate (ls31) at ($(center)+(185:2.3cm)$) {};

\draw (0,0) circle (2cm);

\draw[red,thick,->] (x0)--(x1);
\draw[red,thick,->] (y0)--(y1);
\draw[thick,->] (s0)--(s1);
\draw[thick,->] (s10)--(s11);
\draw[thick,->] (s20)--(s21);
\draw[thick,->] (s30)--(s31);

\tikzstyle{every node}=[inner sep=1pt]
\begin{tiny}
\node at (lx0) {$x^0$};
\node at (lx1) {$x^1$};
\node at (ly0) {$y^0$};
\node at (ly1) {$y^1$};
\node at (ls0) {$r^0$};
\node at (ls1) {$r^1$};
\node at (ls20) {$u^0$};
\node at (ls21) {$u^1$};

\end{tiny}
\draw[white] (-2.5,-2.5)--(-2.5,-2.3);
\draw[white] (2.5,2.5)--(2.5,2.3);
\end{tikzpicture} 
\hspace{0.5cm}
\begin{tikzpicture}[xscale=0.8,yscale=0.8,>=latex]
\coordinate (center) at (0,0);

\coordinate (x0) at ($(center)+(270:2cm)$) {};
\coordinate (x1) at ($(center)+(90:2cm)$) {};
\coordinate (lx0) at ($(center)+(270:2.3cm)$) {};
\coordinate (lx1) at ($(center)+(90:2.3cm)$) {};

\coordinate (y0) at ($(center)+(180:2cm)$) {};
\coordinate (y1) at ($(center)+(0:2cm)$) {};
\coordinate (ly0) at ($(center)+(180:2.3cm)$) {};
\coordinate (ly1) at ($(center)+(0:2.3cm)$) {};

\coordinate (s10) at ($(center)+(40:2cm)$) {};
\coordinate (s11) at ($(center)+(230:2cm)$) {};
\coordinate (ls10) at ($(center)+(40:2.3cm)$) {};
\coordinate (ls11) at ($(center)+(230:2.3cm)$) {};

\coordinate (s20) at ($(center)+(50:2cm)$) {};
\coordinate (s21) at ($(center)+(220:2cm)$) {};
\coordinate (ls20) at ($(center)+(50:2.3cm)$) {};
\coordinate (ls21) at ($(center)+(220:2.3cm)$) {};

\coordinate (t10) at ($(center)+(165:2cm)$) {};
\coordinate (t11) at ($(center)+(135:2cm)$) {};
\coordinate (lt10) at ($(center)+(165:2.2cm)$) {};
\coordinate (lt11) at ($(center)+(135:2.2cm)$) {};

\coordinate (t20) at ($(center)+(-15:2cm)$) {};
\coordinate (t21) at ($(center)+(-45:2cm)$) {};
\coordinate (lt20) at ($(center)+(-15:2.2cm)$) {};
\coordinate (lt21) at ($(center)+(-45:2.2cm)$) {};

\coordinate (t30) at ($(center)+(115:2cm)$) {};
\coordinate (t31) at ($(center)+(-65:2cm)$) {};
\coordinate (lt30) at ($(center)+(115:2.2cm)$) {};
\coordinate (lt31) at ($(center)+(-65:2.2cm)$) {};

\coordinate (s'0) at ($(center)+(45:2.3cm)$) {};
\coordinate (s'1) at ($(center)+(225:2.3cm)$) {};

\draw (0,0) circle (2cm);

\draw[red,thick,->] (x0)--(x1);
\draw[red,thick,->] (y0)--(y1);

\draw[thick,->] (s10)--(s11);
\draw[thick,<-] (s20)--(s21);

\draw[thick,->] (t10)--(t11);
\draw[thick,->] (t20)--(t21);
\draw[thick,->] (t30)--(t31);

\draw[very thick] ([shift=(35:2cm)]0,0) arc (35:55:2cm);
\draw[very thick] ([shift=(215:2cm)]0,0) arc (215:235:2cm);

\tikzstyle{every node}=[inner sep=1pt]
\begin{tiny}
\node at (lx0) {$x^0$};
\node at (lx1) {$x^1$};
\node at (ly0) {$y^0$};
\node at (ly1) {$y^1$};
\node at (lt10) {$t^0$};
\node at (lt11) {$t^1$};
\node at (lt20) {$t^0$};
\node at (lt21) {$t^1$};
\node at (lt30) {$t^0$};
\node at (lt31) {$t^1$};
\node at (s'0) {$R'$};
\node at (s'1) {$R'$};
\end{tiny}
\draw[white] (-2.5,-2.5)--(-2.5,-2.3);
\draw[white] (2.5,2.5)--(2.5,2.3);
\end{tikzpicture} 
\caption{\label{fig:prime_skeleton_unique_model}\ 
}
\end{figure}

Suppose $\phi(x)$ and $\phi(y)$ intersect -- see Figure \ref{fig:prime_skeleton_unique_model} to the right.
Without loss of generality we assume that $\phi \Vert \{x,y\}^* \equiv x^{0}y^{0}x^{1}y^{1}$.
First, note that for every $r \in R_{l-1}$ 
the chord $\phi(r)$ can not have both its ends between $x^{0}$ and $y^{0}$
as otherwise $\phi(r)$ would be on the left side of $\phi(x)$ and on the right side of $\phi(y)$.
For the same reason, $\phi(r)$ can not have both its ends between $x^{1}$ and $y^{1}$.
Let $R'$ be the set of all $r \in R_{l-1}$
such that $\phi(r)$ has one end between $x^{0}$ and $y^{0}$ 
and the other between $x^{1}$ and $y^{1}$.
Clearly, $R' \neq \emptyset$ as $\phi'$ and $\phi''$ are not equivalent.
We claim that $R' \cup \{r_l\}$ is a non-trivial module in $(R_l, {\sim})$.
Indeed, for every $t \in R_{l-1} \setminus R'$ 
the chord $\phi(t)$ has either both ends between $y^{0}$ and $x^{1}$, 
or between $y^{1}$ and $x^{0}$, or has one of its ends between $y^{0}$ and $x^{1}$
and the second between $y^{1}$ and $x^{0}$.
In any case, we have either $t \parallel (R' \cup \{r_l\})$ or $t \sim (R' \cup \{r_l\})$ for every $t \in R_{l-1} \setminus R'$.
This shows that $R' \cup \{r_l\}$ is a module in $(R_l,{\sim})$.
Eventually, note that $R' \cup \{r_l\}$ is strictly contained in $R_l$
as otherwise $r_l \sim R'$ yields $M_i \sim (M \setminus M_i)$ (recall that $r_l \in M_i$), which would contradict that $M$ is prime in $\mathcal{M}(V,{\sim})$.
Now, note that the sets $M_1 \cap R_l,\ldots,M_n \cap R_l$ form
a partition of $R_l$ into $n$ maximal non-trivial modules in $(R_l,{\sim})$.
So, we have $(R' \cup \{r_l\}) \subseteq (M_i \cap R_l)$.
In particular, $M_i$ must be serial as $r_l \sim R'$.
Since for every $u \in U$ such that $u \parallel M_i$ the chord
$\phi(u)$ has both its ends between~$y^{0}$ and~$x^{1}$ or between~$y^{1}$ and~$x^{0}$, we have $r_l K r'$ for every $r' \in R'$ by Claim~\ref{claim:prime-K-relation-U}.\eqref{item:prime-K-relation-U-serial}.
However, this can not be the case as $R$ contains one element from every equivalent class of $K$-relation.

To prove statement \eqref{item:prime_consistent_modules_contiguous_subwords}
assume that $\phi$ is a conformal model of $(Q,{\sim})$.

Suppose $K_j = M_i$, where $M_i$ is a prime child of $Q$.
Then, statement~\eqref{item:prime_consistent_modules_contiguous_subwords}
follows from Lemma~\ref{lemma:circle_models_of_a_proper_prime_module}.

Suppose $K_j \subseteq M_i$, where $M_i$ is a serial child of $Q$.
Assume that $x \in M_{i'}$ is such that $x \sim M_i$ for some $i' \in [n]$ different from $i$.
Suppose $L_1,\ldots,L_p$ are the children of $M_i$ contained in $K_j$, enumerated such that
$$\phi \Vert (\{x\} \cup L_1 \cup \ldots \cup L_p)^* \equiv x^0 \lambda^0_1 \ldots \lambda^0_p x^1 \lambda^1_1 \ldots \lambda^1_p,$$
where $(\lambda^0_t,\lambda^1_t)$ is an oriented permutation model of $(L_t,{\sim})$ for $t \in [p]$.
Lemma~\ref{lemma:circle_models_of_a_serial_module} asserts that $\lambda^0_t, \lambda^1_t$ are contiguous subwords in $\phi$.
In particular, we assume $p \geq 2$ as otherwise statement~\eqref{item:prime_consistent_modules_contiguous_subwords} follows.
Denote by $l^{0}$ and $l^1$ the first and the last letter from $\lambda^0_1$,
by $l^{2}$ and $l^3$ the first and the last letter from $\lambda^0_p$,
by $r^{0}$ and $r^1$ the first and the last letter from $\lambda^1_1$,
and by $r^{2}$ and $r^3$ the first and the last letter from $\lambda^1_p$ --
see Figure \ref{fig:prime_contiguous_subwords} to the left.
\begin{figure}[h!]
\begin{tikzpicture}[xscale=1,yscale=1,>=latex]
\coordinate (center) at (0,0);

\coordinate (x0) at ($(center)+(270:2cm)$) {};
\coordinate (x1) at ($(center)+(90:2cm)$) {};
\coordinate (lx0) at ($(center)+(270:2.3cm)$) {};
\coordinate (lx1) at ($(center)+(90:2.3cm)$) {};

\coordinate (u0) at ($(center)+(120:2cm)$) {};
\coordinate (u1) at ($(center)+(60:2cm)$) {};
\coordinate (lu0) at ($(center)+(115:2.3cm)$) {};
\coordinate (lu1) at ($(center)+(65:2.3cm)$) {};

\coordinate (ly0) at ($(center)+(180:2.3cm)$) {};
\draw[very thick] ([shift=(179:2cm)]0,0) arc (179:181:2cm);

\coordinate (k11) at ($(center)+(20:2cm)$) {};
\coordinate (lk11) at ($(center)+(20:2.3cm)$) {};
\coordinate (k12) at ($(center)+(50:2cm)$) {};
\coordinate (lk12) at ($(center)+(50:2.3cm)$) {};
\coordinate (k13) at ($(center)+(200:2cm)$) {};
\coordinate (lk13) at ($(center)+(200:2.3cm)$) {};
\coordinate (k14) at ($(center)+(230:2cm)$) {};
\coordinate (lk14) at ($(center)+(230:2.3cm)$) {};
\coordinate (k'j) at ($(center)+(215:1.6cm)$) {};

\coordinate (k21) at ($(center)+(-50:2cm)$) {};
\coordinate (lk21) at ($(center)+(-50:2.3cm)$) {};
\coordinate (k22) at ($(center)+(-20:2cm)$) {};
\coordinate (lk22) at ($(center)+(-20:2.3cm)$) {};
\coordinate (k23) at ($(center)+(130:2cm)$) {};
\coordinate (lk23) at ($(center)+(130:2.3cm)$) {};
\coordinate (k24) at ($(center)+(160:2cm)$) {};
\coordinate (lk24) at ($(center)+(160:2.3cm)$) {};
\coordinate (k''j) at ($(center)+(145:1.6cm)$) {};

\draw (0,0) circle (2cm);

\draw[thick,->] (x0)--(x1);
\draw[thick,->] (u0)--(u1);

\draw[very thick,red,<-] ([shift=(20:2cm)]0,0) arc (20:50:2cm);
\coordinate (ll11) at ($(center)+(35:2.3cm)$) {};

\draw[very thick,red,<-] ([shift=(200:2cm)]0,0) arc (200:230:2cm);
\coordinate (ll01) at ($(center)+(215:2.3cm)$) {};

\draw[very thick,red] (k12)--(k13);
\draw[very thick,red] (k14)--(k11);

\draw[very thick,red,<-] ([shift=(-50:2cm)]0,0) arc (-50:-20:2cm);
\coordinate (ll1p) at ($(center)+(-35:2.3cm)$) {};
\draw[very thick,red,<-] ([shift=(130:2cm)]0,0) arc (130:160:2cm);
\coordinate (ll0p) at ($(center)+(145:2.3cm)$) {};
\draw[very thick,red] (k22)--(k23);
\draw[very thick,red] (k24)--(k21);

\tikzstyle{every node}=[inner sep=1pt]
\begin{tiny}
\node at (ll01) {$\lambda^0_1$};
\node at (ll0p) {$\lambda^0_p$};
\node at (ll11) {$\lambda^1_1$};
\node at (ll1p) {$\lambda^1_p$};

\node at (lx0) {$x^0$};
\node at (lx1) {$x^1$};
\node at (lk11) {$r^1$};
\node at (lk12) {$r^0$};
\node at (lk13) {$l^1$};
\node at (lk14) {$l^0$};
\node at (lk21) {$r^3$};
\node at (lk22) {$r^2$};
\node at (lk23) {$l^3$};
\node at (lk24) {$l^2$};
\node at (k'j) {$L_1$};
\node at (k''j) {$L_p$};

\node at (lu0) {$v^0$};
\node at (lu1) {$v^1$};
\node at (ly0) {$y^0$};

\end{tiny}
\draw[white] (-2.5,-2.5)--(-2.5,-2.3);
\draw[white] (2.5,2.5)--(2.5,2.3);
\end{tikzpicture} 
\hspace{0.5cm}
\begin{tikzpicture}[xscale=1,yscale=1,>=latex]
\coordinate (center) at (0,0);

\coordinate (x0) at ($(center)+(270:2cm)$) {};
\coordinate (x1) at ($(center)+(90:2cm)$) {};
\coordinate (lx0) at ($(center)+(270:2.3cm)$) {};
\coordinate (lx1) at ($(center)+(90:2.3cm)$) {};

\coordinate (ly0) at ($(center)+(180:2.3cm)$) {};
\draw[very thick] ([shift=(179:2cm)]0,0) arc (179:181:2cm);

\coordinate (k11) at ($(center)+(20:2cm)$) {};
\coordinate (lk11) at ($(center)+(20:2.3cm)$) {};
\coordinate (k12) at ($(center)+(50:2cm)$) {};
\coordinate (lk12) at ($(center)+(50:2.3cm)$) {};
\coordinate (k13) at ($(center)+(130:2cm)$) {};
\coordinate (lk13) at ($(center)+(130:2.3cm)$) {};
\coordinate (k14) at ($(center)+(160:2cm)$) {};
\coordinate (lk14) at ($(center)+(160:2.3cm)$) {};
\coordinate (k''j) at ($(center)+(140:1.6cm)$) {};

\coordinate (k21) at ($(center)+(-50:2cm)$) {};
\coordinate (lk21) at ($(center)+(-50:2.3cm)$) {};
\coordinate (k22) at ($(center)+(-20:2cm)$) {};
\coordinate (lk22) at ($(center)+(-20:2.3cm)$) {};
\coordinate (k23) at ($(center)+(200:2cm)$) {};
\coordinate (lk23) at ($(center)+(200:2.3cm)$) {};
\coordinate (ll01) at ($(center)+(200:2.3cm)$) {};
\coordinate (k24) at ($(center)+(230:2cm)$) {};
\coordinate (lk24) at ($(center)+(230:2.3cm)$) {};
\coordinate (k'j) at ($(center)+(225:1.6cm)$) {};

\draw (0,0) circle (2cm);

\draw[thick,->] (x0)--(x1);

\draw[very thick,red,<-] ([shift=(20:2cm)]0,0) arc (20:50:2cm);
\coordinate (ll1p) at ($(center)+(35:2.3cm)$) {};
\draw[very thick,red,<-] ([shift=(130:2cm)]0,0) arc (130:160:2cm);
\coordinate (ll0p) at ($(center)+(145:2.3cm)$) {};
\draw[very thick,red] (k12)--(k13);
\draw[very thick,red] (k14)--(k11);

\draw[very thick,red,<-] ([shift=(-50:2cm)]0,0) arc (-50:-20:2cm);
\coordinate (ll11) at ($(center)+(-35:2.3cm)$) {};
\draw[very thick,red,<-] ([shift=(200:2cm)]0,0) arc (200:230:2cm);
\coordinate (ll01) at ($(center)+(215:2.3cm)$) {};
\draw[very thick,red] (k22)--(k23);
\draw[very thick,red] (k24)--(k21);

\tikzstyle{every node}=[inner sep=1pt]
\begin{tiny}
\node at (lx0) {$x^0$};
\node at (lx1) {$x^1$};
\node at (lk11) {$r^1$};
\node at (lk12) {$r^0$};
\node at (lk13) {$l^3$};
\node at (lk14) {$l^2$};
\node at (lk21) {$r^3$};
\node at (lk22) {$r^2$};
\node at (lk23) {$l^1$};
\node at (lk24) {$l^0$};
\node at (k'j) {$L_1$};
\node at (k''j) {$L_p$};
\node at (ly0) {$y^0$};

\node at (ll01) {$\lambda^0_1$};
\node at (ll0p) {$\lambda^0_p$};
\node at (ll11) {$\lambda^1_1$};
\node at (ll1p) {$\lambda^1_p$};

\end{tiny}
\draw[white] (-2.5,-2.5)--(-2.5,-2.3);
\draw[white] (2.5,2.5)--(2.5,2.3);
\end{tikzpicture} 

\caption{\label{fig:prime_contiguous_subwords}
}
\end{figure}

We claim that there is $v \in Q \setminus M_i$ such that
$\phi(v)$ has both its ends either between $l^{3}$ and $r^{0}$ or between $r^{3}$ and $l^{0}$.
Assume otherwise.
Let $T$ be the set of all $t \in Q$ such that $\phi(t)$ has one end between $l^{3}$ and $r^{0}$ and the other end between $r^{3}$ and $l^{0}$.
We show that $M_i \cup T$ is a non-trivial module in $(Q,{\sim})$ strictly containing $M_i$,
which contradicts the fact that $M_i$ is a maximal non-trivial module in $Q$.
Note that $x \in T \setminus M_i$, which shows $M_i \subsetneq (M_i \cup T)$.
For every $w \in Q \setminus (M_i \cup T)$ the chord $\phi(w)$ 
has either both its ends between $l^{1}$ and $l^2$,
or between $r^1$ and $r^2$, or has one end between 
$l^1$ and $l^2$ and the other between $r^1$ and $r^2$.
In particular, for every $w \in Q \setminus (M_i \cup T)$ we have $w \parallel (M_i \cup T)$ or $w \sim (M_i \cup T)$, 
which proves that $M_i \cup T$ is a module in $(Q,{\sim})$.
Since $Q$ is prime, there is $w \in Q \setminus M_i$ such that $w \parallel M_i$.
In particular, $w$ is not in $T \cup M_i$, 
which shows that $T \cup M_i$ is a non-trivial module in $(Q,{\sim})$.
This proves our claim.

Suppose that $K_j$ does not induce a consistent permutation model in $\phi$.
That is, there is $y \in Q \setminus K_j$ 
such that $\phi(y)$ has an end between $l^{1}$ and $l^{2}$
or between $r^{1}$ and $r^{2}$.
Suppose that $y^{0}$ is between $l^{1}$ and $l^{2}$ -- the other case is proved analogously.
First, note that no chord $\phi(w)$ for $w \in Q \setminus M_i$ has its two ends between
$l^1$ and $l^2$ or between $r^1$ or $r^2$.
Otherwise, $w$ and $v$ prove that the vertices from $L_1$ are not in $K$-relation with the vertices from $L_p$.
This shows that $\phi(y)$ has its second end between $r^1$ and $r^2$ and that 
$y$ is not a member of $M_i$ (otherwise, $y K L_1$ and $y K L_p$, which is not the case).
Now, we proceed similarly as earlier end we show that 
$M_i \cup T'$ is a non-trivial module in $(Q,{\sim})$ strictly containing $M_i$,
where $T'$ is the set of all $t \in Q$ such that $\phi(t)$ has one end between $l^{1}$ and $l^2$ and the second end between $r^{1}$ and $r^{2}$.
However, this can not be the case.

Suppose $K_j \subseteq M_i$, where $M_i$ is a parallel child of $Q$.
Assume that $x \in M_{i'}$ is such that $x \sim M_i$ for some $i' \in [n]$ different from $i$.
Suppose $L_1,\ldots,L_p$ are the children of $M_i$ contained in $K_j$, enumerated such that
$$\phi \Vert (\{x\} \cup L_1 \cup \ldots \cup L_p)^* \equiv x^0 \lambda^0_1 \ldots \lambda^0_p x^1 \lambda^1_p \ldots \lambda^1_0,$$
where $(\lambda^0_t,\lambda^1_t)$ is an oriented permutation model of $(L_t,{\sim})$ for $t \in [p]$.
Lemma~\ref{lemma:circle_models_of_a_parallel_module} asserts that $\lambda^0_t, \lambda^1_t$ are contiguous subwords in $\phi$.
In particular, we assume $p \geq 2$ as otherwise statement \eqref{item:prime_consistent_modules_contiguous_subwords} follows easily.
Denote by~$l^{0}$ and~$l^1$ the first and the last letter from~$\lambda^0_1$,
by~$l^{2}$ and~$l^3$ the first and the last letter from~$\lambda^0_p$,
by~$r^{0}$ and~$r^1$ the first and the last letter from~$\lambda^1_p$,
and by~$r^{2}$ and~$r^3$ the first and the last letter from~$\lambda^1_1$ -- see Figure~\ref{fig:prime_contiguous_subwords} to the right.

Suppose that $K_j$ does not induce a consistent permutation model in $\phi$.
That is, there is $y \in Q \setminus K_j$ 
such that $\phi(y)$ has an end between~$l^{1}$ and~$l^{2}$
or between~$r^{1}$ and~$r^{2}$.
Suppose that $y^{0}$ is between $l^{1}$ and $l^{2}$ -- the other case is proved analogously.
First note that $y \parallel K_j$ as $\phi(y)$ can not intersect the chords from $\phi(L_1)$ and $\phi(L_p)$ at the same time. 
Suppose~$\phi(y)$ has both its ends between~$l^{1}$ and~$l^2$.
Then we have $y \notin M_i$ as $\phi(y)$ does not intersect~$\phi(x)$.
Now, let $P$ be a shortest path in $(Q,{\sim})$ between $y$ and $M_i$ with all inner vertices in~$U$.
Then, there must be a vertex $u$ in the path $P$ such that $u \in U \setminus M_i$ and $\phi(u)$ has one of its ends between
$l^1$ and $l^2$ and the second one between $r^1$ and $r^2$.
Then, the vertex $u$ proves that $L_1$ and $L_2$ are in different equivalence classes of $K$-relation, which is not the case.
So, suppose $y^{1}$ is between $r^{1}$ and $r^2$.
Note that $y \notin M_i$ as otherwise $y$ would be in $K$-relation with any vertex from $K_j$, which is not the case.
Then, the vertex $y \in (Q \setminus M_i)$ has $L_1$ and $L_2$ on different sides, which proves that $L_1$ and $L_p$ are not in $K$-relation, which is a contradiction.
\end{proof}

\section{Conformal models for parallel case - appendix} 
\label{sec:parallel_case_properties}
Properties~\ref{prop:PQ_tree} -- \ref{prop:pqtree_P_property}
of the PQS-tree ~$\pqstree$ of $G$ were shown by Hsu~\cite{Hsu95}.
For the sake of completeness, their proofs are also provided below.
\subsection{Properties of $\pqstree$}
\label{subsec:PQ_tree}
We start this section with the claim which proves Property~\ref{prop:PQ_tree} of the graph~$\pqstree$.
We recall that at this stage $\pqstree$ contains only P-nodes and Q-nodes.
\begin{claim}
\label{claim:PQ_tree} The following statements hold:
\begin{enumerate}
\item \label{item:PQ_tree_component_and_neighbors} For every Q-node~$Q$ in~$\pqstree$ and every two P-nodes $P_1,P_2 \in N_{\pqstree}(Q)$ there is a vertex~$v \in Q$ that separates 
the components from~$P_1 \setminus \{Q\}$ and the components from~$P_2 \setminus \{Q\}$.
\item \label{item:PQ_tree} 
The bipartite graph~$\pqstree$ is a tree.
\end{enumerate}
\end{claim}
\begin{proof}
Let $Q$ be a Q-node in $\pqstree$ and let $P_1, P_2$ be two different P-nodes adjacent to $Q$ in~$\pqstree$.
It means that $Q \in P_1$ and $Q \in P_2$.
Since $P_1,P_2$ are different maximal subsets of pairwise neighbouring components from~$\mathcal{Q}$, 
there is a component $Q_1 \in P_1 \setminus P_2$ and a component $Q_2 \in P_2 \setminus P_1$ such that~$Q_1$ and~$Q_2$ are separated by some $v \in V \setminus (Q_1 \cup Q_2)$.
Suppose $Q_1 \subseteq \leftside(v)$ and $Q_2 \subseteq \rightside(v)$.
Note that $v \in Q$.
Otherwise, depending on whether $Q \subseteq \rightside(v)$ or whether $Q \subseteq \leftside(v)$, 
$v$ separates either $Q$ and $Q_1$ or $Q$ and $Q_2$, and either $P_1$ or $P_2$ is not a P-node of~$\pqstree$.
Then, the components from $P_1 \setminus \{Q\}$ are on the left side of $v$ 
and the components from $P_2 \setminus \{Q\}$ are on the right side of $v$ 
as otherwise $P_1$ or $P_2$ is not a P-node of~$\pqstree$.
This proves~\eqref{item:PQ_tree_component_and_neighbors}.

Now, we show that $\pqstree$ is a tree.
First we prove that $\pqstree$ contains no cycles.
Suppose that $Q_1P_1 \ldots Q_kP_k$ is a cycle in $\pqstree$, for some $k \geq 2$.
Since $P_1,P_k$ are neighbors of $Q_1$ in $\pqstree$, there
is $v \in Q_1$ that separates the components in $P_1 \setminus \{Q_1\}$ 
and the components in $P_k \setminus \{Q_1\}$.
So, $v$ separates $Q_{2}$ and $Q_k$. 
In particular, $Q_2$ and $Q_k$ can not be in a same P-node,
and hence $k \geq 3$.
Since $Q_2$ and $Q_k$ are separated by $v$, 
there is $i \in [2,k-1]$ such that $Q_i$ and $Q_{i+1}$ are also separated by $v$.
So, $Q_i$ and $Q_{i+1}$ are not contained in a same P-node,
which is not the case as $P_i$ contains both $Q_i$ and $Q_{i+1}$.
This proves that $\pqstree$ is a forest.
To complete the proof we show that $\pqstree$ is connected.
Suppose $\pqstree$ is not connected and suppose $\phi$ is any conformal model of $G_{ov}$.
Then, there are vertices $u$ and $v$ such that $u^{*}$ and $v^{*}$ are next to each other in $\phi$, 
$u^{*} \in \{u^{0},u^{1}\}$, $v^{*} \in \{v^{0},v^{1}\}$, $u \in Q_u$, $v \in Q_v$, and $Q_u$ and $Q_v$ are components from $\mathcal{Q}$ from different connected components of the graph $\pqstree$.
By the choice of $u$ and $v$, there is no vertex in $V \setminus (Q_u \cup Q_v)$
that separates $Q_u$ and $Q_v$.
So, $Q_u$ and $Q_v$ are contained in some P-node in $\pqstree$,
which contradicts that $Q_u$ and $Q_v$ are in different connected components of~$\pqstree$.
\end{proof}

The next claim shows Property~\ref{prop:pqtree_P_left_right}.
\begin{claim}
\label{claim:P_nodes_in_a_component}
Suppose $Q$ is a Q-node in $\pqstree$, $P$ is a P-node adjacent to $Q$ in $\pqstree$, and $v$ is a vertex in $Q$.
Then, either $V_{\pqstree - Q}(P) \subseteq \leftside(v)$ or $V_{\pqstree - Q}(P) \subseteq \rightside(v)$.
\end{claim}
\begin{proof}
Let $F_P$ be a connected component of $\pqstree \setminus Q$ containing $P$.
Suppose there are two components $Q_1,Q_2 \in F_P$ 
such that $Q_1 \subseteq \leftside(v)$ and $Q_2 \subseteq \rightside(v)$.
Let $p$ be a path between $Q_1$ and $Q_2$ in $F_{P}$.
Clearly, there are three consecutive elements $Q'_1 P' Q'_2$ on the path $p$
such that $Q'_1 \in \leftside(v)$ and $Q'_2 \in \rightside(v)$.
Thus, $Q'_1$ and $Q'_2$ are separated by~$v$,
which contradicts $Q'_1,Q'_2 \in P'$.
\end{proof}

Finally, Claim~\ref{claim:pq_tree_conformal_models_properties_nodes} shows Properties~\ref{prop:pqtree_Q_property} and~\ref{prop:pqtree_P_property}.
\begin{claim}
\label{claim:pq_tree_conformal_models_properties_nodes}
Let $\phi$ be a conformal model of $G_{ov}$.
\begin{enumerate}
\item \label{item:conformal_models_Q_node_property} For every Q-node $Q$ and every $P \in N_{\pqstree}(Q)$ the set $V^*_{\pqstree - Q}(P)$ is contiguous in $\phi$.
Moreover, for every two distinct $P,P' \in N_{\pqstree}(Q)$ there is $v \in Q$ 
such that $\phi(v)$ separates $\phi|V^*_{\pqstree - Q}(P)$
and $\phi|V^*_{\pqstree - Q}(P')$.
\item \label{item:conformal_models_P_node_property} For every P-node $P$ and every $Q \in N_{\pqstree}(P)$ the set $V^*_{\pqstree - P}(Q)$ is contiguous in $\phi$.
\end{enumerate}
\end{claim}
\begin{proof}
Statement~\eqref{item:conformal_models_Q_node_property} follows from
Claim~\ref{claim:P_nodes_in_a_component} and Claim~\ref{claim:PQ_tree}.\eqref{item:PQ_tree_component_and_neighbors}.

Next, we prove statement \eqref{item:conformal_models_P_node_property}.
Since $P$ is a maximal subset of $\mathcal{Q}$ containing pairwise neighbouring components,
the set $Q^*$ is contiguous in the circular word $\phi \Vert (\bigcup P)^*$.
Now, statement \eqref{item:conformal_models_Q_node_property} applied to the neighbours of the component $Q$ different than $P$ proves that $V^*_{\pqstree - P}(Q)$ is a contiguous subword of $\phi$.
\end{proof}

\subsection{The properties of the set $K(Q)$, where $Q$ is a serial non-permutation component}
\label{subsec:parallel_M_serial}
In this subsection we show properties~\ref{prop:parallel_serial_skeleton} and \ref{prop:parallel_serial_contiguous_subwords} of the set $K(Q)$, 
where $Q$ is a serial non-permutation component of $G_{ov}$ and $K(Q)$ is as defined in Definition~\ref{def:serial_K_sets}.
We assume $M_1,\ldots,M_n$ are the children of $Q$.
\begin{lemma}
\label{lemma:serial_consistent_modules}
Let $K_1,\ldots,K_k$ be the members of $K(Q)$ for some $k \leq n$ and let
$R = \{r_1,\ldots,r_k\}$ be such that $r_i \in K_i$ for every $i \in [k]$.
Then:
\begin{enumerate}
 \item \label{item:serial_consistent_modules_skeleton} 
There are two conformal models of $(R,{\sim})$, $\phi^0_R$ and its reflection $\phi^1_R$, 
such that for every extended conformal model $\phi_Q$ of $(Q,{\sim})$ 
we have either $\phi_Q \Vert R^* = \phi^0_R$ or $\phi_Q \Vert R^* = \phi^1_R$. 
 \item \label{item:serial_consistent_modules_contiguous_subwords}  For every extended conformal model $\phi_Q$ of $(Q,{\sim})$ and every $K \in K(Q)$ the set $K$ induces a consistent permutation model in $\phi^Q \equiv \phi \Vert Q^*$.
 Moreover, if $K$ is the union of at least two children of $Q$, the set $K$ induces a consistent permutation model also in $\phi_Q$.
\end{enumerate}
\end{lemma}
\begin{proof}
We assume that $M_1,\ldots,M_n$ are enumerated such that $r_i \in M_i$ for every $i \in [k]$
(note that $k \leq n$).
For $i \in [k]$, let $R_i = \{r_1,\ldots,r_i\}$; note that $(R_i,{\sim})$ is a clique for every $i \in [k]$.
We claim that for every $i \in [2,k]$ there exist two models of $(R_i,{\sim})$, $\phi^{0}_{i}$ 
and its reflection~$\phi^{1}_{i}$, such that for every extended conformal model $\phi_Q$ of $(Q,{\sim})$,
either 
\begin{equation}
\label{eq:serial_module_skeleton}
\phi_Q \Vert R^*_i \equiv \phi^{0}_{i} \quad \text{or} \quad \phi_Q \Vert R^*_i \equiv \phi^{1}_{i}.
\end{equation}
Then, statement \eqref{item:serial_consistent_modules_skeleton} follows from statement \eqref{eq:serial_module_skeleton} for $i=k$.
We prove \eqref{eq:serial_module_skeleton} by induction on~$i$.
Note that $(R_2,{\sim})$ has two conformal models,
$$\phi^{0}_{2} \equiv r^{0}_1r^{0}_2r^{1}_1r^{1}_2 \quad \text{and} \quad \phi^{1}_{2} \equiv r^{0}_1r^{1}_2r^{1}_1r^{0}_2,$$
and $\phi^{0}_{2}$ is the reflection of $\phi^{1}_{2}$.
So, statement \eqref{eq:serial_module_skeleton} holds for $i=2$.

Let $j \in [3,k]$.
Suppose \eqref{eq:serial_module_skeleton} holds for all $i \in [2,j-1]$.
To show \eqref{eq:serial_module_skeleton} for $j$ it suffices to prove there is a unique extension $\phi^{0}_{j}$ of $\phi^{0}_{j-1}$ on the set $R_j$ such that $\phi_Q \Vert R^*_j \equiv \phi^{0}_{j}$ holds for every extended conformal model $\phi_Q$ of $(Q,{\sim})$ such that $\phi_Q \Vert R^*_{j-1} \equiv \phi^{0}_{j-1}$.
Suppose for a contradiction that there are two extended conformal models of $(Q,{\sim})$, say $\phi_Q$ and $\phi'_Q$,
such that $\phi_Q \Vert R^*_{j-1} \equiv \phi'_Q \Vert R^*_{j-1} \equiv \phi^{0}_{j-1}$ and $\phi_Q \Vert R^*_j \not \equiv \phi'_Q \Vert R_j$.
That is, the chords $\phi_Q(r_j)$ and $\phi'_Q(r_j)$ extend $\phi^{0}_{j-1}$ into two non-equivalent models of $(R_j,{\sim})$.
\begin{figure}[!htp]
\begin{tikzpicture}[scale=0.85,>=latex]
    \coordinate (center) at (0,0) {};
    \coordinate (sl0) at ($(center)+(180:2)$) {};
    \coordinate (sl1) at ($(center)+(0:2)$) {};
    \coordinate (lsl0) at ($(center)+(180:2.25)$) {};
    \coordinate (lsl1) at ($(center)+(0:2.25)$) {};

    \coordinate (sk0) at ($(center)+(270:2)$) {};
    \coordinate (sk1) at ($(center)+(90:2)$) {};
    \coordinate (lsk0) at ($(center)+(270:2.25)$) {};
    \coordinate (lsk1) at ($(center)+(90:2.25)$) {};

    \coordinate (x0) at ($(center)+(45:2)$) {};
    \coordinate (x1) at ($(center)+(225:2)$) {};
    \coordinate (lx0) at ($(center)+(45:2.55)$) {};
    \coordinate (lx1) at ($(center)+(225:2.55)$) {};
    
    \coordinate (y0) at ($(center)+(135:2)$) {};
    \coordinate (y1) at ($(center)+(315:2)$) {};
    \coordinate (ly0) at ($(center)+(135:2.55)$) {};
    \coordinate (ly1) at ($(center)+(315:2.55)$) {};
    \tikzstyle{every node}=[inner sep=1pt]
    \begin{tiny}
    \node at (lsl0) {$r^0_q$};
    \node at (lsl1) {$r^1_q$};
    \node at (lsk0) {$r^0_p$};
    \node at (lsk1) {$r^1_p$};

    \node at (lx0) {$\phi_Q(r^1_j)$};
    \node at (lx1) {$\phi_Q(r^0_j)$};
    \node at (ly0) {$\phi'_Q(r^0_j)$};
    \node at (ly1) {$\phi'_Q(r^1_j)$};
    \end{tiny}
    \draw[thick,->] (sl0) -- (sl1);
    \draw[thick,->] (sk0) -- (sk1);
    \draw[thick,->] (x1) -- (x0);
    \draw[thick,->] (y0) -- (y1);

  \draw (0,0) circle (2cm);

  
\end{tikzpicture}
\hspace{1.5cm}
\begin{tikzpicture}[scale=0.85,>=latex]
    \coordinate (center) at (0,0) {};
    \coordinate (sl0) at ($(center)+(180:2)$) {};
    \coordinate (sl1) at ($(center)+(0:2)$) {};
    \coordinate (lsl0) at ($(center)+(180:2.25)$) {};
    \coordinate (lsl1) at ($(center)+(0:2.25)$) {};

    \coordinate (sk0) at ($(center)+(270:2)$) {};
    \coordinate (sk1) at ($(center)+(90:2)$) {};
    \coordinate (lsk0) at ($(center)+(270:2.25)$) {};
    \coordinate (lsk1) at ($(center)+(90:2.25)$) {};

    \coordinate (x0) at ($(center)+(30:2)$) {};
    \coordinate (x1) at ($(center)+(210:2)$) {};
    \coordinate (lx0) at ($(center)+(30:2.55)$) {};
    \coordinate (lx1) at ($(center)+(214:2.55)$) {};
    
    \coordinate (y0) at ($(center)+(60:2)$) {};
    \coordinate (y1) at ($(center)+(240:2)$) {};
    \coordinate (ly0) at ($(center)+(66:2.55)$) {};
    \coordinate (ly1) at ($(center)+(245:2.55)$) {};
    \tikzstyle{every node}=[inner sep=1pt]
    \begin{tiny}
    \node at (lsl0) {$r^0_q$};
    \node at (lsl1) {$r^1_q$};
    \node at (lsk0) {$r^0_p$};
    \node at (lsk1) {$r^1_p$};
    \node at (lx0) {$\phi_Q(r^0_j)$};
    \node at (lx1) {$\phi_Q(r^1_j)$};
    \node at (ly0) {$\phi'_Q(r^1_j)$};
    \node at (ly1) {$\phi'_Q(r^0_j)$};
    \end{tiny}
    \draw[thick,->] (sl0) -- (sl1);
    \draw[thick,->] (sk0) -- (sk1);
    \draw[thick,->] (x0) -- (x1);
    \draw[thick,<-] (y0) -- (y1);

  \draw (0,0) circle (2cm);
\end{tikzpicture}
\caption{\label{fig:skeleton_unique_model_clique}}
\end{figure}

It means that there are two different vertices $r_q,r_p \in R_{j-1}$ such that
$$
\phi'_Q \Vert \{r_q,r_p\}^* \equiv \phi_Q \Vert \{r_q,r_p\}^* \equiv r_p^0r_q^{0}r_p^{1}r_q^{1},
$$
but the chords $\phi_Q(r_j)$ and $\phi'_Q(r_j)$ 
have its endpoints in different sections $r^0_pr^0_q$, $r^0_qr^1_p$, $r^1_pr^1_q$, $r^1_qr^0_q$ of the circular word
$r_p^0r_q^{0}r_p^{1}r_q^{1}$ -- see Figure~\ref{fig:skeleton_unique_model_clique} for an illustration.

First, suppose the case
$$\phi_Q \Vert \{r_q,r_p,r_j\}^* \equiv r_p^0 r_j^0 r^0_q r_p^1 r_j^1 r_q^1 \quad \text{and} \quad 
\phi'_Q \Vert \{r_q,r_p,r_j\}^* \equiv r_p^0 r_q^0 r_j^0 r_p^1 r_q^1 r_j^1,$$
see Figure~\ref{fig:skeleton_unique_model_clique} to the left.
We claim that $\inside(M_q) \neq \emptyset$.
Suppose to the contrary that there is $P \in N_{\pqstree}(Q)$ such that $P \in \inside(M_q)$.
Assume that $P \in \leftside(r_p)$.
Then, in $\phi_Q$ the letter $P$ is on the left side of $\phi_Q(r_j)$
and in the model $\phi'_Q$ the letter $P$ is on the right side of $\phi'_Q(r_j)$,
which can not be the case.
The case when $P \in \rightside(r_p)$ is proven analogously.
Using similar arguments we show that $\inside(M_j) = \emptyset$.
To complete the proof in this case, we show that for every $P \in N_{\pqstree}(Q)$,
$$P \in \leftside(r_q) \iff P \in \leftside(r_j),$$
which shows $r_j K r_q$ and contradicts the fact that $r_j$ and $r_q$ are from two different sets in $K(Q)$.
Suppose that $P \in \leftside(r_q)$ and $P \in \rightside(r_j)$.
Then $P$ is between $r^1_j$ and $r^1_q$ in $\phi_Q$
and between $r^0_q$  and $r^0_j$ in $\phi'_Q$.
Thus, $P$ is on the right side of $\phi_Q(r_p)$ and on the left side of $\phi'_Q(r_p)$, which can not be the case.
The second case is proven analogously.

Next, suppose the case
$$\phi_Q \Vert \{r_q,r_p,r_j\}^* \equiv r_p^0 r_j^1 r^0_q r_p^1 r_j^0 r_q^1 \quad \text{and} \quad
\phi'_Q \Vert \{r_q,r_p,r_j\}^* \equiv r_p^0 r_j^0 r_q^0 r_p^1 r_j^1 r_q^1,$$
see Figure~\ref{fig:skeleton_unique_model_clique} to the right.
First, note that $\inside(M_q) = \inside(M_p) = \emptyset$
as otherwise a P-node $P$ from $\inside(M_q) \cup \inside(M_p)$
would be on different sides of $\phi_Q(r_j)$ and $\phi'_Q(r_j)$.
Now, we show that for every node $P \in N_{\pqstree}(Q)$,
$$P \in \leftside(r_p) \iff P \in \rightside(r_q), $$
which shows $r_p K r_q$ and contradicts the fact that $r_p$ and $r_q$ are from different sets in $K(Q)$.
If $P \in \leftside(r_p) \cap \leftside(r_q)$, then
$P$ is on the right side of $\phi_Q(r_j)$ and on the left side of $\phi'_Q(r_j)$, 
which can not be the case.
The second case is proven analogously.

The other cases corresponding to other placements of the chords $\phi_Q(r_j)$ and $\phi'_Q(r_j)$
in the circular word $r_p^0r_q^{0}r_p^{1}r_q^{1}$ are proven similarly.
This completes the proof of~\eqref{item:serial_consistent_modules_skeleton}.

Let $\phi_Q$ be an extended conformal model of $(Q,{\sim})$.
Statement~\eqref{item:serial_consistent_modules_contiguous_subwords} obviously holds 
when $K = M_i$ and $\inside(M_i) \neq \emptyset$.
Suppose $K \subseteq \bigcup \{M_i: i \in [n] \text{ and } \inside(M_i) = \emptyset\}$.
Let $P$ be any P-node in $N_{\pqstree}(Q)$; such a node exists as $G_{ov}$ is disconnected.
Suppose without loss of generality that $K$ is the union of $M_1,\ldots,M_m$, 
enumerated such that
$$\phi_Q \Vert (\{P\} \cup (M_1 \cup \ldots \cup M_m)^*) \equiv P \lambda^0_1 \ldots \lambda^0_m \lambda^1_1 \ldots \lambda^1_m,$$
where $(\lambda^0_i, \lambda^1_i)$ is an oriented permutation model of $(M_i,{\sim})$ for $i \in [m]$ --
see Figure \ref{fig:serial_consistent_modules_contiguous_subwords} for an illustration.
In particular, the words $\lambda^0_i$ and $\lambda^1_i$ are contiguous in $\phi_Q$ as $\inside(M_i) = \emptyset$.
So, if $m=1$, then statement~\eqref{item:serial_consistent_modules_contiguous_subwords} holds.
Suppose $m \geq 2$.
Denote by $l^0$ and $l^1$ the first and the last letter in $\lambda^0_1$,
by $l^2$ and $l^3$ the first and the last letter in $\lambda^0_m$,
by $r^0$ and $r^1$ the first and the last letter in $\lambda^1_1$,
and by $r^2$ and $r^3$ the first and the last letter in $\lambda^1_m$ --
see Figure \ref{fig:serial_consistent_modules_contiguous_subwords} for an illustration.
Note that there is no letter from $N_{\pqstree}(Q)$ between $l^1$ and $l^2$ or 
between $r^1$ and $r^2$ as otherwise $M_1$ and $M_m$ would not be in $K$-relation.
Suppose for a contrary that there is $v \in Q \setminus K$ such that $\phi_Q(v)$ has 
one end between $l^{1}$ and $l^2$ and the second end between $r^1$ and $r^2$.
Now, our previous observation asserts that we have $v K M_1$ and $v K M_m$, which is a contradiction as 
$v \notin K$.
Hence, the words $\lambda^0_1\ldots \lambda^0_m$ and $\lambda^1_1\ldots \lambda^1_m$ are contiguous in $\phi_Q$.
This completes the proof of statement~\eqref{item:serial_consistent_modules_contiguous_subwords}.
\end{proof}

\begin{figure}[h!]
\begin{tikzpicture}[xscale=1,yscale=1,>=latex]
\coordinate (center) at (0,0);

\coordinate (lN) at ($(center)+(270:2.3cm)$) {};

\coordinate (u0) at ($(center)+(120:2cm)$) {};
\coordinate (u1) at ($(center)+(60:2cm)$) {};
\coordinate (lu0) at ($(center)+(115:2.3cm)$) {};
\coordinate (lu1) at ($(center)+(65:2.3cm)$) {};

\coordinate (y0) at ($(center)+(180:2cm)$) {};
\coordinate (y1) at ($(center)+(0:2cm)$) {};
\coordinate (ly0) at ($(center)+(180:2.3cm)$) {};
\coordinate (ly1) at ($(center)+(0:2.3cm)$) {};
\draw[very thick] ([shift=(179:2cm)]0,0) arc (179:181:2cm);

\coordinate (k11) at ($(center)+(20:2cm)$) {};
\coordinate (lk11) at ($(center)+(20:2.3cm)$) {};
\coordinate (k12) at ($(center)+(50:2cm)$) {};
\coordinate (lk12) at ($(center)+(50:2.3cm)$) {};
\coordinate (k13) at ($(center)+(200:2cm)$) {};
\coordinate (lk13) at ($(center)+(200:2.3cm)$) {};
\coordinate (k14) at ($(center)+(230:2cm)$) {};
\coordinate (lk14) at ($(center)+(230:2.3cm)$) {};
\coordinate (k'j) at ($(center)+(215:1.6cm)$) {};

\coordinate (k21) at ($(center)+(-50:2cm)$) {};
\coordinate (lk21) at ($(center)+(-50:2.3cm)$) {};
\coordinate (k22) at ($(center)+(-20:2cm)$) {};
\coordinate (lk22) at ($(center)+(-20:2.3cm)$) {};
\coordinate (k23) at ($(center)+(130:2cm)$) {};
\coordinate (lk23) at ($(center)+(130:2.3cm)$) {};
\coordinate (k24) at ($(center)+(160:2cm)$) {};
\coordinate (lk24) at ($(center)+(160:2.3cm)$) {};
\coordinate (k''j) at ($(center)+(145:1.6cm)$) {};

\draw (0,0) circle (2cm);

\draw[thick,->] (y0)--(y1);

\draw[very thick,red,<-] ([shift=(20:2cm)]0,0) arc (20:50:2cm);
\coordinate (ll11) at ($(center)+(35:2.3cm)$) {};

\draw[very thick,red,<-] ([shift=(200:2cm)]0,0) arc (200:230:2cm);
\coordinate (ll01) at ($(center)+(215:2.3cm)$) {};

\draw[very thick,red] (k12)--(k13);
\draw[very thick,red] (k14)--(k11);

\draw[very thick,red,<-] ([shift=(-50:2cm)]0,0) arc (-50:-20:2cm);
\coordinate (ll1p) at ($(center)+(-35:2.3cm)$) {};
\draw[very thick,red,<-] ([shift=(130:2cm)]0,0) arc (130:160:2cm);
\coordinate (ll0p) at ($(center)+(145:2.3cm)$) {};
\draw[very thick,red] (k22)--(k23);
\draw[very thick,red] (k24)--(k21);

\draw[very thick,-] ([shift=(265:2.0cm)]0,0) arc (265:275:2.0cm);

\tikzstyle{every node}=[inner sep=1pt]
\begin{tiny}
\node at (ll01) {$\lambda^0_1$};
\node at (ll0p) {$\lambda^0_m$};
\node at (ll11) {$\lambda^1_1$};
\node at (ll1p) {$\lambda^1_m$};

\node at (lk11) {$r^1$};
\node at (lk12) {$r^0$};
\node at (lk13) {$l^1$};
\node at (lk14) {$l^0$};
\node at (lk21) {$r^3$};
\node at (lk22) {$r^2$};
\node at (lk23) {$l^3$};
\node at (lk24) {$l^2$};
\node at (k'j) {$M_1$};
\node at (k''j) {$M_m$};

\node at (ly0) {$v^0$};
\node at (ly1) {$v^1$};
\node at (lN) {$P$};

\end{tiny}
\draw[white] (-2.5,-2.5)--(-2.5,-2.3);
\draw[white] (2.5,2.5)--(2.5,2.3);
\end{tikzpicture} 

\caption{\label{fig:serial_consistent_modules_contiguous_subwords}
}
\end{figure}

\subsection{Refinement procedure}
\label{subsec:refinement_procedure}
Finally, we describe the refinement procedure that produces the CA-modules $\camodules(K)$ for every $K \in K(Q)$
and we show that $\camodules(K)$ satisfies Properties~\ref{prop:K_refinement_partition} -- \ref{prop:K_refinement_contiguous_subwords_order}.

Suppose $K$ is a member of $K(Q)$ for some prime/serial component $Q \in \mathcal{Q}$.
For every strong module $L$ in $\strongModules(K,{\sim})$ and every extended conformal model $\phi_Q$ of $(Q,{\sim})$ we denote by: 
\begin{itemize}
\item $L^0$ and $L^1$ the sets $K^0 \cap L^*$ and $K^1 \cap L^*$, respectively,
\item $\tau(\phi_Q,L^j)$ the shortest contiguous subword of $\phi_Q$ containing
all the letters from $L^j$ and no letter from $L^{1-j}$,
\item $\inside(L)$ the set of all P-nodes from $N_{\pqstree}(Q)$ that appear either in $\tau(\phi_Q,L^0)$ or in 
$\tau(\phi_Q,L^1)$.
\end{itemize}
Since $|K(Q)| \geq 2$, the words $\tau(\phi_Q,K^j)$ are properly defined and satisfy the statements of Claim~\ref{claim:tau_phi_M_K_properties} also for the case when $Q$ is serial.
In particular, it shows that the set $\inside(L)$ is properly defined.

To obtain a partition of $K$ into CA-modules in $\camodules(K)$ we perform the \emph{refinement procedure} on~$K$.
The procedure maintains a partition $\camodules(K)$ of the set $K$ into modules of $(K,{\sim})$.
The procedure marks each member of $\camodules(K)$ either as \emph{active} or \emph{inactive}.
Initially, we set $\camodules(K)=\{K\}$, we mark $K$ as active if $\inside(K) \neq \emptyset$, and as inactive if $\inside(K) = \emptyset$.
The procedure maintains the following invariants:
\begin{itemize}
 \item[\namedlabel{inv:1}{(I1)}] Every active set $L \in S(K)$ is a strong module in $\strongModules(K,{\sim})$ such that $\inside(L) \neq \emptyset$.
 \item[\namedlabel{inv:2}{(I2)}] For every extended conformal model $\phi_Q$ of $(Q,{\sim})$, every $L,T \in \camodules(K)$, and every $j \in \{0,1\}$ the letters from $L^j$ do not overlap with the letters from $T^j$ in the word $\tau(\phi,K^j)$. 
 Moreover, if $\phi'_Q$ and $\phi_Q$ are two extended conformal models of $(Q,{\sim})$ such that $\phi_Q \Vert Q^*$ and $\phi'_Q \Vert Q^*$ are admissible for $\gamma^t(Q)$ for some $t \in \{0,1\}$, then the letters from $L^j$ occur before the letters from $T^j$ in $\tau(\phi_Q,K^j)$ if and only if the letters from $L^j$ occur before the letters from $T^j$ in $\tau(\phi'_Q,K^j)$.
\end{itemize}
The procedure is performed in steps as long as they are active modules in the set $\camodules(K)$.
In a single step, an active module $L$ from $\camodules(K)$ is partitioned into some active/inactive subsets of $L$.
When the procedure is over, all modules in $\camodules(S)$ are inactive; we show that they form the set of all CA-modules of $K$.
Moreover, given~\ref{inv:1} and \ref{inv:2} we show easily that the set $\camodules(K)$ satisfies 
Properties~\ref{prop:K_refinement_partition} -- \ref{prop:K_refinement_contiguous_subwords_order}.

Let $\phi_Q$ and $\phi^Q$ be two extended conformal models of $(Q,{\sim})$ such that 
$\phi_Q \Vert Q^*$ and $\phi'_Q \Vert Q^*$ are admissible for $\gamma^t(Q)$ for some $t \in \{0,1\}$.
Let $\KKK = (K^0,K^1,{<_K})$ be the metachord of~$K$.
Assume that the set $K$ induces in $\phi_Q \Vert Q^*$ and $\phi'_Q \Vert Q^*$ 
consistent permutation models (they are admissible to $\KKK$) that correspond to the transitive orientations ${\prec_{\phi}}$ and ${\prec_{\phi'}}$ of $(K,{\sim})$, respectively.

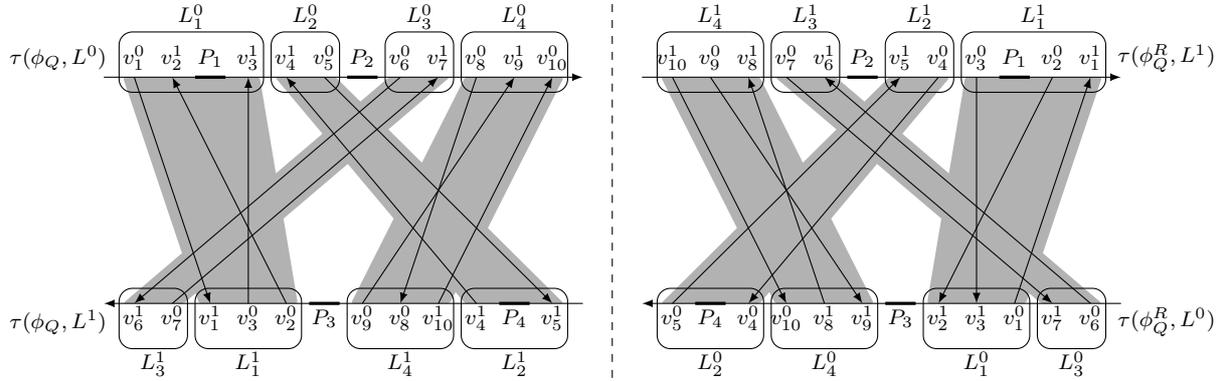
\begin{figure}[!htp]
\begin{tikzpicture}[xscale=0.50,yscale=1,>=latex]
\coordinate (u1) at (0,3) {};
\coordinate (u2) at (1,3) {};
\coordinate (u3) at (2,3) {};
\coordinate (u4) at (3,3) {};
\coordinate (u5) at (4,3) {};
\coordinate (u6) at (5,3) {};
\coordinate (u7) at (6,3) {};
\coordinate (u8) at (7,3) {};
\coordinate (u9) at (8,3) {};
\coordinate (u10) at (9,3) {};
\coordinate (u11) at (10,3) {};
\coordinate (u12) at (11,3) {};

\coordinate (lu1) at (0,3.25) {};
\coordinate (lu2) at (1,3.25) {};
\coordinate (lu3) at (2,3.25) {};
\coordinate (lu4) at (3,3.25) {};
\coordinate (lu5) at (4,3.25) {};
\coordinate (lu6) at (5,3.25) {};
\coordinate (lu7) at (6,3.25) {};
\coordinate (lu8) at (7,3.25) {};
\coordinate (lu9) at (8,3.25) {};
\coordinate (lu10) at (9,3.25) {};
\coordinate (lu11) at (10,3.25) {};
\coordinate (lu12) at (11,3.25) {};

\coordinate (b1) at (0,0) {};
\coordinate (b2) at (1,0) {};
\coordinate (b3) at (2,0) {};
\coordinate (b4) at (3,0) {};
\coordinate (b5) at (4,0) {};
\coordinate (b6) at (5,0) {};
\coordinate (b7) at (6,0) {};
\coordinate (b8) at (7,0) {};
\coordinate (b9) at (8,0) {};
\coordinate (b10) at (9,0) {};
\coordinate (b11) at (10,0) {};
\coordinate (b12) at (11,0) {};

\coordinate (lb1) at (0,-0.2) {};
\coordinate (lb2) at (1,-0.2) {};
\coordinate (lb3) at (2,-0.2) {};
\coordinate (lb4) at (3,-0.2) {};
\coordinate (lb5) at (4,-0.2) {};
\coordinate (lb6) at (5,-0.2) {};
\coordinate (lb7) at (6,-0.2) {};
\coordinate (lb8) at (7,-0.2) {};
\coordinate (lb9) at (8,-0.2) {};
\coordinate (lb10) at (9,-0.2) {};
\coordinate (lb11) at (10,-0.2) {};
\coordinate (lb12) at (11,-0.2) {};

\begin{scope}[fill opacity=0.5]
\draw[rounded corners=1, draw=none, fill=gray!60] (-0.3,3)--(3.3,3) -- (4.3,0) -- (1.7,0)--cycle;
\draw[rounded corners=1, draw=none, fill=gray!60] (6.7,3)--(8.3,3) -- (1.3,0) -- (-0.3,0)--cycle;
\draw[rounded corners=1, draw=none, fill=gray!60] (3.7,3)--(5.3,3) -- (11.3,0) -- (8.7,0)--cycle;
\draw[rounded corners=1, draw=none, fill=gray!60] (8.7,3)--(11.3,3) -- (8.3,0) -- (5.7,0)--cycle;

\end{scope}

\coordinate (tau0) at (-2,3.25);
\coordinate (tau1) at (-2,-0.25);

\coordinate (llu1) at (1.5,3.8) {};
\coordinate (llu2) at (4.5,3.8) {};
\coordinate (llu3) at (7.5,3.8) {};
\coordinate (llu4) at (10,3.8) {};

\coordinate (slu1) at (1.5,4.2) {};
\coordinate (slu2) at (4.5,4.2) {};
\coordinate (slu3) at (7.5,4.2) {};
\coordinate (slu4) at (10,4.2) {};

\coordinate (llb1) at (0.5,-0.8) {};
\coordinate (llb2) at (3,-0.8) {};
\coordinate (llb3) at (7,-0.8) {};
\coordinate (llb4) at (10,-0.8) {};

\coordinate (slb1) at (0.5,-1.2) {};
\coordinate (slb2) at (3,-1.2) {};
\coordinate (slb3) at (7,-1.2) {};
\coordinate (slb4) at (10,-1.2) {};

\tikzstyle{every node}=[inner sep=2pt,fill=white]

\draw[->] (u1)--(b3);
\draw[<-] (u2)--(b5);
\draw[-,very thick] ($(u3)+(-0.4,0)$) -- ($(u3)+(0.4,0)$);
\draw[<-] (u4)--(b4);
\draw[<-] (u5)--(b10);
\draw[->] (u6)--(b12);
\draw[-,very thick] ($(u7)+(-0.4,0)$) -- ($(u7)+(0.4,0)$);
\draw[->] (u8)--(b1);
\draw[<-] (u9)--(b2);
\draw[->] (u10)--(b8);
\draw[<-] (u11)--(b7);
\draw[<-] (u12)--(b9);

\draw[very thick, -] ($(b6)+(-0.4,0)$) -- ($(b6)+(0.4,0)$);
\draw[very thick, -] ($(b11)+(-0.4,0)$) -- ($(b11)+(0.4,0)$);

\draw[->] (-0.8,3) -- (11.8,3);
\draw[<-] (-0.8,0) -- (11.8,0);

\draw[rounded corners=5pt] (-0.4,2.8) rectangle ++(3.8,0.8);
\draw[rounded corners=5pt] (3.6,2.8) rectangle ++(1.8,0.8);
\draw[rounded corners=5pt] (6.6,2.8) rectangle ++(1.8,0.8);
\draw[rounded corners=5pt] (8.6,2.8) rectangle ++(2.8,0.8);

\draw[rounded corners=5pt] (-0.4,-0.6) rectangle ++(1.8,0.8);
\draw[rounded corners=5pt] (1.6,-0.6) rectangle ++(2.8,0.8);
\draw[rounded corners=5pt] (5.6,-0.6) rectangle ++(2.8,0.8);
\draw[rounded corners=5pt] (8.6,-0.6) rectangle ++(2.8,0.8);

\tikzstyle{every node}=[inner sep=1pt]
\begin{tiny}
\node at (llu1) {$L^0_1$};
\node at (llu2) {$L^0_2$};
\node at (llu3) {$L^0_3$};
\node at (llu4) {$L^0_4$};

\node at (llb1) {$L^1_3$};
\node at (llb2) {$L^1_1$};
\node at (llb3) {$L^1_4$};
\node at (llb4) {$L^1_2$};

\node at (lu1) {$v^0_1$};
\node at (lu2) {$v^1_2$};
\node at (lu3) {$P_1$};
\node at (lu4) {$v^1_3$};
\node at (lu5) {$v^1_4$};
\node at (lu6) {$v^0_5$};
\node at (lu7) {$P_2$};
\node at (lu8) {$v^0_6$};
\node at (lu9) {$v^1_7$};
\node at (lu10) {$v^0_8$};
\node at (lu11) {$v^1_9$};
\node at (lu12) {$v^0_{10}$};

\node at (lb1) {$v^1_6$};
\node at (lb2) {$v^0_7$};
\node at (lb3) {$v^1_1$};
\node at (lb4) {$v^0_3$};
\node at (lb5) {$v^0_2$};
\node at (lb6) {$P_3$};
\node at (lb7) {$v^0_9$};
\node at (lb8) {$v^0_{8}$};
\node at (lb9) {$v^1_{10}$};
\node at (lb10) {$v^1_4$};
\node at (lb11) {$P_4$};
\node at (lb12) {$v^1_5$};

\node at (tau0) {$\tau(\phi_Q,L^0)$};
\node at (tau1) {$\tau(\phi_Q,L^1)$};

\end{tiny}
\end{tikzpicture}
\hspace{0.1cm}
\begin{tikzpicture}[xscale=0.50,yscale=1,>=latex]
\draw[dashed] (0,-0.5) -- (0,4.5); 
\end{tikzpicture}
\hspace{0.1cm}
\begin{tikzpicture}[xscale=0.50,yscale=1,>=latex]
\coordinate (u1) at (-0,3) {};
\coordinate (u2) at (-1,3) {};
\coordinate (u3) at (-2,3) {};
\coordinate (u4) at (-3,3) {};
\coordinate (u5) at (-4,3) {};
\coordinate (u6) at (-5,3) {};
\coordinate (u7) at (-6,3) {};
\coordinate (u8) at (-7,3) {};
\coordinate (u9) at (-8,3) {};
\coordinate (u10) at (-9,3) {};
\coordinate (u11) at (-10,3) {};
\coordinate (u12) at (-11,3) {};

\coordinate (lu1) at (-0,3.25) {};
\coordinate (lu2) at (-1,3.25) {};
\coordinate (lu3) at (-2,3.25) {};
\coordinate (lu4) at (-3,3.25) {};
\coordinate (lu5) at (-4,3.25) {};
\coordinate (lu6) at (-5,3.25) {};
\coordinate (lu7) at (-6,3.25) {};
\coordinate (lu8) at (-7,3.25) {};
\coordinate (lu9) at (-8,3.25) {};
\coordinate (lu10) at (-9,3.25) {};
\coordinate (lu11) at (-10,3.25) {};
\coordinate (lu12) at (-11,3.25) {};

\coordinate (b1) at (0,0) {};
\coordinate (b2) at (-1,0) {};
\coordinate (b3) at (-2,0) {};
\coordinate (b4) at (-3,0) {};
\coordinate (b5) at (-4,0) {};
\coordinate (b6) at (-5,0) {};
\coordinate (b7) at (-6,0) {};
\coordinate (b8) at (-7,0) {};
\coordinate (b9) at (-8,0) {};
\coordinate (b10) at (-9,0) {};
\coordinate (b11) at (-10,0) {};
\coordinate (b12) at (-11,0) {};

\begin{scope}[fill opacity=0.5]
\draw[rounded corners=1, draw=none, fill=gray!60] (0.3,3)--(-3.3,3) -- (-4.3,0) -- (-1.7,0)--cycle;
\draw[rounded corners=1, draw=none, fill=gray!60] (-6.7,3)--(-8.3,3) -- (-1.3,0) -- (0.3,0)--cycle;
\draw[rounded corners=1, draw=none, fill=gray!60] (-3.7,3)--(-5.3,3) -- (-11.3,0) -- (-8.7,0)--cycle;
\draw[rounded corners=1, draw=none, fill=gray!60] (-8.7,3)--(-11.3,3) -- (-8.3,0) -- (-5.7,0)--cycle;

\end{scope}

\coordinate (lb1) at (-0,-0.2) {};
\coordinate (lb2) at (-1,-0.2) {};
\coordinate (lb3) at (-2,-0.2) {};
\coordinate (lb4) at (-3,-0.2) {};
\coordinate (lb5) at (-4,-0.2) {};
\coordinate (lb6) at (-5,-0.2) {};
\coordinate (lb7) at (-6,-0.2) {};
\coordinate (lb8) at (-7,-0.2) {};
\coordinate (lb9) at (-8,-0.2) {};
\coordinate (lb10) at (-9,-0.2) {};
\coordinate (lb11) at (-10,-0.2) {};
\coordinate (lb12) at (-11,-0.2) {};

\coordinate (tau0) at (2,3.25);
\coordinate (tau1) at (2,-0.25);

\coordinate (llu1) at (-1.5,3.8) {};
\coordinate (llu2) at (-4.5,3.8) {};
\coordinate (llu3) at (-7.5,3.8) {};
\coordinate (llu4) at (-10,3.8) {};

\coordinate (slu1) at (-1.5,4.2) {};
\coordinate (slu2) at (-4.5,4.2) {};
\coordinate (slu3) at (-7.5,4.2) {};
\coordinate (slu4) at (-10,4.2) {};

\coordinate (llb1) at (-0.5,-0.8) {};
\coordinate (llb2) at (-3,-0.8) {};
\coordinate (llb3) at (-7,-0.8) {};
\coordinate (llb4) at (-10,-0.8) {};

\coordinate (slb1) at (-0.5,-1.2) {};
\coordinate (slb2) at (-3,-1.2) {};
\coordinate (slb3) at (-7,-1.2) {};
\coordinate (slb4) at (-10,-1.2) {};

\tikzstyle{every node}=[inner sep=2pt,fill=white]

\draw[<-] (u1)--(b3);
\draw[->] (u2)--(b5);
\draw[-,very thick] ($(u3)+(-0.4,0)$) -- ($(u3)+(0.4,0)$);
\draw[->] (u4)--(b4);
\draw[->] (u5)--(b10);
\draw[<-] (u6)--(b12);
\draw[-,very thick] ($(u7)+(-0.4,0)$) -- ($(u7)+(0.4,0)$);
\draw[<-] (u8)--(b1);
\draw[->] (u9)--(b2);
\draw[<-] (u10)--(b8);
\draw[->] (u11)--(b7);
\draw[->] (u12)--(b9);

\draw[very thick, -] ($(b6)+(-0.4,0)$) -- ($(b6)+(0.4,0)$);
\draw[very thick, -] ($(b11)+(-0.4,0)$) -- ($(b11)+(0.4,0)$);

\draw[<-] (0.8,3) -- (-11.8,3);
\draw[->] (0.8,0) -- (-11.8,0);

\draw[rounded corners=5pt] (0.4,2.8) rectangle ++(-3.8,0.8);
\draw[rounded corners=5pt] (-3.6,2.8) rectangle ++(-1.8,0.8);
\draw[rounded corners=5pt] (-6.6,2.8) rectangle ++(-1.8,0.8);
\draw[rounded corners=5pt] (-8.6,2.8) rectangle ++(-2.8,0.8);

\draw[rounded corners=5pt] (0.4,-0.6) rectangle ++(-1.8,0.8);
\draw[rounded corners=5pt] (-1.6,-0.6) rectangle ++(-2.8,0.8);
\draw[rounded corners=5pt] (-5.6,-0.6) rectangle ++(-2.8,0.8);
\draw[rounded corners=5pt] (-8.6,-0.6) rectangle ++(-2.8,0.8);

\tikzstyle{every node}=[inner sep=1pt]
\begin{tiny}
\node at (llu1) {$L^1_1$};
\node at (llu2) {$L^1_2$};
\node at (llu3) {$L^1_3$};
\node at (llu4) {$L^1_4$};

\node at (llb1) {$L^0_3$};
\node at (llb2) {$L^0_1$};
\node at (llb3) {$L^0_4$};
\node at (llb4) {$L^0_2$};

\node at (lu1) {$v^1_1$};
\node at (lu2) {$v^0_2$};
\node at (lu3) {$P_1$};
\node at (lu4) {$v^0_3$};
\node at (lu5) {$v^0_4$};
\node at (lu6) {$v^1_5$};
\node at (lu7) {$P_2$};
\node at (lu8) {$v^1_6$};
\node at (lu9) {$v^0_7$};
\node at (lu10) {$v^1_8$};
\node at (lu11) {$v^0_9$};
\node at (lu12) {$v^1_{10}$};

\node at (lb1) {$v^0_6$};
\node at (lb2) {$v^1_7$};
\node at (lb3) {$v^0_1$};
\node at (lb4) {$v^1_3$};
\node at (lb5) {$v^1_2$};
\node at (lb6) {$P_3$};
\node at (lb7) {$v^1_9$};
\node at (lb8) {$v^1_{8}$};
\node at (lb9) {$v^0_{10}$};
\node at (lb10) {$v^0_4$};
\node at (lb11) {$P_4$};
\node at (lb12) {$v^0_5$};

\node at (tau0) {$\tau(\phi^R_Q,L^1)$};
\node at (tau1) {$\tau(\phi^R_Q,L^0)$};

\end{tiny}
\end{tikzpicture}

\caption{\label{fig:slot_refinement_prime} $L$ is prime. 
We replace $L$ in $\camodules(K)$ by $L_1,L_2,L_3,L_4$; $L_1, L_2$ are active, $L_3,L_4$ are inactive.
} 
\end{figure}

Suppose $L$ is an active member of $\camodules(K)$ such that $L$ is prime in $\strongModules(K,{\sim})$.
Suppose $L_1,\ldots,L_n$ are the children of $L$ in $\strongModules(K,{\sim})$ enumerated such that the letters from $L^0_i$ occur before the letters from $L^0_j$ in $\tau(\phi_Q,K^0)$ for $i < j$.
Then we delete $L$ from $\camodules(K)$, add $L_1,\ldots,L_n$ to $\camodules(K)$, we mark $L_i$ such that $\inside(L_i) \neq \emptyset$ as active and as inactive otherwise
-- see Figure~\ref{fig:slot_refinement_prime} for an illustration.
To show that Invariant~\ref{inv:2} is kept it suffices to prove that the transitive orientations ${\prec_{\phi}}$ and ${\prec_{\phi'}}$ of 
$(K,{\sim})$ restricted to the edges of $(L,{\sim}_{L})$ are equal
(recall that $(L,{\sim_L})$ contains all the edges from $(L,{\sim})$ that have both endpoints in different children of $L$).
Since $L$ is active, there is a letter $P$ from $N_{\pqstree}(Q)$ such that $P \in \tau(\phi_Q,L^j)$ for some $j \in \{0,1\}$.
Suppose $P \in \tau(\phi_Q,L^0)$. 
Since $(L,{\sim})$ is prime, there are 
$u \in L_i$ and $v \in L_j$ 
for some $L_i \sim L_j$
such that $u'Pv'$ is a subword of $\tau(\phi_Q,L^0)$ for some $u' \in \{u^0,u^1\}$ and $v' \in \{v^0,v^1\}$. This means $L_i \prec_{\phi} L_j$.
By Claim~\ref{claim:tau_phi_M_K_properties}, $u'Pv'$ is also a subword of $\tau(\phi'_Q,L^0)$, 
and hence we have $L_i \prec_{\phi'} L_j$.
Since $(L,{\sim_L})$ has two transitive orientation, one being the reverse of the other, we deduce that
${\prec_{\phi}}$ and ${\prec_{\phi'}}$ restricted to $(L,{\sim_L})$ are equal.

\begin{figure}[!htp]
\begin{tikzpicture}[xscale=0.52,yscale=1,>=latex]
\coordinate (u1) at (0,3) {};
\coordinate (u2) at (1,3) {};
\coordinate (u3) at (2,3) {};
\coordinate (u4) at (3,3) {};
\coordinate (u5) at (4,3) {};
\coordinate (u6) at (5,3) {};
\coordinate (u7) at (6,3) {};
\coordinate (u8) at (7,3) {};
\coordinate (u9) at (8,3) {};
\coordinate (u10) at (9,3) {};
\coordinate (u11) at (10,3) {};

\coordinate (lu1) at (0,3.25) {};
\coordinate (lu2) at (1,3.25) {};
\coordinate (lu3) at (2,3.25) {};
\coordinate (lu4) at (3,3.25) {};
\coordinate (lu5) at (4,3.25) {};
\coordinate (lu6) at (5,3.25) {};
\coordinate (lu7) at (6,3.25) {};
\coordinate (lu8) at (7,3.25) {};
\coordinate (lu9) at (8,3.25) {};
\coordinate (lu10) at (9,3.25) {};
\coordinate (lu11) at (10,3.25) {};

\coordinate (b1) at (0,0) {};
\coordinate (b2) at (1,0) {};
\coordinate (b3) at (2,0) {};
\coordinate (b4) at (3,0) {};
\coordinate (b5) at (4,0) {};
\coordinate (b6) at (5,0) {};
\coordinate (b7) at (6,0) {};
\coordinate (b8) at (7,0) {};
\coordinate (b9) at (8,0) {};
\coordinate (b10) at (9,0) {};
\coordinate (b11) at (10,0) {};

\coordinate (lb1) at (0,-0.2) {};
\coordinate (lb2) at (1,-0.2) {};
\coordinate (lb3) at (2,-0.2) {};
\coordinate (lb4) at (3,-0.2) {};
\coordinate (lb5) at (4,-0.2) {};
\coordinate (lb6) at (5,-0.2) {};
\coordinate (lb7) at (6,-0.2) {};
\coordinate (lb8) at (7,-0.2) {};
\coordinate (lb9) at (8,-0.2) {};
\coordinate (lb10) at (9,-0.2) {};
\coordinate (lb11) at (10,-0.2) {};

\begin{scope}[fill opacity=0.5]
\draw[rounded corners=1, draw=none, fill=gray!60] (-0.3,3)--(1.3,3) -- (10.3,0) -- (8.7,0)--cycle;
\draw[rounded corners=1, draw=none, fill=gray!60] (1.8,3)--(4.2,3) -- (7.2,0) -- (5.8,0)--cycle;
\draw[rounded corners=1, draw=none, fill=gray!60] (4.8,3)--(6.2,3) -- (5.2,0) -- (3.8,0)--cycle;
\draw[rounded corners=1, draw=none, fill=gray!60] (6.8,3)--(8.2,3) -- (3.2,0) -- (1.8,0)--cycle;
\draw[rounded corners=1, draw=none, fill=gray!60] (9.7,3)--(10.3,3) -- (0.3,0) -- (-0.3,0)--cycle;
\end{scope}

\coordinate (tau0) at (-2,3.25);
\coordinate (tau1) at (-2,-0.25);

\tikzstyle{every node}=[inner sep=2pt,fill=white]

\draw[<-] (u1)--(b10);
\draw[->] (u2)--(b11);
\draw[<-] (u3) -- (b7);
\draw[->] (u5) -- (b8);
\draw[->] (u6)--(b5);
\draw[<-] (u7)--(b6);
\draw[<-] (u8)--(b3);
\draw[->] (u9)--(b4);
\draw[<-] (u11)--(b1);

\draw[-,very thick] ($(u4)+(-0.4,0)$) -- ($(u4)+(0.4,0)$);
\draw[-,very thick] ($(u10)+(-0.4,0)$) -- ($(u10)+(0.4,0)$);
\draw[-,very thick] ($(b2)+(-0.4,0)$) -- ($(b2)+(0.4,0)$);
\draw[-,very thick] ($(b9)+(-0.4,0)$) -- ($(b9)+(0.4,0)$);

\draw[->] (-0.8,3) -- (10.8,3);
\draw[<-] (-0.8,0) -- (10.8,0);

\coordinate (lru1) at (0.7,3.8) {};
\coordinate (lru2) at (3,3.8) {};
\coordinate (lru3) at (5.5,3.8) {};
\coordinate (lru4) at (7.5,3.8) {};
\coordinate (lru5) at (10,3.8) {};

\coordinate (sru1) at (0.7,4.2) {};
\coordinate (sru2) at (3,4.2) {};
\coordinate (sru3) at (6.5,4.2) {};
\coordinate (sru4) at (10,4.2) {};

\coordinate (lrb1) at (0,-0.8) {};
\coordinate (lrb2) at (2.5,-0.8) {};
\coordinate (lrb3) at (4.5,-0.8) {};
\coordinate (lrb4) at (6.5,-0.8) {};
\coordinate (lrb5) at (9.5,-0.8) {};

\coordinate (srb1) at (0,-1.2) {};
\coordinate (srb2) at (3.5,-1.2) {};
\coordinate (srb3) at (6.5,-1.2) {};
\coordinate (srb4) at (9.5,-1.2) {};

\draw[rounded corners=5pt] (-0.4,2.8) rectangle ++(1.8,0.8);
\draw[rounded corners=5pt] (1.6,2.8) rectangle ++(2.8,0.8);
\draw[rounded corners=5pt] (4.6,2.8) rectangle ++(1.8,0.8);
\draw[rounded corners=5pt] (6.6,2.8) rectangle ++(1.8,0.8);
\draw[rounded corners=5pt] (9.6,2.8) rectangle ++(0.8,0.8);

\draw[rounded corners=5pt] (-0.4,-0.6) rectangle ++(0.8,0.8);
\draw[rounded corners=5pt] (1.6,-0.6) rectangle ++(1.8,0.8);
\draw[rounded corners=5pt] (3.6,-0.6) rectangle ++(1.8,0.8);
\draw[rounded corners=5pt] (5.6,-0.6) rectangle ++(1.8,0.8);
\draw[rounded corners=5pt] (8.6,-0.6) rectangle ++(1.8,0.8);

\tikzstyle{every node}=[inner sep=1pt]
\begin{tiny}
\node at (lru1) {$R^0_1$};
\node at (lru2) {$R^0_2$};
\node at (lru3) {$R^0_3$};
\node at (lru4) {$R^0_4$};
\node at (lru5) {$R^0_5$};

\node at (sru1) {$L^0_1$};
\node at (sru2) {$L^0_2$};
\node at (sru3) {$L^0_3$};
\node at (sru4) {$L^0_4$};

\node at (lrb1) {$R^1_5$};
\node at (lrb2) {$R^1_4$};
\node at (lrb3) {$R^1_3$};
\node at (lrb4) {$R^1_2$};
\node at (lrb5) {$R^1_1$};

\node at (srb1) {$L^1_4$};
\node at (srb2) {$L^1_3$};
\node at (srb3) {$L^1_2$};
\node at (srb4) {$L^1_1$};

\node at (lu1) {$v^1_1$};
\node at (lu2) {$v^0_2$};
\node at (lu3) {$v^1_3$};
\node at (lu4) {$P_1$};
\node at (lu5) {$v^0_4$};
\node at (lu6) {$v^0_5$};
\node at (lu7) {$v^1_6$};
\node at (lu8) {$v^1_7$};
\node at (lu9) {$v^0_8$};
\node at (lu10) {$P_2$};
\node at (lu11) {$v^1_9$};

\node at (lb1) {$v^0_9$};
\node at (lb2) {$P_3$};
\node at (lb3) {$v^0_8$};
\node at (lb4) {$v^1_7$};
\node at (lb5) {$v^1_5$};
\node at (lb6) {$v^0_6$};
\node at (lb7) {$v^0_3$};
\node at (lb8) {$v^1_4$};
\node at (lb9) {$P_4$};
\node at (lb10) {$v^0_1$};
\node at (lb11) {$v^1_2$};

\node at (tau0) {$\tau(\phi_Q,L^0)$};
\node at (tau1) {$\tau(\phi_Q,L^1)$};

\end{tiny}
\end{tikzpicture}
\hspace{0.1cm}
\begin{tikzpicture}[xscale=0.50,yscale=1,>=latex]
\draw[dashed] (0,-0.4) -- (0,5.3); 
\end{tikzpicture}
\hspace{0.1cm}
\begin{tikzpicture}[xscale=0.52,yscale=1,>=latex]
\coordinate (u1) at (-0,3) {};
\coordinate (u2) at (-1,3) {};
\coordinate (u3) at (-2,3) {};
\coordinate (u4) at (-3,3) {};
\coordinate (u5) at (-4,3) {};
\coordinate (u6) at (-5,3) {};
\coordinate (u7) at (-6,3) {};
\coordinate (u8) at (-7,3) {};
\coordinate (u9) at (-8,3) {};
\coordinate (u10) at (-9,3) {};
\coordinate (u11) at (-10,3) {};

\coordinate (lu1) at (-0,3.25) {};
\coordinate (lu2) at (-1,3.25) {};
\coordinate (lu3) at (-2,3.25) {};
\coordinate (lu4) at (-3,3.25) {};
\coordinate (lu5) at (-4,3.25) {};
\coordinate (lu6) at (-5,3.25) {};
\coordinate (lu7) at (-6,3.25) {};
\coordinate (lu8) at (-7,3.25) {};
\coordinate (lu9) at (-8,3.25) {};
\coordinate (lu10) at (-9,3.25) {};
\coordinate (lu11) at (-10,3.25) {};

\coordinate (b1) at (-0,0) {};
\coordinate (b2) at (-1,0) {};
\coordinate (b3) at (-2,0) {};
\coordinate (b4) at (-3,0) {};
\coordinate (b5) at (-4,0) {};
\coordinate (b6) at (-5,0) {};
\coordinate (b7) at (-6,0) {};
\coordinate (b8) at (-7,0) {};
\coordinate (b9) at (-8,0) {};
\coordinate (b10) at (-9,0) {};
\coordinate (b11) at (-10,0) {};

\coordinate (lb1) at (-0,-0.2) {};
\coordinate (lb2) at (-1,-0.2) {};
\coordinate (lb3) at (-2,-0.2) {};
\coordinate (lb4) at (-3,-0.2) {};
\coordinate (lb5) at (-4,-0.2) {};
\coordinate (lb6) at (-5,-0.2) {};
\coordinate (lb7) at (-6,-0.2) {};
\coordinate (lb8) at (-7,-0.2) {};
\coordinate (lb9) at (-8,-0.2) {};
\coordinate (lb10) at (-9,-0.2) {};
\coordinate (lb11) at (-10,-0.2) {};

\begin{scope}[fill opacity=0.5]
\draw[rounded corners=1, draw=none, fill=gray!60] (0.3,3)--(-1.3,3) -- (-10.3,0) -- (-8.7,0)--cycle;
\draw[rounded corners=1, draw=none, fill=gray!60] (-1.8,3)--(-4.2,3) -- (-7.2,0) -- (-5.8,0)--cycle;
\draw[rounded corners=1, draw=none, fill=gray!60] (-4.8,3)--(-6.2,3) -- (-5.2,0) -- (-3.8,0)--cycle;
\draw[rounded corners=1, draw=none, fill=gray!60] (-6.8,3)--(-8.2,3) -- (-3.2,0) -- (-1.8,0)--cycle;
\draw[rounded corners=1, draw=none, fill=gray!60] (-9.7,3)--(-10.3,3) -- (-0.3,0) -- (0.3,0)--cycle;
\end{scope}

\coordinate (tau0) at (2,3.25);
\coordinate (tau1) at (2,-0.25);

\tikzstyle{every node}=[inner sep=2pt,fill=white]

\draw[->] (u1)--(b10);
\draw[<-] (u2)--(b11);
\draw[->] (u3) -- (b7);
\draw[<-] (u5) -- (b8);
\draw[<-] (u6)--(b5);
\draw[->] (u7)--(b6);
\draw[->] (u8)--(b3);
\draw[<-] (u9)--(b4);
\draw[->] (u11)--(b1);

\draw[-,very thick] ($(u4)+(-0.4,0)$) -- ($(u4)+(0.4,0)$);
\draw[-,very thick] ($(u10)+(-0.4,0)$) -- ($(u10)+(0.4,0)$);
\draw[-,very thick] ($(b2)+(-0.4,0)$) -- ($(b2)+(0.4,0)$);
\draw[-,very thick] ($(b9)+(-0.4,0)$) -- ($(b9)+(0.4,0)$);

\draw[<-] (0.8,3) -- (-10.8,3);
\draw[->] (0.8,0) -- (-10.8,0);

\coordinate (lru1) at (-0.7,3.8) {};
\coordinate (lru2) at (-3,3.8) {};
\coordinate (lru3) at (-5.5,3.8) {};
\coordinate (lru4) at (-7.5,3.8) {};
\coordinate (lru5) at (-10,3.8) {};

\coordinate (sru1) at (-0.7,4.2) {};
\coordinate (sru2) at (-3,4.2) {};
\coordinate (sru3) at (-6.5,4.2) {};
\coordinate (sru4) at (-10,4.2) {};

\draw[rounded corners=5pt] (0.4,2.8) rectangle ++(-1.8,0.8);
\draw[rounded corners=5pt] (-1.6,2.8) rectangle ++(-2.8,0.8);
\draw[rounded corners=5pt] (-4.6,2.8) rectangle ++(-1.8,0.8);
\draw[rounded corners=5pt] (-6.6,2.8) rectangle ++(-1.8,0.8);
\draw[rounded corners=5pt] (-9.6,2.8) rectangle ++(-0.8,0.8);

\coordinate (lrb1) at (-0,-0.8) {};
\coordinate (lrb2) at (-2.5,-0.8) {};
\coordinate (lrb3) at (-4.5,-0.8) {};
\coordinate (lrb4) at (-6.5,-0.8) {};
\coordinate (lrb5) at (-9.5,-0.8) {};

\coordinate (srb1) at (-0,-1.2) {};
\coordinate (srb2) at (-2.5,-1.2) {};
\coordinate (srb3) at (-5.5,-1.2) {};
\coordinate (srb4) at (-9.5,-1.2) {};

\draw[rounded corners=5pt] (0.4,-0.6) rectangle ++(-0.8,0.8);
\draw[rounded corners=5pt] (-1.6,-0.6) rectangle ++(-1.8,0.8);
\draw[rounded corners=5pt] (-3.6,-0.6) rectangle ++(-1.8,0.8);
\draw[rounded corners=5pt] (-5.6,-0.6) rectangle ++(-1.8,0.8);
\draw[rounded corners=5pt] (-8.6,-0.6) rectangle ++(-1.8,0.8);

\tikzstyle{every node}=[inner sep=1pt]
\begin{tiny}
\node at (lru1) {$R^1_1$};
\node at (lru2) {$R^1_2$};
\node at (lru3) {$R^1_3$};
\node at (lru4) {$R^1_4$};
\node at (lru5) {$R^1_5$};

\node at (sru1) {$L^1_1$};
\node at (sru2) {$L^1_2$};
\node at (sru3) {$L^1_3$};
\node at (sru4) {$L^1_4$};

\node at (lrb1) {$R^0_5$};
\node at (lrb2) {$R^0_4$};
\node at (lrb3) {$R^0_3$};
\node at (lrb4) {$R^0_2$};
\node at (lrb5) {$R^0_1$};

\node at (srb1) {$L^0_4$};
\node at (srb2) {$L^0_3$};
\node at (srb3) {$L^0_2$};
\node at (srb4) {$L^0_1$};

\node at (lu1) {$v^0_1$};
\node at (lu2) {$v^1_2$};
\node at (lu3) {$v^0_3$};
\node at (lu4) {$P_1$};
\node at (lu5) {$v^1_4$};
\node at (lu6) {$v^1_5$};
\node at (lu7) {$v^0_6$};
\node at (lu8) {$v^0_7$};
\node at (lu9) {$v^1_8$};
\node at (lu10) {$P_2$};
\node at (lu11) {$v^0_9$};

\node at (lb1) {$v^1_9$};
\node at (lb2) {$P_3$};
\node at (lb3) {$v^1_8$};
\node at (lb4) {$v^0_7$};
\node at (lb5) {$v^0_5$};
\node at (lb6) {$v^1_6$};
\node at (lb7) {$v^1_3$};
\node at (lb8) {$v^0_4$};
\node at (lb9) {$P_4$};
\node at (lb10) {$v^1_1$};
\node at (lb11) {$v^0_2$};

\node at (tau0) {$\tau(\phi^R_Q,K^1)$};
\node at (tau1) {$\tau(\phi^R_Q,K^0)$};

\end{tiny}
\end{tikzpicture}

\caption{\label{fig:slot_refinement_serial} $L$ is serial, $\camodules(L) = \{R_1,R_2, R_3 \cup R_4, R_5\}$.
We replace $L$ in $\camodules(K)$ by $R_1,R_2,R_3 \cup R_4,R_5$; $R_2$ is active, $R_1,R_3 \cup R_4,R_5$ are inactive.
} 
\end{figure}
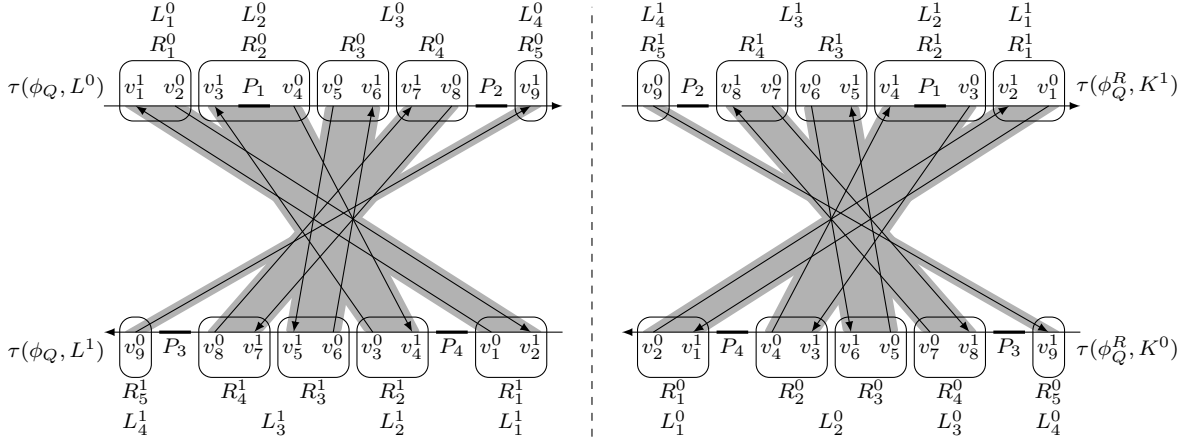

Suppose $L$ is an active member in $\camodules(K)$ such that $L$ is serial.
Suppose $R_1,\ldots,R_t$ are the children of $L$, enumerated such that
for every $i < j$ the letters from $R^0_i$ appear before the letters from $R^0_j$ in $\tau(\phi_Q,K^0)$ --
see Figure \ref{fig:slot_refinement_serial} for an illustration.
Since $L$ is serial, 
for every $i < j$ the letters from $R^1_i$ appear before the letters from $R^1_j$ in $\tau(\phi_Q,K^1)$.
Next, we define a partition $\camodules(L)$ of the set $L$ such that:
\begin{itemize}
\item the set $R_i$ is the member of $S(L)$ if $\inside(R_i) \neq \emptyset$,
\item the set $R_i \cup R_{i+1} \cup \ldots \cup R_j$ is the member of $\camodules(L)$ if $[i,j]$ is a maximal interval in $[t]$ such that $\inside(R_i \cup R_{i+1} \cup \ldots \cup R_j) = \emptyset$. 
\end{itemize}
Suppose $L_1,\ldots,L_s$ are the members of $\camodules(L)$ enumerated such that for every $i < j$ the letters from $L^0_i$ occur 
before the letters from $L^1_j$ in the word $\tau(\phi_Q,K^0)$.
Since $K$ is serial, for every $i < j$ the letters from $L^1_i$ occur before the letters $L^1_j$ in the word $\tau(\phi_M,K^1)$.
We mark $R \in \camodules(L)$ as active if $\inside(R) \neq \emptyset$ and as inactive otherwise.
We delete from $\camodules(K)$ the set $L$ and we add the sets from $\camodules(L)$ to $\camodules(K)$.

To show that Invariant~\ref{inv:2} is kept it is enough to show $L_i \prec_{\phi'} L_j$ for every $i < j$.
Suppose $R_p,R_q$ are children of $L$ such that $R_p \subseteq L_i$ and $R_q \subseteq L_j$.
Note that $p < q$.
Note that either the word $\tau(\phi_Q,R^0_p \cup R^0_{p+1} \cup \ldots \cup R^0_q)$ or the word
$\tau(\phi_Q,R^1_p \cup R^1_{p+1} \cup \ldots \cup R^1_q)$ contains a letter $P$ from $N_{\pqstree}(Q)$
as otherwise $\inside(R_p \cup R_{p+1} \cup \ldots \cup R_q) = \emptyset$ and $R_p$ and $R_q$ would be in the same set of $S(L)$.
It means that there are $a \in R_p$, $b \in R_q$ such that $a' P b'$ is a subword of $\tau(\phi_Q,K^j)$ for some $j \in \{0,1\}$, 
$a' \in \{a^0,a^1\}$, and $b' \in \{b^0,b^1\}$.
By Claim~\ref{claim:tau_phi_M_K_properties}.\eqref{item:inside_definition_1}, 
$a'Pb'$ is also a subword of $\tau(\phi'_Q,K^j)$.
In particular, we have $R_p \prec_{\phi'} R_q$.
Since $R_p$ and $R_q$ are chosen arbitrarily from $L_i$ and $L_j$, we have $L_i \prec_{\phi'} L_j$.

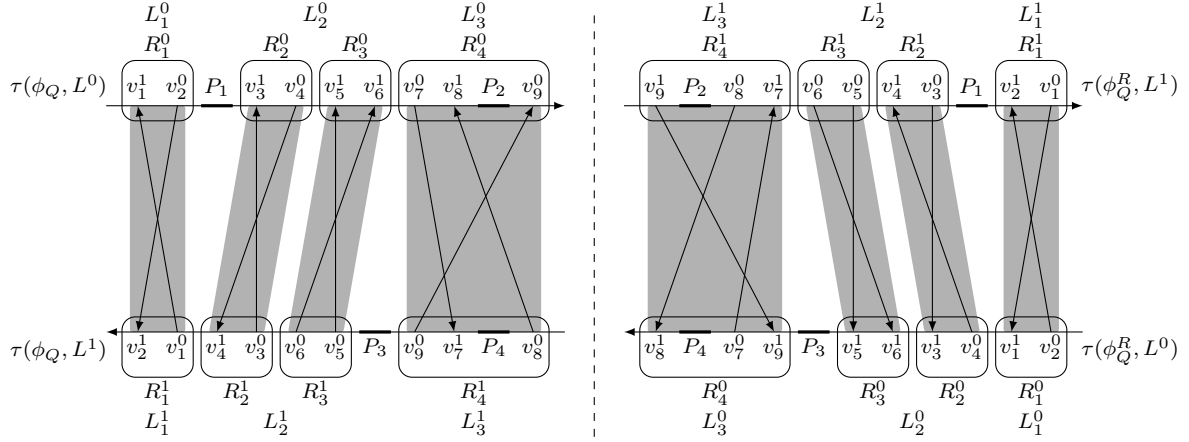
\begin{figure}[!htp]
\begin{tikzpicture}[xscale=0.52,yscale=1,>=latex]
\coordinate (u1) at (0,3) {};
\coordinate (u2) at (1,3) {};
\coordinate (u3) at (2,3) {};
\coordinate (u4) at (3,3) {};
\coordinate (u5) at (4,3) {};
\coordinate (u6) at (5,3) {};
\coordinate (u7) at (6,3) {};
\coordinate (u8) at (7,3) {};
\coordinate (u9) at (8,3) {};
\coordinate (u10) at (9,3) {};
\coordinate (u11) at (10,3) {};

\coordinate (lu1) at (0,3.25) {};
\coordinate (lu2) at (1,3.25) {};
\coordinate (lu3) at (2,3.25) {};
\coordinate (lu4) at (3,3.25) {};
\coordinate (lu5) at (4,3.25) {};
\coordinate (lu6) at (5,3.25) {};
\coordinate (lu7) at (6,3.25) {};
\coordinate (lu8) at (7,3.25) {};
\coordinate (lu9) at (8,3.25) {};
\coordinate (lu10) at (9,3.25) {};
\coordinate (lu11) at (10,3.25) {};

\coordinate (b1) at (0,0) {};
\coordinate (b2) at (1,0) {};
\coordinate (b3) at (2,0) {};
\coordinate (b4) at (3,0) {};
\coordinate (b5) at (4,0) {};
\coordinate (b6) at (5,0) {};
\coordinate (b7) at (6,0) {};
\coordinate (b8) at (7,0) {};
\coordinate (b9) at (8,0) {};
\coordinate (b10) at (9,0) {};
\coordinate (b11) at (10,0) {};

\coordinate (lb1) at (0,-0.2) {};
\coordinate (lb2) at (1,-0.2) {};
\coordinate (lb3) at (2,-0.2) {};
\coordinate (lb4) at (3,-0.2) {};
\coordinate (lb5) at (4,-0.2) {};
\coordinate (lb6) at (5,-0.2) {};
\coordinate (lb7) at (6,-0.2) {};
\coordinate (lb8) at (7,-0.2) {};
\coordinate (lb9) at (8,-0.2) {};
\coordinate (lb10) at (9,-0.2) {};
\coordinate (lb11) at (10,-0.2) {};

\begin{scope}[fill opacity=0.5]
\draw[rounded corners=1, draw=none, fill=gray!60] (-0.2,3)--(1.2,3) -- (1.2,0) -- (-0.2,0)--cycle;
\draw[rounded corners=1, draw=none, fill=gray!60] (2.8,3)--(4.2,3) -- (3.2,0) -- (1.8,0)--cycle;
\draw[rounded corners=1, draw=none, fill=gray!60] (4.8,3)--(6.2,3) -- (5.2,0) -- (3.8,0)--cycle;
\draw[rounded corners=1, draw=none, fill=gray!60] (6.8,3)--(10.2,3) -- (10.2,0) -- (6.8,0)--cycle;
\end{scope}

\coordinate (tau0) at (-2,3.25);
\coordinate (tau1) at (-2,-0.25);

\tikzstyle{every node}=[inner sep=2pt,fill=white]

\draw[<-] (u1)--(b2);
\draw[->] (u2)--(b1);
\draw[<-] (u4) -- (b4);
\draw[->] (u5) -- (b3);
\draw[<-] (u6)--(b6);
\draw[<-] (u7)--(b5);
\draw[->] (u8)--(b9);
\draw[<-] (u9)--(b11);
\draw[<-] (u11)--(b8);

\draw[-,very thick] ($(u3)+(-0.4,0)$) -- ($(u3)+(0.4,0)$);
\draw[-,very thick] ($(u10)+(-0.4,0)$) -- ($(u10)+(0.4,0)$);
\draw[-,very thick] ($(b7)+(-0.4,0)$) -- ($(b7)+(0.4,0)$);
\draw[-,very thick] ($(b10)+(-0.4,0)$) -- ($(b10)+(0.4,0)$);

\draw[->] (-0.8,3) -- (10.8,3);
\draw[<-] (-0.8,0) -- (10.8,0);

\coordinate (lru1) at (0.5,3.8) {};
\coordinate (lru2) at (3.5,3.8) {};
\coordinate (lru3) at (5.5,3.8) {};
\coordinate (lru4) at (8.5,3.8) {};

\coordinate (sru1) at (0.5,4.2) {};
\coordinate (sru2) at (4.5,4.2) {};
\coordinate (sru3) at (8.5,4.2) {};

\coordinate (lrb1) at (0.5,-0.8) {};
\coordinate (lrb2) at (2.5,-0.8) {};
\coordinate (lrb3) at (4.5,-0.8) {};
\coordinate (lrb4) at (8.5,-0.8) {};

\coordinate (srb1) at (0.5,-1.2) {};
\coordinate (srb2) at (3.5,-1.2) {};
\coordinate (srb3) at (8.5,-1.2) {};

\draw[rounded corners=5pt] (-0.4,2.8) rectangle ++(1.8,0.8);
\draw[rounded corners=5pt] (2.6,2.8) rectangle ++(1.8,0.8);
\draw[rounded corners=5pt] (4.6,2.8) rectangle ++(1.8,0.8);
\draw[rounded corners=5pt] (6.6,2.8) rectangle ++(3.8,0.8);

\draw[rounded corners=5pt] (-0.4,-0.6) rectangle ++(1.8,0.8);
\draw[rounded corners=5pt] (1.6,-0.6) rectangle ++(1.8,0.8);
\draw[rounded corners=5pt] (3.6,-0.6) rectangle ++(1.8,0.8);
\draw[rounded corners=5pt] (6.6,-0.6) rectangle ++(3.8,0.8);

\tikzstyle{every node}=[inner sep=1pt]
\begin{tiny}
\node at (lru1) {$R^0_1$};
\node at (lru2) {$R^0_2$};
\node at (lru3) {$R^0_3$};
\node at (lru4) {$R^0_4$};

\node at (sru1) {$L^0_1$};
\node at (sru2) {$L^0_2$};
\node at (sru3) {$L^0_3$};

\node at (lrb1) {$R^1_1$};
\node at (lrb2) {$R^1_2$};
\node at (lrb3) {$R^1_3$};
\node at (lrb4) {$R^1_4$};

\node at (srb1) {$L^1_1$};
\node at (srb2) {$L^1_2$};
\node at (srb3) {$L^1_3$};

\node at (lu1) {$v^1_1$};
\node at (lu2) {$v^0_2$};
\node at (lu3) {$P_1$};
\node at (lu4) {$v^1_3$};
\node at (lu5) {$v^0_4$};
\node at (lu6) {$v^1_5$};
\node at (lu7) {$v^1_6$};
\node at (lu8) {$v^0_7$};
\node at (lu9) {$v^1_8$};
\node at (lu10) {$P_2$};
\node at (lu11) {$v^0_9$};

\node at (lb1) {$v^1_2$};
\node at (lb2) {$v^0_1$};
\node at (lb3) {$v^1_4$};
\node at (lb4) {$v^0_3$};
\node at (lb5) {$v^0_6$};
\node at (lb6) {$v^0_5$};
\node at (lb7) {$P_3$};
\node at (lb8) {$v^0_9$};
\node at (lb9) {$v^1_7$};
\node at (lb10) {$P_4$};
\node at (lb11) {$v^0_8$};

\node at (tau0) {$\tau(\phi_Q,L^0)$};
\node at (tau1) {$\tau(\phi_Q,L^1)$};

\end{tiny}
\end{tikzpicture}
\hspace{0.1cm}
\begin{tikzpicture}[xscale=0.50,yscale=1,>=latex]
\draw[dashed] (0,-0.4) -- (0,5.2); 
\end{tikzpicture}
\hspace{0.1cm}
\begin{tikzpicture}[xscale=0.52,yscale=1,>=latex]
\coordinate (u1) at (0,3) {};
\coordinate (u2) at (-1,3) {};
\coordinate (u3) at (-2,3) {};
\coordinate (u4) at (-3,3) {};
\coordinate (u5) at (-4,3) {};
\coordinate (u6) at (-5,3) {};
\coordinate (u7) at (-6,3) {};
\coordinate (u8) at (-7,3) {};
\coordinate (u9) at (-8,3) {};
\coordinate (u10) at (-9,3) {};
\coordinate (u11) at (-10,3) {};

\coordinate (lu1) at (0,3.25) {};
\coordinate (lu2) at (-1,3.25) {};
\coordinate (lu3) at (-2,3.25) {};
\coordinate (lu4) at (-3,3.25) {};
\coordinate (lu5) at (-4,3.25) {};
\coordinate (lu6) at (-5,3.25) {};
\coordinate (lu7) at (-6,3.25) {};
\coordinate (lu8) at (-7,3.25) {};
\coordinate (lu9) at (-8,3.25) {};
\coordinate (lu10) at (-9,3.25) {};
\coordinate (lu11) at (-10,3.25) {};

\coordinate (b1) at (0,0) {};
\coordinate (b2) at (-1,0) {};
\coordinate (b3) at (-2,0) {};
\coordinate (b4) at (-3,0) {};
\coordinate (b5) at (-4,0) {};
\coordinate (b6) at (-5,0) {};
\coordinate (b7) at (-6,0) {};
\coordinate (b8) at (-7,0) {};
\coordinate (b9) at (-8,0) {};
\coordinate (b10) at (-9,0) {};
\coordinate (b11) at (-10,0) {};

\coordinate (lb1) at (0,-0.2) {};
\coordinate (lb2) at (-1,-0.2) {};
\coordinate (lb3) at (-2,-0.2) {};
\coordinate (lb4) at (-3,-0.2) {};
\coordinate (lb5) at (-4,-0.2) {};
\coordinate (lb6) at (-5,-0.2) {};
\coordinate (lb7) at (-6,-0.2) {};
\coordinate (lb8) at (-7,-0.2) {};
\coordinate (lb9) at (-8,-0.2) {};
\coordinate (lb10) at (-9,-0.2) {};
\coordinate (lb11) at (-10,-0.2) {};

\begin{scope}[fill opacity=0.5]
\draw[rounded corners=1, draw=none, fill=gray!60] (0.2,3)--(-1.2,3) -- (-1.2,0) -- (0.2,0)--cycle;
\draw[rounded corners=1, draw=none, fill=gray!60] (-2.8,3)--(-4.2,3) -- (-3.2,0) -- (-1.8,0)--cycle;
\draw[rounded corners=1, draw=none, fill=gray!60] (-4.8,3)--(-6.2,3) -- (-5.2,0) -- (-3.8,0)--cycle;
\draw[rounded corners=1, draw=none, fill=gray!60] (-6.8,3)--(-10.2,3) -- (-10.2,0) -- (-6.8,0)--cycle;
\end{scope}

\coordinate (tau0) at (2,3.25);
\coordinate (tau1) at (2,-0.25);

\tikzstyle{every node}=[inner sep=2pt,fill=white]

\draw[->] (u1)--(b2);
\draw[<-] (u2)--(b1);
\draw[->] (u4) -- (b4);
\draw[<-] (u5) -- (b3);
\draw[->] (u6)--(b6);
\draw[->] (u7)--(b5);
\draw[<-] (u8)--(b9);
\draw[->] (u9)--(b11);
\draw[->] (u11)--(b8);

\draw[-,very thick] ($(u3)+(-0.4,0)$) -- ($(u3)+(0.4,0)$);
\draw[-,very thick] ($(u10)+(-0.4,0)$) -- ($(u10)+(0.4,0)$);
\draw[-,very thick] ($(b7)+(-0.4,0)$) -- ($(b7)+(0.4,0)$);
\draw[-,very thick] ($(b10)+(-0.4,0)$) -- ($(b10)+(0.4,0)$);

\draw[<-] (0.8,3) -- (-10.8,3);
\draw[->] (0.8,0) -- (-10.8,0);

\coordinate (lru1) at (-0.5,3.8) {};
\coordinate (lru2) at (-3.5,3.8) {};
\coordinate (lru3) at (-5.5,3.8) {};
\coordinate (lru4) at (-8.5,3.8) {};

\coordinate (sru1) at (-0.5,4.2) {};
\coordinate (sru2) at (-4.5,4.2) {};
\coordinate (sru3) at (-8.5,4.2) {};

\coordinate (lrb1) at (-0.5,-0.8) {};
\coordinate (lrb2) at (-2.5,-0.8) {};
\coordinate (lrb3) at (-4.5,-0.8) {};
\coordinate (lrb4) at (-8.5,-0.8) {};

\coordinate (srb1) at (-0.5,-1.2) {};
\coordinate (srb2) at (-3.5,-1.2) {};
\coordinate (srb3) at (-8.5,-1.2) {};

\draw[rounded corners=5pt] (0.4,2.8) rectangle ++(-1.8,0.8);
\draw[rounded corners=5pt] (-2.6,2.8) rectangle ++(-1.8,0.8);
\draw[rounded corners=5pt] (-4.6,2.8) rectangle ++(-1.8,0.8);
\draw[rounded corners=5pt] (-6.6,2.8) rectangle ++(-3.8,0.8);

\draw[rounded corners=5pt] (0.4,-0.6) rectangle ++(-1.8,0.8);
\draw[rounded corners=5pt] (-1.6,-0.6) rectangle ++(-1.8,0.8);
\draw[rounded corners=5pt] (-3.6,-0.6) rectangle ++(-1.8,0.8);
\draw[rounded corners=5pt] (-6.6,-0.6) rectangle ++(-3.8,0.8);

\tikzstyle{every node}=[inner sep=1pt]
\begin{tiny}
\node at (lru1) {$R^1_1$};
\node at (lru2) {$R^1_2$};
\node at (lru3) {$R^1_3$};
\node at (lru4) {$R^1_4$};

\node at (sru1) {$L^1_1$};
\node at (sru2) {$L^1_2$};
\node at (sru3) {$L^1_3$};

\node at (lrb1) {$R^0_1$};
\node at (lrb2) {$R^0_2$};
\node at (lrb3) {$R^0_3$};
\node at (lrb4) {$R^0_4$};

\node at (srb1) {$L^0_1$};
\node at (srb2) {$L^0_2$};
\node at (srb3) {$L^0_3$};

\node at (lu1) {$v^0_1$};
\node at (lu2) {$v^1_2$};
\node at (lu3) {$P_1$};
\node at (lu4) {$v^0_3$};
\node at (lu5) {$v^1_4$};
\node at (lu6) {$v^0_5$};
\node at (lu7) {$v^0_6$};
\node at (lu8) {$v^1_7$};
\node at (lu9) {$v^0_8$};
\node at (lu10) {$P_2$};
\node at (lu11) {$v^1_9$};

\node at (lb1) {$v^0_2$};
\node at (lb2) {$v^1_1$};
\node at (lb3) {$v^0_4$};
\node at (lb4) {$v^1_3$};
\node at (lb5) {$v^1_6$};
\node at (lb6) {$v^1_5$};
\node at (lb7) {$P_3$};
\node at (lb8) {$v^1_9$};
\node at (lb9) {$v^0_7$};
\node at (lb10) {$P_4$};
\node at (lb11) {$v^1_8$};

\node at (tau0) {$\tau(\phi^R_Q,L^1)$};
\node at (tau1) {$\tau(\phi^R_Q,L^0)$};

\end{tiny}
\end{tikzpicture}
\caption{\label{fig:slot_refinement_parallel} $L$ is parallel. 
We replace $L$ in $\camodules(K)$ by $R_1, R_2 \cup R_3, R_4$; $R_4$ is active, $R_1, R_2 \cup R_3$ are inactive.
} 
\end{figure}

Eventually, suppose $L$ is an active member of $\camodules(K)$ such that $L$ is parallel in $\strongModules(K,{\sim})$.
Suppose $R_1,\ldots,R_t$ are the children of $L$, enumerated such that
for every $i < j$ the letters from $R^0_i$ appear before the letters from $R^0_j$ in $\tau(\phi_Q,K^0)$ --
see Figure~\ref{fig:slot_refinement_parallel} for an illustration.
Since $L$ is parallel, 
for every $i < j$ the letters from $R^1_j$ appear before the letters from $R^1_i$ in $\tau(\phi_Q,K^1)$.
Next, we define a partition $\camodules(L)$ of the set $L$ such that:
\begin{itemize}
\item the set $R_i$ is the member of $\camodules(L)$ if $\inside(R_i) \neq \emptyset$,
\item the set $R_i \cup R_{i+1} \cup \ldots \cup R_j$ is the member of $\camodules(L)$ if $[i,j]$ is a maximal interval in $[t]$ such that $\inside(R_i \cup R_{i+1} \cup \ldots \cup R_j) = \emptyset$. 
\end{itemize}
Suppose $L_1,\ldots,L_s$ are the members of $\camodules(L)$ enumerated such that for every $i < j$ the letters from $L^0_i$ occur  before the letters from $L^1_j$ in the word $\tau(\phi_Q,K^0)$.
Since $K$ is parallel, for every $i < j$ the letters from $L^1_j$ occur before the letters $L^1_i$ in the word $\tau(\phi_Q,K^1)$.
We mark $R \in \camodules(L)$ as active if $\inside(R) \neq \emptyset$ and as inactive otherwise.
Now, we delete from $\camodules(K)$ the set $L$ and we add the sets from $\camodules(L)$ to $\camodules(K)$.
Finally, we observe that $L^0_i$ precedes $L^0_j$ in $\tau(\phi_Q,K^0)$ for $i < j$ 
(as $(L,{\sim_L})$ is an empty graph), which shows Invariant~\ref{inv:2}.

First, we prove that the set $\camodules(K)$ satisfies Properties~\ref{prop:K_refinement_partition}--\ref{prop:K_refinement_contiguous_subwords_order}.
Property~\ref{prop:K_refinement_partition} is obviously satisfied.
Property~\ref{prop:K_refinement_contiguous_subwords} follows by Invariant~\ref{inv:2} 
and by the fact that, whenever the set $S$ is marked as inactive, $S^0$ and $S^1$ are contiguous in $\tau(\phi_Q,K^0)$ and $\tau(\phi_Q,K^1)$, respectively.
Property~\ref{prop:K_refinement_contiguous_subwords_order} follows by Invariant~\ref{inv:2}.
Eventually, observe that every inactive module in $S(K)$ is a maximal module in $(K,{\sim})$ whose vertex set admits the same left-right partition of $\inside(K)$, which follows from the way the algorithm refines the strong modules $L$ with $\inside(L) \neq \emptyset$.

\section{Linear-time algorithm constructing the PQSM-tree of a circular-arc graph.}
\label{sec:pqsm_tree_construction}
Let $G = (V,E)$ be a circular-arc graph with no twins and no universal vertices and let $G_{ov}$ be the overlap graph.
In this section we present a linear-time algorithm that 
constructs the PQSM-tree $\pqsmtree$ representing the conformal models of $G_{ov}$.

First, we compute the overlap graph $G_{ov}=(V,{\sim})$ of $G$, 
a conformal model $\phi$ of $G_{ov}$ and its reflection $\phi^R$.
Since the linear-time algorithm recognizing circular-arc graphs by McConnell~\cite{McCon03} 
provides a normalized circular-arc model in the case of yes-instance, 
we can use it to construct a normalized model $\psi$ of $G$.
Then, we compute the overlap graph $(V,{\sim})$ of $G$: for every $u,v \in V$ we set $u \sim v$ if and only if $uv \in E$ and $\psi(u)$ and $\psi(v)$ overlap.
Eventually, we compute the conformal model $\phi$ of $G_{ov}$ corresponding to $\psi$ and its reflection $\phi^R$.
Next, we use a linear-time algorithm of McConnell and Spinrad~\cite{McConnSpin99} to compute the modular decomposition tree $\strongModules(G_{ov})$ of $G_{ov}$.
Clearly, since ${\sim} \subseteq E$, all these steps can be done in linear-time in the size of $G$.

Now we show how we construct PQSM-tree $\pqsmtree$. 
Given $G_{ov}$ and $\phi$, we first compute the set $\camodules$ of CA-modules of $G$.
We leave it to the reader to verify that the set $\camodules$ can be computed using Property~\ref{prop:CA-modules-rule} formulated in Subsection~\ref{sub:ca-modules}.
We proceed the tree~$\strongModules(G_{ov})$ in the bottom-up order and for every strong module~$M$ different than~$V$ we 
compute the set~$\camodules(M)$, as follows.
If~$M$ induces a contiguous permutation model $(\tau'_M,\tau''_M)$ in~$\phi$, 
we set $\camodules(M) = \emptyset$ and we replace $\tau'_M$ and $\tau''_M$ in $\phi$ 
by the letters $M'$ and $M''$.
Otherwise, given that $M_1,\ldots,M_n$ are the children of~$M$:
\begin{itemize}
 \item if $M$ is serial/parallel, then 
 $\camodules(M)$ contains all the sets in $\camodules(M_i)$ for all $M_i$ such that 
 $\camodules(M_i) \neq \emptyset$ and all maximal modules $S \subseteq M$ such that (i) $S$ is the union of some children of $M$ and (ii) $S$ induces a contiguous permutation model in $\phi$,
 \item if $M$ is prime, then 
 $\camodules(M)$ contains all the sets in $\camodules(M_i)$ for $M_i$ such that 
 $\camodules(M_i) \neq \emptyset$ and all the sets $M_i$ for which $\camodules(M_i) = \emptyset$.
\end{itemize}
Finally, if $M$ is a child of $V$ and $\camodules(M)=\emptyset$, we set $\camodules(M) = \{M\}$.
We leave it to the reader to check that the set $\camodules(M)$ comprises all CA-modules contained in $M$.
Since we can process each module in~$\strongModules(G_{ov})$ in time linear in the number of its children, 
we can compute the set $\camodules$ in linear time in the size of $\strongModules(G_{ov})$, and hence in linear time in the size of $G_{ov}$.
Since the set $\camodules$ forms a partition of $V$ into modules in $G_{ov}$, 
we can compute the modular decomposition trees $\strongModules(S,{\sim})$ for all sets $S \in \camodules$ in time linear in the size of~$G_{ov}$.
Further, for all M-nodes $M$ in $\strongModules(S,{\sim})$ we can read the representations of the sets
$\Pi(M)$ from the model $\phi$ -- see Subsection~\ref{sub:admissible_models}.
All these steps can be done in linear time in the size of $G_{ov}$.

Finally, we show how to compute the sets $\Pi(N)$ for every inner node $N$ in the PQS-tree~$\pqstree$. 
We consider only the case when $V$ is parallel as the other cases are trivial.
Recall that the components of $G_{ov}$, whose set was denoted by~$\mathcal{Q}$, 
correspond to the Q-nodes of~$\pqstree$.
Given $\phi$, we first compute the extended conformal model $\phi[Q]$ of $(Q,{\sim})$ induced by $Q$.
For this purpose, for every component $Q$ we compute a circular word $\phi'_Q$, 
where $\phi'_Q$ is obtained from $\phi \Vert Q^*$ by inserting a letter $P'$ 
between every two letters $q',q'' \in Q^*$ such that $q'q''$ is a contiguous subword in $\phi\Vert Q^*$ but not in $\phi$.
We assume all the inserted letters $P'$ are pairwise different
and all belong to the set $\mathcal{P'}$.
Clearly, we can compute $\phi'_Q$ from $\phi \Vert Q^*$ in time linear in~$|Q|$, and
hence, we can compute $\phi'_Q$ for all $Q \in \mathcal{Q}$ in time linear in~$|V|$.
Finally, given $\phi'_Q$ for all $Q \in \mathcal{Q}$, 
in linear time we can compute all P-nodes of $\pqstree$ and the extended conformal models~$\phi[Q]$ for~$Q \in \mathcal{Q}$.
For this purpose we use the following observation - see Figure \ref{fig:P-node} for an illustration.
\begin{observation}
\label{claim:computing_nodes}
Suppose $\mathcal{Q}'=\{Q_1,\ldots,Q_k\}$ is a subset of $\mathcal{Q}$.
Then, $\mathcal{Q}'$ is a P-node
if and only if there are $P'_1,\ldots,P'_k$ in $\mathcal{P}'$ and
$q'_j,q''_j$ in $Q^*_j$ for $j \in [k]$ such that:
\begin{enumerate}
\item \label{item:claim_node_recover_1} $q_j'P'_jq_j''$ is a contiguous subword of $\phi'_{Q_j}$ for every $j \in [k]$,
\item \label{item:claim_node_recover_2} $q_j'q_{j+1}''$ is a contiguous subword of $\phi$ for every $j \in [k]$ (cyclically).
\end{enumerate}
\end{observation}
\begin{proof}
The necessity is obvious. 
We prove the sufficiency.
Since $q_j'q_{j+1}''$ is a contiguous subword of $\phi$, $Q_j$ and $Q_{j+1}$ are neighbouring for all $j \in [k]$ (cyclically).
Hence, $\{Q_1,\ldots,Q_k\}$ are pairwise neighbouring.
Also, $\{Q_1,\ldots,Q_k\}$ is a maximal set of pairwise neighbouring components;
otherwise, $q_j'q_{j+1}''$ for some $j \in [k]$ (cyclically) would not be consecutive in~$\phi$.
\end{proof}
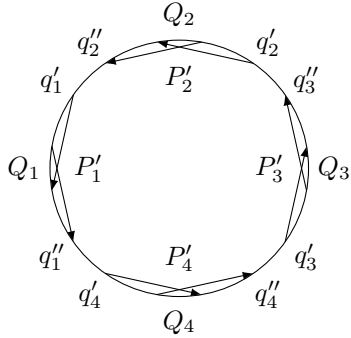
\begin{figure}[htp!]
\centering
\begin{tikzpicture}[xscale=0.85,yscale=0.85,>=latex]
\coordinate (center) at (0,0) {};
\coordinate (lm21) at ($(center)+(125:2.4cm)$) {};
\coordinate (m21) at ($(center)+(125:2cm)$) {};
\coordinate (m22) at ($(center)+(100:2cm)$) {};
\coordinate (m23) at ($(center)+(80:2cm)$) {};
\coordinate (m24) at ($(center)+(55:2cm)$) {};
\coordinate (lm24) at ($(center)+(55:2.4cm)$) {};

\coordinate (lm31) at ($(center)+(35:2.4cm)$) {};
\coordinate (m31) at ($(center)+(35:2cm)$) {};
\coordinate (m32) at ($(center)+(10:2cm)$) {};
\coordinate (m33) at ($(center)+(-10:2cm)$) {};
\coordinate (m34) at ($(center)+(-35:2cm)$) {};
\coordinate (lm34) at ($(center)+(-35:2.4cm)$) {};

\coordinate (lm41) at ($(center)+(305:2.4cm)$) {};
\coordinate (m41) at ($(center)+(305:2cm)$) {};
\coordinate (m42) at ($(center)+(280:2cm)$) {};
\coordinate (m43) at ($(center)+(260:2cm)$) {};
\coordinate (m44) at ($(center)+(235:2cm)$) {};
\coordinate (lm44) at ($(center)+(235:2.4cm)$) {};

\coordinate (lm11) at ($(center)+(215:2.4cm)$) {};
\coordinate (m11) at ($(center)+(215:2cm)$) {};
\coordinate (m12) at ($(center)+(190:2cm)$) {};
\coordinate (m13) at ($(center)+(170:2cm)$) {};
\coordinate (m14) at ($(center)+(145:2cm)$) {};
\coordinate (lm14) at ($(center)+(145:2.4cm)$) {};

\coordinate (lm3) at ($(center)+(0:2.4cm)$) {};
\coordinate (ln3) at ($(center)+(0:1.4cm)$) {};

\coordinate (lm2) at ($(center)+(90:2.4cm)$) {};
\coordinate (ln2) at ($(center)+(90:1.4cm)$) {};

\coordinate (lm1) at ($(center)+(180:2.4cm)$) {};
\coordinate (ln1) at ($(center)+(180:1.4cm)$) {};

\coordinate (lm4) at ($(center)+(270:2.4cm)$) {};
\coordinate (ln4) at ($(center)+(270:1.4cm)$) {};

\draw (0,0) circle (2cm);
\draw[->] (m24)--(m22);
\draw[->] (m23)--(m21);

\draw[->] (m14)--(m12);
\draw[->] (m13)--(m11);

\draw[->] (m34)--(m32);
\draw[->] (m33)--(m31);

\draw[->] (m44)--(m42);
\draw[->] (m43)--(m41);

\tikzstyle{every node}=[inner sep=1pt]
\begin{footnotesize}
\node at (lm11) {$q''_1$};
\node at (lm14) {$q'_1$};
\node at (lm21) {$q''_2$};
\node at (lm24) {$q'_2$};
\node at (lm31) {$q''_3$};
\node at (lm34) {$q'_3$};
\node at (lm41) {$q''_4$};
\node at (lm44) {$q'_4$};

\node at (lm1) {$Q_1$};
\node at (ln1) {$P'_1$};

\node at (lm2) {$Q_2$};
\node at (ln2) {$P'_2$};

\node at (lm3) {$Q_3$};
\node at (ln3) {$P'_3$};

\node at (lm4) {$Q_4$};
\node at (ln4) {$P'_4$};
\end{footnotesize}

\end{tikzpicture}
\caption{\label{fig:P-node} The letters $P'_1,P'_2,P'_3,P'_4$ are merged to a single P-node $P = \{Q_1,Q_2,Q_3,Q_4\}$.}
\end{figure}

Given the set of P-nodes, the extended conformal models $\phi[Q]$, and the CA-modules of~$G$, 
we can compute easily~$\pqstree$ and the sets $\Pi(\cdot)$ for all inner nodes of~$\pqstree$.

\section{Linear-time canonization of a circular-arc graph}
\label{sec:canonization}
Let $G'$ be a circular-arc graph.
Let $U(G')$ denote the set of universal vertices of $G'$ and let
$u = |U(G')|$.
For every vertex $v \in V(G') \setminus U(G')$ let $T_{G'}(v)$ be the set of twins of $v$ in $G'$, that is,
$$T_{G'}(v) = \{w \in V(G'): N_{G'}[v] = N_{G'}[w]\}.$$
Clearly, $\{T_{G'}(v): v \in V(G') \setminus U(G')\}$ forms a partition of $V(G) \setminus U(G')$.
Let $V$ be the set containing a vertex from every set $\{T_{G'}(v): v \in V(G') \setminus U(G')\}$, 
let $G$ be a subgraph of $G'$ induced by $V$, and let $m(v)$ for every $v \in V$ be the size of the set $T_{G'}(v)$.
Clearly, $G$ is a circular-arc graph with no universal vertices and no twins.
The triple $(G,m,u)$ is called the \emph{representation} of $G'$.
\begin{claim}
Let $G'$ be a circular-arc graph.
The representation $(G,m,u)$ of $G'$ can be computed in linear time in the size of $G'$.   
\end{claim}
\begin{proof}
Let $\strongModules(G')$ be the modular decomposition of $G'$.
The following observations hold for every $v \in V(G')$ and every $T \subseteq V(G')$ such that $|T| \geq 2$:
\begin{itemize}
 \item $v \in U(G')$ if and only if $V(G')$ is serial in $\strongModules(G')$ and  $\{v\}$ is the child of $V(G')$ in~$\strongModules(G')$.
 \item $T$ is a set of twins in $G'$ if and only if $T \subseteq T'$ for some serial module $T'$ in $\strongModules(G')$ 
 and $T = \{u \in V(G'): \{u\} \text{ is a child of $T'$ in $\strongModules(G')$}\}$.
\end{itemize}
Since the modular decomposition tree $\strongModules(G')$ can be computed in linear time in the size of $G'$, in the same time we can compute the representation $(G,m,u)$ of $G'$.
\end{proof}

Suppose $G'$ and $H'$ are circular arc graphs represented by $(G,m_G,u_G)$ and $(H,m_H,u_H)$.
We say that $(G,m_G,u_G)$ and $(H,m_H,u_H)$ are \emph{isomorphic} if 
$u_G = u_H$ and there is an isomorphism $\alpha$ from $G$ to $H$ that satisfies $m_G(v) = m_H(\alpha(v))$ for every $v \in V(G)$.
Clearly, all representations of $G'$ (corresponding to different choices of the vertices in~$V$) are isomorphic.
\begin{claim}
\label{claim:isomorphism_between_graphs_and_representants}
$G'$ and $H'$ are isomorphic if and only if $(G,m_G,u_G)$ and $(H,m_H,u_H)$ are isomorphic.
\end{claim}

Our goal is to present a linear time algorithm that computes a string representation $\canon(G')$ of a circular-arc graph $G'$.
For this purpose we compute the representation $(G,m_G,u_G)$ of $G'$, the overlap
graph $G_{ov}$ of $G$, and the PQSM-tree $\pqsmtree$ representing the conformal models of $G_{ov}$.
For the reason that became clear later, for every node $L$ in $\strongModules(S,{\sim})$ for $S \in \camodules$ we define two metachords $\LLL^0$ and $\LLL^1$: $\LLL^0 = (L^0,L^1,{<^0_L})$ and $\LLL^1 = (L^1,L^0,{<^1_L})$, where ${<^0_L} = {<_L}$ and $<^1_L$ is the reverse of ${<_L}$.
In particular, for every $S \in \camodules$ model $\tau = (\tau^0,\tau^1)$ is admissible for $\SSS^0$ if and only if model $(\tau^1,\tau^0)$
is admissible for $\SSS^1$.
We set 
$$\Pi(\LLL^0) = \Pi(L) \quad \text{and} \quad \Pi(\LLL^1) = \{(\pi^1, \pi^0): (\pi^0, \pi^1) \in \Pi(L)\}.$$
Also, we let 
$$
\begin{array}{ccl}
\mathcal{L}^0 &=& \{\mathbb{L}^0: L \text{ is a node in } \strongModules(S,{\sim}) \text{ for some } S \in \camodules\}, \\
\mathcal{L}^1 &=& \{\mathbb{L}^1: L \text{ is a node in } \strongModules(S,{\sim}) \text{ for some } S \in \camodules\},\\
\mathcal{L} &=& \{L: L \text{ is a node in } \strongModules(S,{\sim}) \text{ for some } S \in \camodules\}.
\end{array}
$$
We have shown in Section~\ref{sec:pqsm_tree_construction} that we can compute these components in time linear in the size of~$G$.

We partition the members of~$\mathcal{L}$ into~\emph{levels}: 
for every $S \in \camodules$ and every $L \in \strongModules(S,{\sim})$ we have
$\level(L) = l$ if the distance between $L$ and $S$ in $\strongModules(S,{\sim})$
equals to $l$. 
In particular, $\level(S)=0$ for every $S \in \camodules$.
We assume $\level(\mathbb{L}^0) = \level(\mathbb{L}^1) = \level(L)$ for every $L \in \mathcal{L}$.
By $\mathcal{L}_l$, $(\mathcal{L}^0 \cup \mathcal{L}^1)_{l}$ 
we denote the restrictions of the sets $\mathcal{L}$ and $\mathcal{L}^0 \cup \mathcal{L}^1$, respectively, to the elements from level $l$.
We assume $\mathcal{L}_{l,l+1} = \mathcal{L}_{l} \cup \mathcal{L}_{l+1}$ and similarly for the set $(\mathcal{L}^0 \cup \mathcal{L}^1)_{l,l+1}$. 

Similarly, we partition the nodes of the PQS-tree $\pqstree$ into levels with respect to the \emph{root} of $\pqstree$.
To define the root of $\pqstree$ we need some definitions.
A~\emph{center} of a tree is a vertex with the minimum maximum distance to a leaf.
It is commonly known that any tree has either one center (and then the tree is called \emph{centered}) or has two centers joined with an edge (and then the tree is called \emph{bicentered}).
Next, we root the tree~$\pqstree$ such that:
\begin{itemize}
\item if $\pqstree$ is centered, we root $\pqstree$ in the center of $\pqstree$,
\item if $\pqstree$ is bicentered, we add a special node $R$ on the edge joining two centers of~$\pqstree$ (one is a Q-node and one is a P-node) and we root $\pqstree$ in the node $R$.
\end{itemize}
Let $N(\pqstree)$ denote the set of the nodes of $\pqstree$.
For $N \in N(\pqstree)$ by $\level(N)$ we denote the distance of $N$ to the root of 
$\pqstree$.
We let $N(\pqstree)_{l}$ and $N(\pqstree)_{l,l+1}$ to contain the nodes of $\pqstree$ 
whose distance to the root is in the sets~$\{l\}$ and~$\{l,l+1\}$, respectively.
Eventually, for every inner node $N$ in $\pqstree$, by $V_{\pqstree}(N)$ 
we denote the vertices of $V$ from the components of $G_{ov}$ 
corresponding to Q-nodes contained in the subtree of $\pqstree$ rooted in $N$ 
(the set $V_{\pqstree}(N)$ contains also the vertices from $N$ if $N$ is a Q-node).

Our goal is to compute a string representation $\canon(G,m_G,u_G)$ of $(G,m_G,u_G)$ such that for every other circular arc graph $H'$ represented by $(H,m_H,u_H)$:
$$
\begin{array}{c}
\canon(G,m_G,u_G) = \canon(H,m_H,u_H) \quad \iff \\
\text{$(G,m_G,u_G)$ and $(H,m_H,u_H)$ are isomorphic.}
\end{array}
$$
Finally, we set $\canon(G') = \canon(G,m_G,u_G)$ and we note that, by Claim \ref{claim:isomorphism_between_graphs_and_representants}, $\canon(G') = \canon(H')$ if and only if $G'$ and $H'$ are isomorphic.

To compute $\canon(G,m_G,u_G)$ we first calculate the number $\num(X)$ for every
object~$X$ in $\mathcal{L}^0 \cup \mathcal{L}^1 \cup N(\pqstree)$.
We compute $\num(\cdot)$ in two steps, first for the objects in $\mathcal{L}^0 \cup \mathcal{L}^1$, and then for the nodes in $\pqstree$.
In each step, we process the objects in the decreasing order of their levels; 
that is, the objects at level $l+1$ are processed
before the objects at level~$l$.
Two objects $X,Y$ on the same level will satisfy $\num(X) = \num(Y)$ if and only if $X$ and $Y$ are ``locally isomorphic'', where the meaning of a `` local isomorphism'' depends on the type of the objects
(informally, two objects $X,Y$ are locally isomorphic if the parts of the graph $G$ induced by the vertices from the subtrees rooted at $X$ and at $Y$ are isomorphic).
Given computed the numbers $\num(\cdot)$ for the objects from level $l+1$, we 
process the objects from level $l$.
First, we compute a tuple $\canon(X)$ for every object $X$ from level~$l$.
The tuple $\canon(X)$ encodes the object $X$ (in particular, it uses the numbers $\num$ encoding the local isomorphism type of the children of $X$ from level $l+1$)
so as for two objects $X$ and $Y$ at level $l$ we have $\canon(X) = \canon(Y)$ if and only if $X$ and $Y$ are locally isomorphic.
Roughly speaking, $\canon(X)$ is the lexicographically smallest tuple among
appropriately defined linear representations of the members of the set $\Pi(X)$.
We sort all the tuples $\canon(X)$ for the objects $X$ from level $l$ 
and we identify those that share $\canon(\cdot)$ (those that are locally isomorphic).   
All the tuples $\canon(\cdot)$ are stored in the table $\canon$, 
indexed with the consecutive natural numbers starting from $Num = 0$.
The variable $Num$ always indicates the first free index in the $\canon$ table.
The tuple $\canon(X)$ for the object $X$ is stored at the position $\num(X)$, that is,
$$\canon(X) = \canon[\num(X)] \quad \text{holds for every object } X \in \mathcal{L}^0 \cup \mathcal{L}^1 \cup N(\pqstree).$$
Summing up, the group of locally isomorphic objects is represented by one entry in the table $\canon$, 
stored at the position $\num$ which is common for the objects of this group
(that is, $\num = \num(X)$ for every object $X$ in this group).
The tuples $\canon$ for objects from the same level are stored in a contiguous block of the table $\canon$.
Eventually, $\canon(G,m_G,u_G)$ is defined as a linearisation of the table $\canon$. 

We start by showing how the algorithm computes the values $\canon(\mathbb{L}')$ and $\num(\mathbb{L}')$ for the metachords $\mathbb{L}'$ in the set $(\mathcal{L}^0 \cup \mathcal{L}^1)_l$.
We assume that $\num(\mathbb{L}')$ and $\canon(\mathbb{L}')$ are computed for all $\mathbb{L}' \in (\mathcal{L}^0 \cup \mathcal{L}^1)_{l+1}$.
We assume also that $\num(\mathbb{L}') = \num(\mathbb{R}')$ holds for every two metachords $\mathbb{L}',\mathbb{R}' \in (\mathcal{L}^0 \cup \mathcal{L}^1)_{l+1}$ 
if and only if $\mathbb{L}'$ and $\mathbb{R}'$ are locally isomorphic according to the following definition.
\begin{definition}
\label{def:metachords_loc_iso}
Suppose $\mathbb{L}' = (L',L'',{<'_L})$ and $\mathbb{R}' = (R',R'',{<'_R})$ are two metachords such that $\level(\mathbb{L}') = \level(\mathbb{R}')$, where  
$$ \mathbb{L}' = \left\{ 
\begin{array}{rcl}
(L^0,L^1,{<^0_L}) & \text{ if } & \mathbb{L'} \in \mathcal{L}^0 \\  
(L^1,L^0,{<^1_L}) & \text{ if } & \mathbb{L'} \in \mathcal{L}^1 \\  
\end{array}
\right.
\text{ and }
\quad 
\mathbb{R}' = \left\{ 
\begin{array}{rcl}
(R^0,R^1,{<^0_R}) & \text{ if } & \mathbb{R}' \in \mathcal{L}^0 \\  
(R^1,R^0,{<^1_R}) & \text{ if } & \mathbb{R}' \in \mathcal{L}^1 \\  
\end{array}
\right.
$$
Let $\alpha$ be a bijection from $L$ to $R$. 
We say that $\alpha$ is a \emph{local isomorphism} from $\mathbb{L}'$ to $\mathbb{R}'$ if for every $u,v \in L$:
\begin{enumerate}
 \item \label{def:metachords_loc_iso_left_right} $u <'_{L} v \iff \alpha(u) <'_{R} \alpha(v)$,
 \end{enumerate}
 and for every $u \in L$:
 \begin{enumerate}[resume]
 \item \label{def:metachords_loc_iso_orientation} $u$ is oriented from $L'$ to $L''$ $\iff$ $\alpha(u)$ is 
oriented from $R'$ to $R''$,
 \item \label{def:metachords_loc_iso_mul} $m_G(u) = m_G(\alpha(u))$.
 \end{enumerate}
We say $\mathbb{L}'$ and $\mathbb{R}'$ are \emph{locally isomorphic} if there is a local isomorphism from $\mathbb{L}'$ to $\mathbb{R}'$. 
\end{definition}

Let~$\mathbb{L}' = (L',L'',{<'_{L}})$ is a metachord from $(\mathcal{L}^0 \cup \mathcal{L}^1)_l$ and let $S \in \camodules$ be such that $L \subseteq S$.
The tuple $\canon(\mathbb{L}')$ is defined as follows:
\begin{itemize}
\item If $L = \{u\}$ is a leaf in $\strongModules(S,{\sim})$, then 
$$\canon(\mathbb{L}') = \big([u^0 \in L'], m_G(u) + Num \big),$$
where $[u^0 \in L'] = 1$ if $u^0 \in L'$ and $[u^0 \in L'] = 0$ if $u^0 \notin L'$.
\item If $L$ in a non-leaf in $\strongModules(S,{\sim})$ with children $L_1,\ldots,L_k$, 
then $\canon(\mathbb{L}')$ is the lexicographically smallest tuple 
$\canon\big(\mathbb{L}', \pi(\mathbb{L}') \big)$ over all members $\pi(\mathbb{L}')$ of $\Pi(\mathbb{L}')$,
where for $\pi(\mathbb{L}') \in \Pi(\mathbb{L}')$ of the form $\pi(\mathbb{L}') = \big(L'_{\delta'(1)},\ldots, L'_{\delta'(k)}), (L''_{\delta''(1)},\ldots, L''_{\delta''(k)})\big)$ for some permutations $\delta', \delta''$ of $[k]$ we set
$$
\begin{array}{lcc}
\canon\big(\mathbb{L}', \pi(\mathbb{L}')\big) &=&
\left( \begin{array}{c}
\ \num(\mathbb{L}'_{\delta'(1)}), \ pos\big(L''_{\delta'(1)},(L''_{\delta''(1)},\ldots, L''_{\delta''(k)})\big), \\
 \vdots\\
 \num(\mathbb{L}'_{\delta'(k)}), \ pos\big(L''_{\delta'(k)},(L''_{\delta''(1)},\ldots, L_{\delta''(k)})\big),
\end{array} \right),
\end{array}$$
where $pos\big(L''_{\delta'(i)},(L''_{\delta''(1)},\ldots, L''_{\delta''(k)})\big)$ is the position of $L''_{i}$ in $(L''_{\delta''(1)},\ldots, L''_{\delta''(k)})$ increased by the current value of $Num$.
See Figure~\ref{fig:Pi_prime_canon} for an example.
\end{itemize}
Note that the type of the module $L$ can be easily recovered from the entries $pos(\cdot)$.
\begin{figure}[!htp]
\begin{tikzpicture}[xscale=0.50,yscale=1,>=latex]
\coordinate (u1) at (0,3) {};
\coordinate (u2) at (1,3) {};
\coordinate (u3) at (2,3) {};
\coordinate (u4) at (2,3) {};
\coordinate (u5) at (3,3) {};
\coordinate (u6) at (4,3) {};
\coordinate (u7) at (5,3) {};
\coordinate (u8) at (5,3) {};
\coordinate (u9) at (6,3) {};
\coordinate (u10) at (7,3) {};
\coordinate (u11) at (8,3) {};
\coordinate (u12) at (9,3) {};

\coordinate (lu1) at (0,3.25) {};
\coordinate (lu2) at (1,3.25) {};
\coordinate (lu3) at (2,3.25) {};
\coordinate (lu4) at (2,3.25) {};
\coordinate (lu5) at (3,3.25) {};
\coordinate (lu6) at (4,3.25) {};
\coordinate (lu7) at (5,3.25) {};
\coordinate (lu8) at (5,3.25) {};
\coordinate (lu9) at (6,3.25) {};
\coordinate (lu10) at (7,3.25) {};
\coordinate (lu11) at (8,3.25) {};
\coordinate (lu12) at (9,3.25) {};

\coordinate (b1) at (0,0) {};
\coordinate (b2) at (1,0) {};
\coordinate (b3) at (2,0) {};
\coordinate (b4) at (3,0) {};
\coordinate (b5) at (4,0) {};
\coordinate (b6) at (5,0) {};
\coordinate (b7) at (5,0) {};
\coordinate (b8) at (6,0) {};
\coordinate (b9) at (7,0) {};
\coordinate (b10) at (8,0) {};
\coordinate (b11) at (9,0) {};
\coordinate (b12) at (9,0) {};

\coordinate (lb1) at (0,-0.2) {};
\coordinate (lb2) at (1,-0.2) {};
\coordinate (lb3) at (2,-0.2) {};
\coordinate (lb4) at (3,-0.2) {};
\coordinate (lb5) at (4,-0.2) {};
\coordinate (lb6) at (5,-0.2) {};
\coordinate (lb7) at (5,-0.2) {};
\coordinate (lb8) at (6,-0.2) {};
\coordinate (lb9) at (7,-0.2) {};
\coordinate (lb10) at (8,-0.2) {};
\coordinate (lb11) at (9,-0.2) {};
\coordinate (lb12) at (9,-0.2) {};

\begin{scope}[fill opacity=0.5]
\draw[rounded corners=1, draw=none, fill=gray!60] (-0.3,3)--(2.3,3) -- (4.3,0) -- (1.7,0)--cycle;
\draw[rounded corners=1, draw=none, fill=gray!60] (4.7,3)--(6.3,3) -- (1.3,0) -- (-0.3,0)--cycle;
\draw[rounded corners=1, draw=none, fill=gray!60] (2.7,3)--(4.3,3) -- (9.3,0) -- (7.7,0)--cycle;
\draw[rounded corners=1, draw=none, fill=gray!60] (6.7,3)--(9.3,3) -- (7.3,0) -- (4.7,0)--cycle;

\end{scope}

\coordinate (tau0) at (-1.5,3.25);
\coordinate (tau1) at (-1.5,-0.25);

\coordinate (llu1) at (1,3.8) {};
\coordinate (llu2) at (3.5,3.8) {};
\coordinate (llu3) at (5.5,3.8) {};
\coordinate (llu4) at (8,3.8) {};

\coordinate (llb1) at (0.5,-0.8) {};
\coordinate (llb2) at (3,-0.8) {};
\coordinate (llb3) at (6,-0.8) {};
\coordinate (llb4) at (8.5,-0.8) {};

\tikzstyle{every node}=[inner sep=2pt,fill=white]

\draw[->] (u1)--(b3);
\draw[<-, very thick] (u2)--(b5);
\draw[<-] (u4)--(b4);
\draw[<-, very thick] (u5)--(b10);
\draw[->] (u6)--(b12);
\draw[->, very thick] (u8)--(b1);
\draw[<-] (u9)--(b2);
\draw[->] (u10)--(b8);
\draw[<-] (u11)--(b7);
\draw[<-, very thick] (u12)--(b9);

\draw[->] (-0.8,3) -- (9.8,3);
\draw[<-] (-0.8,0) -- (9.8,0);

\draw[rounded corners=5pt] (-0.4,2.8) rectangle ++(2.8,0.8);
\draw[rounded corners=5pt] (2.6,2.8) rectangle ++(1.8,0.8);
\draw[rounded corners=5pt] (4.6,2.8) rectangle ++(1.8,0.8);
\draw[rounded corners=5pt] (6.6,2.8) rectangle ++(2.8,0.8);

\draw[rounded corners=5pt] (-0.4,-0.6) rectangle ++(1.8,0.8);
\draw[rounded corners=5pt] (1.6,-0.6) rectangle ++(2.8,0.8);
\draw[rounded corners=5pt] (4.6,-0.6) rectangle ++(2.8,0.8);
\draw[rounded corners=5pt] (7.6,-0.6) rectangle ++(1.8,0.8);

\tikzstyle{every node}=[inner sep=1pt]
\begin{tiny}
\node at (llu1) {$L^0_1$};
\node at (llu2) {$L^0_2$};
\node at (llu3) {$L^0_3$};
\node at (llu4) {$L^0_4$};

\node at (llb1) {$L^1_3$};
\node at (llb2) {$L^1_1$};
\node at (llb3) {$L^1_4$};
\node at (llb4) {$L^1_2$};

\node at (lu1) {$v^0_1$};
\node at (lu2) {$v^1_2$};
\node at (lu4) {$v^1_3$};
\node at (lu5) {$v^1_4$};
\node at (lu6) {$v^0_5$};
\node at (lu8) {$v^0_6$};
\node at (lu9) {$v^1_7$};
\node at (lu10) {$v^0_8$};
\node at (lu11) {$v^1_9$};
\node at (lu12) {$v^0_{10}$};

\node at (lb1) {$v^1_6$};
\node at (lb2) {$v^0_7$};
\node at (lb3) {$v^1_1$};
\node at (lb4) {$v^0_3$};
\node at (lb5) {$v^0_2$};
\node at (lb7) {$v^0_9$};
\node at (lb8) {$v^0_{8}$};
\node at (lb9) {$v^1_{10}$};
\node at (lb10) {$v^1_4$};
\node at (lb12) {$v^1_5$};


\end{tiny}
\end{tikzpicture}

\caption{\label{fig:Pi_prime_canon} 
The pair $\big( (L^0_1,L^0_2,L^0_3,L^0_4),(L^1_2,L^1_4,L^1_1,L^1_3)\big)$ is in $\Pi(\mathbb{L}^0)$.
Assuming $Num=0$ we have $canon\Big(\mathbb{L}, \big( (L^0_1,L^0_2,L^0_3,L^0_4),(L^1_2,L^1_4,L^1_1,L^1_3)\big)\Big) = \big(num(\mathbb{L}^0_1),3,num(\mathbb{L}^0_2),1,num(\mathbb{L}^0_3),4,num(\mathbb{L}^0_4),2 \big).$}
\end{figure}

\begin{claim}
\label{claim:metachord_local_isomorphism}
Let $\mathbb{L}'$ and $\mathbb{R}'$ be two metachords from $(\mathcal{L}^0 \cup \mathcal{L}^1)_l$.
Then, $\mathbb{L}'$ and $\mathbb{R}'$ are locally isomorphic if and only if $\canon(\mathbb{L}') = canon(\mathbb{R}')$.
\end{claim}
\begin{proof}
Suppose $L$ is a strong module in $\strongModules(S_L,{\sim})$ and $R$ is a strong module in $\strongModules(S_R,{\sim})$ for some $S_L,S_R \in \camodules$.

Assume that $\canon(\mathbb{L}') = \canon(\mathbb{R}')$.
Suppose $L$ and $R$ are leaves, say, $L = \{u\}$ and $R = \{u'\}$.
Since $\canon(\mathbb{L}') = \canon(\mathbb{R}')$, the mapping $\alpha:L \to R$ such that $\alpha(u) = u'$ establishes a local isomorphism between $\mathbb{L}'$ and $\mathbb{R}'$.
Suppose $L$ and $R$ are non-leaves.
Let $L_1,\ldots,L_k$ be the children of $L$, enumerated such that 
$$\canon(\mathbb{L}') = \canon\Big(\mathbb{L}', \big((L'_1,\ldots,L'_k),(L''_{\delta(1)},\ldots,L''_{\delta(k)})\big)\Big)$$ 
for some $\big((L'_1,\ldots,L'_k),(L''_{\delta(1)},\ldots,L''_{\delta(k)})\big) \in \Pi(\mathbb{L}')$. 
Let $R_1,\ldots,R_k$ be the children of $R$, enumerated such that 
$$\canon(\mathbb{R}') = \canon\Big(\mathbb{R}', \big((R'_1,\ldots,R'_k),(R''_{\sigma(1)},\ldots,R''_{\sigma(k)})\big)\Big)$$ 
for some $\big((R'_1,\ldots,R'_k),(R''_{\sigma(1)},\ldots,R''_{\sigma(k)})\big) \in \Pi(\mathbb{R}')$. 
Since $\canon(\mathbb{L}')=\canon(\mathbb{R}')$, we have
$\num(\mathbb{L}'_{i}) = \num(\mathbb{R}'_{i})$ and $\delta = \sigma$. 
Since $\num(\mathbb{L}'_{i}) = \num(\mathbb{R}'_{i})$, $\mathbb{L}'_{i}$ is locally isomorphic to $\mathbb{R}'_{i}$ for every $i \in [k]$.
Suppose that $\alpha_i$ establishes a local isomorphism between $\mathbb{L}'_{i}$ and $\mathbb{R}'_{i}$.
We claim that $\alpha:L \to R$, where $\alpha = \bigcup_{i=1}^k \alpha_i$, establishes a local isomorphism between 
$\mathbb{L}'$ and $\mathbb{R}'$.
Clearly, $\alpha$ satisfies conditions \ref{def:metachords_loc_iso}.\eqref{def:metachords_loc_iso_orientation}--\eqref{def:metachords_loc_iso_mul} as $\alpha_i$ satisfies   
\ref{def:metachords_loc_iso}.\eqref{def:metachords_loc_iso_orientation}--\eqref{def:metachords_loc_iso_mul} for every $i \in [k]$.
Also, $\alpha$ satisfies condition \ref{def:metachords_loc_iso}.\eqref{def:metachords_loc_iso_left_right} as
$\alpha_i$ satisfies \ref{def:metachords_loc_iso}.\eqref{def:metachords_loc_iso_left_right} and $\delta = \sigma$.

Suppose $\alpha$ is a local isomorphism between $\mathbb{L}'$ and $\mathbb{R}'$.
If $L$ and $R$ are leaves, then
$\canon(\mathbb{L}') = \canon(\mathbb{R}')$ as $\alpha$ satisfies conditions \ref{def:metachords_loc_iso}.\eqref{def:metachords_loc_iso_orientation}--\eqref{def:metachords_loc_iso_mul}.
If $L$ and $R$ are non-leaves, then $L$ and $R$ have the same number of children in $\strongModules(S_L,{\sim})$ and in $\strongModules(S_R,{\sim})$, respectively.
Suppose $\mathbb{L}'_1,\ldots,\mathbb{L}'_k$ are the restrictions of $\mathbb{L}'$ to the sets $L_1,\ldots,L_k$ and 
$\mathbb{R}'_1,\ldots,\mathbb{R}'_k$ are the restrictions of $\mathbb{R}'$ to the sets $R_1,\ldots,R_k$, where
$L_1,\ldots,L_k$ and $R_1,\ldots,R_k$ are the children of $L$ and $R$, respectively, enumerated such that $\alpha(L_i) = R_i$ for every $i \in [k]$.
Clearly, $\mathbb{L}'_1,\ldots,\mathbb{L}'_k$ and $\mathbb{R}'_1,\ldots,\mathbb{R}'_k$ are from level $l+1$.
Now, note that the mapping 
$$\big((L'_{\delta'(1)}, \ldots, L'_{\delta'(k)}),(L''_{\delta''(1)}, \ldots, L''_{\delta''(k)})\big) \to \big((R'_{\delta'(1)}, \ldots, R'_{\delta'(k)}),(R''_{\delta''(1)}, \ldots, R''_{\delta''(k)})\big)$$
establishes a bijection between the members of $\Pi(\mathbb{L}')$ and the members of $\Pi(\mathbb{R}')$.
Since $\alpha|L_i$ establishes a local isomorphism between $\mathbb{L}'_i$ and $\mathbb{R}'_i$, we must have $\canon(\mathbb{L}_i) = \canon(\mathbb{R}_i)$ and $\num(\mathbb{L}_i) = \num(\mathbb{R}_i)$ for every $i \in [k]$.
Thus, we have $\canon(\mathbb{L}') = \canon(\mathbb{R}')$.
\end{proof}
Next, we claim we can compute $\canon(\mathbb{L}')$ and $\num(\mathbb{L}')$ for all $\mathbb{L}' \in (\mathcal{L}^0 \cup \mathcal{L}^1)_l$
in total time linear in the size of the set  $(\mathcal{L}^0 \cup \mathcal{L}^1)_{l,l+1}$.  
Clearly, if $L$ is prime or parallel, then $|\Pi(\mathbb{L}')| \leq 2$, and $\canon(\mathbb{L}')$ can be computed in linear time in the number of the children of $L$.
Suppose $L$ is serial and suppose $L_1,\ldots,L_k$ are the children of $L$.
To compute $\canon(\mathbb{L}')$ it suffices to sort the numbers in the tuple $(\num(\mathbb{L}'_{1}), \ldots, \num(\mathbb{L}'_{k}))$.
Note that the numbers $\num(\mathbb{L}'_i)$ are integers from the interval whose length is bounded by the size of $(\mathcal{L}^0 \cup \mathcal{L}^1)_{l+1}$.  
Since the total number of the entries in the tuples $(\num(\mathbb{L}'_{1}), \ldots, \num(\mathbb{L}'_{k}))$ for all $\mathbb{L}' \in (\mathcal{L}^0 \cup \mathcal{L}^1)_l$ 
is bounded by the size of $(\mathcal{L}^0 \cup \mathcal{L}^1)_{l+1}$, 
we can sort the entries in all such tuples in time linear in the size of $(\mathcal{L}^0 \cup \mathcal{L}^1)_{l+1}$, which follows from the following proposition.
\begin{proposition}
\label{prop:sorting_entries_in_tuples}
Let $\mathcal{T}$ be a set of arbitrarily length tuples, $t$ be the total number of the entries in $\mathcal{T}$, and $d$ be the difference between the maximum and the minimum entry among all the tuples from $\mathcal{T}$.
We can sort the entries of every tuple from $\mathcal{T}$ in time $\Oh{t+d}$.
\end{proposition}
\begin{proof}
We use a modified \emph{counting-sort} algorithm.
First, we find the minimum entry $Min$ in all the tuples from $\mathcal{T}$ and we subtract $Min$ from every entry of every tuple in $\mathcal{T}$.
Now, every entry is in the interval $[0,d]$.
Next, for every $x \in [0,d]$ we compute the set $p(x)$ containing the pointers to the tuples $T$ from $\mathcal{T}$ which contain~$x$.
We clear all the tuples in~$\mathcal{T}$. 
Then, we proceed the table~$p$ from~$d$ down to~$0$, 
and for every pointer to the tuple $T$ in the set $p(x)$ 
we insert the entry $(x + Min)$ to the tuple~$T$ at the first position.
Clearly, the algorithm works in time $O(t+d)$ and sorts the entries of all tuples in~$\mathcal{T}$.
\end{proof}
Suppose that the tuples $\canon(\mathbb{L}')$ are computed for all $\mathbb{L}' \in (\mathcal{L}^0 \cup \mathcal{L}^1)_l$.
Note that the total number of the entries in all those tuples, excluding those corresponding to the leaves~$L$ in $\mathcal{L}_{l}$ (one entry in $\canon(\mathbb{L}^0)$ and $\canon(\mathbb{L}^1)$ encodes the number of twins of the vertex in $L$), is linearly bounded by the size of the set $(\mathcal{L}^0 \cup \mathcal{L}^1)_{l,l+1}$ 
and every such entry is an integer from the interval whose length is linearly bounded by the size of $(\mathcal{L}^0 \cup \mathcal{L}^1)_{l,l+1}$.
The next proposition asserts we can sort all those tuples in time linear in the size of $(\mathcal{L}^0 \cup \mathcal{L}^1)_{l,l+1}$.
\begin{proposition}[\cite{AHU74}]
Let $\mathcal{T}$ be a set of arbitrarily length tuples, $t$ be the total number of the entries in $\mathcal{T}$, and $d$ be the difference between the maximum and the minimum entry among all the tuples from $\mathcal{T}$.
We can lexicographically sort the tuples from $\mathcal{T}$ in time $\Oh{t+d}$.
\end{proposition}
Also, we can sort the tuples $\canon(\mathbb{L}')$ for leaf metachords $\mathbb{L}' \in (\mathcal{L}^0 \cup \mathcal{L}^1)_l$ in time linear in 
$\sum \{ m(v) : \{v\} \text{ is a leaf module from } (\mathcal{L})_l \}$.

Let $n = |V(G')|$. 
Since $|\mathcal{L}^0 \cup \mathcal{L}^1| \leq 4n$, we deduce that we can perform the first step of the canonization procedure in time $\Oh{n}$.

We proceed to the second step of the canonization procedure.
Our goal is to compute $\canon(N)$ and $\num(N)$ for every inner node of the PQS-tree $\pqstree$; 
for leaves $S^0$ and $S^1$ (which are slots of $G$) 
in $\pqstree$ we set $\canon(S^j) = \canon(\SSS^j)$ and 
$\num(S^j) = \num(\SSS^j)$ for $j \in \{0,1\}$.
We assume that $\num(N_1) = \num(N_2)$ for every two locally isomorphic nodes $N_1, N_2$ in $N(\pqstree)_{l+1}$, 
where the local isomorphism between the inner nodes of $\pqstree$ is defined as follows.
\begin{definition}
Suppose $N_1,N_2$ are two inner nodes of the PQS-tree $\pqstree$ such that $\level(N_1) = \level(N_2)$.
Let $\alpha$ be a bijection from $V_{\pqstree}(N_1)$ to $V_\pqstree(N_2)$.
We say $\alpha$ is a \emph{local isomorphism} between $N_1$ and $N_2$ if for every $u,v \in V_{\pqstree}(N_1)$:
\begin{enumerate}
 \item $m_G(u) = m_G(\alpha(u))$, 
 \item $u \in \leftside(v) \iff \alpha(u) \in \leftside(\alpha(v))$,
 \item $u \in \rightside(v) \iff \alpha(u) \in \rightside(\alpha(v))$.
\end{enumerate}
Moreover, if $N'_1$ and $N'_2$ are the parents of $N_1$ and $N_2$ in $\pqstree$, 
for every $v \in V_{\pqstree}(N_1)$:
\begin{enumerate}[resume]
 \item $N'_1 \in \leftside(v) \iff N'_2 \in \leftside(\alpha(v))$.
 \item $N'_1 \in \rightside(v) \iff N'_2 \in \rightside(\alpha(v))$.
\end{enumerate}
\end{definition}
Observe that if $\alpha$ establishes a local isomorphism from $N_1$ to $N_2$, then:
\begin{itemize}
 \item if $N_1,N_2 \in N(\pqstree)_l$ for some $l \geq 1$, 
 then both $N_1$ and $N_2$ are either P-nodes or Q-nodes,
 \item if $N_1=N_2$ is the root of $\pqstree$,
 then $\alpha$ is an isomorphism from $(G, m_G, u_G)$ to $(G, m_G, u_G)$.
\end{itemize}
Suppose $\alpha$ is a local isomorphism between $N_1$ and $N_2$.
For convenience, we introduce the concept of the ``image by $\alpha$'' for words and subsets of $V^*_{\pqstree}(N_1) \cup V_{\pqstree}(N_1)$.
Suppose $w_1$ is a word consisting of some letters from $V^*_{\pqstree}(N_1) \cup V_{\pqstree}(N_1)$.
The \emph{image of $\tau$ by $\alpha$}, denoted by $\alpha(\tau)$, is a word
that arises from $\tau$ by replacing every letter $u^i$ of $\tau$ from $V^*_{\pqstree}(N_1)$ by~$\alpha(u)^i$ and every letter $u$ of $\tau$ from $V_{\pqstree}(N_1)$ by~$\alpha(u)$.
We use an analogous notation for subsets of $V^*_{\pqstree}(N_1) \cup V_{\pqstree}(N_1)$.

Now, we define $\canon(\cdot)$ for every node in $N(\pqstree)_l$.
Suppose $Q$ is a Q-node in $N(\pqstree)_l$.
We transform every circular word $\pi(Q)$ in $\Pi(Q)$ into a tuple $\canon(Q,\pi(Q))$, as follows.
First, we let $\pi'(Q)$ to be a circular tuple that arises from the circular word $\pi(Q)$ such that:
\begin{itemize}
 \item for every $S \in \camodules(Q)$ and every $i \in \{0,1\}$ we replace the slot $S^i$ in $\pi(Q)$ by two entries, $\num(\mathbb{S}^i)$ and $dist(S^i,S^{1-i},\pi(Q))$,
 where $dist(S^i,S^{1-i},\pi(Q))$ is the number of the letters between $S^i$ and $S^{1-i}$ in $\pi(Q)$ increased by the current value of $Num$.
 \item for every P-node $P$ neighbouring $Q$ from level $(l+1)$ we 
 replace $P$ in $\pi(Q)$ by the entry $\num(P)$.
\end{itemize}
If $Q$ has a parent $P$ in $\pqstree$, we set $\canon(Q,\pi(Q))$ such that the circular word $P \cdot \canon(Q,\pi(Q))$ equals to $\pi'(Q)$ (note that $\canon(Q,\pi(Q))$ is a non-circular word).
Otherwise ($Q$ is the root of $\pqstree$), 
we set $\canon(Q,\pi(Q))$ as the lexicographically smallest word 
that satisfies $\pi'(Q) \equiv \canon(Q,\pi(Q))$ (i.e. $\canon(Q,\pi(Q))$
is the lexicographically smallest word which made circular gives~$\pi'(Q)$).
Eventually, we set $\canon(Q)$ as the lexicographically smallest tuple in the set $\{\canon(Q,\pi(Q)): \pi(Q) \in \Pi(Q) \}$.

We claim that the tuple $\canon(Q)$  
can be computed in time linear in the size of the set $\camodules(Q)$.
If $Q$ has a parent in the PQS-tree $\pqstree$, 
$\Pi(Q)$ has exactly two admissible orders, each of size at most $4|\camodules(Q)|$,
and hence $\canon(Q)$ can be computed in linear time in $|\camodules(Q)|$.
Suppose $Q$ is the root of $\pqstree$.
Again, the set $\Pi(Q)$ contains exactly two elements, 
each of size at most $4|\camodules(Q)|$, for the cases where $V$ is parallel/prime in $\strongModules(G_{ov})$.
Moreover, each entry in the tuple $\pi'(Q)$ is contained in the interval whose length is linearly bounded in $|N(\pqstree)|$ (and hence in time linear in the size of $G$).
The next proposition asserts we can compute $\canon(Q)$ in linear time in $|N(\pqstree)|$.
\begin{proposition}[\cite{Booth80}]
Suppose $\pi'$ is a circular word of size $t$ whose letters are integers from the interval of length at most $d$.
Then, in time $\Oh{t+d}$ we can compute the lexicographically smallest (simple, non-circular) word $\pi$ 
such that $\pi \equiv \pi'$.
\end{proposition}
Finally, assume that $Q$ is serial in $\strongModules(G_{ov})$.
That is, we have $Q=V$.
In this case we have
$$\canon(V) = 
\begin{array}{l}
\Big(\num(\mathbb{S}'_1),Num + |\camodules|-1, \ldots, \canon(\mathbb{S}'_k), Num + |\camodules| -1, \\
\ \ \num(\mathbb{S}''_1),Num + |\camodules|-1, \ldots, \canon(\mathbb{S}''_k), Num + |\camodules| -1
\Big),
\end{array}$$
where $\mathbb{S}'_i, \mathbb{S}''_i$ are such that $\{\mathbb{S}',\mathbb{S}''\} = \{\mathbb{S}^0,\mathbb{S}^1\}$ and $\mathbb{S}'_i = min \{\num(\mathbb{S}^0_i),\num( \mathbb{S}^1_i)\}$,
and the sets $S_1,\ldots,S_k$ in $\camodules$ are enumerated such that we have $\num(\mathbb{S}'_i) \leq \num(\mathbb{S}'_j)$ for every $i < j$.
Hence, to compute $\canon(V)$ it suffices to sort the numbers from the set $\{\num(\mathbb{S}'_1),\ldots, \num(\mathbb{S}'_k)\}$, which can be done in linear time in $|\camodules|$.

\begin{claim}
\label{claim:module_local_isomorphism}
Suppose $Q_1$ and $Q_2$ are two Q-nodes such that $\level(Q_1) = \level(Q_2)$ with parents $P_1$ and $P_2$ in $\pqstree$.
Then, $Q_1$ is locally isomorphic to $Q_2$ if and only if $\canon(Q_1) = \canon(Q_2)$.
\end{claim}
\begin{proof}
Suppose $\alpha$ is a local isomorphism between $Q_1$ and $Q_2$.
Note that $\alpha$ maps the vertices of $Q_1$ into the vertices of $Q_2$.
Moreover, $\phi \Vert Q^*_1$  is a conformal model of $(Q_1,{\sim})$ and the image $\alpha(\phi\Vert Q^*_1)$ of $\phi \Vert Q^*_1$ by $\alpha$ is a conformal model of $(Q_2,{\sim})$.
Moreover, observe that:
\begin{itemize}
 \item For every $S \in \camodules(Q_1)$, the image $\alpha(S)$ of a CA-module $S \in \camodules(Q_1)$ is a CA-module in $\camodules(Q_2)$ and 
 the mapping $S \to \alpha(S)$ for $S \in \camodules(Q_1)$ establishes a bijection between the sets in $\camodules(Q_1)$ and the sets in $\camodules(Q_2)$.
 \item For every $S \in \camodules(Q_1)$ the images $\alpha(S^0),\alpha(S^1)$ of the slots $S^0,S^1$ of $S$ are the slots of $\alpha(S)$.
 Moreover, $\alpha$ restricted to $S$ establishes a local isomorphism between the metachords $\mathbb{S}^0, \mathbb{S}^1$ 
 and $\alpha(\mathbb{S}^0), \alpha(\mathbb{S}^1)$, respectively.
 Hence we have $\num(\mathbb{S}^0) = \num(\alpha(\mathbb{S}^0))$
 and $\num(\mathbb{S}^1) = \num(\alpha(\mathbb{S}^1))$.
 \item For every $P \in N_\pqstree(Q) \setminus \{P_1\}$, $\alpha$ maps the set $V_{\pqstree}(P)$ into the set $V_{\pqstree}(P')$ for some $P' \in N_{\pqstree}(Q_2) \setminus \{P_2\}$ and $\alpha$ restricted to $V_{\pqstree}(P)$ establishes a local isomorphism between $P$ and $P'$. Hence we have $\num(P) = \num(P')$.
\end{itemize}
Given the above observations we easily check that $\canon(Q_1) = \canon(Q_2)$.

Suppose $\canon(Q_1) = \canon(Q_2)$.
Let $\pi(Q_1) \in \Pi(Q_1)$ and $\pi(Q_2) \in \Pi(Q_2)$ be such that
$$\canon(Q_1) = \canon(Q_1,\pi(Q_1)) \quad \text{and} \quad \canon(Q_2) = \canon(Q_2,\pi(Q_2)).$$
Let $P,P'$ be nodes from $N_{\pqstree}(Q_1) \setminus \{P_1\}$ and $N_{\pqstree}(Q_2) \setminus \{P_2\}$, respectively, 
such that $\num(P)$ and $\num(P')$ appear in $\canon(Q_1,\pi(Q_1))$ and $\canon(Q_2,\pi(Q_2))$ at the same position.
Since $\canon(Q_1,\pi(Q_1)) = \canon(Q_2,\pi(Q_2))$, $\num(P) = \num(P')$, and hence $P$ is locally isomorphic to $P'$.
Suppose $\alpha_P:V_{\pqstree}(P) \to V_{\pqstree}(P')$ establishes a local isomorphism between $P$ and $P'$.

Let $S$ be a CA-module from $\camodules(Q_1)$.
Suppose $\mathbb{R}', \mathbb{R}''$ are the metachords in $(\mathcal{L}^0 \cup \mathcal{L}^1)_0$ 
such that $\{\mathbb{R}', \mathbb{R}''\} = \{\mathbb{R}^0,\mathbb{R}^1\}$ for some $R \in \camodules(Q_2)$ and $\num(\mathbb{R}'), \num(\mathbb{R}'')$ 
are at the same positions in $\canon(Q_2,\pi(Q_2))$
as $\num(\mathbb{S}^0), \num(\mathbb{S}^1)$ in $\canon(Q_1,\pi(Q_1))$.
Clearly, such $\mathbb{R}', \mathbb{R}''$ exists as $\canon(Q_1,\pi(Q_1)) = \canon(Q_2,\pi(Q_2))$.
In particular, it means that $\mathbb{S}^0$ and $\mathbb{R}'$ are locally isomorphic.
Suppose $\alpha_S: S \to R$ establishes a local isomorphism between $\mathbb{S}^0$ and $\mathbb{R}'$.
Now, we can easily check that the mapping 
$$\alpha = \bigcup  \{ \alpha_P: P \in N_{\pqstree}(Q_1) \setminus \{P_1\} \} \cup \bigcup \{ \alpha_S: S \in \camodules(Q_1)\}$$
establishes a local isomorphism between $Q_1$ and $Q_2$.  
\end{proof}

Suppose $P$ is a P-node in $\pqstree$.
We arrange the children of $P$ in $\pqstree$ into a sequence $Q_1,\ldots,Q_k$ 
such that $\num(Q_i) \leq \num(Q_j)$ for every $i < j$.
We set $$\canon(P) = (\num(Q_1),\ldots,\num(Q_k)).$$
Clearly, by Proposition \ref{prop:sorting_entries_in_tuples}, we can compute $\canon(P)$ for all P-nodes  
$P \in N(\pqstree)_l$ in linear time in the size of $N(\pqstree)_{l+1}$.
\begin{claim}
\label{claim:node_local_isomorphism}
Suppose $P_1$ and $P_2$ are two nodes such that $\level(P_1) = \level(P_2)$ with the parents $Q_1,Q_2$ in $\pqstree$.
Then, $P_1$ is locally isomorphic to $P_2$ if and only if $\canon(P_1) = \canon(P_2)$.
\end{claim}
\begin{proof}
Suppose $\alpha$ is a local isomorphism between $P_1$ and $P_2$.
Note the $\alpha$ maps a child $Q'$ of $P_1$ into a child $Q''$ of $P_2$ and 
the restriction of $\alpha$ to $V_{\pqstree}(Q')$ establishes a local isomorphism between $Q'$ and $Q''$ (and hence $\num(Q') = \num(Q'')$).
Suppose $Q'_1,\ldots,Q'_k$ are the children of $P_1$ in $\pqstree$ and 
$Q''_1, \ldots, Q''_k$ are the children of $P_2$ in $\pqstree$ such that $\alpha$ restricted to $V_{\pqstree}(Q'_i)$ is a local isomorphism between $Q'_i$ and $Q''_i$.
So, we have $\num(Q'_i) = \num(Q''_i)$ for every $i \in [k]$, and hence $\canon(P_1) = \canon(P_2)$.

Suppose $\canon(P_1) = \canon(P_2)$.
Suppose $\canon(P_1) = (\num(Q'_1),\ldots,\num(Q'_k))$ and $\canon(P_2) = (\num(Q''_1),\ldots,\num(Q''_k))$, where 
$Q'_1,\ldots,Q'_k$ and $Q''_1,\ldots,Q''_k$ are the children of $P_1$ and $P_2$, respectively, in $\pqstree$.
Since $\canon(P_1) = \canon(P_2)$, we have $\num(Q'_i) =\num(Q''_i)$ for every $i \in [k]$.
It means that $Q'_i$ and $Q''_i$ are locally isomorphic.
Suppose that $\alpha_{i}: V_{\pqsmtree}(Q'_i) \to V_{\pqsmtree}(Q''_i)$ is a local isomorphism between $Q'_i$ and $Q''_i$ for $i \in [k]$.
Now, one can easily check that the mapping $\alpha = \bigcup_{i=1}^k \alpha_i$
establishes a local isomorphism between $P_1$ and $P_2$.
\end{proof}
Eventually, suppose $\pqstree^{PQ}$ is bicentered and $R$ is the root of $\pqstree^{PQ}$.
In this case we set $$\canon(R) = (\num(P),\num(Q)),$$
where $P$ and $Q$ are the children of $R$.

Summing up, the observations made above assert that we can compute $\canon(N)$ and $\num(N)$ for all nodes in $N(\pqstree)_l$ in time linear in the size of 
$$|N(\pqstree)_{l,l+1}| + \sum \big{\{} |S(Q)|: Q \text{ is a Q-node in } N(\pqstree)_l\big{\}}.$$
So, the second step of the canonization procedure can be performed in linear time in the size of $G$.
Eventually, we set
$$\canon(G,m_G,u_G) = (u_G,Num-1,\canon(Num-1),\ldots,2,\canon(2),1,\canon(1))$$
and 
$$\canon(G') = \canon(G,u_G,m_G).$$
As we have argued, $\canon(G')$ can be computed in time $O(n)$, where $n = |V(G')|$.
Note that the size of $\canon(G')$ is linear in $n$, 
and each entry in $\canon(G')$ is a natural number in the range $O(n)$ .
So, to prove Theorem~\ref{thm:main_canonization_theroem} it remains to show
$$G' \text{ and } H' \text{ are isomorphic if and only if } \canon(G') = \canon(H')$$
for every two circular-arc graphs $G'$ and $H'$.
\begin{proof}[Proof of Theorem~\ref{thm:main_canonization_theroem}]
Suppose $G'$ and $H'$ are isomorphic circular-arc graphs, 
represented by $(G,m_G,u_G)$ and $(H,m_H,u_H)$, respectively.
Suppose $\alpha$ establishes an isomorphism between $(G,m_G,u_G)$ and $(H,m_H,u_H)$.
Hence, $\alpha$ establishes a local isomorphism between the root of the PQS-tree $\pqstree_G$
of $G$ and the root of the PQS-tree $\pqstree_H$ of $H$.
Since $\alpha$ establishes a local isomorphism between the metachords of $G$ and of $H$, 
and between the nodes of~$\pqstree_G$ and~$\pqstree_H$, 
the tables $\canon_{G'}$ and $\canon_{H'}$ computed for $G'$ and $H'$ contain 
the tuples with the same entries. 
Hence, we have $\canon(G,m_G,u_G) = \canon(G,m_H,u_H)$.

Suppose that $\canon(G,m_G,u_G) = \canon(G,m_H,u_H)$.
To show that $(G,m_G,u_G)$ and $(H,m_H,u_H)$ are isomorphic, 
we traverse the tables $\canon_G$ and $\canon_H$ from the first to the last entry and we prove, 
as in Claim~\ref{claim:metachord_local_isomorphism} and Claims~\ref{claim:module_local_isomorphism}--\ref{claim:node_local_isomorphism}, that the tuples
stored at the same position in tables $\canon_G$ and $\canon_H$ encode the objects (metachords or nodes of the PQS-trees) of $G$ and of $H$ that are locally isomorphic.
In particular, the roots of $\pqstree_G$ and of $\pqstree_H$ are locally isomorphic, 
which shows that $(G,m_G,u_G)$ and $(H,m_H,u_H)$, and hence $G'$ and $H'$, are isomorphic.
\end{proof}

\section{Acknowledgments}

We sincerely thank Janek Derbisz and Stefan Felsner for their valuable comments on this manuscript.

\bibliographystyle{plain}
\bibliography{lit_short}

\end{document}